\documentclass[english]{article}
\usepackage[T1]{fontenc}
\usepackage[latin9]{inputenc}
\usepackage{geometry}
\usepackage{xcolor}
\usepackage{babel}
\usepackage{amsmath}
\usepackage{mathtools}
\usepackage{amssymb}
\usepackage{amsthm}
\usepackage{subcaption}
\usepackage{qcircuit}
\usepackage{graphicx}
\usepackage{adjustbox}
\usepackage[normalem]{ulem}
\usepackage{enumitem}
\usepackage{makecell}
\usepackage{soul}
\usepackage{authblk}

\usepackage[unicode=true]{hyperref}
\hypersetup{colorlinks=true,linkcolor=blue,citecolor=blue,urlcolor=blue}

\newcommand{\el}{\textrm{el}}
\newcommand{\val}{\textrm{val}}
\newcommand{\ion}{\textrm{ion}}
\newcommand{\loc}{\mathrm{loc}}
\newcommand{\NL}{\textrm{NL}}
\newcommand{\BE}{\text{BE}}
\newcommand{\PREP}{\text{PREP}}
\newcommand{\SEL}{\text{SEL}}
\newcommand{\SWUP}{\text{SWUP}}

\renewcommand{\v}[1]{\mathbf{{#1}}}


\begin{document}
\global\long\def\brat#1{\langle#1|}
\global\long\def\kett#1{|#1\rangle}
\global\long\def\d{\partial}
\global\long\def\s#1{\mathcal{#1}}
\global\long\def\p#1{\left(#1\right)}
\global\long\def\dag{\dagger}

\global\long\def\ex#1{\bra#1\ket}
\global\long\def\qb#1#2{\langle#1|#2\rangle}
\global\long\def\mel#1#2#3{\langle#1|#2|#3\rangle}

\newcommand{\burak}[1]{\textcolor{red}{[BS:#1]}}
\newcommand{\ks}[1]{\textcolor{blue}{KS: #1}}
\newcommand{\sam}[1]{\textcolor{violet}{SP: #1}}
\newcommand{\mat}[1]{\textcolor{orange}{ML: #1}}
\newcommand{\FHJ}[1]{\textcolor{magenta}{FJ: #1}}
  
\newcommand{\ketbra}[2]{\ensuremath{\left|#1\right\rangle\!\!\left\langle#2\right|}}
\newcommand{\braket}[2]{\ensuremath{\!\left\langle#1|#2\right\rangle}\!}

\theoremstyle{plain}
\newtheorem{thm}{Theorem}
\newtheorem{remark}[thm]{Remark}

\setcounter{tocdepth}{2}

\title{A comprehensive framework to simulate real-time chemical dynamics on a fault-tolerant quantum computer}

\author[1]{Felipe H. da Jornada}
\author[2]{Matteo Lostaglio}
\author[2]{Sam Pallister}
\author[2]{Burak \c{S}ahino\u{g}lu}
\author[2]{Karthik I. Seetharam\thanks{\texttt{Lead author, karthik@psiquantum.com}}}
\affil[1]{\emph{Department of Materials Science and Engineering, Stanford University, Stanford, CA 94305, USA}}
\affil[2]{\emph{PsiQuantum, 700 Hansen Way, Palo Alto, CA 94304}}

\date{\today}
\maketitle

\begin{abstract}
We present a comprehensive end-to-end framework for simulating the real-time dynamics of chemical systems on a fault-tolerant quantum computer, incorporating both electronic and nuclear quantum degrees of freedom. An all-particle simulation is nominally efficient on a quantum computer, but practically infeasible.
Hence, central to our approach is the construction of a first-quantized plane-wave algorithm making use of
\emph{pseudoions}. 
The latter consolidate chemically inactive electrons and the nucleus into a single effective dynamical ionic entity, extending the well-established concept of pseudopotentials in quantum chemistry to a two-body interaction. 
We explicitly describe efficient quantum circuits for initial state preparation across all 
degrees of freedom, as well as for block-encoding the Hamiltonian describing interacting pseudoions and chemically active electrons, by leveraging recent advances in quantum rejection sampling to optimize the implementations. 
To extract useful chemical information, we first design molecular fingerprints by combining density-functional calculations with machine learning techniques, and subsequently validate them through surrogate classical molecular dynamics simulations. These fingerprints are then coherently encoded on a quantum computer for efficient molecular identification via amplitude estimation. 
We provide an extensive analysis of the cost of running the algorithm on a fault-tolerant quantum computer for several chemically interesting systems. 
As an illustration, simulating the interaction between $\mathrm{NH_3}$ and $\mathrm{BF_3}$ (a 40-particle system) requires 808 logical qubits to encode the problem, and approximately $10^{11}$ Toffoli gates per femtosecond of time evolution. Our results establish a foundation for further quantum algorithm development targeting chemical and material dynamics.
\end{abstract}

\tableofcontents{}

\part*{Framework}
\addcontentsline{toc}{part}{I - Framework}

\section{Introduction}

\subsection{Modeling catalysis on a classical computer}
\label{sec:modeling_catalysis}

Our ability to analyze and manipulate chemical reaction pathways underpins modern technology, from material science to chemical engineering and biochemistry.
In particular, chemical reactions utilizing
a catalyst - a substance that facilitates a
 reaction without being consumed - is crucial to the global economy, playing a pivotal
role in the production of the majority of all manufactured goods today~\cite{catlow2016catalysis}.
Industrial catalysts are heavily used for petroleum
refining, synthesis of base chemicals (e.g., ammonia, ethylene, methanol),
and petrochemicals production. Faced with growing energy demand from
industrial catalysts and the growing need for clean fuels to replace
fossil-based fuels, major research efforts are underway to pursue lower-cost,
earth-abundant, and environmentally sustainable classes of catalysts,
including electrochemical catalysts (e.g. for electrolysis, fuel cells,
etc.), and photocatalysts and plasmonic photocatalysts (e.g. for green
$\textrm{H}_2$ production, pollutant degradation, etc.). Organocatalysts and biocatalysts
are also being investigated for both biotechnological applications
(e.g. synthetic biology) and non-biological applications such as fuels
production, plastics degradation, and pharmaceutical production. Catalyst
formulations are diverse, ranging from metals and metal oxides to
zeolites, transition metal complexes, metal-organic frameworks, and
other organic compounds~\cite{hu2021heterogenous,guajardo2021production,bell2021biocatalysis,hassaan2023principles,schmal2014heterogenous, xiang2020advances}.

In the ubiquitous case of heterogeneous catalysis~\cite{norskov2014fundamental}, a reaction pathway typically involves
three elementary processes at the atomic level: (1) Reactant molecules are adsorbed onto
the catalyst surface (2) Bonds of the reactant molecules are reconfigured
through key intermediate states (often called \emph{transition states}) (3)
Newly reconfigured products desorb from the surface. Each of these
processes individually, along with surface diffusion to and from the
catalytically active surface sites, are constantly occuring and modeled
at the ensemble-level by a microkinetic model of the system, whose
rate constants are typically derived from the underlying statistical-mechanics
of the system, e.g. the Eyring rate from transition state theory which
relates the reaction temperature and the free energy difference of an elementary
process to a kinetic rate.\footnote{In transition state theory, the Eyring rate $k$ for
an elementary process involves the Gibbs free energy difference $\Delta G_{\mathrm{TS}}$
between the initial state and the transition state of that process,
$
k=\kappa(T)\frac{k_{\mathrm{B}}T}{h}e^{-\Delta G_{\mathrm{TS}}/k_{\mathrm{B}}T}
$
where $\kappa(T)\leq1$ is a transmission coefficient (Eq. 4.26 in~\cite{norskov2014fundamental}) dependent on the temperature $T$, $h$ is Planck's constant, and $k_{\mathrm{B}}$ is Boltzmann's constant.}

The design and development of
catalysts that are both reactive and selective is  an extremely
challenging problem. The physics involved spans a wide range of spatiotemporal scales and requires a delicate interplay of angstrom/nanometer
scale quantum dynamics for elementary atomic-level processes over
fast timescales, and micro/mesoscopic scale statistical mechanics/kinetics over slow timescales. Conventional studies~\cite{bruix2019first,norskov2014fundamental} of heterogenous catalysis rely on chemical intuition  to construct reaction networks that specify the relevant chemical species and
their elementary reaction processes, with the energy barriers required to compute the corresponding rate constants given by approximate electronic
structure calculations. Microkinetic models are then developed for
the reaction network and either evolved in time to predict ensemble
dynamics or solved for steady-state ensemble properties. This First-Principles
Micro Kinetic (FPMK)~\cite{bruix2019first} modeling procedure 
is both partially heuristic and cumbersome. To complicate matters
further, for a full reactor model of a catalytic reaction, local FPMK
models must be coupled to macroscopic transport models that govern
the heat and mass flows determining the local environment
conditions throughout the reactor.

At the atomic scale, the fundamental challenge is accurate computational modeling of the quantum behavior of electrons
and ions~\cite{perez-ramirez2019strategies}. The current catalysis simulation pipeline gives an atomic-scale description of elementary chemical processes utilizing various
computational tools. For example, most catalytic studies utilize density-functional
theory (DFT), a formally exact formalism for computing the total ground-state electronic energy which depends primarily on the charge density $\rho(\mathbf{r})$. DFT maps the real, strongly-interacting electronic problem to an effective weakly-interacting problem -- the Kohn-Sham system. However, this mapping critically depends on a quantity, the exchange-correlation functional, which can only be approximated in practice~\cite{martin2020electronic}. While DFT can accurately give relative structural energies, vibrational frequencies, and formation energies, it is typically less accurate when dealing with heterogeneous systems and bond reconfiguration, which involve a higher degree of electronic correlation. 
More accurate methods that explicitly capture the correlated nature of the electronic wavefunctions, such as quantum chemistry expansions and quantum Monte Carlo, can address some of these shortcomings and are sometimes utilized for catalytic studies, although are often limited to smaller systems due to their higher computational complexities and difficulty in describing chemical bond reconfiguration ~\cite{fanta2025resolution,kowalski2000method}.
Besides the challenges of accurately computing the electronic structure, one must couple electronic motion with atomic motion to accurately describe realistic chemical physics. A common approach is to utilize the Born-Oppenheimer (BO) approximation, which decouples these degrees of freedom, and even allows one to treat the atomic motion classically. This allows one to more straightforwardly perform BO molecular dynamics (MD) simulations, which address the evolution of atoms when electrons are approximately always in an instantaneous lowest-energy configuration. Clearly, this approximation fails precisely when there is feedback between the electronic and atomic degrees of freedom. For instance, BOMD simulations cannot qualitatively capture how electrons may transition from one set of electronic states (known as a BO surface) to another as the ions move -- critical for simulating many catalysts, and necessary for even qualitatively describing photocatalytic processes~\cite{worth2004beyond,hammesSchiffer1994proton,subotnik2016understanding,meng2024firstprinciples}.
Tools for studying coupled ion and electron dynamics, such as real-time time-dependent
DFT (RT-TDDFT), have been recently developed and are under investigation,
but suffer nonetheless from significant drawbacks such as the incorrect prediction of thermalization, the inherit inaccuracies given typical approximations to the exchange-correlation functionals employed, and large computational costs, which limits such methods to small systems~\cite{shepard2021simulating,tully2023ehrenfest,lian2019indirect}. 

In summary, the construction of a physically consistent, multi-scale model, from atomic scale to macroscopic flow, to
accelerate research and development of new catalysts presents several formidable challenges. A major one, at its atomistic foundation, is that we lack a satisfactory techniques to capture the electronic system including its environment-dependent correlations, 
that treats ions and
electrons on the same footing, does not use uncontrolled approximations,
retains all correlations up to finite discretization effects, and can study the dynamics (and hence excited states) in real-time. Here, we put forth a quantum computational framework for this atomistic foundation.

\subsection{Quantum computing for chemical dynamics}
\label{subsec:QCforChemicalDynamics}

\begin{figure}
\centering
\includegraphics[scale=0.07]{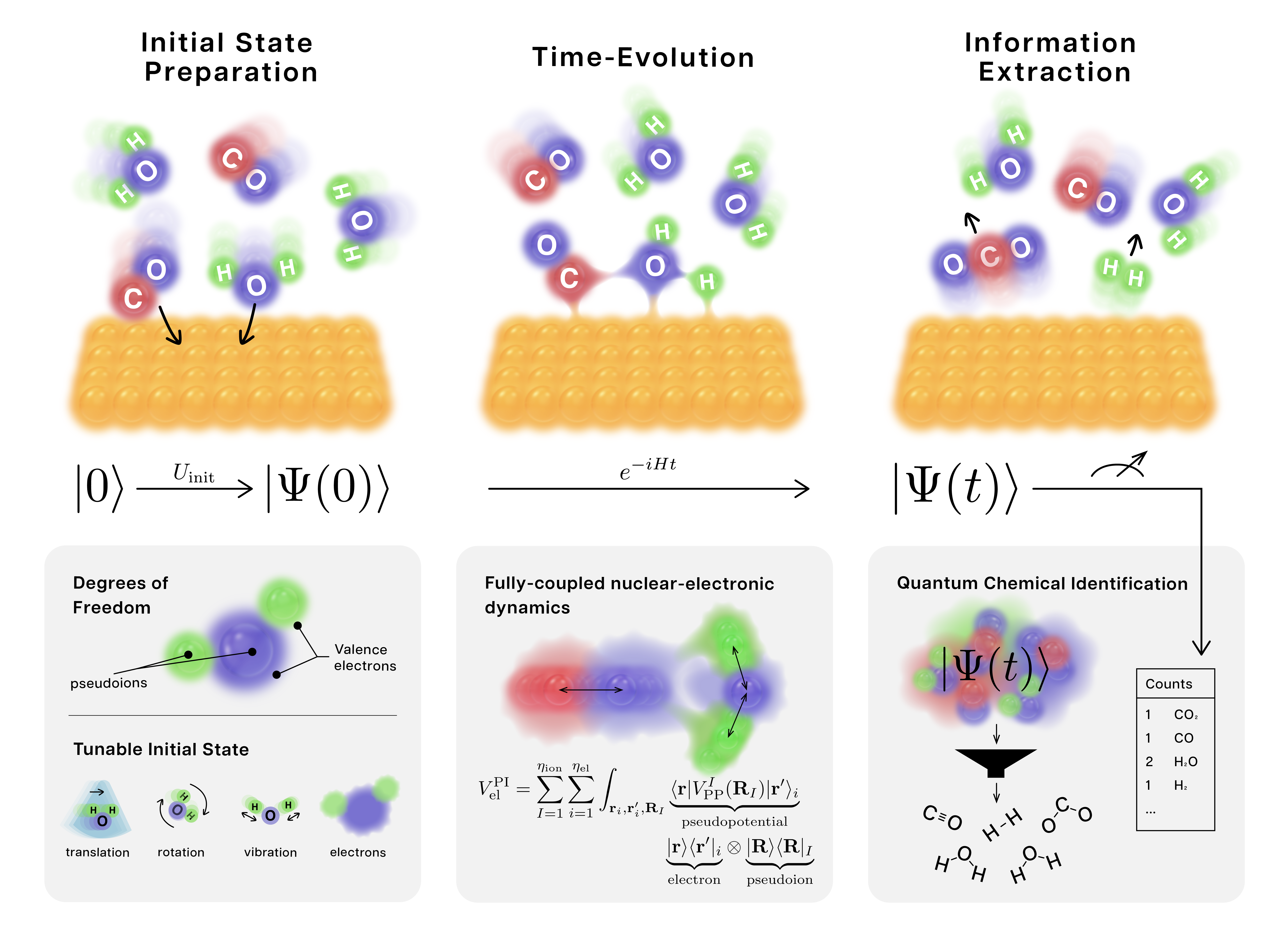}
\caption{The algorithm consists of three main steps: Initial state preparation of a physically relevant quantum state of pseudoions and electrons, followed by time-evolution until a desired final time, and finally information extraction which includes the identification of chemical species present in the wavefunction. We define a compact description of the chemically--relevant degrees of freedom - pseudoions and electrons - and
propose a flexible initial state construction by considering the important physical motions - molecular translations, rotations, vibrations, and electronics. Our time evolution procedure involves the construction of an efficient block-encoding of a Hamiltonian describing interacting pseudoions and electrons. Finally, we develop a new method to identify chemical species by using classical computational methods to develop chemical fingerprints that are efficient to implement coherently on a quantum computer.}
\label{fig:overview}
\end{figure}

Exact simulation of chemical dynamics on classical computers incurs exponential computational costs with the number of constituents and/or the target simulation time. 
Hence such simulations are currently inaccessible, and most likely will remain out of reach for the foreseeable future. 
In contrast, fault-tolerant quantum computers (FTQC) present a novel computing
paradigm that can efficiently and exactly time-evolve wavefunctions present within an exponentially large Hilbert space.
Significant prior work has focused on increasingly efficient
methods for this problem~\cite{childs2021theory,berry2015simulating,low2017optimal,low2019hamiltonian,kalev2021quantum}. Among other techniques, quantum signal processing
(QSP)~\cite{low2017optimal,low2019hamiltonian,motlagh2024generalized,berry2024doubling} applied to Hamiltonian evolution efficiently outputs a quantum state encoding the solution to the Schr\"odinger equation at a final time $t$, given an efficient unitary encoding of the Hamiltonian and an efficient protocol to prepare an initial state. For example, evolution up to precision $\epsilon$ of an initial state of $\eta$ particles under a molecular Hamiltonian in first quantization, discretized with $N$ plane waves per particle,\footnote{The simulation cell is assumed to have volume scaling linearly with $\eta$.} may be performed with $\tilde{O}(\eta \log N)$ qubits and $\tilde{O}(N^{1/3} \eta^{8/3}|t|\log \frac{1}{\epsilon})$ elementary quantum operations (gates) -- using the algorithm of Ref.~\cite{babbush2019sublinear} -- or $\tilde{O}((N^{2/3}\eta^{4/3}+N^{1/3} \eta^{8/3})(|t|+ \log \frac{1}{\epsilon}))$ elementary quantum operations -- using QSP and the encoding in Ref.~\cite{berry2023quantum}.
Nonetheless, quantum algorithms for quantum chemistry and materials
science have mostly focused on the electronic structure problem
under the BO approximation, proposing algorithms to compute ground state energies (and sometimes
a few excited states, see~\cite{bauman2021toward})
with controlled precision using quantum phase estimation
(QPE)
~\cite{motta2021emerging,liu2022prospects,bauer2020quantum,mcardle2020quantum,burg2021quantum,dalzell2023quantum,lee2021even,nirmal2022modular}.
Ground state energy computations
using FTQC circumvent many of the heuristics and approximations used
in DFT and other conventional techniques.
While valuable and conceptually straightforward to integrate as a reliable, high-precision replacement for
(static) electronic structure calculations in the aforementioned computational pipeline for catalyst design, this quantum computational paradigm can address dynamical properties only indirectly via a costly reconstruction of potential energy surfaces/force-fields, and is still limited to the BO approximation.
Performing time-evolution of quantum systems which is exact up to a user-specified precision and finite basis size is particularly efficient with an FTQC. 
The question then becomes, how do we make use of this capability of FTQC to study chemical dynamics?

Several prior works have considered using FTQC to simulate quantum dynamics in a chemical setting. Chan et al.~\cite{chan2024grid} classically emulated
a short (variant of) Trotterized quantum time-evolution in a few scenarios, including
single electron ionization under an applied field in 2D, two-electron
collision in 2D, and the evolution of two electrons in a He atom (with no
nuclear dynamics) in 3D. Kale and Kais~\cite{kale2024simulation} developed
a time-dependent formulation of quantum scattering of wavepackets
based on the M{\o}ller operator formulation to compute scattering matrix
elements, which requires one to specify the molecular scattering channel,
i.e. the reactant and product of interest. Closer to the present work, Kassal et al.~\cite{kassal2008polynomial} first suggested the possibility
to use FTQC for direct grid-based simulation
of non-BO chemical dynamics via a first-order Trotter method. Ideas were sketched for how to initialize
appropriate molecular states and compute reaction probabilities. Following that, Schleich et al.~\cite{schleich2024chemically}
outlined a ``molecular factory'' approach using time-evolution to
prepare good initial molecular states for subsequent use in a dynamics simulation. The idea of exact dynamical evolution has also been applied to other physically interesting problems. McArdle~\cite{mcardle2021learning} proposed time-evolution to simulate a muon-resonance experiment and determine Hamiltonian interaction parameters that characterize a material. Rubin et al.~\cite{rubin2024quantum} proposed and costed the quantum dynamical simulation of a nuclear projectile moving through a material which is modeled under the BO approximation, in order to determine the stopping power for inertial fusion target design by measuring the projectile kinetic energy over time. Motlaugh et al.~\cite{motlagh2024quantum} proposed and costed a quantum dynamical evolution of a specific vibronic Hamiltonian relevant to a class of photochemistry.

Despite encouraging proposals and initial results, the study of FTQC algorithms for real-time quantum chemical dynamics has remained a fairly niche direction in quantum algorithmic research so far. Compared to quantum algorithms for ground state energy estimation, detailed algorithm instantiations and costing are generally unavailable. As discussed above, the multi-scale modeling required in the catalysis pipeline involves formidable  challenges, with or without quantum computers, and there is little hope of providing a solution to the entire problem at once. Here we target specifically the most quantum-mechanical aspects of the pipeline, the study of reaction mechanisms involving bond reconfiguration. With this application in mind, we introduce a framework to target this problem, and a detailed analysis that includes all three core components of an end-to-end description: the preparation of initial states, the time-evolution, and the information extraction from a final state, schematically represented in Fig.~\ref{fig:overview}. A summary of the challenges and our contributions is as follows:

\begin{enumerate}
    \item \emph{Initial state preparation} pertinent to chemical dynamics invokes protocols like thermal state preparation to account for finite temperature effects. Algorithms for thermal state preparation are inefficient in the worst-case~\cite{poulin2009sampling, chiang2010quantum, chowdhury2016quantum} or have a complexity that depends on parameters whose behavior has not been analyzed in chemical scenarios~\cite{temme2011quantum,chen2021fast, holmes2022quantum, chen2023quantum, rall2023thermal, gilyen2024quantum, jiang2024quantum}. In fact, substantial work is needed to extend these algorithms from finite-dimensional/lattice systems to a chemical context. Ref.~\cite{kassal2008polynomial} suggested to initialize the quantum state at the initial time in Born-Oppenheimer approximation, exploiting the representation of nuclear motion in normal mode coordinates to prepare a nuclear wavefunction, but did not give an explicit construction. Here we provide such construction, while also discussing the role that kinematic degrees of freedom play, and introducing physically-motivated adjustments on the state preparation to simulate useful reaction processes, e.g., by mitigating the statistical rarity of reaction events via biasing. 
    \item For \emph{time-evolution}, two general strategies have been proposed. A brute-force approach involves evolution under the Hamiltonian of all electrons and nuclei~\cite{kassal2008polynomial}, which is nominally efficient but extremely impractical whenever bulk material properties of a catalyst or the role of a thermal bath must be included in the model, due to the large number of electrons and ions involved. That is, the number of degrees of freedom required to represent such physics of interest is overwhelming despite the efficient algorithmic scaling. Alternatively, it has been suggested that thermalization be included via dissipative Markovian dynamics~\cite{schleich2024chemically}, although explicit constructions are not provided. It is nontrivial how bulk material properties and a physically representative thermalization model can be included via a Markovian master equation, and without an efficient construction of unitaries encoding the Lindbladian (e.g. via its jump operators~\cite{cleve2016efficient}), one cannot claim efficiency. Here, we follow an alternative path to reducing degrees of freedom and introduce a partially tunable method to replace atoms with \emph{pseudoions}, a new dynamical object that we propose, built using the well-known technique of pseudopotentials~\cite{austin1962general, hamann1979norm, kresse1999ultrasoft,hartwigsen1998relativistic}.
    \item To \emph{extract information} about reaction products encoded inside a quantum state, it was previously assumed that one has access to a unitary that, given access to any nuclear configuration, outputs a list of products~\cite{kassal2008polynomial}, but the construction of such unitary was not specified. In the worst case, encoding information about how to classify nuclear configurations according to products compromises the efficiency of the algorithm. Here, we propose an efficient, practical algorithm to identify and count a limited number of chemical species (molecules, radicals, adsorbed species, etc.) in
an exponentially large wavefunction via \emph{fingerprints}. Fingerprints are simple functions of the nuclear coordinates, constructed through classical machine learning techniques to efficiently classify molecular specifies. Weight information from the classically trained model is loaded onto the quantum computer and fingerprints are computed coherently, accumulating the amplitude associated to a given chemical species in an auxiliary register. This amplitude, which carries information about a specific reaction rate, is reconstructed via amplitude estimation. We have validated that this procedure works well with a restricted classical dynamical simulation. Ensemble averaging over initial conditions is performed coherently via amplitude amplification, so that information is extracted at Heisenberg scaling. 
\end{enumerate}
In summary, we present an end-to-end algorithmic framework for optimized simulation of quantum chemical dynamics. 
In the conventional chemical modeling parlance, we put forth a trajectory-less
approach evolving the entire wavefunction under exact, non-Born-Oppenheimer dynamics up to a controlled, user-specified precision, with polynomial space (qubits)
and time (gates) cost.
As well as proposing novel solutions, we present detailed implementations and compute the dominant cost of the routines we use. The results are showcased in Sec.~\ref{sec:sequence} for a sequence of increasingly challenging applications. Our study provides evidence that algorithms for time dynamics are likely to become a complementary to energy estimation algorithms in quantum chemistry. However, we note that these algorithms have not yet been the subject of a concerted optimization effort like those for ground state energy estimation, and so expect that further cost reductions are possible in the future. Furthermore, we stress that performing exact time evolution is substantially more challenging on classical machines, and so FTQC may provide a unique scientific advantage already with rather small problem instances. In specific instances (e.g. charge migration in $\mathrm{C_2HI}$) one can find an efficient matrix-product state description which recovers the results of full configuration-interaction calculations up to a small error~\cite{lars-hendrik2019ultrafast}, but these approaches are fundamentally limited to small correlation regimes. In a general setting, strongly correlated states that might occur during bond reconfiguration/charge transfer, especially when considering photochemistry, are expected to be challenging to capture even with such state-of-the-art classical methods. Our FTQC algorithm can address these challenging cases.

\section{Designing in-silico quantum dynamical simulation of chemical reactions} \label{sec:framework_design}

In this overview, we begin by discussing the dynamical degrees of freedom and the Hamiltonian required for time-evolution (Sec.~\ref{sec:dynamic_DoF}). We then discuss how to construct a tunable and physically salient initial state for simulating reaction mechanisms (Sec.~\ref{sec:generalconsiderationstateprep}). Next, we develop a flexible chemical species identification protocol assisted by classical computational pre-processing for information extraction after time-evolution (Sec.~\ref{sec:info_extraction}).
We then propose a sequence of increasingly challenging applications and give a summary of the corresponding quantum computing resource estimates (Sec.~\ref{sec:sequence}).

\subsection{Dynamical degrees of freedom and  pseudoions}\label{sec:dynamic_DoF}

The non-relativistic Hamiltonian of interacting electrons and ions,
written in terms of atomic units with position coordinates in units
of the Bohr radius $a_{0}= 4 \pi \epsilon_0 \hbar^2/(m_e e^2)$, spatial derivatives (i.e. momenta) in
units $a_{0}^{-1}$, masses in units of the electron mass $m_{e}$,
and energies in units of $1\;\mathrm{Hartree} =\frac{\hbar^{2}}{m_e a^2_0}$, is given by
\begin{align}
H & =\underset{T_{\mathrm{el}}}{\underbrace{-\frac{1}{2}\sum_{i=1}^{\eta_{\mathrm{el}}}\nabla_{i}^{2}}}\underset{T_{\mathrm{ion}}}{\underbrace{-\sum_{I=1}^{\eta_{\mathrm{ion}}}\frac{1}{2M_{I}}\nabla_{I}^{2}}}\underset{V_{\mathrm{el}}}{\underbrace{+\frac{1}{2}\sum_{i\ne j}^{\eta_{\mathrm{el}}}\frac{1}{|\mathbf{r}_{i}-\mathbf{r}_{j}|}}}\underset{V_{\mathrm{ion}}}{\underbrace{+\frac{1}{2}\sum_{I\ne J}^{\eta_{\mathrm{ion}}}\frac{Z_{I}Z_{J}}{|\mathbf{R}_{I}-\mathbf{R}_{J}|}}}\underset{V_{\mathrm{el-ion}}}{\underbrace{-\sum_{i=1}^{\eta_{\mathrm{el}}}\sum_{I=1}^{\eta_{\mathrm{ion}}}\frac{Z_{I}}{|\mathbf{r}_{i}-\mathbf{R}_{I}|}}},\label{eq:H_bare_position}
\end{align}
where $i=1,\dots\eta_{\mathrm{el}}$ and $I=1,\dots,\eta_{\mathrm{ion}}$
label the electrons and ions, respectively, and the ions have atomic
masses and charges $M_{I},Z_{I}$, respectively.

The Hamiltonian in Eq.~\eqref{eq:H_bare_position}, when expressed in a finite plane-wave basis, can be used to construct block-encodings, and
to implement polynomials of $H$ via quantum signal processing~\cite{gilyen2018quantum}, e.g., for the purpose of implementing
time-evolution or eigenstate filtering. 
The gate complexity will be
polynomial in the total number of particles ($\eta=\eta_{\mathrm{el}}+\eta_{\mathrm{ion}}$) for systems in condensed phase (such as those that involve a metal slab),
specifically $\mathcal{O}(\eta^{8/3})$ and sublinear dependence in the
number of plane-waves $|G|$ in the worst case, when using a plane-wave basis.
While this is still considered efficient, the scaling with $\eta^{8/3}$ will lead to a very large cost in practice. Modeling a typical scenario of a catalytic chemical reaction, involving small molecules and an extended (condensed phase or large cluster) system, requires a far too large number of degrees of freedom (DoF) to brute-force simulate all of the electrons and ions, and a correspondingly large basis size per particle to maintain fine resolution of the highly energetic core electrons. In particular, the extended system poses the largest difficulty in a dynamical situation since a finite size approximation to a fully extended system, even with periodic boundary conditions, needs to be large enough to model realistic physics, i.e. there must be sufficient DoF to act as both a surface of a bulk material and a thermal bath for the catalytic reaction. A full dynamical simulation of such a large number of degrees of freedom is prohibitively expensive even with state-of-the-art current quantum Hamiltonian simulation techniques. 

As a simple yet effective way of eliminating
chemically less relevant electronic degrees of freedom, particularly those of the subsurface layers of the catalyst, is to employ pseudopotentials. 
Pseudopotential methods are a mainstay in conventional quantum chemistry calculations~\cite{austin1962general, hamann1979norm, kresse1999ultrasoft}. The main idea is to replace core electron degrees of freedom by an effective 
modification of the electron-ion interaction
- the so-called pseudopotential term. The number of removed electronic DoF is a partially tunable knob interpolating between an all-electron simulation, and one where all but the outermost valence electrons have been replaced by an effective potential. For quantum algorithms, employing pseudopotentials reduces
the complexity in two ways: first, by reducing the number of electrons
in the system;
and secondly by reducing the number
of plane-waves needed for convergence
since core electrons tend to have
highly oscillatory wavefunctions that are no longer required to be
captured.
A possible complication is whether we can efficiently implement the
new electron-interaction term on a quantum computer. The answer is
affirmative, and Sec.~\ref{sec:time_evolution} is devoted to that analysis. 

Pseudopotentials  have been extensively used 
in classical algorithms for static electronic structure  calculations, 
and most recently they have also been employed in quantum algorithms for the same problem~\cite{zini2023quantum,berry2023quantum}. 
Here we extend the idea by generalizing pseudopotentials to quantum dynamical evolution, for which the notion of a simple one-body potential is no longer applicable. 
We define a new object - a \emph{pseudoion} - as a point-like dynamical object constructed from combining chemically-inactive \emph{core} electrons and a nucleus.
Specifically, for each atom $I$, we
treat $\eta_{\textrm{core},I}$ core electrons together with the positively
charged nucleus (of atomic number $Z_{I}$) as a single ion with positive
charge 
\begin{align}
    Z_{I}^{\mathrm{PI}}=Z_{I}-\eta_{\textrm{core},I}.
    \label{eq:Z_PI}
\end{align}
In total, we have $\eta_{\textrm{val}}=\eta_{\textrm{el}}-\eta_{\textrm{core}}$
valence electrons, where $\eta_{\textrm{core}}=\sum_{I=1}^{\eta_{\textrm{ion}}}\eta_{\textrm{core},I}$. Note that here \emph{valence} refers to the residual electronic degrees of freedom, which is a parameter rather than a chemically defined property. Hence, for each atom, electrons are divided into two sets - valence and core electrons.
This modifies the Hamiltonian into the following form,
\begin{align}
H^{\mathrm{PI}} & =\underbrace{-\frac{1}{2}\sum_{i=1}^{\eta_{\mathrm{val}}}\nabla_{i}^{2}-\sum_{I=1}^{\eta_{\mathrm{ion}}}\frac{1}{2M_{I}}\nabla_{I}^{2}+\frac{1}{2}\sum_{i\ne j}^{\eta_{\mathrm{val}}}\frac{1}{|\mathbf{r}_{i}-\mathbf{r}_{j}|}}_{ \textrm{same as Eq.~\eqref{eq:H_bare_position}, but only valence electrons}}+\underset{V_{\mathrm{ion}}^{\mathrm{PI}}}{\underbrace{\frac{1}{2}\sum_{I\ne J}^{\eta_{\mathrm{ion}}}\frac{Z_{I}^{\mathrm{PI}}Z_{J}^{\mathrm{PI}}}{|\mathbf{R}_{I}-\mathbf{R}_{J}|}}}+V_{\el-\ion}^{\textrm{PI}},\label{eq:H_PI}
\end{align}
where $V_{\el-\ion}^{\textrm{PI}}$ is a modified interaction
term between the electrons and the newly-defined pseudoions, which, as we shall see shortly, is constructed from a pseudopotential. 

There are different classes of pseudopotentials, obtained by the procedures described
in Refs.~\cite{hamann1979norm, bachelet1982pseudopotentials, bachelet1982relativistic, goedecker1996separable, hartwigsen1998relativistic}. We follow the same kind of norm-conserving pseudopotentials recently introduced into quantum algorithms~\cite{zini2023quantum, berry2023quantum}, specifically the HGH pseudopotentials~\cite{goedecker1996separable, hartwigsen1998relativistic}. However, as stated earlier, we consider the problem
of simulating dynamics instead. To this end, crucially, we do \emph{not}
make the ubiquitous Born-Oppenheimer approximation and consider both
the electrons and pseudoions as \emph{dynamical} and interacting objects. 
Accordingly, we promote the pseudopotentials to operators on the full Hilbert
space of electrons and pseudoions as follows,
\begin{align}
V_{\el-\ion}^{\mathrm{PI}} & :=\sum_{I=1}^{\eta_{\mathrm{ion}}}\int_{\mathbf{R}_I}V_{\mathrm{PP}}^{I}(\mathbf{R}_I)\otimes\kett{\mathbf{R}}\brat{\mathbf{R}}_{I}\label{eq:pseudoion_lifting_procedure}
\end{align}
where $V^{I}_{\mathrm{PP}}(\mathbf{R}_I)$ is the pseudopotential, i.e. the sum of potential energy terms imposed on electrons by the ion labeled by $I$ at position $\mathbf{R}_I$.
For simplicity, we take the HGH form from Ref.~\cite{hartwigsen1998relativistic}, which expands the total pseudopotential
into `local' and `nonlocal' components as
\begin{align}
\label{eq:HGHPPexpansion}
    V^{I}_{\mathrm{PP}}(\mathbf{R}_I) = \sum_{i=1}^{\eta_{\mathrm{val}} }V^{i,I}_{\mathrm{PP, loc}}(\mathbf{R}_I) + \sum_{i=1}^{\eta_{\mathrm{val}}} V_{\mathrm{PP, NL}}^{i,I}(\mathbf{R}_I),
\end{align}
with
\begin{align}
V^{i,I}_{\mathrm{PP, loc}}(\mathbf{R}_I)& =\int_{\mathbf{r}} \p{\frac{-Z_{I}}{|\mathbf{r}-\mathbf{R}_I|}\mathrm{erf}(\bar{\lambda}_{\mathrm{loc}}^{I}|\mathbf{r}-\mathbf{R}_I|)+e^{-(\bar{\lambda}_{\mathrm{loc}}^{I}|\mathbf{r}-\mathbf{R}_I|)^{2}}\sum_{c=1}^{4}C_{c}^{I}(\sqrt{2}\bar{\lambda}_{\mathrm{loc}}^{I}|\mathbf{r}-\mathbf{R}_I|)^{2(c-1)}}\kett{\mathbf{r}}\brat{\mathbf{r}}_{i},\label{eq:localPP_term_position}\\
V_{\mathrm{PP, NL}}^{i,I}(\mathbf{R}_I) & =\int_{\mathbf{r},\mathbf{r}'}\sum_{l=0}^{l_{\mathrm{max}}}\sum_{m=-l}^{l}\sum_{a,b=1}^{3}\p{\qb{\mathbf{r},\mathbf{R}_I}{\zeta_{a}^{I,l,m}}B_{a,b}^{I,l}\qb{\zeta_{b}^{I,l,m}}{\mathbf{r}',\mathbf{R}_I}}\kett{\mathbf{r}}\brat{\mathbf{r}'}_{i},\label{eq:NLPP_term_position}
\end{align}
where $\mathbf{r},\mathbf{r}'$ are the positions of electron $i$, $\mathbf{R}_I$ is the position of pseudoion $I$, and 
\begin{align}
\bar{\lambda}_{\mathrm{loc}}^{I}:=\frac{1}{\sqrt{2}\bar{r}_{\mathrm{loc}}^{I}}, \quad \qb{\mathbf{r},\mathbf{R}}{\zeta_{a}^{I,l,m}}=\zeta_{a}^{I,l}(|\mathbf{r}-\mathbf{R}|)Y_{l}^{m}(\widehat{\mathbf{r}-\mathbf{R}}),\label{eq:NL_position_elements}
\end{align}
where, for any vector $\mathbf{v}$, $\hat{\mathbf{v}}$ is the unit vector $\mathbf{v}/\| \mathbf{v}\|$.
Here $Y_{l}^{m}$ are complex spherical harmonics (see App. \ref{app:basis_change}
for convention), $\zeta_{a}^{I,l}(r)$ are radial
functions defined as
\begin{align}
\zeta_{a}^{I,l}(r) & =A_{a}^{I,l}r^{l+2(a-1)}e^{-\frac{1}{2}(\frac{r}{\bar{r}_{l}^{I}})^{2}}, \quad \quad 
A_{a}^{I,l} =\frac{\sqrt{2}}{(\bar{r}_{l}^{I})^{(l+\frac{4a-1}{2})}\sqrt{\Gamma(l+\frac{4a-1}{2})}}, \nonumber 
\end{align}
where $\bar{r}_{\mathrm{loc}}^{I},C_{1}^{I},C_{2}^{I},C_{3}^{I},C_{4}^{I},\bar{r}_{l}^{I},B_{a,b}^{I,l}$
are the HGH fitting parameters that depend on the ion type found by matching various quantities in the valence region with an all-electron
or DFT calculation.\footnote{The variables $r_{\mathrm{loc}}^{I},C_{1}^{I},C_{2}^{I},C_{3}^{I},C_{4}^{I},\bar{r}_{l}^{I},B_{a,b}^{I,l}$
are, respectively, the $r_{\mathrm{loc}},C_{1},C_{2},C_{3},C_{4},r_{l},h_{i,j}^{l}$
parameters from Ref.~\cite{hartwigsen1998relativistic}
(up to index relabeling $i,j\rightarrow a,b$). We append the superscript
here and elsewhere to show explicit dependence on the ion $I$. The radial functions $\zeta_{a}^{I,l}$ are denoted as 
$p_{i}^{l}$ in \cite{hartwigsen1998relativistic}.} {Eq.~\eqref{eq:NLPP_term_position} expresses
 $V_{\mathrm{PP,NL}}^{i,I}$ as a sum of  projectors on states 
with low angular momentum -- typically,
$l_{\mathrm{max}}\leq2$.

We see from Eq.~\eqref{eq:pseudoion_lifting_procedure} that the $1$-body pseudopotentials are ``lifted'' to bona-fide 2-body electron-pseudoion
interactions which are diagonal in
the
pseudoion position (hence the traditional 1-body pseudopotential term is no longer present).
In contrast to the Born-Oppenheimer approximation,
this treatment includes the quantum mechanical motion of all chemically relevant degrees of freedom, i.e. the dynamics of both pseudoions and electrons, and their mutual interactions.
Besides the
interactions with electrons, the pseudoions themselves evolve via their own kinetic terms $T_{\mathrm{ion}}$
and the pseudoion-pseudoion interactions $V_{\mathrm{ion}}^{\mathrm{PI}}$
as seen in Eq.~\eqref{eq:H_PI}. The simple form of $V_{\mathrm{ion}}^{\mathrm{PI}}$
stems from the intuition that in non-relativistic chemical
and material systems, pseudoions are sufficiently spatially separated
such that, to leading order in the multipole expansion, they can be
treated as effective point charges interacting via Coulomb repulsion, with $Z_{I}$ replaced by $Z_{I}^{\mathrm{PI}}$ defined in Eq.~\eqref{eq:Z_PI}.
More sophisticated models could be introduced in the future.

Using Eq.~\eqref{eq:HGHPPexpansion} and Eq.~\eqref{eq:pseudoion_lifting_procedure}, explicitly, we have
\begin{equation}
V_{\el-\ion}^{\mathrm{PI}}=\underset{V_{\mathrm{loc}}^{\mathrm{PI}}}{\underbrace{\sum_{I=1}^{\eta_{\mathrm{ion}}}\sum_{i=1}^{\eta_{\mathrm{val}}}V_{\mathrm{loc}}^{i,I}}}+\underset{V_{\mathrm{NL}}^{\mathrm{PI}}}{\underbrace{\sum_{I=1}^{\eta_{\mathrm{ion}}}\sum_{i=1}^{\eta_{\mathrm{val}}}V_{\mathrm{NL}}^{i,I}}},\label{eq:VPI_elion}
\end{equation}
where
\begin{align}
    V_{\mathrm{loc}}^{i,I} &= \int_{\mathbf{R}_I}V_{\mathrm{PP, loc}}^{i,I}(\mathbf{R}_I)\otimes\kett{\mathbf{R}}\brat{\mathbf{R}}_{I},
    \label{eq:local_term_position} \\
     V_{\mathrm{NL}}^{i,I} &= \int_{\mathbf{R}_I}V_{\mathrm{PP, NL}}^{i,I}(\mathbf{R}_I)\otimes\kett{\mathbf{R}}\brat{\mathbf{R}}_{I}.
\label{eq:NL_term_position}
\end{align}

The Hamiltonian in Eq.~\eqref{eq:H_PI} is written in a general way. That is, for each atom of the catalyst and, if desired, for each atom of the reactants, one can decide how many of the electrons are removed, if any.\footnote{The HGH pseudopotentials often give a couple of options for a given atom, usually corresponding to natural division of core and valence electrons based on the orbital structure and filling of the atom.}
A conservative approach would be to employ the pseudoion description only for the atoms in the extended catalyst, particularly for those atoms away from a chemically active site, whose main purpose in the catalyst is to provide increased electronic delocalization (as would appear in a realistic extended system) and behave as an effective (small) thermal bath.
However, one may liberally and flexibly use pseudoions as replacements to all-electron atoms, and check the convergence of the results when varying the number of DoF combined into pseudoions.
Such quantum dynamical simulations are completely out of reach for classical computers, and could produce valuable scientific insight. 
For example, relative reaction rates between configurations (e.g., computed from two different catalyst surfaces/motifs) may  
display systematic error cancellation giving rise to more physically accurate numerical results than one might apriori expect, and be sufficiently accurate to rank their relative reactivity. 
In any case, under reasonable choices the coarse-grained treatment of bulk catalytic properties leads to drastic savings for the quantum algorithm compared to the full simulation under the Hamiltonian in Eq.~\eqref{eq:H_bare_position},  with the number of electrons cut by a factor of $10$ or more, and lower momentum cutoffs for the
plane-wave basis set by a factor of up to $1000$~\cite{gygi2023allelectron}, at a relatively small price of a more
complicated form of the modified electron-ion interaction term. 

\subsection{Physically salient and algorithmically efficient quantum state preparations}
\label{sec:generalconsiderationstateprep}

The choice of the initial state is consequential and it is informed by a range of considerations. We want our choice to capture salient physical features, but only to the extent that they impact the physics of interest, while leaving any remaining freedom to improve algorithmic efficiency. This ensures that we can make maximal use of limited quantum resources to glean useful physical insight. 

As discussed in the introduction (Sec.~\ref{sec:modeling_catalysis}), a typical catalytic reaction involves adsorption of reactant molecules onto a catalyst surface -- possibly with in-situ surface modifications -- bond reconfiguration of the reactants usually through key intermediate (often rate-determining) states, and desorption of the newly reconfigured products.
Here we specifically target the study of reaction mechanisms, i.e. bond reconfiguration, the ``quantum heart'' of catalytic reactions. The DoF of reactant molecules (or catalyst species) can be approximately decomposed \emph{at the initial time} into four independent classes: the translational motion of the center of mass (CoM), the overall rotational motion of the molecule, internal vibrational motions of the bonds, and the electronic motion. The total state for each chemical species at the initial time is taken to be the product of the wavefunctions for each of its motional classes.
Details including algorithmic implementations will be given in Sec.~\ref{sec:quantum_state_preparation}, while here we only introduce the main conceptual ideas.

\subsubsection*{Translational Motion}
Statistical mechanics considerations might suggest preparing the CoM translational degrees of freedom of each reactant as a Gaussian wavepacket (e.g., of minimal uncertainty), with velocity taken from a Maxwell distribution at the reaction temperature~$T$. However, it is typically the case that most initial kinematic configurations will be unlikely to undergo bond reconfigurations -- the reactants might not reach the catalyst surface at the same space or time, might spend considerable time hopping on the surface of the catalyst to/away from the active sites, or might reach the surface at speeds or angles of incidence unsuitable for adsorption, a prerequisite for a catalytic reaction to occur. We wish to remove these inactive initial configurations from our simulation as much as prior knowledge and chemical intuition allows, while still capturing relevant physical features:

\begin{enumerate}
\item \emph{Chemically active kinematic configurations}: 
For optimal use of quantum computational resources, sufficiently strong mutual interactions between
species (reactants and catalyst) should develop within a suitably short time, i.e., we would like the reactants to reach the catalyst surface at similar times, and fairly quickly. 
On the other hand, if the reactants approach the surface with sufficiently large velocities, adsorption is unlikely. 
Hence, while ideally we would initialize the reactants far from each other and the surface to ensure small initial interactions -- thus justifying independent state preparation of each chemical species as per commonly-invoked molecular chaos hypothesis -- for the sake of optimal simulation it is preferable to deviate from this ideal and focus attention solely on the reaction mechanism/bond reconfiguration step.\footnote{If we insisted in preparing reactants far enough from the surface that they are approximately non-interacting at the initial time, we would easily find ourselves spending the majority of the overall simulation time waiting for the reactants to reach the surface.} We then initialize the reactants close to the catalyst surface with minimal velocity, giving rise to a ``soft-landing'' to facilitate adsorption. The preparation of independent states for each reactant and the catalyst
then remains justified to the extent that it does not significantly influence the reaction mechanism.
This can be deemed acceptable based on heuristic considerations, or alternatively the initial mutual separation between the chemical species could also be treated as a convergence parameter subject to numerical experimentation.

\item \emph{Kinematic modeling requirements}: These refer to further kinematic constraints imposed by the single reaction-event physics we focus on. For example, to faithfully model a small time
window of a single reaction event of a larger chemical reaction, the wavepackets should not significantly delocalize before strong interactions between all species develop, and the time for strong interactions to develop should be smaller than the average time between reactant collisions, e.g. in the case of the gas, the mean free time.
\end{enumerate}

\subsubsection*{Rotational Motion}
We may approach the rotational motion with a similar spirit. Statistical mechanics again suggests preparing these DoF in a thermal state. A collection of atoms forming metal clusters/slabs, or most molecules for that matter, 
have large equilibrium principal moments of inertia, which are inversely proportional to the energy scale of rotations. If one associates a rotational temperature $\Theta_{\mathrm{rot}}$
to this rotational energy scale, for typical reaction temperatures $T$
we have $T\gg\Theta_{\mathrm{rot}}$. Barring H$_2$, with $\Theta_{\mathrm{rot}}\sim 10-40K$, small molecules have rotational temperatures $\Theta_{\mathrm{rot}}\lesssim0.1-10K$ (see Table 11C.1 in~\cite{atkins2022physical} and Table 6.1 in~\cite{gordy1984microwave}). Hence, naively we
would have to prepare a thermal state of rotational eigenstates
(e.g., eigenstates of the rigid rotor Hamiltonian under typical approximations discussed later)
to very high angular momentum for molecules. This is a non-trivial state preparation that could be carried out by a similar procedure as that discussed for the vibrational modes in the next section. In practice, however, such a rotational state preparation is not desirable. For small molecules, larger angular momenta, and therefore larger angular velocities, are associated with initial configurations that are less likely to adsorb to the catalysts. Hence, these configurations are less kinetically relevant for simulations. For larger molecules, even though one naively expects the angular momentum to acquire a large quantum number, the effective linear velocities of extremal atoms decreases with increasing molecular size, and are asymptotically small compared to the contribution from the vibrational states.  Accordingly, we construct the molecular reactants in the rotational ground state.
This choice carries a potential additional benefit. Assume that a uniform superposition in the rotational degrees of freedom \emph{effectively} behaves as a uniform ensemble for the purposes of chemical bonding and determining the output reaction rates. Then, initializing the molecule in the spherically symmetric rotational ground state effectively captures uniform averaging over 
classical molecular orientations relative to the catalytic surface, which would instead require several distinct classical calculations.
Finally, note that substrates (slabs) are approximated in our approach as being periodic, and hence do not have rotational modes associated with them due to being confined in a simulation super cell. Isolated atoms also do not have rotational modes.

\subsubsection*{Vibrational Motion}
The vibrational modes, extensive in the number of atoms for any molecule, cluster, or other extended phase, typically carry the majority of the thermal energy and are mechanistically critical for bond reconfiguration. After all, vibrationally stretching a bond is highly conducive to eventually breaking it, a precursor to forming new bond configurations.
At typical reaction temperatures, for most small and medium size molecules, only a few excitations of vibrational modes occur, since they typically involve energies scales between 10 to 300~meV ($\sim$ 80 to 2500~cm$^{-1}$). Therefore, at the initial time, we prepare the vibrational modes of molecules in a thermal state with a finite cutoff on the number of excitations per mode, i.e. a \emph{truncated thermal state}. 
Similarly, we build a truncated thermal state for the vibrational modes of the catalyst, although the cutoffs for a large cluster or extended phase are higher and therefore will consume more computational resources. 
Furthermore, note that the initial thermal state can be biased, e.g., by removing certain vibrational energies/states which are known from independent arguments or simulations to lead to little or no products.

\subsubsection*{Electronic Motion}

From known physical reasoning about energy scales for molecules and extended phases including metals, semiconductors, and insulators, electrons
can be considered to be approximately in their ground state for a fixed equilibrium configuration of the ions (see App.~\ref{app:detailed_physical_justification}). 
That is, at the initial time the electronic excitations for reactants and the catalyst are not thermally relevant.
This is in line with most conventional quantum chemistry approaches which, for the most part, focus on electronic ground states. We stress again that this assumption is only made at the initial time - electronic excitations are included at all $t>0$. Hence, we initialize the electronic ground state in the equilibrium ionic positions for all chemical species.\footnote{We have implicitly assumed the electronic ground state of the
chemical species is non-degenerate. As per Sec. 14.5.4 in~\cite{jensen2017introduction},
the large majority of stable molecules have non-degenerate ground
states. For exceptional cases with degeneracy, the user may specify
what appropriate state from the ground state manifold to prepare.}

\subsection{Chemical species identification} \label{sec:info_extraction}

 In order to define an end-to-end quantum algorithm, we need to decide what data of interest is to be extracted from the coherently encoded output of the simulation. A common approach is that this is chosen to be one of a handful of observables (most commonly, just energy) which are the same across a large swathe of use-cases. Here, instead, we suggest to construct observables tailored to each specific instantiation of the problem at hand.

In this spirit, we propose \emph{quantum chemical identification} (QCI), a methodology for identifying a set of chemical species
in a wavefunction of a chemical dynamics simulation. Examples of chemical species can include free molecules, adsorbed surface species, free radicals, etc., depending on the chemical scenario of interest. The protocol consists of three steps. First, fingerprint functions for each chemical species are classically pre-computed. Second, the collection of fingerprint functions for a reaction are classically compiled to form species counter functions. Third, the species counter functions are efficiently implemented on a quantum computer using coherent arithmetic and logic. The output is chemical species counts entangled to each ionic configuration in the wavefunction, which may be easily measured or further processed. Notably, the proposed QCI protocol only utilizes ionic information to indicate and count chemical species. However, it is most certainly a fruitful line of future work to consider species identification and extraction of other observables based on joint electronic and ionic quantum information.

\subsubsection*{Classification}

Consider chemical species (e.g. molecules or ions) $X_1, \dots, X_M$ consisting of  $n^{X_1}, \dots n^{X_M}$ ions, respectively. To each $X_\alpha$ we associate the configuration space of its constituent ions, $\mathbb{R}^{3n^{X_\alpha}}$. We define an associated \emph{configuration indicator}
function for $X_\alpha$ as, 
\[
I_{X_\alpha}:\mathbb{R}^{3n^{X_\alpha}}\rightarrow\{0,1\}
\]
that takes configurational information and outputs a binary classification --
whether the configuration represents molecule $X_\alpha$ or not. 

This indicator is obtained using classical computational techniques. As a practical example, we could compute
\begin{align}
    I_{X_\alpha}(x)=\Theta(\Delta^{>}_{X_\alpha}-E_{X_\alpha}(x))\Theta(E_{X_\alpha}(x) - \Delta^{<}_{X_\alpha})
    \label{eq:conf_indic}
\end{align} 
where $\Theta$ is the Heaviside step function, $E_{X_\alpha}(x)$
is the energy of a configuration $x\in\mathbb{R}^{3n^{X_\alpha}}$ and $\Delta^{<}_{X_\alpha}$, $\Delta^{>}_{X_\alpha}$ are suitably chosen energy thresholds such that energy values
within the interval $[\Delta^{<}_{X_\alpha}, \Delta^{>}_{X_\alpha}]$ indicate the presence of $X_\alpha$. Appropriate
classical computational techniques (e.g. DFT) may
be used to compute the configuration energy over a relevant subset
of $\mathbb{R}^{3n^{X_\alpha}}$. The lower threshold $\Delta^{<}_{X_\alpha}$ can, for example, be a local minimum in the potential energy surface, while the upper bound $\Delta^{>}_{X_\alpha}$ can, for example, be the approximate activation energy required to escape the local minimum. Note that the accuracy of the energy estimates must only be sufficiently high to discriminate among a set of potential products. There is an inherent robustness in the choice of accuracy since typically
only after a significant distortion do we want to identify the collection of atoms as a separate chemical species.
Ultimately, the exact region of configuration space we identify as the chemical species of interest depends to an extent on the chemistry we wish to capture, but we expect some inherent stability under small changes of this definition.\footnote{On an abstract note, the system only consists of interacting electrons and ions, and so what to identify as a molecule/chemical species must, in part, by up to our interpretation of the collective/bound structures that appear in the wavefunction. Also note that one may consider exchanging the above binary
(``hard'') indicator for a non-binary (e.g. real number) ``soft'' indicator.}

\subsubsection*{Compilation of counting logic}

It is convenient to view $I_{X_\alpha}$ as acting on a subspace of the configuration space of all ions, and so given $x \in \mathbb{R}^{3 \eta_{\mathrm{ion}}}$, we have that $I_{X_\alpha}$ acts on the appropriate $n^{X_\alpha}$ components of $x$. Then, for a given reaction or other dynamical process being simulated, we specify a list of chemical species
$(X_{1},...,X_{M})$ to identify, and for $x \in \mathbb{R}^{3 \eta_{\mathrm{ion}}}$, compute a corresponding set of
configuration indicators $\{I_{X_1}(x) = i_{X_1}, \dots, I_{X_M}(x) = i_{X_M}\}$. The indicators are then mapped to molecular species counts using a \emph{compiler} $\s{C}$ to produce \emph{species counters},
\[
\s C_{X_{1},...,X_{M}}: \{0,1\}^{\otimes M} 
 \rightarrow \mathbb{N}^{M}, \quad  (i_{X_{1}},...,i_{X_{M}})\mapsto \left(C_{X_{1}}(i_{X_{1}},...,i_{X_{M}}),...,C_{X_{M}}(i_{X_{1}},...,i_{X_{M}}) \right)
\]
where $C_{X_{\alpha}}$ counts the number of chemical species $X_\alpha$.
The compiler processes the configuration indicators to (1) ensure that there is no double-counting or overlapping of indicators, and (2) for the collection of atoms in the simulation, check occurrence and count all (combinatorially many) possible ways a species $X_\alpha$ can appear as a combination of the available atoms. As an example of the first property, suppose hypothetically that we have prior knowledge that the CO configuration indicator does not sufficiently discern CO as an isolated species and CO as a substructure within $\textrm{CO}_2$. Then, positive indicators for both CO and $\textrm{CO}_2$ sharing a common C atom do not indicate the presence of both species. There is a single $\textrm{CO}_2$, but it triggers both indicators. To avoid such double-counting, the compiler function performs additional logic to ensure that CO is only flagged when it is not a derivative of $\textrm{CO}_2$.
The compiling logic is also pre-determined classically given knowledge of the fidelity and interplay of indicators.\footnote{One might fairly worry that checking combinatorially-many instances results in poor scaling for identification. However, for a small handful of chemical species of interest, as we envision for any reasonably early to mid generation FTQC, the practical costs of checking all instances (involving coherent arithmetic and additional simple logic) are negligible compared to that of performing quantum time-evolution.}

\subsubsection*{Fingerprinting for an individual species}
Direct coherent implementation of the classical indicators can be impractically expensive. For the energetic classification we suggested, one would need to either directly load the indicator functions that partition the ionic configuration space into the quantum computer, or run a quantum phase estimation, an expensive subroutine introducing further large multiplicative costs. All of these solutions have serious drawbacks. Therefore, we must construct a more efficient strategy for chemical species identification. 
To this end, we propose a classical pre-processing algorithm to identify discerning fingerprints of the chemical species, i.e., we develop an efficient substitute for $I_{X_\alpha}$ which, after compilation, we efficiently implement on a quantum computer. Using machine-learning parlance, for each chemical species $X_\alpha$, we identify a set of \emph{features},
\begin{align}
\label{eq:feature_map}
F_{X_\alpha}:\mathbb{R}^{3n^{X_\alpha}}\rightarrow\mathbb{R}^{Q^{X^\alpha}}.
\end{align}
Here, $F_{X_\alpha}$ maps the $3n^{X_\alpha}$ configurational coordinates of molecule $X_\alpha$ to a smaller $Q^{X^\alpha}$-dimensional \emph{feature space}. 
Features must be efficiently computable on a quantum computer, ideally only involving simple arithmetic and a small number of free parameters to be loaded. 
Examples of features include functions of bond distances, bond angles, or other chemically justified quantities. 
A \emph{feature indicator} map 
\begin{align}
\label{eq:feature_indicator}
I^{\textrm{feat}}_{X_\alpha}:\mathbb{R}^{Q^{X^\alpha}}\rightarrow\{0,1\}
\end{align}
is constructed such that $\tilde{I}_{X_\alpha} = I^{\textrm{feat}}_{X_\alpha} \circ F_{X_\alpha}$ closely approximates the indicator $I_{X_\alpha}$. 
The feature indicator map may be developed agnostically and scalably by using an appropriately-chosen
machine-learning model (e.g. logistic regression or support vector
machines with linear kernels), trained on the information encoded by $I_{X_\alpha}$
over a relevant subset of the configuration space. In other words, $I^{\textrm{feat}}_{X_\alpha}$ is the model trained on the classical data $\{(F_{X_\alpha}(x),I_{X_\alpha}(x))\}$. We refer to $\tilde{I}_{X_\alpha}$ as the \emph{fingerprint} for molecule $X_\alpha$. A set of $M$ fingerprints are compiled to form the fingerprint-based species counters via $C_{X_{1},...,X_{M}}$ defined above (viewing the fingerprints acting on the appropriate subspace of $R^{3\eta_\mathrm{ion}}$).

By construction, machine learning models involve layered compositions of linear functions and a simple nonlinear function (e.g. $\tanh$). As before, the compiler appends additional logic onto the fingerprints functions, which can be also be implemented coherently, possibly with the use of extra ancilla (to coherently implement classically irreversible logic). Hence, coherent implementation of both the feature map and the feature indicator map (and therefore $\tilde{I}_{X_\alpha}$), and additional compiler logic only utilizes straightforward coherent arithmetic and logic. The full expression for the proposed form of the fingerprinting function is detailed in Sec.~\ref{sec:chemical_species_id}. Critically, it is obtained from a simple logistic regression model, involves only additions and multiplications, and requires only ionic distances as input. 

\subsubsection*{Readout}

We may either directly measure the chemical species counts
or we may implement amplitude estimation. For the latter, we first flag a subset of the count registers indicating a subspace
of interest, e.g. all pieces of the wavefunction that contain a desired
set of product molecules, perform amplitude amplification on that subspace, and hence measure
the amplitude for which such products occur as a result of the reaction,
with quadratic improvement over naive sampling in terms of the total end-to-end quantum resources..

\bigskip

In summary, in this framework, collections of simple ionic configuration features, such as bond lengths or bond angles, can be used to classify chemical species via a classical machine learning model, and the resulting feature indicator maps are tabulated for all desired chemical species. Composition of the feature maps and the learned feature indicator gives chemical fingerprints, which are then mapped to species counts via a reaction-specific compiler. These fingerprint-based species counters are coherently implemented using arithmetic and logic in order to identify and
evaluate production rates for all chemical species of interest in the wavefunction of a chemical reaction.
Algorithmic implementation of the fingerprints, along with a classical simulation to exemplify and validate the QCI protocol, are discussed in detail in Sec.~\ref{sec:chemical_species_id}.

\subsection{Results for a sequence of increasingly challenging applications}
\label{sec:sequence}
We consider 3 classes of chemical reactions to exemplify our approach.
In each of the classes, there are one or more problem instances, where
an instance is an exact specification of pseudoions, supercell, and
basis. Table~\ref{tab:problem_instances} displays all specifications for the problem instances. Here, $|G|,|\overline{G}|$ denote the sizes of the plane-wave bases for electrons and pseudoions, determined by the respective momentum cutoff vectors $\mathbf{\Lambda},\mathbf{\overline{\Lambda}}$. Note that $\mathbf{\overline{\Lambda}_\mathrm{trunc}}$ is a hard-cutoff vector for the pseudoion-pseudoion interaction (for details, see Sec.~\ref{sec:Hamiltonian_in_plane_waves} and Sec.~\ref{sec:PI_interactions}). 

\begin{table}[h!]
\begin{adjustbox}{width=1.05\textwidth,center}
\begin{tabular}{|c|c|c|c|c|c|c|c|}
\hline 
Instance & Pseudoion Counts & $\eta_{\mathrm{tot}}^T$ & $\mathbf{\Lambda}, \overline{\mathbf{\Lambda}}, \overline{\mathbf{\Lambda}}_{\mathrm{trunc}}$ & $\mathbf{n},\overline{\mathbf{n}}$ & $(|G|,|\overline{G}|)^{T}$ & Super cell & Qubits\tabularnewline
\hline 
\hline 
$\mathrm{NH_{3}BF_{3}}$ & $\begin{array}{c}
\mathrm{N^{5}:1,B^{3}:1}\\
\mathrm{F^{7}:3,H^{1}:3}
\end{array}$ & 
$\small \begin{pmatrix}
    32\\ 8 \\ 40
\end{pmatrix}$ &
$\small \begin{pmatrix}
    10.31 \\ 10.31 \\ 13.96
\end{pmatrix}$,
$\small  \begin{pmatrix}
42.22 \\ 42.22 \\ 56.52
\end{pmatrix}$, 
$\small  \begin{pmatrix}
1.66 \\ 1.66 \\ 1.55
\end{pmatrix}$ & 
$\small \small \begin{pmatrix}
6\\6\\7
\end{pmatrix}$,
$\small \begin{pmatrix}
8\\8\\9
\end{pmatrix}$ & 
$\small \begin{pmatrix}
504\;063\\
33\;227\;775
\end{pmatrix}$ & $\begin{array}{c}
(18.90,18.90,28.35)\\
\mathrm{Cuboid}
\end{array}$ & $808$\tabularnewline
\hline 

\centering
\makecell{DMTM \\ Molecular}
 & $\begin{array}{c}
\mathrm{N^{5}:3,C^{4}:1,O^{6}:1}\\
\mathrm{H^{1}:10,Pd^{18}:1}
\end{array}$ & $\small \begin{pmatrix}
53\\
16\\
69
\end{pmatrix}$ & 
$\small \begin{pmatrix}
12.32\\12.32\\12.32 
\end{pmatrix}$,
$\small \begin{pmatrix}
49.87\\49.87\\49.87
\end{pmatrix}$,
$\small \begin{pmatrix}
1.56\\1.56\\1.56
\end{pmatrix}$
&
$\small \begin{pmatrix}
7\\7\\7
\end{pmatrix}$,
$\small \begin{pmatrix}
9\\9\\9
\end{pmatrix}$ &
$\small \begin{pmatrix}
2\;048\;383\\
133\;432\;831
\end{pmatrix}$ & $\begin{array}{c}
(32.13,32.13,32.13)\\
\mathrm{Cuboid}
\end{array}$ & $1545$\tabularnewline
\hline 

\centering
 
\makecell{DMTM \\ 3$\times$3}
& $\begin{array}{c}
\mathrm{B^{3}:8,N^{5}:9,C^{4}:1}\\
\mathrm{O^{6}:1,H^{1}:4,Pd^{18}:1}
\end{array}$ & $\small \begin{pmatrix}
101\\
24\\
125
\end{pmatrix}$ & 
$\small \begin{pmatrix}
15.87\\15.87\\10.47
\end{pmatrix}$,
$\small \begin{pmatrix}
32.25\\32.25\\42.39
\end{pmatrix}$,
$\small \begin{pmatrix}
1.54\\1.54\\1.50
\end{pmatrix}$ & 
$\small \begin{pmatrix}
6\\6\\7
\end{pmatrix}$,
$\small \begin{pmatrix}
7\\7\\9
\end{pmatrix}$ & 
$\small \begin{pmatrix}
504\;063\\
8\;241\;919
\end{pmatrix}$ & $\begin{array}{c}
(14.18,14.18,37.80)\\
\mathrm{Rhombohedron\;120{^\circ}}
\end{array}$& $2471$\tabularnewline
\hline 

\centering
\makecell{DMTM \\ 5$\times$5}
& $\begin{array}{c}
\mathrm{B^{3}:24,N^{5}:25,C^{4}:1}\\
\mathrm{O^{6}:1,H^{1}:4,Pd^{18}:1}
\end{array}$ & $\small \begin{pmatrix}
229\\
56\\
285
\end{pmatrix}$ &
$\small \begin{pmatrix}
9.52\\9.52\\10.47
\end{pmatrix}$,
$\small \begin{pmatrix}
39.00\\39.00\\42.39
\end{pmatrix}$,
$\small \begin{pmatrix}
1.54\\1.54\\1.50
\end{pmatrix}$ &
$\small \begin{pmatrix}
6\\6\\7
\end{pmatrix}$,
$\small \begin{pmatrix}
8\\8\\9 
\end{pmatrix}$
& $\small \begin{pmatrix}
504\;063\\
33\;227\;775
\end{pmatrix}$ & $\begin{array}{c}
(23.63,23.63,37.80)\\
\mathrm{Rhombohedron\;120{^\circ}}
\end{array}$& $5751$\tabularnewline
\hline 

\centering
\makecell{DMTM \\ 9$\times$9}
& $\begin{array}{c}
\mathrm{B^{3}:80,N^{5}:81,C^{4}:1}\\
\mathrm{O^{6}:1,H^{1}:4,Pd^{18}:1}
\end{array}$ & $\small \begin{pmatrix}
677\\
168\\
845
\end{pmatrix}$ & $\small \begin{pmatrix}
10.75\\10.75\\10.47
\end{pmatrix}$,
$\small \begin{pmatrix}
43.51\\43.51\\42.39
\end{pmatrix}$
$\small \begin{pmatrix}
1.54\\1.54\\1.50
\end{pmatrix}$ &
$\small \begin{pmatrix}
7\\7\\7
\end{pmatrix}$,
$\small \begin{pmatrix}
9\\9\\9
\end{pmatrix}$
& $\small \begin{pmatrix}
2\;048\;383\\
133\;432\;831
\end{pmatrix}$ & $\begin{array}{c}
(42.53,42.53,37.80)\\
\mathrm{Rhombohedron\;120{^\circ}}
\end{array}$& $18753$\tabularnewline
\hline 

\centering
 
\makecell{WGS \\ 2$\times$3$\times$3}

 & $\begin{array}{c}
\mathrm{C^{4}:1,O^{6}:2,H^{1}:2}\\
\mathrm{Cu^{1}:85,Cu^{11}:5}
\end{array}$ & $\small \begin{pmatrix}
158\\
95\\
253
\end{pmatrix}$ &
$\small \begin{pmatrix}
9.42\\9.50\\9.50
\end{pmatrix}$,
$\small \begin{pmatrix}
38.13\\38.93\\38.93
\end{pmatrix}$,
$\small \begin{pmatrix}
1.50\\1.53\\1.53
\end{pmatrix}$ &
$\small \begin{pmatrix}
7\\6\\6
\end{pmatrix}$,
$\small \begin{pmatrix}
9\\8\\8
\end{pmatrix}$ &
$\small \begin{pmatrix}
504\;063\\
33\;227\;775
\end{pmatrix}$ & $\begin{array}{c}
(42.01,20.50,20.50)\\
\mathrm{Cuboid}
\end{array}$& $5377$\tabularnewline
\hline 

\centering
\makecell{WGS \\ 2$\times$5$\times$5}
& $\begin{array}{c}
\mathrm{C^{4}:1,O^{6}:2,H^{1}:2}\\
\mathrm{Cu^{1}:245,Cu^{11}:5}
\end{array}$ & $\small \begin{pmatrix}
318\\
255\\
573
\end{pmatrix}$ & 
$\small \begin{pmatrix}
9.42\\11.59\\11.59
\end{pmatrix}$,
$\small \begin{pmatrix}
38.13\\46.90\\46.90
\end{pmatrix}$,
$\small \begin{pmatrix}
1.50\\1.47\\1.47
\end{pmatrix}$ &
$\small \begin{pmatrix}
7\\7\\7
\end{pmatrix}$,
$\small \begin{pmatrix}
9\\9\\9
\end{pmatrix}$ &
$\small \begin{pmatrix}
2\;048\;383\\
133\;432\;831
\end{pmatrix}$ & $\begin{array}{c}
(42.01,34.16,34.16)\\
\mathrm{Cuboid}
\end{array}$& $13563$\tabularnewline
\hline 
\end{tabular}
\end{adjustbox}

\caption{Instances for all 3 classes of problems considered. Superscripts on
the HGH pseudoions indicate the number of valence electrons. $\eta_{\mathrm{tot}}=(\eta_\mathrm{val},\eta_\mathrm{ion},\eta)$
indicates the total particle counts for electrons, pseudoions, the sum total of both. The vectors $\mathbf{\Lambda}=(\Lambda_{1},\Lambda_{2},\Lambda_{3})$
($\overline{\mathbf{\Lambda}}=(\overline{\Lambda}_{1},\overline{\Lambda}_{2},\overline{\Lambda}_{3})$)
indicate the anisotropic momentum cutoffs for electrons (pseudoions),
and $\overline{\mathbf{\Lambda}}_{\mathrm{trunc}}$ is the anisotropic
momentum cutoff for the pseudoion-pseudoion interaction. The vectors
$\mathbf{n}=(n_{1},n_{2},n_{3})$ ($\overline{\mathbf{n}}=(\overline{n}_{1},\overline{n}_{2},\overline{n}_{3})$)
indicate the number of qubits in each reciprocal lattice direction for
electrons (pseudoions), and $|G|$ ($|\overline{G}|$)
indicates the the basis size for electrons (pseudoions). The super
cell types are ``Cuboid'' with lattice
vectors $a_{1}(1,0,0),a_{2}(0,1,0),a_{3}(0,0,1)$ and ``Rhombohedron 120${^\circ}$'' with lattice vectors
$a_{1}(1,0,0),a_{2}(\text{-\ensuremath{\frac{1}{2}}},\text{\ensuremath{\frac{\sqrt{3}}{2}}},0),a_{3}(0,0,1)$,
with $(a_{1},a_{2},a_{3})$ given in the table. The last column denotes the total number of qubits (space cost) required for the system, excluding all ancilla needed in the computation.}
\label{tab:problem_instances}
\end{table}

\begin{figure}[h!]
    \centering
    \includegraphics[width=0.9\textwidth]{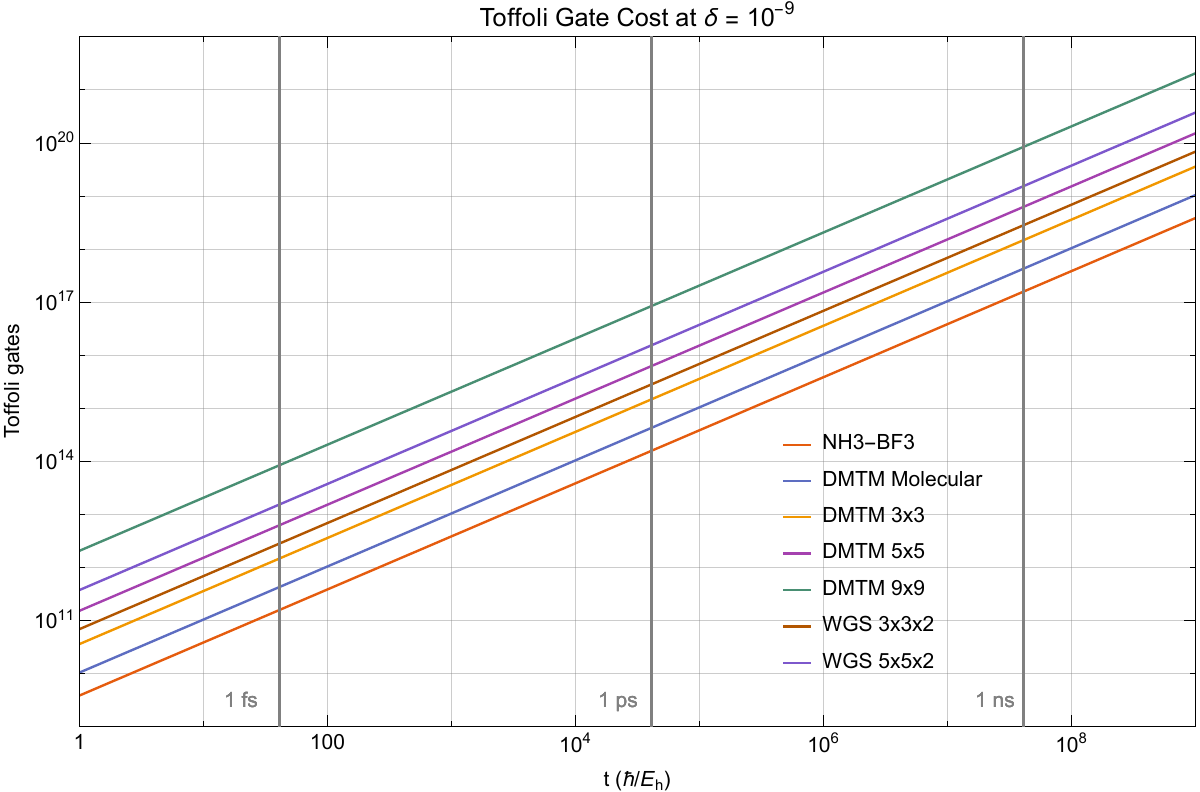}
    \caption{Total Toffoli gate cost for all problem instances in Table~\ref{tab:problem_instances} as a function of time (in atomic units, the unit time step is $\hbar/E_h=2.42\times10^{-17}s$ where $E_h$ is 1 Hartree) for total time-evolution error $\delta=10^{-9}$. Larger values of $\delta$ may be sufficient for several applications but does not decrease the resource costs substantially for the studied time regimes.}
            \label{fig:timeEvoCost}
\end{figure}
The $3$ classes represent:
\begin{enumerate}
\item Ammonia - Boron Trifluoride: We consider a typical Lewis acid-base
interaction $\mathrm{NH}_{3}+\mathrm{BF_{3}}\rightleftharpoons\mathrm{NH}_{3}\mathrm{BF_{3}}$,
where $\mathrm{NH}_{3}$ donates its $\mathrm{sp}^{3}$ lone-pair
to the vacant $\mathrm{p}$ orbital of the $\mathrm{BF_{3}}$ to form
a dative covalent bond. This simple system is fundamental in understanding
electron donation, charge transfer, and bond stabilization, key processes
in catalysis and molecular recognition. There is only one instance
in this class.
\item Direct Methane to Methanol (DMTM): Two-dimensional transition metal
doped Boron-Nitride (BN) is considered an interesting class of catalysts
that can be used to convert methane to methanol, a highly significant
process in scientific and industrial applications to remove and economically
repurpose (into fuel, chemical feedstock, etc.) the potent greenhouse
gas, methane. We consider 4 increasingly demanding representations
of such a system divided into 1 molecular instance and 3 extended
instances. The molecular system is a Pd-O complex bonded to 3 $\mathrm{NH}_2$
groups. The extended systems are of the form of $m\times m$ conventional
unit cells with 1 B and 1 N atom each, and a single B atom replaced
by a Pd-O complex. All instances have a single $\mathrm{CH}_4$ molecule.
\item Water Gas Shift (WGS): The WGS reaction $\mathrm{CO}+\mathrm{H_{2}O}\rightleftharpoons\mathrm{CO_{2}}+\mathrm{H}_{2}$
plays a pivotal role in energy production, the hydrogen economy, and environmental
sustainability, making its study essential for both scientific and
industrial advancements. We consider 2 increasingly demanding representations
of bilayer Cu(100) systems of the form $2\times m\times m$ conventional
unit cells, where the the center 5 atoms on the top surface are $\textrm{Cu}^{11}$
and the rest of the copper atoms are $\textrm{Cu}^{1}$. Both instances have a single
CO and a single $\textrm{H}_2\textrm{O}$ molecule.
\end{enumerate}

In Fig.~\ref{fig:timeEvoCost}, we compute the quantum resource estimate (QRE) in terms
of Toffoli gates for time-evolving the pseudoion Hamiltonian
(Eq.~\eqref{eq:H_PI}) for each of the 7 instances above for time $t$ (in atomic
units) with evolution error $\delta$. We perform a detailed cost
analysis of the time-evolution circuit, neglecting the one-time and
significantly sub-leading costs of initial state preparation and information
extraction. Note that QSP-based time-evolution gives a linear scaling
in the simulation time and an additive cost in the log of the inverse error (details in Sec.~\ref{subsec:TimeEvolution}). For molecular identification, a further multiplicative cost in the square root of the inverse of the (potentially biased) chemical rate has to be included. A detailed analysis of asymptotic scaling and a discussion of practical quantum resource estimates is given in Sec.~\ref{sec:resource_estimates}.

For the smallest instance of $\mathrm{NH}_3 \mathrm{BF}_3$ with $\eta=40$, we find the cost per time-step to be $\sim4\times10^{9}$ Toffolis with a spatial cost of $808$ qubits (excluding ancillae). For
the largest instance of DMTM $9\times 9$ with $\eta=845$, we find the cost per time step to be $\sim2\times10^{12}$ Toffolis with a spatial cost of $18753$ qubits (excluding ancilla). Elementary chemical processes
are likely captured within $\lesssim1$ps and so a useful simulation
likely requires $\lesssim4 \times 10^4$ time steps, adding an additional
few orders of magnitude overhead. Deploying our framework on
the limited resources available on early generation FTQCs requires
careful consideration of the appropriate problem instances and evolution
time, in order to extract scientifically useful information. Addressing
this issue is a subject of future work, perhaps involving co-development
of classical methods to pair with quantum dynamical simulations, as well as further improvements to the quantum algorithm. We see the above estimates as a starting point for further development.

\part*{Algorithm}
\addcontentsline{toc}{part}{II - Algorithm}

\section{Hamiltonians in plane-wave basis \label{sec:problem_statement}}

\subsection{Electrons and nuclei}
\label{sec:Hamiltonian_in_plane_waves}
In this section, we define notation and derive the full Hamiltonian describing electrons and nuclei in
the plane-wave basis, which we shall then modify in Sec.~\ref{sec:PI_interactions} to include a pseudoion description. Let $\kett{\mathbf{r}}_{i}\ensuremath{,}\kett{\mathbf{R}}_{I}$
be the position eigenstates of electron $i=1,\dots,\eta_{\mathrm{el}}$
and ion $I=1,\dots,\eta_{\mathrm{ion}}$, respectively, labeled by
the position operators eigenvalues $\mathbf{r},\mathbf{R}$. Define
the finite periodic plane wave basis,

\begin{equation}
\kett{\mathbf{p}}_{i}:=\frac{1}{\sqrt{\Omega}}\int_{{\bf r}_i}e^{-i{\bf k_{p}}\cdot{\bf r}_i}\kett{{\bf r}}_{i},\quad\kett{\mathbf{P}}_{I}:=\frac{1}{\sqrt{\Omega}}\int_{{\bf R_I}}e^{-i{\bf K_{P}}\cdot{\bf R}_I}\kett{{\bf R}}_{I},\label{eq:plane_wave_def}
\end{equation}
for electron and ion $i,I$, respectively.  The integral is over the
real-space simulation cell defined by the real-space lattice vectors
$\{{\bf a}_{\alpha}\}_{\alpha=1}^3$. The volume of the simulation cell is 
\begin{align}
    \Omega=|\det(\s A)|=|{\bf a}_{1}\cdot({\bf a}_{2}\times{\bf a}_{3})|,
\end{align}
where $\s A$ is a matrix with columns $\{{\bf a}_{\alpha}\}_{\alpha=1}^3$. Note
that we label the electronic momentum eigenstates by integers \mbox{${\bf p}=(p_{1},p_{2},p_{3})$}, where each $p_{\alpha}\in \mathbb{Z}$
(for $\alpha=1,2,3$) lives in the set 
\begin{align}
    G_{\alpha}:=[-p_{\alpha}^{\max},p_{\alpha}^{\max}], \quad G=G_{1}\times G_{2}\times G_{3},
\end{align}
with $p_{\alpha}^{\max}\in\mathbb{Z}$ denoting the electronic momentum cutoff. Similarly, we label the nuclear momentum eigenstates by $\mathbf{P}=(P_{1},P_{2},P_{3})$, where each $P_{\alpha}\in\mathbb{Z}$ lives in the set 
\begin{align}
    \overline{G}_{\alpha}:=[-P_{\alpha}^{\max},P_{\alpha}^{\max}], \quad \overline{G}=\bar{G}_{1}\times \overline{G}_{2}\times \overline{G}_{3},
    \label{eq:Galphabar}
\end{align}
with $P_{\alpha}^{\max}\in\mathbb{N}$ denoting the nuclear momentum cutoff.

The integer labels $\mathbf{p}, \mathbf{P}$ index
the electron and ion momentum eigenvalues $\mathbf{k_{p}}$, $\mathbf{K_{P}}$ given by
\begin{equation}
\mathbf{k_{p}}=\sum_{\alpha=1}^{3}p_{\alpha}\mathbf{b}_{\alpha},\quad\mathbf{K_{P}}=\sum_{\alpha=1}^{3}P_{\alpha}\mathbf{b}_{\alpha},\label{eq:physical_to_comp_momenta}
\end{equation}
where $\mathbf{b}_{\alpha}$ are the reciprocal lattice vectors along
directions $\alpha=1,2,3$, that define the full computational reciprocal
lattice. The reciprocal lattice vectors $\{{\bf b}_{\alpha}\}$ are
constructed from the real-space lattice vectors~$\{{\bf a}_{\alpha}\}$:

\begin{equation}
{\bf b}_{1}=\frac{2\pi}{\Omega}{\bf a}_{2}\times{\bf a}_{3},\quad{\bf b}_{2}=\frac{2\pi}{\Omega}{\bf a}_{3}\times{\bf a}_{1},\quad{\bf b}_{3}=\frac{2\pi}{\Omega}{\bf a}_{1}\times{\bf a}_{2},\label{eq:reciprocol_lattice_vectors}
\end{equation}
where the reciprocal lattice volume is $\bar{\Omega}=|\det(\s B)|=\frac{(2\pi)^{3}}{\Omega}$,
with $\s B$ as a matrix with columns $\{{\bf b}_{\alpha}\}_{\alpha=1}^3$. Hence, the
simulation takes place in the plane-wave basis labeled by $\{{\bf p}\}_{{\bf p}\in G}$, $\{{\bf P}\}_{{\bf P}\in G}$.

The elements of $G_{\alpha}$ are stored using $n_{\alpha}$ qubits
in a signed momentum representation giving the max value of the momentum
integers as $p_{\alpha}^{\max}=2^{n_{\alpha}-1}-1$.\footnote{The signed representation has $p_{\alpha}=(-1)^{p_{\alpha,n_{\alpha}-1}}\sum_{r=0}^{n_{\alpha}-2}2^{r}p_{\alpha,r}$
where $p_{\alpha,r} \in \{0,1\}$ is the $r$-th bit of $p_{\alpha}$.
The label $p_{\alpha}=-0$ indexes an unphysical redundant state not
present in $G_{\alpha}$. Hence, $|G_{\alpha}|=2^{n_{\alpha}}-1=2p_{\alpha}^{\max}+1$
and $|G|=\prod_{\alpha=1}^{3}|G_{\alpha}|$.} The total number of qubits per electron is therefore 
\begin{align}
n=\sum_{\alpha=1}^{3}n_{\alpha}, \quad n_{\alpha} = \lceil \log_2( p_{\alpha}^{\textrm{max}} +1)\rceil +1.\label{eq:qubit_number}  
\end{align}
We choose an approximate maximum momentum $\Lambda$, and set each $p^{\textrm{max}}_{\alpha}$ such that\footnote{In practice, we choose $\Lambda$
and compute $ p^{\textrm{max}}_{\alpha}=\Lambda/|\mathbf{b}_{\alpha}|$ which, in general, is a floating point number. Using Eq.~\eqref{eq:qubit_number}, we compute the integer number of qubits and then recompute the integer $p^{\textrm{max}}_{\alpha}$ as shown above Eq.~\eqref{eq:qubit_number}. This then gives the ``true'' cutoffs, which are in general anisotropic, and larger than $\Lambda$ due to the ceiling in Eq.~\eqref{eq:qubit_number}. When quoting the instances in Sec.~\ref{sec:sequence}, we list these true cutoffs.} 
\begin{align}
    \Lambda \approx p^{\textrm{max}}_{\alpha} |\mathbf{b}_{\alpha}|, \quad \textrm{for } \alpha =1,2,3.
    \label{eq:electroncutoff}
\end{align}
One may choose anisotropic cutoffs, i.e., $\Lambda_1, \Lambda_2, \Lambda_3$, if desired (as we do in Sec.~\ref{sec:sequence}). 
The interaction terms in the Hamiltonian of interest will involve
a momentum exchange that has twice the range of the momentum values
defined in $G$. For this we define a new set 
\begin{align}
    {\bf q}\in G^{0}:=G_{1}^{0}\times G_{2}^{0}\times G_{3}^{0}, \quad G_{\alpha}^{0}:=[-2p_{\alpha}^{\max},2p_{\alpha}^{\max}]\backslash\{0\},
\end{align}
where zero-momentum differences are neglected to avoid singularities
in the interaction.

A completely analogous discussion holds for nuclei. We shall repeat it to introduce the relevant notation. The elements of $\overline{G}_{\alpha}$ are stored using $\overline{n}_{\alpha}$ qubits
in a signed momentum representation giving the max value of the momentum
integers as $P_{\alpha}^{\max}=2^{\bar{n}_{\alpha}-1}-1$. The total number of qubits per nucleus is therefore 
\begin{align}
\bar{n}=\sum_{\alpha=1}^{3}\bar{n}_{\alpha}, \quad \bar{n}_{\alpha} = \lceil \log_2( P_{\alpha}^{\textrm{max}} +1)\rceil +1.  
\label{eq:qubit_number2}
\end{align}
Each $P_{\alpha}^{\max}$ is chosen with the approximate maximum momentum as before, 
\begin{align}
\label{eq:nuclearcutoff}
    \bar{\Lambda}\approx P_\alpha^{\max}|{\bf b}_{\alpha}|.
\end{align}
Similarly, the nuclear momentum exchange is given by,
\begin{align}
    {\bf Q}\in \bar{G}^{0}:=\bar{G}_{1}^{0}\times \bar{G}_{2}^{0}\times \bar{G}_{3}^{0}, \quad \bar{G}_{\alpha}^{0}:=[-2P_{\alpha}^{\max},2P_{\alpha}^{\max}]\backslash\{0\},
\end{align}
In the plane-wave
basis, the Hamiltonian terms given in Eq.~\eqref{eq:H_bare_position}
take the following form:
\begin{align}
T_{\mathrm{el}} & =\sum_{i=1}^{\eta_{\mathrm{el}}}\left(\sum_{\mathbf{p}\in G}\frac{|{\bf k_{p}}|^{2}}{2}\kett{\mathbf{p}}\brat{\mathbf{p}}_{i}\right),\label{eq:Tel_plane-wave-elements}\\
T_{\mathrm{ion}} & =\sum_{I=1}^{\eta_{\mathrm{ion}}}\left(\sum_{\mathbf{P}\in \bar{G}}\frac{|{\bf k_{P}}|^{2}}{2M_{I}}\kett{\mathbf{P}}\brat{\mathbf{P}}_{I}\right),\label{eq:Tion_plane-wave-elements}\\
V_{\mathrm{el}} & =\frac{2\pi}{\Omega}\sum_{i\ne j=1}^{\eta_{\mathrm{el}}}\left(\sum_{\substack{\mathbf{p},\mathbf{p'}\in G,{\bf q}\in G^{0}\\
\mathbf{p}-\mathbf{q},\mathbf{p'}+\mathbf{q}\in G
}
}\frac{1}{|{\bf k_{q}}|^{2}}\kett{\mathbf{p-q}}\brat{\mathbf{p}}_{i}\otimes\kett{{\bf p'+q}}\brat{{\bf p'}}_{j}\right),\label{eq:Vel_plane-wave-elements}\\
V_{\mathrm{ion}} & =\frac{2\pi}{\Omega}\sum_{I\ne J=1}^{\eta_{\mathrm{ion}}}\left(\sum_{\substack{\mathbf{P},\mathbf{P'}\in \bar{G},{\bf Q}\in \bar{G}^{0}\\
\mathbf{P}-\mathbf{Q},\mathbf{P'}+\mathbf{Q}\in \bar{G}
}
}\frac{Z_{I}Z_{J}}{|{\bf k_{Q}}|^{2}}\kett{\mathbf{P-Q}}\brat{\mathbf{P}}_{I}\otimes\kett{{\bf P'+Q}}\brat{{\bf P'}}_{J}\right),\label{eq:Vion_plane-wave-elements}\\
V_{\mathrm{el-ion}} & =-\frac{4\pi}{\Omega}\sum_{i=1}^{\eta_{\mathrm{el}}}\sum_{I=1}^{\eta_{\mathrm{ion}}}\left(\sum_{\substack{\mathbf{p}\in G,\mathbf{P}\in \bar{G},\mathbf{q}\in G^{0}\\
\mathbf{p}-\mathbf{q}\in G,\mathbf{P}+\mathbf{q}\in \bar{G}
}
}\frac{Z_{I}}{|{\bf k_{q}}|^{2}}\kett{\mathbf{p-q}}\brat{{\bf p}}_{i}\otimes\kett{{\bf P+q}}\brat{{\bf P}}_{I}\right),\label{eq:Velion_plane-wave-elements}
\end{align}
where the expressions are obtained by calculating matrix elements
via integrals involving plane-wave basis states given in Eq.~\eqref{eq:plane_wave_def},
and where 
\begin{align}
|\mathbf{k_{p}}|^{2}=\sum_{\alpha,\beta=1}^{3}p_{\alpha}p_{\beta}(\mathbf{b}_{\alpha}\cdot\mathbf{b}_{\beta})
\label{eq:ksq_formula}
\end{align}
is obtained by expressing the physical momenta $\mathbf{k_{p}},\mathbf{K_{P}}$
in terms of the integer labels $\mathbf{p},\mathbf{P}$ via Eq.~\eqref{eq:physical_to_comp_momenta}.
Note that, for the special case of a cubic lattice, $\mathbf{b}_{\alpha}=\frac{2\pi}{\Omega^{1/3}}\hat{\mathbf{a}}_{\alpha}$
and so $|\mathbf{k_{p}}|=\frac{2\pi|\mathbf{p}|}{\Omega^{1/3}}$.

\subsection{Valence electrons and pseudoions}\label{sec:PI_interactions}

In Sec.~\ref{sec:dynamic_DoF} we described how the quantum simulation of dynamics with pseudoions is governed by the modified Hamiltonian in Eq.~\eqref{eq:H_PI}. Here we describe that modified Hamiltonian in the plane-wave basis. Moving to the pseudoion description:
\begin{itemize}
    \item The electronic kinetic energy and electron-electron interaction (Eq.~\eqref{eq:Tel_plane-wave-elements} and Eq.~\eqref{eq:Vel_plane-wave-elements}) are simply modified by replacing the number of electrons $\eta_{\mathrm{el}}$ with the chosen number of valence electrons $\eta_\mathrm{val}$. 
    \item The nuclear kinetic energy (Eq.~\eqref{eq:Tion_plane-wave-elements}) is left unchanged.
    \item The ion-ion interaction in Eq.~\eqref{eq:Vion_plane-wave-elements} is modified in two ways. First, for each ion $I$  we replace the atomic numbers $Z_I$ with $Z_I^{\mathrm{PI}} = Z_{I}-\eta_{\textrm{core},I}$, as described in Eq.~\eqref{eq:Z_PI}, where $\eta_{\textrm{core},I} = \eta_{\textrm{el},I} - \eta_{\textrm{val},I}$. Second, we introduce a cutoff $\overline{\Lambda}_\mathrm{trunc}$ and the associated basis set $\overline{G}_{\textrm{trunc}}$ (and momentum exchange set $\overline{G}_{\textrm{trunc}}^0$) which is constructed identically to $\overline{G}$ (and $\bar{G}^0$) around Eq.~\eqref{eq:Galphabar} with $P^{\textrm{max}}_{\alpha}$ replaced by $P^{\textrm{trunc}}_{\alpha} < P^{\textrm{max}}_{\alpha}$. This is consistent with the assumption that pseudoions interact among themselves as point-charges of atomic number $Z_I^{\mathrm{PI}}$, which is only valid as a far-field approximation. It would be inconsistent to consider pseudoion-pseudoion interactions in the near-field (i.e. resolved by high momenta), as then the notion of pseudoions itself breaks down. 
    \item Finally, the electron-ion interaction in Eq.~\eqref{eq:Velion_plane-wave-elements} has a highly nontrivial modification. As we have seen in Eq.~\eqref{eq:VPI_elion}, $V_{\mathrm{el-ion}}$ is replaced by 
$V_{\el-\ion}^{\mathrm{PI}}=\sum_{I=1}^{\eta_{\mathrm{ion}}}\sum_{i=1}^{\eta_{\mathrm{val}}}(V_{\mathrm{loc}}^{i,I}+V_{\mathrm{NL}}^{i,I})$,  with the definitions in Eq.~\eqref{eq:local_term_position}-\eqref{eq:NL_term_position}. 
 \end{itemize} In App.~\ref{app:basis_change} we show how to express these terms in the plane wave basis. In summary, the pseudoion Hamiltonian in Eq.~\eqref{eq:H_PI} is characterized by the following terms:
\begin{align}
T_{\mathrm{el}} & =\sum_{i=1}^{\eta_{\mathrm{val}}}\left(\sum_{\mathbf{p}\in G}\frac{|{\bf k_{p}}|^{2}}{2}\kett{\mathbf{p}}\brat{\mathbf{p}}_{i}\right),\label{eq:Tel_plane-wave-elements-valence}\\
T_{\mathrm{ion}} & =\sum_{I=1}^{\eta_{\mathrm{ion}}}\left(\sum_{\mathbf{P}\in \overline{G}}\frac{|{\bf k_{P}}|^{2}}{2M_{I}}\kett{\mathbf{P}}\brat{\mathbf{P}}_{I}\right),\label{eq:Tion_plane-wave-elements-repeated}\\
V_{\mathrm{el}} & =\frac{2\pi}{\Omega}\sum_{i\ne j=1}^{\eta_{\mathrm{val}}}\left(\sum_{\substack{\mathbf{p},\mathbf{p'}\in G,{\bf q}\in G^{0}\\
\mathbf{p}-\mathbf{q},\mathbf{p'}+\mathbf{q}\in G
}
}\frac{1}{|{\bf k_{q}}|^{2}}\kett{\mathbf{p-q}}\brat{\mathbf{p}}_{i}\otimes\kett{{\bf p'+q}}\brat{{\bf p'}}_{j}\right),\label{eq:Vel_plane-wave-elements-valence}
\\
V_{\mathrm{ion}}^{\mathrm{PI}}  = &\frac{2\pi}{\Omega}\sum_{I\ne J=1}^{\eta_{\mathrm{ion}}}\left(\sum_{\substack{\mathbf{P},\mathbf{P'}\in \overline{G}_{\textrm{trunc}},{\bf Q}\in \overline{G}^{0}_{\textrm{trunc}}\\
\mathbf{P}-\mathbf{Q},\mathbf{P'}+\mathbf{Q}\in \overline{G}_{\textrm{trunc}}
}
}\frac{Z_{I}^{\mathrm{PI}}Z_{J}^{\mathrm{PI}}}{|{\bf k_{Q}}|^{2}}\kett{\mathbf{P-Q}}\brat{\mathbf{P}}_{I}\otimes\kett{{\bf P'+Q}}\brat{{\bf P'}}_{J}\right), \label{eq:VPI_ion_plane_waves}\\ \nonumber
\\
 V_{\el-\ion}^{\mathrm{PI}} = & \sum_{I=1}^{\eta_{\mathrm{ion}}}
 \sum_{i=1}^{\eta_{\mathrm{val}}}  (V^{i,I}_{\loc} + V^{i,I}_{\mathrm{NL}}), \label{eq:VPI_el-ion_plane_waves_total}\\
V_{\mathrm{loc}}^{i,I} = & \sum_{\substack{\mathbf{p}\in G,\mathbf{P}\in \overline{G}, \mathbf{q}\in G^{0}\\
\mathbf{p-q} \in G,\mathbf{P}+\mathbf{q}\in \overline{G}
}
}\frac{4\pi(\bar{r}_{\mathrm{loc}}^{I})^{3}}{\Omega}\sqrt{\frac{\pi}{2}}e^{-(|\mathbf{k_{q}}|\bar{r}_{\mathrm{loc}}^{I})^{2}/2}\sum_{s=-1}^{3}c_{s}^{I}(|\mathbf{k_{q}}|\bar{r}_{\mathrm{loc}}^{I})^{2s}\kett{\mathbf{p-q}}\brat{\mathbf{p}}_{i}\otimes\kett{\mathbf{P}+\mathbf{q}}\brat{\mathbf{P}}_{I},\label{eq:local_term_plane_waves}
\end{align}
\small
\begin{align}
V_{\mathrm{NL}}^{i,I} & = \sum_{\substack{\mathbf{p_{1}},\mathbf{p_{2}} \in G,\mathbf{P}\in \overline{G}\\
\mathbf{P}+\mathbf{p_{1}}-\mathbf{p_{2}}\in \overline{G}
}
}\sum_{a,b=1}^{3}\sum_{l=0}^{l_{\mathrm{max}}}\frac{4\pi}{\Omega}(\bar{r}_{l}^{I})^{3}(2l+1)\mathrm{g}_{a}^{l}(|\mathbf{k_{p_{2}}}|\bar{r}_{l}^{I})B_{a,b}^{I,l}\mathrm{g}_{b}^{l}(|\mathbf{k_{p_{1}}}|\bar{r}_{l}^{I})P_{l}(\mathbf{\hat{k}_{p_{1}}}\cdot\mathbf{\hat{k}_{p_{2}}})\kett{\mathbf{p_{2}},\mathbf{P}+\mathbf{p_{1}}-\mathbf{p_{2}}}\brat{\mathbf{p_{1}},\mathbf{P}}_{i,I}.
\label{eq:nl_legendre}
\end{align}
\normalsize
where 
\begin{align}
c_{-1}^{I} & =-\sqrt{\frac{2}{\pi}}\frac{Z_{I}^{\mathrm{PI}}}{\bar{r}_{\mathrm{loc}}^{_{I}}},  \quad 
& c_{0}^{I} 
 & =C_{1}^{I}+3C_{2}^{I}+15C_{3}^{I}+105C_{4}^{I}, \nonumber \\
c_{1}^{I} & =-C_{2}^{I}-10C_{3}^{I}-105C_{4}^{I}, \quad 
& c_{2}^{I} & =C_{3}^{I}+21C_{4}^{I}, \quad 
c_{3}^{I}  =-C_{4}^{I}, \label{eq:c_coeffs}
\end{align}
with $\bar{r}_{\mathrm{loc}}^{I},C_{1}^{I},C_{2}^{I},C_{3}^{I},C_{4}^{I},\bar{r}_{l}^{I},B_{a,b}^{I,l}$
as the HGH fitting parameters, $\mathrm{g}_{a}^{l}(x)$ the radial functions 
\begin{align}
\mathrm{g}_{a}^{l}(x) =e^{-x^2/2}x^{l}\frac{\sqrt{\pi}2^{a-1}(a-1)!}{\sqrt{\Gamma(l+2a-\frac{1}{2})}}\mathrm{L}_{a-1}^{l+\frac{1}{2}}(x^2/2), \label{eq:g_function}
\end{align}
with $\mathrm{L}_{n}^{m}(x)$ generalized Laguerre polynomials and $P_{l}(\hat{\mathbf{x}})$ Legendre polynomials of degree $l$.\footnote{Our notation differs from that of Ref.~\cite{berry2023quantum}
in that we provide more general expressions without any reference
to a table of coefficients as is done in their Table 13. However, the connection
is simple in that, for the definition of $F$ functions in
their Eq.~15,
we have $F_{l,I}^{a}(|\mathbf{k_{p}}|)=(4\pi)(\bar{r}_{l}^{I})^{3/2}\mathrm{g}_{a}^{l}(|\mathbf{k_{p}}|\bar{r}_{l}^{I})$.}
This is the form that we will use to compile the block encoding in Sec.~\ref{subsubsec:BENonlocalPP} after a minor additional step. As with $V_{\mathrm{el-ion}}$ in Eq.~\eqref{eq:Velion_plane-wave-elements},
the modified interaction $V_{\el-\ion}^{\mathrm{PI}}$ involves a momentum
exchange between electrons and pseudoions, while the main difference arises in the matrix elements, which are significantly
more involved for the non-local pseudoion interactions than a simple Coulomb interaction.

In summary, our proposal
involves extending successful constructions for the (static) electronic
structure problem in condensed matter and chemical systems to the
simulation of fully-interacting quantum dynamics, through lifting $1$-body pseudopotentials to 2-body electron-pseudoion
interactions. We may also expect that pseudoion interactions
better suited to describe dynamical problems may be introduced in
the future. The plan ahead is as follows:
\begin{itemize}
    \item Prepare a suitable initial state for reactants and catalyst, justified by appropriate physical and chemical arguments, as well as considerations of algorithmic efficiency (Sec.~\ref{sec:quantum_state_preparation}).
    \item Describe the algorithm for the evolution under the pseudoion Hamiltonian in Eq.~\eqref{eq:H_PI} (Sec.~\ref{sec:time_evolution}).
    \item Describe the molecular identification algorithm to be used to extract information about the reaction products (Sec.~\ref{sec:chemical_species_id}).
\end{itemize}

\subsubsection*{Remark on cutoffs}
We have introduced three different momentum cutoffs: $\Lambda,\overline{\Lambda}$
for the electrons, pseudoions, and $\overline{\Lambda}_{\mathrm{trunc}}<\overline{\Lambda}$
only for the pseudoion-pseudoion interaction $V_{\mathrm{ion}}^{\mathrm{PI}}$.
The first two cutoffs determine the size of the qubit registers associated
to each electron and pseudoion and hence appear in all the Hamiltonian
terms except $V_{\mathrm{ion}}^{\mathrm{PI}}$. For this latter case,
we replace the naive $\overline{\Lambda}$ with $\overline{\Lambda}_{\mathrm{trunc}}$
which functionally truncates the momentum sums to exclude unphysical
parts of the Hamiltonian (that would never be relevant in physical
situations) in order to reduce the resources. As is the convention
in quantum-chemistry simulations with plane waves, one may associate an
energy, in Hartree, to the electron cutoff $E_{\Lambda}=\frac{1}{2}\Lambda^{2}$.
Typically, electronic structure problems converge well with $E_{\Lambda}\sim50\mathrm{Ha}$, and so $\Lambda\sim10$~\cite{lejaeghere2016reproducibility}. For pseudoion wavefunctions to be resolved spatially, since
they are significantly more massive than electrons (and hence more
localized), we estimate -- by approximating low-lying quantum harmonic
oscillator states with plane waves as a rough surrogate model for bound ions (not shown) --  that $\overline{\Lambda}\sim3\Lambda$
is required as a conservative estimate to ensure critical nuclear phenomena, such
as proton delocalization, are not missed in the dynamics. Finally, in physical
situations we expect pseudoions to be separated $\gtrsim 2a_0$. Inspired by this, we choose to resolve interactions between pseudoions up to $\lesssim 1 a_0$, although a different choice could be made based on the desired resolution. Converting to a momentum cutoff, this gives $\overline{\Lambda}_{\mathrm{trunc}} \sim 1$. These cutoff considerations are used to define our problem instances in Sec.~\ref{sec:sequence}.

In this work, the hard cutoffs abruptly truncate the momentum set
for the Coulomb interactions ($V_{\mathrm{el}},V_{\mathrm{ion}}^{\mathrm{PI}}$).
There is no fundamental obstacle in alternatively considering replacing $1/r$ with a softened
interaction.\footnote{Any radially-symmetric interaction $f(r)$, can be moved into the
plane-wave basis with a 3d Fourier transformation, which, for computational ease, can be expressed as a Hankel transform of order $1/2$ denoted $\s H_{1/2}\{\cdot\}(k)$
with $k=|\mathbf{k}|$, $
f(k)=4\pi\sqrt{\frac{\pi}{2}}\frac{1}{\sqrt{k}}\s H_{1/2}\{f(r)\sqrt{r}\}(k)$.} Consider for example $\mathrm{erf}(r/a)/r$  where $a$
determines the scale where the short-distance divergence is softened,
just as is seen in the $s=-1$ term of $V_{\mathrm{loc}}^{\mathrm{PI}}$.
In effect,
the hard
cutoff $\overline{\Lambda}_{\mathrm{trunc}}$ is replaced by the softened
interaction with parameter $a$ while keeping $\overline{\Lambda}$
as the momentum set for pseudoions. 
Intuition suggests that numerical convergence
for physical
scenarios might be easier with such a softened interaction as well.
This investigation is left for future work.

\section{Initial state construction}
\label{sec:quantum_state_preparation}
We construct an initial state that is both physically representative in the chemical context of interest (e.g. understanding bond reconfiguration/reaction mechanism) and efficient
to construct on an quantum computer. Towards this end, we will be inspired by a Hamiltonian constructed by following the standard approach that incorporates the BO approximation.  However, we will evolve the initial state under the fully quantum-mechanical pseudoion Hamiltonian in Eq.~\eqref{eq:H_PI} for all $t>0$. We leave exploring settings where the initial state preparation requires
inclusion of some non-BO terms to future work. We shall also follow the guideline drawn in Sec.~\ref{sec:generalconsiderationstateprep} (and App.~\ref{app:detailed_physical_justification}). With this premise, we start with the full Hamiltonian in Eq.~\eqref{eq:H_bare_position}.
\begin{align}
H  ={-\frac{1}{2}\sum_{i=1}^{\eta_{\mathrm{el}}}\nabla_{i}^{2}}{-\sum_{I=1}^{\eta_{\mathrm{ion}}}\frac{1}{2M_{I}}\nabla_{I}^{2}}{+\frac{1}{2}\sum_{i\ne j}^{\eta_{\mathrm{el}}}\frac{1}{|\mathbf{r}_{i}-\mathbf{r}_{j}|}}{+\frac{1}{2}\sum_{I\ne J}^{\eta_{\mathrm{ion}}}\frac{Z_{I}Z_{J}}{|\mathbf{R}_{I}-\mathbf{R}_{J}|}}{-\sum_{i=1}^{\eta_{\mathrm{el}}}\sum_{I=1}^{\eta_{\mathrm{ion}}}\frac{Z_{I}}{|\mathbf{r}_{i}-\mathbf{R}_{I}|}}.
\end{align}

\subsection{Kinematics}
\label{sec:kinematics}

\subsubsection{Species Partitioning}
\label{sec:speciespartitioning}

We partition the degrees of freedom (electrons and nuclei)
into disjoint sets, each of which forms a chemical ``species'' defined by spatial proximity and chemical bonding at $t=0$. 
For example, consider simulating a single event of the water gas-shift (WGS) reaction
\begin{align}
    \mathrm{CO + H_2 O \rightarrow H_2 + CO_2}
\end{align}
on a copper catalyst, with 1 C, 2 O, 2 H, and a collection of Cu atoms. We lump all the metal $\mathrm{Cu}$ nuclei and their associated electrons into one set representing the metal catalyst species, labeled $s=0$. Similarly, we lump 1 C and 1 O together representing the $\mathrm{CO}$ species, labeled $s=1$, and lump the remaining 2 H and 1 O together representing $\mathrm{H_2 O}$, labeled $s=2$. For our purposes of a single catalyst and $S$
reactants (e.g. $S=2$ above), we label all $S+1$ species from $s=0,...,S$, where $s=0$
refers to the catalyst and the rest index the reactants. For
a species $s$, the labels $\mathcal{E}'_{s},\s I_{s}$ denote a
corresponding set of associated electrons and nuclei, where $|\s E'_{s}|=\eta_{\mathrm{el}}^{s},|\s I_{s}|=\eta_{\mathrm{ion}}^{s}$.
With this, let us define,
    \begin{align}
    \label{eq:Hsoriginal}
H^{s} & =-\frac{1}{2}\sum_{i\in\mathcal{E}'_{s}}\nabla_{i}^{2}-\sum_{I\in\mathcal{I}_{s}}\frac{1}{2M_{I}}\nabla_{I}^{2}+\frac{1}{2}\sum_{i\ne j\in\mathcal{E}'_{s}}\frac{1}{|\mathbf{r}_{i}-\mathbf{r}_{j}|}+\sum_{I\ne J\in\mathcal{I}_{s}}\frac{Z_{I} Z_{J}}{|\mathbf{R}_{I}-\mathbf{R}_{J}|}-\sum_{i \in \mathcal{E}'_s} \sum_{I \in \mathcal{I}_s} \frac{Z_{I}}{|\mathbf{r}_{i}-\mathbf{R}_{I}|} ,\nonumber \\
H & =\sum_{s=0}^{S}H^{s}+H^{\textrm{INT}},
\end{align}
where
$H^{\textrm{INT}}$ is implicitly defined by collecting
all mutual interactions between different chemical species (e.g. a
reactant with another reactant, a reactant with the catalyst, etc.). We neglect interspecies interactions $H^{\mathrm{INT}}$ for the purpose of creating an initial state ansatz and further assume
that all species are initially mutually uncorrelated. This invokes
the same spirit as in the usual treatment of quantum-mechanical scattering,
wherein initial states are product states of spatially well-separated non-interacting particles, although for efficiency purposes, we might choose to prepare species in close proximity to each other as discussed in Sec.~\ref{sec:generalconsiderationstateprep}. Altogether, this yields a tensor product ansatz for the initial
state, described by the density operator
\begin{equation}
\rho(t=0)=\otimes_{s=0}^{S}\rho_{s}(t=0), \label{eq:species_partition}
\end{equation}
where $\rho_{s}(t=0)$ is the initial state of species $s$.

\subsubsection{Reaction geometry} \label{sec:reactiongeometry}

Fixing a species $s$, for each of its constituents labeled by $i \in \mathcal{E}_s'$ and $I \in \mathcal{I}_s$, we move to the corresponding center-of-mass (CoM) coordinates,
\begin{equation}
\tilde{\mathbf{r}}_{i}=\mathbf{r}_{i}-\mathbf{R}_{\textrm{CoM}}^s,\quad\tilde{\mathbf{R}}_{I}=\mathbf{R}_{I}-\mathbf{R}_{\textrm{CoM}}^s,\quad\mathbf{R}_{\textrm{CoM}}^s=\sum_{I\in\mathcal{I}_s}\frac{M_{I}}{M_{\textrm{tot}}^{s}}\mathbf{R}_{I},
\end{equation}
where $M_{\textrm{tot}}^{s}=\sum_{I\in\mathcal{I}_{s}}M_{I}$. Ignoring
the typically small \emph{mass polarization} term coupling internal
(non-CoM) and CoM DoF (Ref.~\cite{jensen2017introduction}),
we get the separation
 
\small
\begin{align}
H^{s} \approx -\frac{1}{2M_{\textrm{tot}}^{s}}\nabla_{\textrm{CoM},s}^{2} -\frac{1}{2}\sum_{i\in\mathcal{E}'_{s}}\tilde{\nabla}_{i}^{2}-\sum_{I\in\mathcal{I}_{s}}\frac{1}{2M_{I}}\tilde{\nabla}_{I}^{2}+\frac{1}{2}\sum_{i\ne j\in\mathcal{E}'_{s}}\frac{1}{|\tilde{\mathbf{r}}_{i}-\tilde{\mathbf{r}}_{j}|}+\sum_{I\ne J\in\mathcal{I}_{s}}\frac{Z_{I}Z_{J}}{|\tilde{\mathbf{R}}_{I}-\tilde{\mathbf{R}}_{J}|}-\sum_{i \in \mathcal{E}'_s} \sum_{I \in \mathcal{I}_s} \frac{Z_{I}}{|\tilde{\mathbf{r}}_{i}-\tilde{\mathbf{R}}_{I}|}
\label{eq:HPIseparation}
\end{align}
\normalsize
where $\nabla_{\textrm{CoM},s}$ is the gradient with respect to $\mathbf{R}_{\textrm{CoM}}^s$,
$\tilde{\nabla}_{i,I}$ are gradients in relative coordinates with respect to the CoM. 
 Note that the CoM coordinates obey $\sum_{I\in\mathcal{I}_s}M_{I}\tilde{\mathbf{R}}_{I}=0$
by construction and hence only $3\eta_{\mathrm{ion}}^{s}-3$ of the
CoM coordinates are independent.

We choose the CoM for each species to be prepared in an isotropic minimal uncertainty Gaussian wavepacket. Such a wavepacket has $7$ free parameters per species, $3$ fixing the mean position, $3$ fixing the mean momentum, and $1$ for the spatial standard deviation
of the wavepacket at $t=0$. Hence, for $S+1$ species, we have $7(S+1)$
free parameters to specify which we define as the \emph{reaction geometry}.
The reaction geometry is user-specified and based on the
kinematic scenario desired for simulation. While in principle any
configuration is acceptable, certain desiderata exist
to bolster a combination of physical realism, scientific utility, and algorithmic efficiency, as discussed in Sec.~\ref{sec:generalconsiderationstateprep}. 

\subsection{Approximate molecular Hamiltonian for initial state preparation}
\label{sec:internaldof}

We invoke a standard series of approximations in molecular physics literature \cite{bunker2006molecular,wilson1980molecular,littlejohn1997gauge}:
\begin{itemize}
    \item Born-Oppenheimer approximation (Sec.~\ref{sec:BO}),
    \item Rigid-rotor approximation (Sec.~\ref{sec:rigidrotorapproximation}),
    \item Harmonic oscillator approximation (Sec.~\ref{sec:harmonicoscillatorapproximation}).
\end{itemize}

\subsubsection{Born-Oppenheimer (BO) approximation}
\label{sec:BO}

We assume that the degrees
of freedom of the molecules are well-described by
the Born-Oppenheimer (BO) approximation at $t=0$. This means separating
the electronic and nuclear Schr\"odinger equations as per the standard
procedure.\footnote{The BO approximation, as per usual, entails the adiabatic approximation,
where potential energy surfaces (PES) of all electronic energy manifolds are mutually well-separated
such that the \emph{non-adiabatic} coupling terms between PES are
ignored; and it entails neglecting the \emph{diagonal correction}, such that the 
Hamiltonian simply reduces to kinetic terms with the ground
state PES as an effective potential. See Sec.~3.1 in Ref.~\cite{jensen2017introduction} for details.}  We solve for the electronic ground state with parametric dependence
on the nuclear positions to construct the ground state potential
energy surface (PES). Specifically, we start with the operator $H^{s}$ in Eq.~\eqref{eq:Hsoriginal}, drop the nuclear kinetic term and fix nuclear positions for species $s$
 \begin{align}
\mathbf{R}^s = \{  \mathbf{R}_I \}_{I \in \mathcal{I}_s},
 \end{align}
obtaining the operator
\begin{align}
H_{\textrm{BO}}^{s}(\mathbf{R}^s) & =-\frac{1}{2}\sum_{i\in\mathcal{E}'_{s}}\nabla_{i}^{2}+\frac{1}{2}\sum_{i\ne j\in\mathcal{E}'_{s}}\frac{1}{|\mathbf{r}_{i}-\mathbf{r}_{j}|}+\sum_{I\ne J\in\mathcal{I}_{s}}\frac{Z_{I}Z_{J}}{|\mathbf{R}_{I}-\mathbf{R}_{J}|}-\sum_{i \in \mathcal{E}'_s} \sum_{I \in \mathcal{I}_s} \frac{Z_{I}}{|\mathbf{r}_{i}-\mathbf{R}_{I}|}.\label{eq:BO_electronic_hamiltonian}
\end{align}
We now introduce the pseudopotential approximation~\cite{hartwigsen1998relativistic}. That is, we remove some of the electron DoF (non-valence) and replace the Coulomb potential felt by the electrons due to the fixed nuclei with a pseudopotential:
\begin{align}
H_{\textrm{BO}}^{\mathrm{PP},s}(\mathbf{R}^s) & =-\frac{1}{2}\sum_{i\in\mathcal{E}_{s}}\nabla_{i}^{2}+\frac{1}{2}\sum_{i\ne j\in\mathcal{E}_{s}}\frac{1}{|\mathbf{r}_{i}-\mathbf{r}_{j}|}+\sum_{I\ne J\in\mathcal{I}_{s}}\frac{Z_{I}^{\mathrm{PI}}Z_{J}^{\mathrm{PI}}}{|\mathbf{R}_{I}-\mathbf{R}_{J}|}+\sum_{I \in \mathcal{I}_s} V^{I}_{\mathrm{PP,s}}(\v{R}_I),
\label{eq:BOPP_electronic_hamiltonian}
\end{align}
where  $Z_{I}^{\mathrm{PI}}=Z_{I}-\eta_{\textrm{core},I}$ as given in Eq.~\eqref{eq:Z_PI}, $\mathcal{E}_{s}$ is a set of labels for the `valence' electrons of molecular species $s$, with $|\mathcal{E}_{s}|= \eta^s_{\mathrm{val}}$ and $V_{\mathrm{PP},s}^I(\v{R}_I)$ is defined as in Eq.~\eqref{eq:HGHPPexpansion}, just restricted to species $s$:
\begin{align}
    V^{I}_{\mathrm{PP,s}}(\mathbf{R}_I) = \sum_{i \in \mathcal{E}_s}V^{i,I}_{\mathrm{PP, loc}}(\mathbf{R}_I) + \sum_{i \in \mathcal{E}_s} V_{\mathrm{PP, NL}}^{i,I}(\mathbf{R}_I),
\end{align}
with $V^{i,I}_{\mathrm{PP, loc}}(\mathbf{R}_I), V_{\mathrm{PP, NL}}^{i,I}(\mathbf{R}_I)$ given in Eq.~\eqref{eq:localPP_term_position} and Eq.~\eqref{eq:NLPP_term_position}, respectively.
The Hamiltonian $H_{\textrm{BO}}^{\mathrm{PP},s}$ is, in essence, the BO version of the pseudoion Hamiltonian of Eq.~\eqref{eq:H_PI} for a single molecular species.

We then consider the eigenproblem for $H_{\textrm{BO}}^{\mathrm{PP},s}(\mathbf{R}^s)$ for fixed nuclear coordinates.
We make the assumption that the electronic degrees of freedom are in
the ground state as discussed in Sec.~\ref{sec:generalconsiderationstateprep}. Let $E^{s}(\mathbf{R}^s)$ be an approximation to the ground state PES, i.e., the ground state energy  of $H_{\textrm{BO}}^{\mathrm{PP},s}(\mathbf{R}^s)$ as a function of $\mathbf{R}^s$. The corresponding ground state for species $s$ and fixed nuclear coordinates $\mathbf{R}^s$ is denoted by $\kett{g^{\mathrm{el}}_s(\mathbf{R}^{s})}$. The nuclei then obey a Hamiltonian describing particles experiencing a potential given by the PES, 
\begin{equation}
H_{\mathrm{ion}}^{s}=-\sum_{I\in\mathcal{I}_{s}}\frac{1}{2M_{I}}\nabla_{I}^{2}+E^{s}(\mathbf{R}^s).\label{eq:BO_nuclear_hamiltonian}
\end{equation}

\subsubsection{Rigid-rotor approximation}
\label{sec:rigidrotorapproximation}

The ion Hamiltonian in Eq.~\eqref{eq:BO_nuclear_hamiltonian} is
historically the starting point of a lengthy analysis of
translational, rotational, and vibrational modes~\cite{wilson1980molecular, littlejohn1997gauge, bunker2006molecular}. Here we primarily follow the geometrical
construction of Ref.~\cite{littlejohn1997gauge} and incorporate
some aspects of Ref.~\cite{bunker2006molecular}. Conceptually,
our goal is to perform, for each species $s$, a coordinate transformation in $\mathbb{R}^{3\eta_{\mathrm{ion}}^{s}}$:
\begin{align}
\mathbf{R}^{s}\mapsto(\mathbf{R}_{\mathrm{CoM}}^{s},\mathbf{S}^{s},\underset{\mathbf{Q}^{s}}{\underbrace{Q_{1}^{s},...,Q_{3\eta_{\mathrm{ion}}^{s}-6}^{s}}})
\end{align}
such that the operator Eq.~\eqref{eq:BO_nuclear_hamiltonian} factorizes
into three approximately independent sets of terms on: 
\begin{itemize}
    \item The $3$ translational CoM DoF denoted by $\mathbf{R}^s_{\mathrm{CoM}}$,
    \item The $3$ (or $2$ for linear molecules, $0$ for periodic substrates or isolated atoms) rotational DoF around axes
$\mathbf{S}^s$ passing through the CoM,
\item The remaining $3\eta_{\mathrm{ion}}^{s}-6$
(or $3\eta_{\mathrm{ion}}^{s}-5$ for linear molecules, $3\eta_{\mathrm{ion}}^{s}-3$ for periodic substrates, $0$ for isolated atoms), which represent the remaining DoF denoted by $\mathbf{Q}^s$.
\end{itemize}
 The space of $\mathbf{Q}^s$, formally defined as the quotient of $\mathbb{R}^{3\eta_{\mathrm{ion}}^{s}}$ by the special Euclidean group for non-linear molecules,\footnote{For linear molecules, the symmetry group of rigid body rotations and translations is the subgroup of $SE(3)$ that fixes an axis of rotation.} will be called \emph{shape space}, even though it also a includes a degree of freedom measuring the overall size or scale of the system.
For example, shape space for a water molecule consists of the bond angle, the ratio of OH bond lengths, and the overall scale or size (e.g. the CoM moment of inertia). Alternatively, we may choose the bond angle and the two OH bond lengths.
While separating out the $3$ CoM translational
coordinates is straightforward, as discussed in Sec.~\ref{sec:reactiongeometry},\footnote{We can also use $3\eta_{\mathrm{ion}}^{s}-3$ Jacobi coordinates $\rho_{\alpha}$
such that the kinetic energy becomes,
\[
T_{\mathrm{ion}}=-\sum_{I\in\mathcal{I}_{s}}\frac{1}{2M_{I}}\nabla_{I}^{2}=-\frac{1}{2M_{\textrm{tot}}^{s}}\nabla_{\textrm{CoM}}^{2}-\sum_{\alpha=1}^{3\eta_{\mathrm{ion}}^{s}-3}\frac{1}{2\mu_{\alpha}}\nabla_{\alpha}^{2},
\]
where $\mu_{\alpha}$ are suitably defined reduced masses based on
the choice of $\rho_{\alpha}$. The Jacobi coordinates are theoretically
cleaner since they explicitly separate the CoM with no mass polarization term and leave the form
of the kinetic energy unchanged.
See Ref.~\cite{littlejohn1997gauge}.} fully separating the rotations and shape
degrees of freedom  (molecular vibrations) is not possible.

The ``translation-reduced configuration space'',  $\mathbb{R}^{3\eta_{\mathrm{ion}}^{s}-3}$, obtained after separation of the CoM degrees of freedom, can be
mathematically endowed with the structure of a principal fiber bundle, with the base space being shape space and the fiber space being the 
$SO(3)$ group manifold,\footnote{Recall this is $\mathbb{RP}^{3}$, the 3-sphere with antipodal
points identified.} parameterized by $3$ Euler angles that specify the rotational orientation
of the molecule.
The Hamiltonian
is a Hermitian operator defined using a section of the fiber bundle,
where a section is a suitably continuous choice of reference orientation for rotations at each shape point, i.e. a choice of a
fiber point at each base point, often called the \emph{body frame}.
This is a gauge choice which determines the form of the Hamiltonian, although the total energy is gauge-invariant.

We want to use the fiber bundle structure with a well-chosen gauge
in order to maximally decouple rotation and shape interactions. In
this spirit, the Hamiltonian can be thought of as a sum of three terms:
 a term that depends on the base point/shape and its change, measuring the vibrational energy of the molecule;
a term that depends on the change in the displacement along the fiber from the reference section, measuring the energy in overall rotation; and a coupling term between the base and fiber, measuring the
the rotational-vibrational interaction energy. We want to choose
a gauge/section in which the latter term is small relative to the
former terms, particularly near the equilibrium configuration of the molecule. A historically successful choice is the Eckart gauge
(or Eckart frame), which is a choice of section given by $3$ linear
conditions,
\begin{equation}
F_{\mathrm{Eck}}(\mathbf{R}^s)=\sum_{I\in \mathcal{I}_s} \mathbf{R}_{I}\times \mathbf{R}_{I}^{s,0}=0, \label{eq:Eckart_conditions}
\end{equation}
where $\mathbf{R}_{I}^{s,0}$ is the value of the section at the base point corresponding to the equilibrium molecular shape~$\mathbf{Q}^{s,0}$, with the latter defined as the base point projection from a suitable minimum of the PES. This encodes the choice of orientation when the nuclear coordinates are at equilibrium.
The Eckart frame is specified everywhere once a choice of $\mathbf{R}_{I}^{s,0}$ is made. Often, this equilibrium reference orientation is taken to be
a principal axis frame for the equilibrium shape, i.e., a frame  which diagonalizes the moment of inertia tensor at equilibrium. This is the choice we make in this work. The Eckart
section is not, in general, perpendicular to the fibers, except at the equilibrium base point. This means that, in general, the rotational-vibrational coupling vanishes only at equilibrium. However, if the molecule is semi-rigid, as is typically the case, and we do not depart too much from the equilibrium shape, these couplings will be small.  
This is the scenario we envision for all the species at $t=0$.
Capitalizing on this, we further ignore the rotational-vibrational
coupling terms to obtain the rigid-rotor Hamiltonian,\footnote{See Eq.~(5.68) in Ref.~\cite{littlejohn1997gauge} for the fully coupled rotational-vibrational Hamiltonian
in Eckart gauge. We ignore all rovibrational coupling terms
involving the variable $\mathbf{K}$ and also the so-called ``Watson
term'' $V_{2}(q)$. The error in the rovibrational decoupling
is often only of the order of a few percent in practice (see Sec.~14.5 in~\cite{jensen2017introduction}). Finally, we approximate the inverse inertia tensor 
to leading order  in the displacement around equilibrium as the inverse of the equilibrium inertia tensor. As we discussed earlier,  with our choice of $\mathbf{R}^{s,0}_I$, the moment of inertia tensor is a diagonal matrix. This leads to the rigid-rotor Hamiltonian given
below, which is constructed in the Eckart frame and so is a gauge-dependent
approximation of the full rovibrational Hamiltonian. The full gauge-invariant rovibrational Hamiltonian is given in Eq.~(4.146) in Ref.~\cite{littlejohn1997gauge}.} displaying the desired separation of translations, rotations, and
vibrations:\footnote{Here and below, we show the DoF count for the general non-linear molecule case but note the DoF counts are straightforwardly different for linear molecules, periodic substrates, and isolated atoms as discussed earlier.}

\begin{equation}
H_{\mathrm{ion}}^{s}\approx-\frac{1}{2M_{\textrm{tot}}^s}\nabla_{\textrm{CoM},s}^{2}+\frac{1}{2}\sum_{\alpha=1}^{3}\mu_{\alpha}^{s,0}(J_{\alpha}^s)^{2}-\frac{1}{2}\sum_{r=1}^{3\eta_{\mathrm{ion}}^{s}-6}\frac{\d^{2}}{\d (Q_{r}^s)^{2}}+ E^{s}(\mathbf{Q}^s),
\label{eq:RR_hamiltonian}
\end{equation}
where $\mu_{\alpha}^{s,0}$ are the inverse eigenvalues of the equilibrium moment of inertia tensor
\begin{equation} 
I^{s,0}_{ij}:=\sum_{I \in \mathcal{I}_s} M_{I}(|\mathbf{R}_{I}^{s,0}|^{2}\delta_{ij}-R_{i,I}^{s,0}R_{I,j}^{s,0}), \label{eq:principle_inertia_tensor}
\end{equation}
 with $\alpha$ labeling the principal axis directions, $J_{\alpha}^s$
are the associated components of the body-fixed rigid-rotor angular
momentum operator obeying the commutation relation (note the minus sign)
\begin{align}
    [J_{\alpha}^s,J_{\beta}^s]=-i\epsilon_{\alpha\beta\gamma}J_{\gamma}^s,
\end{align}
where $\epsilon_{\alpha\beta\gamma}$ is the Levi-Civita symbol.

\subsubsection{Harmonic oscillator approximation}
\label{sec:harmonicoscillatorapproximation}

Let us now focus on the \emph{vibrational Hamiltonian} component of Eq.~\eqref{eq:RR_hamiltonian} defined on shape space
\begin{align}
   H^s_{\mathrm{vib}}:= -\frac{1}{2}\sum_{r=1}^{3\eta_{\mathrm{ion}}^{s}-6}\frac{\d^{2}}{\d (Q_{r}^s)^{2}}+ E^{s}( \mathbf{Q}^s).
\end{align}
We expand the above close to the equilibrium shape $\mathbf{Q}^{s,0}$ for species $s$, up to second order:
\begin{align}
   H^s_{\mathrm{vib}}\approx -\frac{1}{2}\sum_{r=1}^{3\eta_{\mathrm{ion}}^{s}-6}\frac{\d^{2}}{\d (Q_{r}^s)^{2}}+ E^{s}( \mathbf{Q}^{s,0}) + \frac{1}{2} \sum_{r,r'=1}^{3\eta_{\mathrm{ion}}^{s}-6} \frac{\partial^2 E^{s}}{\partial (Q_{r}^s) \partial (Q_{r'}^s)}|_{\mathbf{Q}= \mathbf{Q}^{s,0}}. 
\end{align}
By diagonalizing the Hessian $\frac{\partial^2 E^{s}}{\partial (Q_{r}^s) \partial (Q_{r'}^s)}|_{\mathbf{Q}= \mathbf{Q}^{s,0}}$ at the equilibrium shape we obtain eigenvalues $f_{s,k}$ and eigenvectors $q^{s,k}$ corresponding to the normal modes corresponding to $3\eta^s_{\textrm{ion}}-6$ decoupled oscillators. We hence obtain the rigid-rotor-harmonic-oscillator (RRHO) Hamiltonian approximating $H_{\mathrm{ion}}^{s}$ in Eq.~\eqref{eq:RR_hamiltonian}:
\begin{equation}
H_{\mathrm{ion}}^{s}\approx\underset{H_{\textrm{CoM},s}}{\underbrace{-\frac{1}{2M_{\textrm{tot}}^s}\nabla_{\textrm{CoM},s}^{2}}}\underset{H_{\mathrm{rot},s}}{\underbrace{+\frac{1}{2}\sum_{\alpha=1}^{3}\mu_{\alpha}^{s,0}(J_{\alpha}^s)^{2}}}\underset{H_{\mathrm{vib},s}}{\underbrace{ +E^{s}( \mathbf{Q}^{s,0}) +\sum_{k=1}^{3\eta_{\mathrm{ion}}^{s}-6}(-\frac{1}{2}\frac{\d^{2}}{\d (q^{s,k})^{2}}+\frac{1}{2}f_{s,k}(q^{s,k})^{2})}}.
\label{eq:RRHO_hamiltonian}
\end{equation}
The RRHO Hamiltonian in Eq.~\eqref{eq:RRHO_hamiltonian} is the desired factorized description of the ionic motion approximating Eq.~\eqref{eq:BO_nuclear_hamiltonian}. Together with the BO Hamiltonian with pseudopotentials in Eq.~\eqref{eq:BOPP_electronic_hamiltonian}, that separately describes the electronic motion, we obtain an approximate description of molecular interactions at $t=0$ decoupled  into four classes: electronics, translations, rotations, and vibrations.

\subsection{Form of the initial state}

\label{sec:form_of_initial_state}
The decoupled structure of the RRHO Hamiltonian admits product eigenstates over motional classes and as such we construct an initial product state, with justifications for the choice of each factor presented in Sec.~\ref{sec:generalconsiderationstateprep}. Recall again that although the RRHO Hamiltonian is constructed in the Eckart frame and motivates the initial state ansatz, we do not actually use the RRHO Hamiltonian for evolution.

The full initial state is taken to be,
\begin{equation}
\rho(t=0)=\otimes_{s=0}^{S}\rho_{s}(t=0),
\label{eq:species_partition2}
\end{equation}
where
\begin{align}
\rho_{s}(t=0) & =\ketbra{\psi^{\textrm{CoM}}_{\bar{\mathbf{R}}^s,\bar{\mathbf{P}}^s,\sigma_s}}{\psi^{\textrm{CoM}}_{\bar{\mathbf{R}}^s,\bar{\mathbf{P}}^s,\sigma_s}}\otimes \ketbra{g^{\mathrm{el}}_s(\mathbf{R}^{s,0})}{g^{\mathrm{el}}_{s}(\mathbf{R}^{s,0})}\otimes\ketbra{g^{\mathrm{rot}}_s}{g^{\mathrm{rot}}_s}\otimes \rho_{s}^{\mathrm{vib}},\label{eq:ansatz}
\end{align}
is the initial state of chemical species $s$ with the following parts, 
\begin{itemize}
    \item $\kett{\psi^{\textrm{CoM}}_{\bar{\mathbf{R}}^s,\bar{\mathbf{P}}^s,\sigma_s}}$ is a minimal uncertainty, isotropic Gaussian wavepacket over the center-of-mass degrees of freedom, with mean position, mean momentum, and spatial standard deviation $\bar{\mathbf{R}}^s,\bar{\mathbf{P}}^s,\sigma_s$, respectively, specified according to the reaction geometry, as described in Sec.~\ref{sec:reactiongeometry}.
    \item  $\kett{g^{\mathrm{el}}_s(\mathbf{R}^{s,0})}$ is the ground
state, at fixed nuclear equilibrium
positions~$\mathbf{R}^{s,0}$, of the BO electronic Hamiltonian with pseudopotentials  in Eq.~\eqref{eq:BOPP_electronic_hamiltonian}.
\item $\kett{g^{\mathrm{rot}}_s}$ is the ground state of zero total body-fixed angular momentum,
i.e., a uniform superposition over all rotational configurations parameterized by,
for example, by the three Euler angles. 
\item $\rho_{s}^{\mathrm{vib}}$ is a truncation of the thermal state $e^{- \beta H_{\textrm{vib},s}}/\textrm{Tr}(e^{- \beta H_{\textrm{vib},s}})$:
\begin{align}
\rho_{s}^{\mathrm{vib}}= \bigotimes_{k=1}^{3\eta_{\mathrm{ion}}^{s}-6}\left(\sum_{l=0}^{l^{\max}_{s,k}}\mathrm{Pr}_{s,k}(l)\ketbra{\phi_{l}^{s,k}}{\phi_{l}^{s,k}}\right),
\end{align}
where $\beta=\frac{1}{k_{B}T}$ is the inverse of the desired (e.g. reaction) temperature and
\begin{align}
    Z^{\mathrm{vib}}_s= \Pi_{k=1}^{3\eta_{\mathrm{vib}}^{s}-6}  Z^{\mathrm{vib}}_{s,k}, \quad
    Z^{\mathrm{vib}}_{s,k} = \sum_{l=0}^{l^{\max}_{s,k}} e^{-\beta\omega_{s,k}(l+\frac{1}{2})}, \quad \mathrm{Pr}_{s,k}(l)=\frac{1}{Z^{\mathrm{vib}}_{s,k}}e^{-\beta\omega_{s,k}(l+\frac{1}{2})},
\end{align}
where $l$ labels the excitations, $k$ labels the modes,
$\omega_{s,k}=\sqrt{f_{s,k}}$ are the angular frequencies, $l^{\max}_{s,k}$ is the excitation
number cutoff for mode $k$ of species $s$, and $\kett{\phi_{l}^{s,k}}=\sum_{q^{s,k}}\qb{q^{s,k}}{\phi_{l}^{s,k}}\kett{q^{s,k}}$
is the $l$-th quantum harmonic oscillator wavefunction with amplitudes
along the $k$-th shape coordinate of species $s$,

\begin{equation}
\qb{q^{s,k}}{\phi_{l}^{s,k}}=\frac{1}{\s{N}^{s,k,l}_\mathrm{vib}}e^{-\frac{\omega_{s,k}}{2}{q^{s,k}}^{2}}\mathrm{H}_{l}\left(\sqrt{{\omega_{s,k}}}q^{s,k}\right),\label{eq:QHO_eigenstate}
\end{equation}

with Hermite polynomials $\mathrm{H}_{l}(x)$. 
\end{itemize}
Note that, for the purposes of initial state preparation, we can neglect the species-dependent energy shift $E^{s}( \mathbf{Q}^{s,0})$ since the product state over all species aggregates a global phase (which anyway also vanishes at $t=0$). Often, it may be of interest to sample over a range of initial configurations and compute average results, for which we discuss how standard quantum algorithmic techniques yield quadratically improved complexity scaling in the sampling error in Sec.~\ref{sec:coherentsampling}.

\subsection{Algorithm for initial state preparation}
\label{sec:algorithm for initial state preparation}

\subsubsection{Quantum algorithm for initial state preparation}
\label{sec:qalgo_initial_state_prep}
We present a quantum algorithm to prepare the initial state ansatz
in Eq.~\eqref{eq:ansatz}. The classical precomputations required for the input are reviewed in Sec.~\ref{sec:classicalprecomp}. For each of the $S$ molecular species, the pseudoion DoFs are described in their respective
natural product basis 
\begin{align}
\kett{\mathbf{R}^s_{\mathrm{CoM}}}\otimes\kett{\mathbf{S}^s}\otimes\kett{\mathbf{q}^s}, \quad s=0,\dots, S
\end{align}
where $\kett{\mathbf{R}^s_{\mathrm{CoM}}}$ are eigenstates of the CoM position operator, $\kett{\mathbf{S}^s}$  is a basis of $L^2(\mathbb{R} \mathbb{P}^3)$ (square integrable functions with, for example, coordinates being Euler angles), and $\kett{\mathbf{q}^s}=\kett{q^{s,1},...,q^{s,3\eta_{\mathrm{ion}}^{s}-6}}$
is the basis of
vibrational normal modes for each species~$s$ (for linear molecules there are $2$ rotations, and a shape basis $\kett{\mathbf{q}^s}=\kett{q^{s,1},...,q^{s,3\eta_{\mathrm{ion}}^{s}-5}}$).  All of these basis components are suitably discretized. The discretization used in the initial state preparation is not the same as the plane wave discretization used in the simulation, and so requires a discussion of ``grid-matching'' in Sec.~\ref{sec:classicalprecomp}.

\subsubsection*{Translations}

The translational state is the launched Gaussian wavepacket,
 
\begin{align}
\kett{\psi^{\textrm{CoM}}_{\bar{\mathbf{R}}^s,\bar{\mathbf{P}}^s,\sigma_s}}& =\sum_{\mathbf{R}_{\mathrm{CoM}}^s}\underset{f_{\mathrm{trans}}(\mathbf{R}_{\mathrm{CoM}}^s)}{\underbrace{\frac{1}{(2\pi\sigma_s^{2})^{3/4}}e^{i\bar{\mathbf{P}}^{s}\cdot\mathbf{R}_{\mathrm{CoM}}^s}e^{-\frac{1}{4\sigma^{2}_s}\sum_{i=1}^{3}(\mathbf{R}_{\mathrm{CoM}}^s-\bar{\mathbf{R}}^s)^{2}}}}\kett{\mathbf{R}_{\mathrm{CoM}}^s},\label{eq:translation_state}
\end{align}
where, as before, $\bar{\mathbf{R}}^{s},\bar{\mathbf{P}}^{s},\sigma_s$
are the reaction-geometry-specified mean position, mean momentum,
and spatial uncertainty of the wavepacket.  We have taken the spatial
uncertainty $\sigma_s$ to be equal in all directions. Each of the three components of $\mathbf{R}_{\mathrm{CoM}}^s$ takes values on a uniform cubic grid with $n_{\mathrm{trans}}$ qubit discretization per dimension (corresponding to $N_{\mathrm{trans}} = 2^{n_{\mathrm{trans}}}$ points per direction). This state can
be created using $O(n_{\mathrm{trans}})$ gates by known methods such as quantum rejection
sampling as discussed in Ref.~\cite{lemieux2024quantum}. In practice, this cost will be dwarfed by the rest of the algorithm and is hence neglected. 

\subsubsection*{Rotations}

The ground state of the rigid rotor Hamiltonian is the uniform superposition
over all rotational configurations (Ref.~\cite{bunker2006molecular}, Problem 11-2) and hence it is rotationally invariant. We use an
extrinsic Euler angle representation of $SO(3)$ rotations with angles
$\alpha,\gamma\in[-\pi,\pi),\beta\in[0,\pi]$ such that a general
rotation can be written as $S(\alpha,\beta,\gamma)=S_{\hat{a}}(\alpha)S_{\hat{b}}(\beta)S_{\hat{a}}(\gamma)$
where $S_{\hat{n}}(\theta)$ is the 3d matrix for a rotation of angle
$\theta$ about axis $\hat{n}$, and $\hat{a},\hat{b}$ are any two
independent axis. Classically, uniform sampling over $SO(3)$ can
be accomplished by generating uniform random variables $u_{1},u_{2},u_{3}\in[-1,1]$
and creating $(\alpha,\beta, \gamma)=(\pi u_{1},\cos^{-1}(u_{2}), \pi u_{3})$. We mimic this spirit for state preparation. Let $\mathrm{USP}(a,b;n)$ be a unitary that creates from the $n$ qubit all zeroes state a uniform superposition over $2^{n}$ states labeling a uniform discretization of the interval $[a,b)$:
\begin{align}
    \mathrm{USP}(a,b;n) \kett{0} = \frac{1}{\sqrt{2^n}} \sum_{j=0}^{2^n-1} \left|\frac{j(b-a)}{2^n}\right\rangle.
\end{align}
For a $n_{\mathrm{rot}}$ qubit resolution in each Euler angle we obtain,
\begin{align}
\kett{g^{\mathrm{rot}}_s} & =\mathrm{USP}(-\pi,\pi;n_{\mathrm{rot}})\kett 0\otimes\mathrm{USP}(-1,1;n_{\mathrm{rot}})\kett 0\otimes\mathrm{USP}(-\pi,\pi;n_{\mathrm{rot}})\kett 0\nonumber \\
 & = \sum_{\cos\beta\in u_{n_{\mathrm{rot}}} [-1,1]}\sum_{\alpha,\gamma\in u_{n_{\mathrm{rot}}}[-\pi,\pi)}\underset{f_{\mathrm{rot}}(\alpha,\cos\beta,\gamma)}{\underbrace{\frac{1}{\s N_{\mathrm{rot}}}}}\kett{\alpha,\cos\beta,\gamma}, \label{eq:rotation_state}
\end{align}
where $\s N_{\mathrm{rot}} = 2^{3n_{\mathrm{rot}}/2}$, and $u_{n}[a,b)$ denotes the set of $2^n$ points obtained by uniform discretization of the corresponding interval $[a,b)$. For linear molecules,
there are only $2$ unique angles and so we just omit preparation
of the $\gamma$ angle above. This state preparation only involves $3n_{\mathrm{rot}}$ Hadamard gates, whose cost is completely negligible.

\subsubsection*{Vibrations}

\paragraph*{Truncated thermal state.} 
The truncated thermal state over $3\eta_{\mathrm{ion}}^{s}-6$ ($3\eta_{\mathrm{ion}}^{s}-5$ for linear molecules) independent
quantum harmonic oscillator (QHO) states for each species $s$ is,
\begin{align}
\rho_{s}^{\mathrm{vib}}= \bigotimes_{k=1}^{3\eta_{\mathrm{ion}}^{s}-6}\left(\sum_{l=0}^{l^{\max}_{s,k}}\mathrm{Pr}_{s,k}(l)\ketbra{\phi_{l}^{s,k}}{\phi_{l}^{s,k}}\right).
\end{align}
We will prepare a purification
of $\rho_{s}^{\mathrm{vib}}$, specifically
\begin{align}
\kett{\psi_{s}^{\mathrm{vib}}} & =\bigotimes_{k=1}^{3\eta_{\mathrm{ion}}^{s}-6}\left(\sum_{l_k=0}^{l^{\max}_{s,k}}\sqrt{\mathrm{Pr}_{s,k}(l_k)}\kett{l_{k}}\kett{\phi_{l_k}^{s,k}}\right)\nonumber \\
 & =\sum_{q^{s,1},...,q^{s,3\eta_{\mathrm{ion}}^{s}-6}}\underset{f_{\mathrm{vib}}(\textbf{q}^{s},\textbf{l}^{s})}{\sum_{l_{1}=0}^{l^{\max}_{s,1}}...\sum_{l_{3\eta_{\mathrm{ion}}-6}=0}^{l^{\max}_{s,3\eta_{\mathrm{ion}}^{s}-6}}\underbrace{\p{\sqrt{\mathrm{Pr}_{s,1}(l_{1})...\mathrm{Pr}_{s,3\eta_{\mathrm{ion}}^{s}-6}(l_{3\eta_{\mathrm{ion}}^{s}-6})}\qb{q^{s,1}}{\phi^{s,1}_{l_{1}}}...\qb{q^{s,3\eta_{\mathrm{ion}}^{s}-6}}{\phi^{s,3\eta_{\mathrm{ion}}^{s}-6}_{l_{3\eta_{\mathrm{ion}}^{s}-6}}}}}}\nonumber\\ &\kett{l_{1},...,l_{3\eta_{\mathrm{ion}}^{s}-6}}\kett{q^{s,1},...,q^{s,3\eta_{\mathrm{ion}}^{s}-6}}.\label{eq:purified_vibration_state}
\end{align}
where $\mathbf{l}^s=(l_1,...,l_{3\eta_{\mathrm{ion}}^{s}-6})$ label computational basis states of an ancilla and $\mathbf{q}^s=(q^{s,1},...,q^{s,3\eta_{\mathrm{ion}}^{s}-6})$ as before.
A physically motivated cutoff is to choose the smallest $l$ such
that $\textrm{Pr}_{k}(l)\leq\epsilon$ for all $l\geq l^{\max}_{s,k}$,
for some fixed $\epsilon$, although other choices are possible.\footnote{Note that since $\mathrm{Pr}_{s,k}(l)$ is temperature dependent, so
too will be the choice of $l^{\max}_{s,k}$, since higher temperatures
mean higher cutoffs to achieve the same $\epsilon$. An alternative choice is to fix $l$ so that we have a trace norm error $\epsilon$ to the exact thermal state.} We obtain,
\begin{equation}
l^{\max}_{s,k}=\left\lceil \frac{1}{\beta\omega_{k}^s}\log\p{\frac{1}{\epsilon}(1-e^{-\beta\omega_{k}^s})}\right\rceil =O(\log(1/\epsilon))\label{eq:QHO_cutoff}
\end{equation}
For each $k$, introduce $\lceil\log_{2}l^{\max}_{s,k}\rceil=O(\log\log(1/\epsilon))$
ancillary qubits for the purification register.
As desired, performing the trace over the first register $\kett{l_{1},...,l_{3\eta_{\mathrm{ion}}^{s}-6}}$
yields $\rho_{s}^{\mathrm{vib}}$. We create this purification in the following manner.

\paragraph*{Preparation of QHO eigenstates.}
With slight abuse of notation, we define $\phi_{l}^{s,k}(q^{s,k})$
as a vector encoding $\qb{q^{s,k}}{\phi_{l}^{s,k}}$ discretized
using $n_{\mathrm{vib}}$ qubits, where we keep the same number of
discretization points for all modes, although other choices are possible. On the $k$-th mode,

\begin{equation}
\kett{\phi_{l}^{s,k}}=\frac{1}{\s N_{\mathrm{vib}}^{s,k,l}}\sum_{b=1}^{2^{n_{\mathrm{vib}}}}\phi_{l}^{s,k}(q^{s,k}(b))\kett{q^{s,k}(b)}\label{eq:QHO_discretized_eigenstate}
\end{equation}
where $q^{s,k}(1)$ and $q^{s,k}(2^{n_{\mathrm{vib}}})$ are determined by the relevant range in shape space, $q^{s,k}(b)$ linearly interpolates between the two with $N_{\textrm{vib}} = 2^{n_{\mathrm{vib}}}$ discretization points, and $\s N_{\mathrm{vib}}^{s,k,l}$
is a normalization that depends on the excitation number $l$ and
on the number of discretization points.

Let $U_{l}^{s,k}$ be a unitary that prepares the state in Eq.~\eqref{eq:QHO_discretized_eigenstate} (the $l=0,\dots, l^{\max}_{s,k}$-th discretized QHO eigenstate of mode $k$, species $s$)
from the all zero state $\kett{0^{n_{\mathrm{vib}}}}$, given the
vibrational mode frequencies,
\begin{align}
U_{l}^{s,k}\kett{0^{n_{\mathrm{vib}}}}=\kett{\phi_{l}^{s,k}}
\end{align}
The unitary $U_{l}^{s,k}$ may be, for example, efficiently realized by
quantum rejection sampling discussed in Ref.~\cite{lemieux2024quantum} with the Type I reference states detailed in App.~\ref{app:reference_states_QRS}, with complexity scaling with the logarithm of the number discretization points, $O(n_{\mathrm{vib}})$.

\paragraph*{Purification.}
Consider the unitary $V^{s,k}$ for each of the $3\eta_{\mathrm{ion}}^{s}-6$
modes, 
\begin{equation}
V^{s,k}\kett{0^{\lceil\log_{2}l^{\max}_{s,k}\rceil}}=\sum_{l=0}^{l^{\max}_{s,k}}\sqrt{\textrm{Pr}_{k}(l)}\kett l.
\end{equation}
A generic state preparation algorithm can achieve this unitary with a number of gates scaling linearly with the Hilbert space dimension in the worst case, hence as
$O(\log(1/\epsilon))$ due to Eq.~\eqref{eq:QHO_cutoff}~\cite{grover2002creating, mottonen2004transformation}. We apply the following protocol to each
of the $k=1,...,3\eta_{\mathrm{ion}}^{s}-6$ modes, to obtain the state in Eq.~\eqref{eq:purified_vibration_state}:
\begin{itemize}
\item Initialize all qubits in zero: $\kett{0^{\lceil\log_{2}l^{\max}_{s,k}\rceil}}\otimes\kett{0^{n_{\mathrm{vib}}}}$.
\item Perform the unitary $V^{s,k}$ on the $\kett{0^{\lceil\log_{2}l^{\max}_{s,k}\rceil}}$
qubits. 
\item Perform the controlled unitary $U_{C}^{s,k}=\sum_{l=0}^{l^{\max}_{s,k}}\kett l\brat l\otimes U_{l}^{s,k}$.
\end{itemize}
 The total cost of this procedure
across all ions scales as $O(\eta_{\mathrm{ion}} n_{\mathrm{vib}}\log(1/\epsilon)))$ in the worst case, and in practice it can be neglected compared to the rest of the algorithm.

Note that the protocol
we have described can be generalized to other non-thermal mixtures
as desired. For example, we might create a mixture of modes in an
specific energy band that is relevant to a reaction mechanism of interest,
or rather understand a reaction mechanism itself by creating an energy
band of vibrational excitations and moving the energy band higher/lower
while observing the resulting products.

\paragraph*{Electrons.}
We prepare the electrons in an approximate ground state of the
BO Hamiltonian (Eq.~\eqref{eq:BOPP_electronic_hamiltonian}) as computed
by classical computational methods such as DFT. Explicitly, for each particle on a grid $\mathbf{r}$ that is the inverse Fourier transform of the plane wave basis, we wish
to create,
\[
\kett{g^{\mathrm{el}}_s(\mathbf{R}^{s,0})}=\s A\sum_{i_{1},...,i_{\eta_{\mathrm{val}}^{s}}}\sum_{\mathbf{r}_{1},...,\mathbf{r}_{\eta_{\mathrm{val}}^{s}}}c_{i_{1},...i_{\eta_{\mathrm{val}}^{s}}}(\mathbf{R}^{s,0})\phi_{i_{1}}(\mathbf{r}_{1};\mathbf{R}^{s,0})...\phi_{i_{\eta_{\mathrm{val}}^{s}}}(\mathbf{r}_{\eta_{\mathrm{val}}^{s}};\mathbf{R}^{s,0})\kett{\mathbf{r}_{1},...,\mathbf{r}_{\eta_{\mathrm{val}}^{s}}}
\]
where $\{\phi_{i}(\mathbf{r};\mathbf{R}^{s,0})\}$ are a suitably chosen
set of normalized spin orbitals in which the classical calculation
is performed, and $\s A$ denotes the anti-symmetrization operator.
Since classical methods often give a single (single-reference/single-determinant,
e.g. the Hartree-Fock state) or at most a superposition (multi-reference/multi-determinant)
of a few product states,\footnote{Increasing the amount of determinants in the state comes with significant
classical computational overhead. Often a single determinant is sufficient
for practical situations but, on occasion, one might want to refine
this approximation with additional terms. Note again that very high
accuracy in the initial state should not be needed for our purposes,
since it is only an approximate initial state from which significant
entanglement can arise through full Hamiltonian dynamics.} most of the coefficients $c_{i_{1},...i_{\eta_{\mathrm{val}}^{s}}}(\mathbf{R}^{s,0})$ vanish.

For the typical case of a single product state given by single set
$\alpha=\{i_{1},...,i_{\eta_{\mathrm{val}}^{s}}\}$, we load the $\eta_{\mathrm{val}}^{s}$
single-particle states in parallel and use the efficient anti-symmetrization
circuit of Ref.~\cite{berry2018improved} for $\s A$, to prepare a single-reference state,

\[
\kett{\phi_{\alpha}^s}=\s A\bigotimes_{j=1}^{\eta_{\mathrm{val}}^{s}}\sum_{\mathbf{r}_{j}}\phi_{i_{j}}(\mathbf{r}_{j};\mathbf{R}^{s,0})\kett{\mathbf{r}_{j}}
\]
Recalling that
$|G|$ is the size of the basis, the loading procedure has a worst case cost $O(\eta_{\mathrm{val}}^{s}|G|)$, and the
anti-symmetrization circuit costs $O(\eta_{\mathrm{val}}^{s}\log^{c}\eta_{\mathrm{val}}^{s}\log|G|)$ (with $c\geq1$). More refined techniques than simple loading could be used to lower the cost to logarithmic in the number of plane waves and polynomial in the number of primitive Gaussian functions approximating the Slater determinant~\cite{huggins2024efficient}, but in practice including these is not currently a priority in our setting. For example, with $|G|\sim 10^6$, $\eta_{\textrm{val}}\sim100$, $S\sim10$ one can expect an overall gate cost $\sim 10^{9}$, which as we shall see is a sub-leading contribution to the overall cost of the algorithm.

Multi-reference states of $n_{\mathrm{det}}$ products can be prepared
using an additional LCU procedure. Defining $U_{\alpha}^s\kett 0=\kett{\phi_{\alpha}^s}$
for the sets of determinants $\alpha_{1},...,\alpha_{n_{\mathrm{det}}}$,
$V^s\kett 0=\sum_{\alpha=1}^{n_{\mathrm{det}}}\sqrt{c_{\alpha}^s(\mathbf{R}^{s,0})}\kett{\alpha}$
for their associated coefficients, and the select operator $S=\sum_{\alpha}\kett{\alpha}\brat{\alpha}\otimes U_{\alpha}^s$,
we obtain,
\[
(V^{s})^\dag SV^s\kett 0\kett 0=\frac{1}{\lambda_{\mathrm{det}}^s}\kett 0\kett{g^{\mathrm{el}}_s(\mathbf{R}^{s,0})}+\kett{0^{\perp}}
\]
where $\kett{0^{\perp}}$ is an unnormalized state orthogonal to the projection onto the zero state of the first qubit. The rescaling factor is $\lambda_{\mathrm{det}}^s=\sum_{\alpha}|c_{\alpha}^s(\mathbf{R}^{s,0})|$
and the state can be amplified to near unity. Since there are usually only a few
non-zeros coefficients, this rescaling is fairly benign
in practical settings ($\lambda_{\mathrm{det}}^s \leq \sqrt{n_{\mathrm{det}}}$ in the worst case) and the number of amplitude amplification rounds, which scales as $O(\lambda_{\mathrm{det}})$, is limited. A final quantum Fourier transform (QFT) can be used to move to the plane-wave basis.

\subsection*{Pseudoion coordinate transformation}
Defining the full set of pseuodion normal coordinates $\mathbf{\Xi}^s=(\mathbf{R}_{\mathrm{CoM}}^s,\mathbf{S}^s,\mathbf{q}^s)$ for a species $s$ and the function
\begin{align}
f(\mathbf{\Xi}^s, \textbf{l}^{s})= f_{\mathrm{trans}}(\mathbf{R}_{\mathrm{CoM}}^s) f_{\mathrm{rot}}(\alpha^s,\cos\beta^s,\gamma^s)f_{\mathrm{vib}}(\textbf{q}^{s},\textbf{l}^{s}),
\end{align}
for ease, we have so far constructed the initial state (suppressing
the species superscript),
\[
\kett{\Psi_{0}^{\mathrm{PI}}}=\sum_{\mathbf{\Xi},\textbf{l}} f(\mathbf{\Xi}, \mathbf{l})\kett{\mathbf{\Xi}} \kett{\textbf{l}}.
\]
We must transform from the normal mode basis $\kett{\mathbf{\Xi}}$
to the plane-wave basis $\kett{\mathbf{P}}=\kett{\mathbf{P}_{1},...,\mathbf{P}_{\eta_{\mathrm{ion}}^{s}}}$.
Denoting the inverse Fourier transform of the plane wave basis (for pseudoions) as $\kett{\mathbf{R}}=\kett{\mathbf{R}_{1},...,\mathbf{R}_{\eta_{\mathrm{ion}}^{s}}}$, we use a sparse oracle $O_{\mathrm{coord}}$ that provides an injective map $\mathbf{\Xi}\rightarrow\mathbf{R}$ via $O_{\mathrm{coord}}\kett{\mathbf{\Xi}}=\kett{\mathbf{R}}$,
constructed from a classically-computed table that utilizes a grid matching description (see Sec.~\ref{sec:classicalprecomp}). This requires data loading with a cost  $O(\eta_{\mathrm{ion}}^{s}|G|)$ in the worst case. By an argument similar to the one made for electrons, this cost is a subleading contribution in the overall cost of the algorithm.  Performing
the coordinate change, we obtain
\[
(O_{\mathrm{coord}} \otimes I) \sum_{\mathbf{\Xi}, \mathbf{l}}f(\mathbf{\Xi}, \mathbf{l})\kett{\mathbf{\Xi}}\kett{\textbf{l}}=\sum_{\mathbf{\Xi},\mathbf{l}}f(\mathbf{\Xi}(\mathbf{R}), \mathbf{l})\kett{\mathbf{R}(\mathbf{\Xi})}\kett{\textbf{l}}\equiv\sum_{\mathbf{R}, \mathbf{l}}\tilde{f}(\mathbf{R}, \mathbf{l})\kett{\mathbf{R}}\kett{\textbf{l}}
\]
where $\tilde{f}(\mathbf{R}, \mathbf{l}):= f(\mathbf{\Xi}(\mathbf{R}),\mathbf{l})$. We then perform a Quantum Fourier Transform
to move to the plane-wave basis~$\kett{\mathbf{P}}$,
\[
\kett{\Psi_{0}^{\mathrm{PI}}}=\sum_{\mathbf{P}, \mathbf{l}}f_0(\mathbf{P}, \mathbf{l})\kett{\mathbf{P}}\kett{\mathbf{l}},
\]
where $f_0$ and $\tilde{f}$ are related by Fourier transform. 
This completes the construction of
the approximate initial state in the desired basis as needed for time-evolution.

\subsubsection{Classical precomputations}
\label{sec:classicalprecomp}
We describe the classical pre-processing that must be performed in
order to provide the requisite information to construct the initial
state. 

\subsubsection*{Potential Energy Surface analysis and shape coordinates}

We first find an appropriate minimum of the ground-state potential
energy surface (PES) for fixed $s$, $E^{s}(\mathbf{R}^{s})$, where $\mathbf{R}^{s}=\{\mathbf{R}_{I}\}_{I\in\s I_{s}}$ represents configurations of the chemical species described under the BO approximation
Eq.~\eqref{eq:BOPP_electronic_hamiltonian}. This can be accomplished, for example,
by constructing the appropriate electronic energy gradients by repeatedly
computing approximate electronic energies for fixed pseudoion positions
with a conventional method such as DFT, and using gradient descent
or other standard optimization techniques to find a minimum $\mathbf{R}^{s,0}=\{\mathbf{R}_{I}^{s,0}\}_{I\in\s I_{s}}$.
Expanding the PES to second order around the equilibrium configuration (the minimum),

\begin{align*}
E^{s}(\mathbf{R}^{s}) & \approx E^{s}(\mathbf{R}^{s,0})+\frac{1}{2}\sum_{j,j'=1}^{3\eta_{\mathrm{ion}}^{s}}(R_{j}^{s}-R_{j}^{s,0})\underset{\mathrm{F}_{jj'}^s}{\underbrace{\frac{\d^{2}E^{s}(\mathbf{R}^{s})}{\d R_{j}^{s}\d R_{j'}^{s}}}}|_{\mathbf{R}^{s}=\mathbf{R}^{s,0}}(R_{j'}^{s}-R_{j'}^{s,0}),
\end{align*}
where $j=(\alpha,I)$ is a composite index for the $\alpha$-th Cartesian
component of $I$-th pseudoion position, such that $\mathbf{R}^{s}$
is a $3\eta_{\mathrm{ion}}^{s}$ element vector with elements $R_{j}^{s}$
(and similarly $\mathbf{R}^{s,0}$ with elements $R_{j}^{s,0}$). The second term
contains the $3\eta_{\mathrm{ion}}^{s}\times3\eta_{\mathrm{ion}}^{s}$
Hessian matrix, aka force matrix, denoted by $\mathbf{F}^s$ with elements $\mathrm{F}_{jj'}^s$.

Let
$\Delta\mathbf{R}^{s}=\mathbf{R}^{s}-\mathbf{R}^{s,0}$ denote a vector of displacements.
Reexpressing the nuclear Hamiltonian of Eq.~\eqref{eq:BO_nuclear_hamiltonian}
in terms of mass-weighted coordinate displacements $\Delta\tilde{\mathbf{R}}^{s}=\{\sqrt{M_{I}}\Delta\mathbf{R}_{I}^{s}\}_{I\in\s I_{s}}$
with elements $\Delta \tilde{R}_{j}^{s}$ and using the PES expansion,

\[
H_{\mathrm{ion}}^{\mathrm{PI},s}\approx E^{s}(\mathbf{R}^{s,0})-\sum_{j=1}^{3\eta_{\mathrm{ion}}^{s}}\frac{1}{2}\frac{\d^{2}}{\d(\Delta \tilde{R}_{j})^{2}}+\frac{1}{2}\sum_{j,j'=1}^{3\eta_{\mathrm{ion}}^{s}}\Delta \tilde{R}_{j}^{s}\mathrm{\tilde{F}}_{jj'}^s{}_{|\Delta \tilde{\mathbf{R}}^s=\textbf{0}} \Delta \tilde{R}_{j'}^{s}
\]
where $\tilde{\mathrm{F}}_{jj'}^s=\frac{1}{\sqrt{M_{j}M_{j'}}}\mathrm{F}_{jj'}^s$
are the elements of the mass-weighted Hessian $\tilde{\mathbf{F}}^s$, which
is real and symmetric since the PES is real.
In this standard form, diagonalization of $\tilde{\mathbf{F}}^s$ can be realized by an orthogonal transformation that decouples the quadratic potential term and leaves the form of
the kinetic term unchanged, thereby fully decoupling the harmonic
oscillators.

Note, however, that the PES has the global Euclidean symmetry $SE(3)$,
\[
E^{s}(\mathbf{S}\mathbf{R}_{1}^{s}+\mathbf{R},...,\mathbf{S}\mathbf{R}_{\eta_{\mathrm{ion}}^{s}}^{s}+\mathbf{R})=E^{s}(\mathbf{R}_{1}^{s},...,\mathbf{R}_{\eta_{\mathrm{ion}}^{s}}^{s})
\]
where $\mathbf{R},\mathbf{S}$ are an arbitrary translation and rotation.
As a result, the force matrix $\mathbf{F}^{s}$ has exactly 6 (or
5 for linear molecules) zero eigenvalues that correspond to the global
Euclidean symmetry generators, since the eigenvectors corresponding
to infinitesimal global translations and rotations of the pseudoions
incur no energy cost.

For numerical accuracy, it is important to remove the
global Euclidean symmetry before computing the remaining eigenmodes
associated to the internal degrees of freedom. This can be achieved
with the following procedure. First, explicitly define the global translation
and rotation vectors in mass-weighted coordinates as
\begin{align*}
\tilde{t_{x}^{s}} & =\{\sqrt{M_{I}}(1,0,0)_{I}^{s}\}_{I\in\s I_{s}},\;\tilde{t_{y}^{s}}=\{\sqrt{M_{I}}(0,1,0)_{I}^{s}\}_{I\in\s I_{s}},\;\tilde{t_{z}^{s}}=\{\sqrt{M_{I}}(0,0,1)_{I}^{s}\}_{I\in\s I_{s}},\\
\tilde{r_{x}^{s}} & =\{\mathbf{\hat{x}}\times\Delta\tilde{\mathbf{R}}_{I}^{s}\}_{I\in\s I_{s}},\;
\tilde{r_{y}^{s}}=\{\mathbf{\hat{y}}\times\Delta\tilde{\mathbf{R}}_{I}^{s}\}_{I\in\s I_{s}},\;
\tilde{r_{z}^{s}}=\{\mathbf{\hat{z}}\times\Delta\tilde{\mathbf{R}}_{I}^{s}\}_{I\in\s I_{s}},
\end{align*}
which respectively correspond to the action of the generators of translations and rotations on a generic displacement vector $\Delta \tilde{\mathbf{R}}_I^s$.
Perform a QR decomposition on the $3\eta_{\mathrm{ion}}^{s}\times3\eta_{\mathrm{ion}}^{s}$
matrix 
$$
[\tilde{t_{x}^{s}}^{T},\tilde{t_{y}^{s}}^{T},\tilde{t_{z}^{s}}^{T},\tilde{r_{x}^{s}}^{T},\tilde{r_{y}^{s}}^{T},\tilde{r_{z}^{s}}^{T}|*]
$$
where $*$ indicates $3\eta_{\mathrm{ion}}^{s}\times3\eta_{\mathrm{ion}}^{s}-6$
random real entries, and define $\tilde{Q}$ as the last $3\eta_{\mathrm{ion}}^{s}-6$
(or $3\eta_{\mathrm{ion}}^{s}-5$) columns of the unitary part of
the decomposition which, by construction, spans the subspace orthogonal
to the Euclidean transformations. Construct the $3\eta_{\mathrm{ion}}^{s}-6\times3\eta_{\mathrm{ion}}^{s}-6$
(or $3\eta_{\mathrm{ion}}^{s}-5\times3\eta_{\mathrm{ion}}^{s}-5$)
matrix $\bar{\tilde{\mathrm{F}}}^{s}=\tilde{Q}^{T}\tilde{\mathrm{F}}^{s}\tilde{Q}$
in which the global symmetry has been removed. Diagonalization of
$\bar{\tilde{\mathrm{F}}}^{s}$ gives eigenvalues and eigenstates
$f_{s,k},\bar{e}{}_{\lambda,k}^{s}$ from which we obtain $e_{j,k}^{s}=\sum_{\lambda=1}^{3\eta_\mathrm{ion}^s-6}\tilde{Q}_{j\lambda}\bar{e}_{\lambda,k}$
as the components of the ``polarization'' eigenvectors $\hat{\mathbf{e}}_{k}^{s}$
that obey the orthonormality condition $\sum_{j=1}^{3\eta_{\mathrm{ion}}^{s}}e_{j,k}^{s}e_{j,k'}^{s}=\delta_{kk'}$
for $k,k'=1,...,3\eta_{\mathrm{ion}}^{s}-6$ (or $3\eta_{\mathrm{ion}}^{s}-5$). These are the normal modes in mass-weighted Cartesian coordinates.

Defining new variables $q^{s,k}=\tilde{\Delta\mathbf{R}^{s}}\cdot\hat{\mathbf{e}}_{k}^{s}=\sum_{j=1}^{3\eta_{\mathrm{ion}}^{s}}\sqrt{M_{j}}\Delta R_{j}^{s}e_{j,k}^{s}$
(and conjugate $\tilde{p}^{s,k}\equiv i\frac{\d}{\d q^{s,k}}=i\frac{\d}{\d(\tilde{\Delta\mathbf{R}^{s}})}\cdot\hat{\mathbf{e}}_{k}^{s}$)
we obtain decoupled harmonic oscillators for normal modes with angular
frequency $\omega_{s,k}=\sqrt{f_{s,k}}$ and unit mass as,

\[
H_{\mathrm{vib},s}=E^{s}(\mathbf{R}^{s,0})+\sum_{k=1}^{3\eta_{\mathrm{ion}}^{s}-6}\left(-\frac{1}{2}\frac{\d^{2}}{(\d q^{s,k})^{2}}+\frac{1}{2}f_{s,k}(q^{s,k})^{2}\right)
\]
which is the same term as in Eq.~\eqref{eq:RRHO_hamiltonian}.\footnote{With slight abuse of notation, the constant $E^{s}(\mathbf{R}^{s,0})$
is the same as the constant $E^{s}(\mathbf{Q}^{s,0})$
since indeed the PES only depends non-trivially on the $3\eta_{\mathrm{ion}}^{s}-6$
shape coordinates.} Note that the computational procedure above works in the redundant
description of $3\eta_{\mathrm{ion}}^{s}$ coordinates $\mathbf{R}^{s}$
rather than the abstract shape coordinates used
for theoretical clarity in Sec.~\ref{sec:internaldof}, giving rise
to explicit formulae for the shape coordinates.\footnote{As mentioned in Sec.~\ref{sec:generalconsiderationstateprep}, substrates (slabs) do not have rotational modes and so for these species, we just omit the rotation vectors and perform the calculation as before. The result will be $3\eta_\mathrm{ion}^s-3$ shape coordinates instead.}

The dominant (classical) cost in the PES analysis is finding the equilibrium configuration, since QR decomposition and diagonalization of force matrix for a range of typical molecules/clusters $\eta_{\mathrm{ion}}^{s}\sim1-100$ is negligible (scaling as $O((\eta_{\mathrm{ion}}^{s})^{3})$ classically). However, equilibrium configurations for many molecules are either already known, or very good guesses are known leading to fast convergence. In practice, these classical computations are routine and are not of practical concern.

\subsubsection*{Grid Matching}

The initial state is prepared in the set of normal coordinates $\mathbf{\Xi}^{s}$
discretized as discussed in Sec.~\ref{sec:qalgo_initial_state_prep}. We must
perform a coordinate transformation mapping coordinates $\mathbf{\Xi}^{s}\rightarrow\mathbf{R}^{s}$
where $\mathbf{R}^{s}$ are the original $3\eta_{\mathrm{ion}}^{s}$
Cartesian coordinates of the chemical species discretized on a spatial
grid given by the inverse Fourier transformation of the finite plane
wave basis defined in Sec.~\ref{sec:problem_statement}. Noting that $\Delta R_{\alpha I}^{s}=\frac{1}{\sqrt{M_{I}}}\sum_{k=1}^{3\eta_{\mathrm{ion}}^{s}-6}e_{\alpha I,k}^{s}q^{s,k}$,
we explicitly write the Euclidean transformation,
\[
\bar{R}_{\alpha I}^{s}:=R_{\mathrm{CoM},\alpha I}^{s}+\sum_{\alpha}S_{\alpha\beta}^{s}\left(R_{\alpha I}^{s,0}+\frac{1}{\sqrt{M_{I}}}\sum_{k=1}^{3\eta_{\mathrm{ion}}^{s}-6}e_{\beta I,k}^{s}q^{s,k}\right)
\]
where $S_{\alpha\beta}^{s}$ are the elements of the 3D rotation matrix
formed from the Euler angles in $\mathbf{S}^{s}$. The coordinates
$\bar{\mathbf{R}}_{I}^{s}$ take values on a non-Euclidean spatial
grid distinct from the uniform spatial grid $\mathbf{R}_{I}^{s}$.
We classically precompute a map $\s R$ to suitably assign each value of $\bar{\mathbf{R}}_{I}^{s}$
to a unique value $\mathbf{R}_{I}^{s}$, in order to injectively
match every point of the discretized $\mathbf{\Xi}^{s}$ coordinates to a point of the discretized $\mathbf{R}^{s}$ coordinates. This $\mathcal{R}$ is required as part of the definition of
the preparation oracle $O_{\mathrm{coord}}$ used in the initial state
preparation. Given a sufficient resolution on each grid we expect the
physical realism of the initial state to be not very sensitive to this
choice of assignment.\footnote{For example, we could (classically) map each point from normal to Cartesian coordinates, divide each coordinate by the lattice spacing and round up or down to find a representative on the lattice. This may map two shape coordinates into the same lattice representative, so a separate subroutine should be introduced to resolve the clashes.}
Once a solution is obtained, it is loaded on the quantum computer at a cost of $O(\eta^s_{\mathrm{ion}} |G|)$ quantum operations. More broadly, the quantum
state preparation is agnostic to the grid matching prescription and ultimately
the choice is left to the user.

In summary, the classical pre-processing required for the initial
state preparation are as follows.
\begin{itemize}
\item The classically-computed (e.g. using DFT) approximate electronic wavefunction
at equilibrium is required to prepare the electronic state.
\item The angular frequencies $\omega_{s,k}=\sqrt{f_{s,k}}$ are required
to prepare the truncated thermal state of vibrations.
\item The motional information about the vibrational modes $\hat{\mathbf{e}}_{k}^{s}$
and the grid-matching prescription $\s R:\mathbf{\bar{R}}_{I}^{s}\rightarrow\mathbf{R}_{I}^{s}$
are required to classically compute the map $\mathbf{\Xi}^{s}\rightarrow\mathbf{R}^{s}$
that is loaded as the oracle $O_{\mathrm{coord}}$.
\end{itemize}

\section{Quantum simulation of time evolution using pseudoions}
\label{sec:time_evolution}

In this section, we describe the quantum algorithm for simulating time-evolution using the Hamiltonian with pseudoions. The core piece of time evolution algorithm is the (unitary) \emph{block-encoding} of the pseudoion Hamiltonian~$H$.
This is a unitary $U_{sa}$, acting on the extended Hilbert space $\mathcal{H}_{sa}= \mathcal{H}_s \otimes \mathcal{H}_a$, where the subscripts $s$ and $a$ denote the system and the required ancilla space, respectively. $U_{sa}$ is constructed such that it satisfies
\begin{align}
\frac{H}{\lambda}= \brat{G} U_{sa} \kett{G}_a
\end{align}
where $\lambda >0$ is the rescaling factor, and $\kett{G}_a$ is a normalized quantum state in the ancillary Hilbert space~$\mathcal{H}_a$.
We construct the block-encodings such that $U_{sa}$ has the additional property of being self-inverse, $U_{sa}^2 = I$.
Then, one can directly construct a so-called \emph{walk operator} or \emph{iterate} $W_{sa}$ as
\begin{align}
    W_{sa} = [I_s \otimes (2 \ketbra{G}{G}_a - I_a)] U_{sa}.
\end{align}
This is the basic building block of Hamiltonian simulation algorithms based on quantum signal processing (QSP), which have optimal asymptotic scaling with simulation time $t$ and error $\delta$~\cite{low2019hamiltonian, gilyen2018quantum}:
\begin{align}
O( \lambda t + \log 1/\delta).
\label{eq:QSPscaling}
\end{align}
Most of the challenge in constructing the algorithm is then finding a block-encoding $U_{sa}$ of $H$ with (a) Scaling $\lambda$ as close as possible to the theoretical optimal value $\lambda = \| H\|$ and (b) An efficient quantum circuit implementation. In Sec.~\ref{subsec:CircuitsForBE} we present the block-encoding constructions, while in Sec.~\ref{subsec:TimeEvolution} we review how to use the block-encodings to devise a state-of-the art QSP-based Hamiltonian simulation algorithm.

\subsection{Quantum circuits for block-encodings}\label{subsec:CircuitsForBE}

In this section, we give detailed block-encoding circuits for all of the terms of the pseudoion Hamiltonian in the plane wave basis: the kinetic terms $T_{\el}+T_{\ion}$ of Eqs.~\eqref{eq:Tel_plane-wave-elements-valence},\eqref{eq:Tion_plane-wave-elements-repeated} in Sec.~\ref{subsubsec:BEKinetic}; Coulomb terms $V_{\el}+V^\mathrm{PI}_{\ion}$ of Eqs.~\eqref{eq:Vel_plane-wave-elements-valence},\eqref{eq:VPI_ion_plane_waves} in Sec.~\ref{subsubsec:BECoulomb}; the local term $V_{\loc}^\mathrm{PI}$ with elements given in Sec.~\ref{subsubsec:BELocalPP} (Eq.~\eqref{eq:local_term_plane_waves}); and the non-local term $V_{\NL}^\mathrm{PI}$ with elements  given in Sec.~\ref{subsubsec:BENonlocalPP} (Eq.~\eqref{eq:nl_legendre}). 

Combining these block-encodings with a linear combination of unitaries (LCU) approach yields a block-encoding of the full Hamiltonian, leading to a total rescaling factor given as 
\begin{align}
\lambda= \lambda_{T_{\el}+T_{\ion}} + \lambda_{V_{\el} + V_{\ion}} + \lambda_{V_{\loc}} + \lambda_{V_{\NL}}. 
\end{align}
where the rescaling factors for each of the terms are given by Eq.~\eqref{eq:lambdaT},\eqref{eq:lambdaV},\eqref{eq:lambdaLoc}, and \eqref{eq:lambdaVNL} respectively, and where we drop the superscript $\mathrm{PI}$ for brevity. In the simplest LCU setting, the cost of block-encoding the full Hamiltonian is the sum of the cost of individual block-encodings with some additional two-qubit gates for the controlled logic used in combining the block-encodings.
However, we further optimize the compilation of these quantum circuits by identifying common subroutines between the block-encodings of different terms and
invoking them only once with appropriate control logic gates, thereby reducing the overall resource cost. We make heavy use of quantum rejection sampling as discussed in Ref.~\cite{lemieux2024quantum} for several underlying state preparations.

\subsubsection{Combining block-encodings of individual terms}
The block-encoding circuit proceeds as shown in Fig.~\ref{fig:BE_highlevel}:
\begin{figure}
\includegraphics[width=\textwidth]{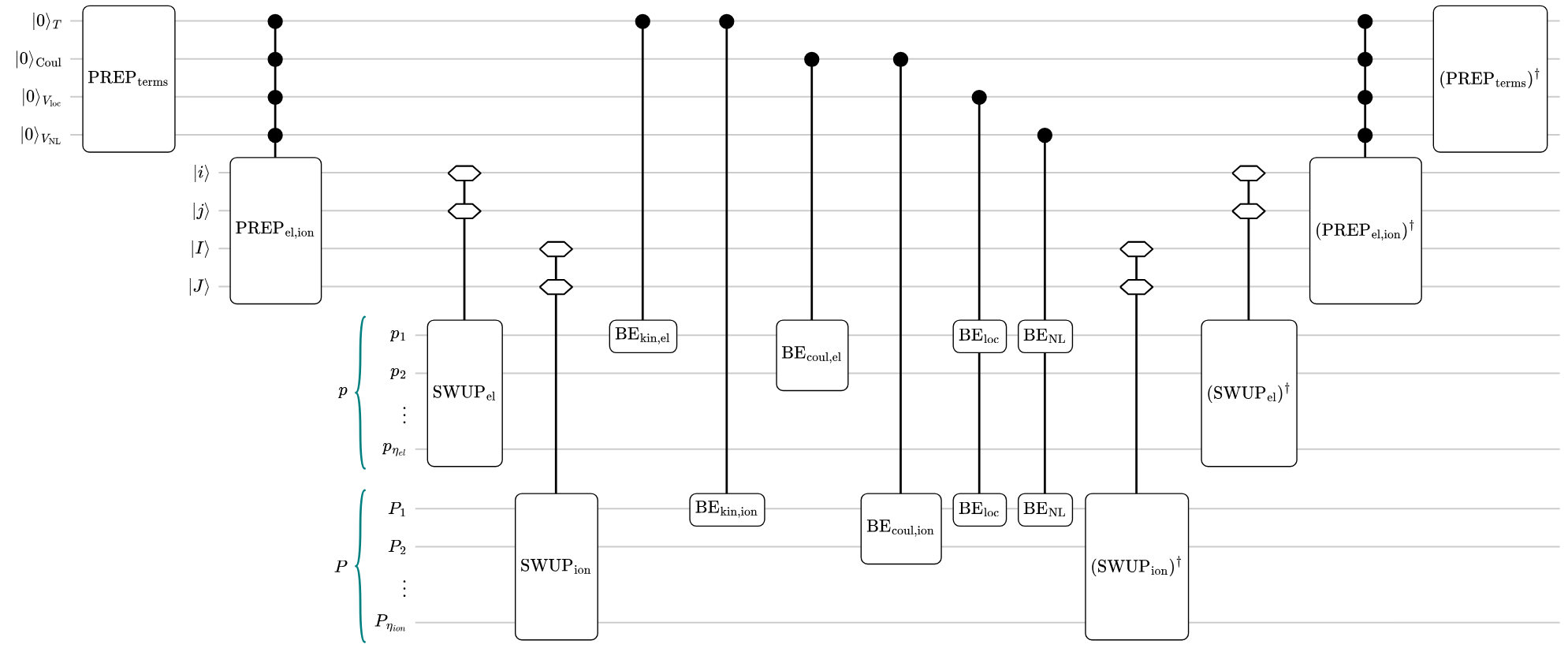}
   \caption{High level circuit for block-encoding the Hamiltonian $H^\mathrm{PI}$ in Eq.~\eqref{eq:H_PI}. Note that we have shown the block-encoding of the kinetic terms individually for conceptual clarity in viewing the target registers, but they are jointly block-encoded (and equivalently for the Coulomb terms).}
   \label{fig:BE_highlevel}
\end{figure}

\begin{enumerate}
\item A first $4$-qubit register is introduced, with a one-hot unary encoding of $4$ possible ``branches' ($\kett{0}$ for the kinetic terms, $\kett{1}$ for the Coulomb terms, $\kett{2}$ for the local term and $\kett{3}$ for the non-local term). We apply a unitary $\PREP_{\textrm{terms}}$ on this register, that sets up an LCU to correct for the term-dependent block-encoding prefactors and sum over the terms:
\begin{align}\label{eq:CombiningPREP}
\PREP_{\textrm{terms}} \kett{0}= \frac{1}{\sqrt{\lambda}} \left(\sqrt{\lambda_{T_{\el}+T_{\ion}}} \kett{0} + \sqrt{\lambda_{V_\el + V_\ion}} \kett{1} + \sqrt{\lambda_{V_\loc}} \kett{2} + \sqrt{\lambda_{V_{\NL}}} \kett{3}\right).
\end{align}

\item  A second register is introduced, labeling valence electrons $i,j$ and pseudoions $I,J$. We introduce unitaries $\PREP_p$, for $p=0,1,2,3$, acting on this register. These unitaries set up an LCU to adjust $(i,j)$ and $(I,J)$-dependent weights for each term $p$ and sum over particles (or particle pairs). Specifically, we apply 
\begin{align}
\label{eq:PREPelion}
    \PREP_{\el,\ion}= \sum_{p=0}^3  \ketbra{p}{p} \otimes \PREP_p,
\end{align} 
where the control is on the first register, and $\PREP_p$ acts on the second register. Expressions for each $\PREP_p$ are given in the following sections, where we focus on each term $p$ separately.

\item Apply $\mathrm{SWUP}_{\mathrm{el}}$, a unitary that controlled  on the $i,j$ labels of the second register, swaps the $i$th and $1$st electron momentum registers, and swaps the $j$th and $2$nd electron momentum registers. Then apply  $\mathrm{SWUP}_{\mathrm{ion}}$, a unitary that controlled on the $I,J$ labels of the second register swaps the $I$th and $1$st pseudoion momentum registers, and swaps the $J$th and $2$nd pseudoion momentum registers.

\item Conditioned on label $p$ from the first register, apply on the first two electron and the first two pseudoion momentum registers, the block-encoding of the single or two-particle versions of the kinetic, Coulomb, local and non-local terms, as indicated in Fig.~\ref{fig:BE_highlevel}. Note that these block-encodings bring in further ancilla qubits that are not represented.

\item Uncompute the controlled SWUPs, and all of the state preparation unitaries.
\end{enumerate}
The resulting unitary acts as $H/\lambda$ on the momentum registers, when conditioned on the qubits of the first two registers as well as the ancilla flags in each of the block-encoding pieces being $\kett{0}$. The costs for the shared subroutines across all block-encoding terms are given in Table~\ref{tab:qre_swap}.

\begin{table}[h!]
\begin{center}
\begin{tabular}{|c|c|c|c|}
\hline
Routine & Toffoli gates & Ancilla qubits & Reference \\
\hline
$\PREP_{\textrm{terms}}$  & $3{\color{red}b_P}+3$ & $2$ & App.~\ref{sec:comp_tu_rotn}\\
$\text{SWUP}_{\text{el}}$    & $2 n \eta_{\text{val}} + 2 \eta_{\text{val}} - 4$ & 0 & App.~\ref{sec:comp_swaps}\\
$\text{SWUP}_{\text{el}}^\dagger$    & $2 n \eta_{\text{val}} + 2 \eta_{\text{val}} - 4$ & 0 & App.~\ref{sec:comp_swaps}\\
$\text{SWUP}_{\text{ion}}$    & $2 \bar{n} \eta_{\text{ion}} + 2 \eta_{\text{ion}} - 4$ & 0 & App.~\ref{sec:comp_swaps}\\
$\text{SWUP}_{\text{ion}}^\dagger$    & $2 \bar{n} \eta_{\text{ion}} + 2 \eta_{\text{ion}} - 4$ & 0 & App.~\ref{sec:comp_swaps}\\
$\PREP_{\textrm{terms}}^\dagger$  & $3{\color{red}b_P}+3$ & $2$ & App.~\ref{sec:comp_tu_rotn}\\
\hline
\end{tabular}
\caption{The resource costs for the subroutines shared across all terms in the block encoding, assuming the compilation scheme in App.~\ref{sec:comp_shared}. The parameters $n$ and $\bar{n}$ were introduced in Eq.~\eqref{eq:qubit_number} and \eqref{eq:qubit_number2}, respectively, $\eta_{\text{ion}}$ is the total number of pseudoions and $\eta_{\text{val}}$ the total number of valence electrons. The parameters in red (defined in Appendix~\ref{sec:comp}) are tunable, and must be chosen to satisfy an overall error budget for the block encoding.}
\label{tab:qre_swap}
\end{center}
\end{table}

\subsubsection{Kinetic term: $T_\mathrm{el}+T_\mathrm{ion}$}\label{subsubsec:BEKinetic} 
We begin from Eq.~\eqref{eq:Tel_plane-wave-elements-valence},\eqref{eq:Tion_plane-wave-elements-repeated} and express the kinetic term in
a convenient per-particle form,

\begin{align}
T=T_{\mathrm{el}}+T_{\mathrm{ion}}=\sum_{i=1}^{\eta_{\mathrm{val}}}T^{i}+\sum_{I=1}^{\eta_{\mathrm{ion}}}T^{I}
\end{align}
where the operator $T^{i}$ (or $T^{I}$) acting on the $i$-th electron
register (or $I$-th pseudoion register) is defined to be
\begin{align}
T^{i}=\sum_{\mathbf{p}\in G}\frac{|\mathbf{k_{p}}|^{2}}{2}\kett{\mathbf{p}}\brat{\mathbf{p}}, \quad T^{I}=\sum_{\mathbf{P}\in \overline{G}}\frac{|\mathbf{k_{P}}|^{2}}{2}\kett{\mathbf{P}}\brat{\mathbf{P}}.
\label{eq:T_perParticle}
\end{align}
The $\mathrm{BE}_{{\mathrm{kin, el}}},\mathrm{BE}_{{\mathrm{kin, ion}}}$ routines in Fig.~\ref{fig:BE_highlevel} utilize explicit
quantum rejection sampling~\cite{lemieux2024quantum} to block-encode
the diagonal matrix in Eq.~\eqref{eq:T_perParticle} with rescaling factors
for electrons and ions respectively 
\begin{align}
    \lambda_{\tilde{T}_{\mathrm{el}}}:= \max_{\mathbf{p} \in G}\frac{|\mathbf{k_{p}}|^{2}}{2}, \quad  \quad \lambda_{\tilde{T}_{\mathrm{ion}}} := \max_{\mathbf{P} \in \overline{G}}\frac{|\mathbf{k_{P}}|^{2}}{2}.
\end{align}
The particle index register preparation routine is
\begin{align}
\PREP_0\kett 0=\frac{1}{\sqrt{\lambda_{T_{\mathrm{el}}+T_{\mathrm{ion}}}}}\p{\sqrt{\lambda_{\tilde{T}_{\mathrm{el}}}} \sum_{i=1}^{\eta_{\mathrm{val}}}\kett i+\sqrt{\lambda_{\tilde{T}_{\mathrm{ion}}}} \sum_{I=1}^{\eta_{\mathrm{ion}}}\sqrt{\frac{1}{M_{I}}}\kett{I}},
\label{eq:PREPelion_T}
\end{align}
where $\lambda_{T_{\mathrm{el}}+T_{\mathrm{ion}}}=\lambda_{T_\mathrm{el}}+\lambda_{T_\mathrm{ion}}$ with $\lambda_{T_\mathrm{el}}=\lambda_{\tilde{T}_{\mathrm{el}}} \eta_{\mathrm{val}},\;\lambda_{T_\mathrm{ion}}=\lambda_{\tilde{T}_{\mathrm{ion}}} \sum_{I=1}^{\eta_{\mathrm{ion}}}\frac{1}{M_{I}}$. The simplest (albeit perhaps not the cheapest) approach implementing $\PREP_0$ is to also use rejection sampling, but with a classically-precomputed set of coefficients proportional to $M_{I}^{-\frac{1}{2}}$ loaded as part of the QROM discussed as part of the interaction term (see App.~\ref{sec:comp_alprep}). Given the loaded data, rejection sampling with a uniform reference state is good enough; 
the state in Eq.~\eqref{eq:PREPelion_T} is close to uniform given that most of the particles are electrons and that their corresponding amplitudes are all equal and significantly larger than the nuclear terms. 
While the precise success probability, and therefore the number of rounds of amplitude amplification, depends on the particle masses, we make the conservative assumption that a single round of amplitude amplification will suffice. We must also include the cost of adding a control to this preparation from the term selection register in Eq.~\eqref{eq:PREPelion}. The construction is similar to~\cite{berry2023quantum}:
\begin{enumerate}
\item Controlled on the momentum register $\kett{\mathbf{p_{1}}}$ (or
$\kett{\mathbf{P_{1}}}$ for the pseudoion) that has been appropriately
selected by the SWUP, compute $|\mathbf{k_{p_1}}|^{2}$ (or $|\mathbf{k_{P_1}}|^{2}$
for the ions) to a second register. 
\item Create a uniform state $\frac{1}{\sqrt{M}}\sum_{m=1}^{M}\kett m$
and perform an inequality test that checks $m   \max_{\mathbf{p} \in G} |\mathbf{k_{p}}|^{2} \leq M|\mathbf{k_{p_1}}|^{2}$ ($m \max_{\mathbf{P} \in \overline{G}} |\mathbf{k_{P}}|^{2} \leq M |\mathbf{k_{P_1}}|^{2}$ for the pseudoion)
and for all $m$ passing the test outputs a flag $\kett 0$ indicating the desired subspace.
\item Uncompute the uniform state. As the kinetic term is implemented before the interaction term, do not uncompute $|\mathbf{k_{p_1}}|^{2}$; we reuse it as part of that block encoding.
\end{enumerate}
 The total rescaling factor for the kinetic
term becomes, 
\begin{align}
\lambda_{T_{\mathrm{el}}+T_{\mathrm{ion}}}=\eta_{\mathrm{val}}  \max_{\mathbf{p} \in G}\frac{|\mathbf{k_{p}}|^{2}}{2} +\sum_{I=1}^{\eta_{\mathrm{ion}}}\frac{1}{M_{I}}\max_{\mathbf{P} \in \overline{G}}\frac{|\mathbf{k_{P}}|^{2}}{2}.
\label{eq:lambdaT}
\end{align}
The cost for preparing $|\mathbf{k_{p}}|^{2}$ is included in Table~\ref{tab:qre_nl} (and App.~\ref{sec:comp_G}). All costs specific to the kinetic term are included in Table~\ref{tab:qre_T}.

\begin{table}[h!]
\begin{center}
\begin{adjustbox}{width=1.05\textwidth,center}
\begin{tabular}{|c|c|c|c|c|}
\hline
Routine & Subroutine        & Toffoli gates & Ancilla qubits & Reference \\
\hline
$\text{PREP}_0$ &  & $6\lceil \log \frac{2}{{\color{red}\epsilon_T}} \rceil + 13\lceil \log(\eta) \rceil + 8 {\color{red} b_T} - 30$ & $\max\{\lceil \log \frac{2}{{\color{red}\epsilon_T}} \rceil, \lceil \log(\eta) \rceil, {\color{red}b_T}\}$ & App.~\ref{sec:comp_T_prep_eta} \\
$\text{BE}_{T, \el (\ion)}$ & Ref. state for $|\mathbf{k_p}|^2$ ($|\mathbf{k_P}|^2$) & $\lceil \log(\eta) \rceil + 7{\color{red}b} + 7{\color{red}\bar{b}} + 4{\color{red}b_k} - 12$ & $\max\{{\color{red}\bar{b}} + 1, {\color{red}b_k}\}$ & App.~\ref{sec:comp_T_usp} \\
 & Comp. $|\v{k_P}|^2$ & $\frac{5}{2}\tilde{\bar{n}} + 2\bar{n}^2 + 4{\color{red}\bar{b}}\bar{n} - 2\bar{n}_{\max}(\bar{n}_{\max}+{\color{red}\bar{b}})$ & ${\color{red}\bar{b}}$ & App.~\ref{sec:comp_G}\\
  & Ineq. test & ${\color{red}\bar{b}} + {\color{red}b}$ & ${\color{red}\bar{b}} + {\color{red}b}$ & App.~\ref{sec:comp_T_sel} \\
 & Uncomp. $|\v{k_P}|^2$ & $\frac{5}{2}\tilde{\bar{n}} + 2\bar{n}^2 + 4{\color{red}\bar{b}}\bar{n} - 2\bar{n}_{\max}(\bar{n}_{\max}+{\color{red}\bar{b}})$ & ${\color{red}\bar{b}}$ & App.~\ref{sec:comp_G}\\
 & Ref. state for $|\mathbf{k_p}|^2$ ($|\mathbf{k_P}|^2$) & $\lceil \log(\eta) \rceil + 7{\color{red}b} + 7{\color{red}\bar{b}} + 4{\color{red}b_k} - 12$ & $\max\{{\color{red}\bar{b}} + 1, {\color{red}b_k}\}$ & App.~\ref{sec:comp_T_usp} \\
$\text{PREP}_{0}^{\dagger}$ & & $6\lceil \log \frac{2}{{\color{red}\epsilon_T}} \rceil + 13\lceil \log(\eta) \rceil + 8 {\color{red}b_T} - 30$ & $\max\{\lceil \log \frac{2}{{\color{red}\epsilon_T}} \rceil, \lceil \log(\eta) \rceil, {\color{red}b_T}\}$ & App.~\ref{sec:comp_T_prep_eta} \\
\hline
\end{tabular}
\end{adjustbox}
\caption{The resource costs for the kinetic term $T$, assuming the compilation scheme in App.~\ref{sec:comp_T}. The parameters in red are tunable, and must be chosen to satisfy an overall error budget for the block encoding. Here, $\eta$ is the total number of particles (electrons plus ions).}
\label{tab:qre_T}
\end{center}
\end{table}

\subsubsection{Coulomb term: $V_{\el} + V_{\ion}^\mathrm{PI}$}\label{subsubsec:BECoulomb}

We begin from
Eqs.~\eqref{eq:Vel_plane-wave-elements-valence},\eqref{eq:VPI_ion_plane_waves} and rewrite the Coulomb term to easily handle momentum conservation, 
\begin{align}
V_\el = \sum^{\eta_\mathrm{val}}_{i\ne j=1} V_{\el}^{i,j}, \; \quad
V_\ion^\mathrm{PI}&= \sum^{\eta_\ion}_{I\ne J=1} (V_{\ion}^\mathrm{PI})^{I,J},
\end{align}
with
\small
\begin{align}
V_{\mathrm{el}}^{i,j} & =\sum_{\substack{{\bf q}\in G^{0}}
}\sum_{c\in\{0,1\}}\frac{\pi}{\Omega|{\bf k_{q}}|^{2}}\left(\sum_{\substack{\mathbf{p},\mathbf{p'}\in G}
}(-1)^{c([\mathbf{p}-\mathbf{q}\notin G]\vee[\mathbf{p'}+\mathbf{q}\notin G])}\kett{\mathbf{p-q}}\brat{\mathbf{p}}_{i}\otimes\kett{{\bf p'+q}}\brat{{\bf p'}}_{j}\right), \label{eq:Vel_LCU-expansion}\\
(V_{\ion}^\mathrm{PI})^{I,J} & =\sum_{\substack{{\bf Q}\in \overline{G}^{0}_{\textrm{trunc}}}
}\sum_{c\in\{0,1\}}\frac{\pi Z_{I}^{\mathrm{PI}}Z_{J}^{\mathrm{PI}}}{\Omega|{\bf k_{Q}}|^{2}}\left(\sum_{\substack{\mathbf{P},\mathbf{P'}\in \overline{G}_{\textrm{trunc}}}
}(-1)^{c([\mathbf{P}-\mathbf{Q}\notin \overline{G}_{\textrm{trunc}}]\vee[\mathbf{P'}+\mathbf{Q}\notin \overline{G}_{\textrm{trunc}}])}\kett{\mathbf{P-Q}}\brat{\mathbf{P}}_{I}\otimes\kett{{\bf P'+Q}}\brat{{\bf P'}}_{J}\right).\label{eq:VPI_ion_LCU-expansion}
\end{align}
\normalsize
where we have used the condition $\frac{1}{2}\sum_{c\in\{0,1\}}(-1)^{c([\mathbf{p}-\mathbf{q}\notin G]\vee[\mathbf{p'}+\mathbf{q}\notin G])}$
to impose that the final state momenta are in $G$ to generate a valid matrix element (and similarly for $\overline{G}_{\textrm{trunc}}$). The particle index register preparation routine is,
\begin{align}
   \mathrm{PREP}_{1}\kett 0=\frac{1}{\sqrt{\lambda_{V_{\mathrm{el}}+V_{\mathrm{ion}}}}}\p{\sqrt{\lambda_{\tilde{V}_{\mathrm{el}}}}\sum_{i\ne j=1}^{\eta_{\mathrm{val}}}\kett{i,j}+\sqrt{\lambda_{\tilde{V}_{\mathrm{ion}}}}\sum_{I\ne J=1}^{\eta_{\mathrm{ion}}}\sqrt{Z_{I}^{\mathrm{PI}}Z_{J}^{\mathrm{PI}}}\kett{I,J}} 
\end{align} where $\lambda_{V_{\mathrm{el}}+V_{\mathrm{ion}}}=\lambda_{V_\mathrm{el}}+\lambda_{V_\mathrm{ion}}$, with $\lambda_{V_\mathrm{el}}=\lambda_{\tilde{V}_{\mathrm{el}}}\eta_{\mathrm{val}}(\eta_{\mathrm{val}}-1),\;\lambda_{V_\mathrm{ion}}=\lambda_{\tilde{V}_{\mathrm{ion}}}\sum_{I\ne J=1}^{\eta_{\mathrm{ion}}}Z_{I}^{\mathrm{PI}}Z_{J}^{\mathrm{PI}}$ where $\lambda_{\tilde{V}_\mathrm{el}}=\frac{2\pi}{\Omega}\sum_{\mathbf{q}\in G^{0}}\frac{1}{|\mathbf{k_{q}}|^{2}}$ and $\lambda_{\tilde{V}_\mathrm{ion}}=\frac{2\pi}{\Omega}\sum_{\mathbf{Q}\in \overline{G}^{0}_{\textrm{trunc}}}\frac{1}{|\mathbf{k_{Q}}|^{2}}$.
We achieve this preparation by first preparing the state
\begin{equation}
\propto \p{\lambda_{V_{\mathrm{el}}}^{1/4}\sum_{i=1}^{\eta_{\mathrm{val}}}\kett i+\lambda_{V_{\mathrm{ion}}}^{1/4}\sum_{I=1}^{\eta_{\mathrm{ion}}}\sqrt{Z_{I}^{\mathrm{PI}}}\kett I}\otimes\p{\lambda_{V_{\mathrm{el}}}^{1/4}\sum_{j=1}^{\eta_{\mathrm{val}}}\kett j+\lambda_{V_{\mathrm{ion}}}^{1/4}\sum_{J=1}^{\eta_{\mathrm{ion}}}\sqrt{Z_{J}^{\mathrm{PI}}}\kett J}
\end{equation}
by data-loading and then using inequality tests to flag discard parts
of the state that have: (i) $i=j$; (ii) $I=J$; (iii) the cross terms
$\kett{i,J}$; (iv) the cross terms $\kett{j,I}$. Amplifying the
unflagged part of the state and uncomputing the flag leads to the
desired result. See App.~\ref{sec:comp_Coul_V} for details.

For the block-encoding of the electron-electron and pseudoion-pseudoion interactions, denoted by $\mathrm{BE}_{{\mathrm{coul, el}}}$ and $\mathrm{BE}_{{\mathrm{coul, ion}}}$ in Fig.~\ref{fig:BE_highlevel}, we perform an LCU with the preparations respectively
\small
\begin{align}
\PREP_{\mathrm{coul,el}}\kett 0  =\frac{1}{\sqrt{\lambda_{\tilde{V}_\mathrm{el}}}}\sum_{\mathbf{q}\in G^{0}}\sum_{c \in \{0,1\}} \sqrt{\frac{\pi}{\Omega|\mathbf{k_{q}}|^{2}}} \kett{\mathbf{q},c}, \quad 
\PREP_{\mathrm{coul,ion}}\kett 0  =\frac{1}{\sqrt{\lambda_{\tilde{V}_\mathrm{ion}}}}\sum_{\mathbf{Q}\in \overline{G}^{0}_{\mathrm{trunc}}}\sum_{c \in \{0,1\}} \sqrt{\frac{\pi}{\Omega|\mathbf{k_{Q}}|^{2}}} \kett{\mathbf{Q},c},
\label{eq:PREP_V}
\end{align}
\normalsize
using quantum rejection sampling as in Ref.~\cite{lemieux2024quantum} (see Type II reference state in App.~\ref{app:reference_states_QRS}, here the same as in Ref.~\cite{berry2023quantum}) as well as a trivial
$\frac{1}{\sqrt{2}}\sum_{c\in\{0,1\}}\kett c=\kett +$ state. The SELECT for the electron-electron and pseudoion-pseuodion Coulomb terms are respectively given by
\begin{align}
\SEL_{\mathrm{coul,el}}=\sum_{\mathbf{q}\in G^{0}}\sum_{c\in\{0,1\}}\kett{\mathbf{q},c}\brat{\mathbf{q},c}\otimes U_{(\mathbf{q},c)}^{\mathrm{coul,el}},\quad\SEL_{\mathrm{coul,ion}}=\sum_{\mathbf{Q}\in \overline{G}^{0}_\mathrm{trunc}}\sum_{c\in\{0,1\}}\kett{\mathbf{Q},c}\brat{\mathbf{Q},c}\otimes U_{(\mathbf{Q},c)}^{\mathrm{coul,ion}},
\end{align}
where the unitaries are 
\begin{align}
U_{(\mathbf{q},c)}^{\mathrm{coul,el}} & =\sum_{\substack{\mathbf{p},\mathbf{p'}\in G}
}(-1)^{c([\mathbf{p}-\mathbf{q}\notin G]\vee[\mathbf{p'}+\mathbf{q}\notin G])}\kett{\mathbf{p-q}}\brat{\mathbf{p}}\otimes\kett{{\bf p'+q}}\brat{{\bf p'}}, \\
U_{(\mathbf{Q},c)}^{\mathrm{coul,ion}} & =\sum_{\substack{\mathbf{P},\mathbf{P'}\in \overline{G}_\mathrm{trunc}}
}(-1)^{c([\mathbf{P}-\mathbf{Q}\notin \overline{G}_\mathrm{trunc}]\vee[\mathbf{P'}+\mathbf{Q}\notin \overline{G}_\mathrm{trunc}])}\kett{\mathbf{P-Q}}\brat{\mathbf{P}}\otimes\kett{{\bf P'+Q}}\brat{{\bf P'}}. \label{eq:SEL_Vel_VPI_ion}
\end{align}
The total rescaling factor is then,
\begin{align}
\lambda_{V_{\mathrm{el}}+V_{\mathrm{ion}}}=\frac{2\pi}{\Omega}\left(\eta_{\mathrm{val}}(\eta_{\mathrm{val}}-1) \sum_{\mathbf{q}\in G^{0}}\frac{1}{|\mathbf{k_{q}}|^{2}} +\sum_{I\ne J=1}^{\eta_{\mathrm{ion}}}Z_{I}^{\mathrm{PI}}Z_{J}^{PI} \sum_{\mathbf{Q}\in \overline{G}^{0}_{\mathrm{trunc}}}\frac{1}{|\mathbf{k_{Q}}|^{2}}\right),
\label{eq:lambdaV}
\end{align}
where we may consider the first/second term in the parenthesis as the rescaling factors for the block-encoding of $V_{\mathrm{el}},V_{\ion}^\mathrm{PI}$, respectively, as given above by $\lambda_{V_\mathrm{el}},\lambda_{V_\mathrm{ion}}$.

The cost for $\PREP_1$, tabulated in App.~\ref{sec:comp_Coul_V}, is included in Table~\ref{tab:qre_Coul}. We tabulate the most significant costs for the preparation of $\PREP_{\text{coul}, \el}$ and $\PREP_{\text{coul}, \ion}$ in App.~\ref{sec:comp_Coul_q}. The cost of SELECT is calculated in App.~\ref{sec:comp_Coul_sel}.

\begin{table}[h!]
\begin{center}
\begin{tabular}{|c|c|c|c|}
\hline
Routine & Toffoli gates & Ancilla qubits & Reference \\
\hline
$\PREP_{1}$ & $6(\eta_{\text{val}} +5 \lceil \log \eta \rceil + 2\lceil \log 2\eta_{\text{val}} \rceil + 2{\color{red}b_\kappa} - 8)$ & $\lceil \log(2\eta_{\text{val}}) \rceil$ & App.~\ref{sec:comp_Coul_V}\\
    $\PREP_{\text{coul}, \el}$ & $5\tilde{n} + 4n^2 + 8{\color{red}b_g} n$ &  $\tilde{n}$ & App.~\ref{sec:comp_Coul_q} \\
    $\PREP_{\text{coul}, \ion}$ & $5\tilde{\bar{n}} + 4\bar{n}^2 + 8{\color{red}b_g} \bar{n}$ &  $\tilde{\bar{n}}$ & App.~\ref{sec:comp_Coul_q} \\
$\text{SEL}_{\text{coul}, \el}$ & $8n$ & $n_{\text{max}}$ & App.~\ref{sec:comp_Coul_sel} \\
$\text{SEL}_{\text{coul}, \ion}$ & $8\bar{n}$ & $\bar{n}_{\text{max}}$ & App.~\ref{sec:comp_Coul_sel} \\
    $\PREP_{\text{coul}, \ion}^\dagger$ & $\frac{5}{2}\tilde{\bar{n}} + 2\bar{n}^2 + 4 {\color{red}b_g} \bar{n}$ & 0 & App.~\ref{sec:comp_Coul_q} \\
    $\PREP_{\text{coul}, \el}^\dagger$ & $\frac{5}{2}\tilde{n} + 2n^2 + 4 {\color{red}b_g} n$ & 0 & App.~\ref{sec:comp_Coul_q} \\
  $\PREP^\dagger_{1}$ & $2(\eta_{\text{val}} +5 \lceil \log \eta \rceil + 2\lceil \log 2\eta_{\text{val}} \rceil + 2{\color{red}b_\kappa} - 8)$ & $\lceil \log(2\eta_{\text{val}}) \rceil$ & App.~\ref{sec:comp_Coul_V} \\
\hline
\end{tabular}
\caption{The resource costs for the Coulomb terms $V_{\text{el}}$ and $V_{\text{ion}}^\mathrm{PI}$, assuming the compilation scheme in App.~\ref{sec:comp_Coul}. The parameters in red are tunable, and must be chosen to satisfy an overall error budget for the block encoding. Here, $\tilde{n}= \sum_{i=1}^3 n_i^2$ and $n_{\text{max}} = \max_i n_i$. The corresponding quantities for pseudoions are indicated with an overbar. The resource costs assume that the system is charge-neutral; see App.~\ref{sec:comp_Coul_V}.}
\label{tab:qre_Coul}
\end{center}
\end{table}

\subsubsection{Electron-pseudoion interaction: The local term $V_{\loc}^\mathrm{PI}$}\label{subsubsec:BELocalPP}

We begin from Eq.~\eqref{eq:local_term_plane_waves} and rewrite the local term to simply handle momentum conservation,

\begin{align}
V_{\loc}^{i,I}  =\sum_{s=-1}^{3}& \sum_{\substack{\mathbf{q}\in G^{0}}
}\sum_{c\in\{0,1\}}\frac{2\pi(\bar{r}_{\mathrm{loc}}^{\zeta_I})^{3}}{\Omega}\sqrt{\frac{\pi}{2}}c_{s}^{\zeta_I}e^{-(|\mathbf{k_{q}}|\bar{r}_{\mathrm{loc}}^{\zeta_I})^{2}/2}(|\mathbf{k_{q}}|\bar{r}_{\mathrm{loc}}^{\zeta_I})^{2s}\nonumber \\
 & \p{\sum_{\substack{\mathbf{p}\in G,\mathbf{P}\in \overline{G}}
}(-1)^{c([\mathbf{p}-\mathbf{q}\notin G]\vee[\mathbf{P}+\mathbf{q}\notin \overline{G}])}\kett{\mathbf{p-q}}\brat{\mathbf{p}}_{i}\otimes\kett{\mathbf{P}+\mathbf{q}}\brat{\mathbf{P}}_{I}},\label{eq:Vloc_LCU-expansion}
\end{align}
where the expression in parenthesis is a unitary operator with the
same form as that in the Coulomb term (c.f. Eq.~\eqref{eq:Vel_LCU-expansion},\eqref{eq:VPI_ion_LCU-expansion}). 
The particle index register preparation routine is
\begin{align}
\label{eq:LocalPREP}
\PREP_{2} \kett{0} = \frac{1}{\sqrt{\eta_\mathrm{val}}} \sum^{\eta_\mathrm{val}}_{i=1} \kett{i} \otimes \frac{1}{\sqrt{\sum^{\eta_\ion}_{I=1}  
 \lambda_{\tilde{V}^{I}_{\loc}}}} \sum^{\eta_\ion}_{I=1}  \sqrt{\lambda_{\tilde{V}^I_\loc}}\kett{I}
\end{align}
where
\begin{align}
\lambda_{\tilde{V}_{\mathrm{loc}}^{I}} & =\sum_{s=-1}^{3}\sum_{c\in\{0,1\}}\frac{2\pi(\bar{r}_{\mathrm{loc}}^{\zeta_{I}})^{3}}{\Omega}\sqrt{\frac{\pi}{2}}|c_{s}^{\zeta_{I}}|\lambda_{\mathrm{loc}}^{\zeta_{I},s}, \\ 
\lambda_{\mathrm{loc}}^{\zeta_{I},s} & =\sum_{\mathbf{q}\in G^{0}}e^{-(|\mathbf{k_{q}}|\bar{r}_{\mathrm{loc}}^{\zeta_{I}})^{2}/2}(|\mathbf{k_{q}}|\bar{r}_{\mathrm{loc}}^{\zeta_{I}})^{2s}.
\label{eq:LambdaLocPerElectronPerIon}
\end{align}
The expression before the parenthesis in Eq.~\eqref{eq:Vloc_LCU-expansion}, and therefore the amplitudes necessary for $\PREP$, only depend on $I$ through the pseudoion type $\zeta_I$. For both the local and non-local interaction terms, we first load the ion type, indexed on $I$, using a QROM $\kett{I} \kett{0} \mapsto \kett{I} \kett{\zeta_I}$ (this shared cost is accounted for in the cost for the non-local term). The state preparations below are then controlled on the register encoding $\zeta_I$.
For the $\mathrm{BE}_\loc$ routine we perform an LCU with the preparation,
\begin{equation*}
\PREP_{\mathrm{loc}}\kett 0 \kett{\zeta_{I}}=\frac{1}{\sqrt{\lambda_{\tilde{V}_{\mathrm{loc}}^{I}}}}\sum_{s=-1}^{3}\sum_{\mathbf{q}\in G^{0}}\sum_{c\in\{0,1\}}\sqrt{\frac{2\pi(\bar{r}_{\mathrm{loc}}^{\zeta_{I}})^{3}}{\Omega}\sqrt{\frac{\pi}{2}}|c_{s}^{\zeta_{I}}|e^{-(|\mathbf{k_{q}}|\bar{r}_{\mathrm{loc}}^{\zeta_{I}})^{2}/2}(|\mathbf{k_{q}}|\bar{r}_{\mathrm{loc}}^{\zeta_{I}})^{2s}}\kett{s,\mathrm{sgn}(c_{s}^{\zeta_{I}}),\mathbf{q},c} \kett{\zeta_{I}},
\end{equation*}
using a sequence of controlled unitaries 
 $\PREP_{\mathrm{loc}}=\PREP_{\mathrm{loc}, 2}\cdot\PREP_{\mathrm{loc},1}$ where
their actions are
\begin{align}
\PREP_{\mathrm{loc,1}}\kett 0  \kett{\zeta_{I}}& =\frac{1}{\sqrt{\lambda_{\tilde{V}_{\mathrm{loc}}^{I}}}}\sum_{s=-1}^{3}\sum_{c\in\{0,1\}}\sqrt{\frac{2\pi(\bar{r}_{\mathrm{loc}}^{\zeta_{I}})^{3}}{\Omega}\sqrt{\frac{\pi}{2}}|c_{s}^{\zeta_{I}}|\lambda^{\zeta_{I},s}_\loc}\kett{s,\mathrm{sgn}(c_{s}^{\zeta_{I}}),c} \kett{\zeta_{I}}, \\
\PREP_{\mathrm{loc}, 2} \kett 0 \kett{s} \kett{\zeta_{I}} & =\frac{1}{\sqrt{\lambda_{\mathrm{loc}}^{\zeta_{I},s}}}\sum_{\mathbf{q}\in G^{0}}\sqrt{e^{-(|\mathbf{k_{q}}|\bar{r}_{\mathrm{loc}}^{\zeta_{I}})^{2}/2}(|\mathbf{k_{q}}|\bar{r}_{\mathrm{loc}}^{\zeta_{I}})^{2s}}\kett{\mathbf{q}}\kett{s}\kett{\zeta_{I}}. \label{eq:s_prep}
\end{align}
The $\mathrm{SELECT}$ for the local term is given by,
\begin{align}
\mathrm{SEL}_{{\mathrm{loc}}} & =\sum_{s=-1}^{3}\sum_{\mathbf{q}\in G^{0}}\sum_{c\in\{0,1\}}\kett{s,\mathrm{sgn}(c_{s}^{\zeta_I}),\mathbf{q},c}\brat{s,\mathrm{sgn}(c_{s}^{\zeta_I}),\mathbf{q},c}\otimes U_{(s,\mathrm{sgn}(c_{s}^{\zeta_I}),\mathbf{q},c)}^{\loc},\\
U_{(s,\mathrm{sgn}(c_{s}^{\zeta_I}),\mathbf{q},c)}^{\loc} & =\sum_{\mathbf{p}\in G,\mathbf{P}\in \overline{G}}(-1)^{c(\mathbf{p}-\mathbf{q}\notin G\lor\mathbf{P}+\mathbf{q}\notin \overline{G})+\mathrm{sgn}(c_{s}^{\zeta_I})}\kett{\mathbf{p}-\mathbf{q},\mathbf{P}+\mathbf{q}}\brat{\mathbf{p},\mathbf{P}},
\end{align}
where the unitaries $U^\loc$ are very similar to that of the Coulomb case (c.f. Eq.~\eqref{eq:SEL_Vel_VPI_ion}), but additionally include the accumulation of
$\mathrm{sgn}(c_{s}^{\zeta_I})$ in the phase. The total rescaling factor of then becomes,
\begin{align}
\lambda_{V_\mathrm{loc}}=\eta_{\mathrm{val}} \sum_{I=1}^{\eta_{\mathrm{ion}}}\lambda_{\tilde{V}_{\mathrm{loc}}^{I}}=\eta_{\mathrm{val}}\sum_{I=1}^{\eta_{\mathrm{ion}}}\sum_{s=-1}^{3}\frac{4\pi(\bar{r}_{\mathrm{loc}}^{\zeta_{I}})^{3}}{\Omega}\sqrt{\frac{\pi}{2}}|c_{s}^{\zeta_{I}}|\sum_{\mathbf{q}\in G^{0}}e^{-(|\mathbf{k_{q}}|\bar{r}_{\mathrm{loc}}^{\zeta_{I}})^{2}/2}(|\mathbf{k_{q}}|\bar{r}_{\mathrm{loc}}^{\zeta_{I}})^{2s}.
\label{eq:lambdaLoc}
\end{align}
The subroutine $\PREP_{\mathrm{loc},1}$ is a state preparation of 5 elements (the $c$ variable sum just gives a $\kett{+}$ state). We achieve this with coherent alias sampling; see App.~\ref{sec:comp_loccoeff}. The dominant preparation cost is from $\PREP_{\mathrm{loc},2}$, which invokes quantum rejection sampling as described in Ref.~\cite{lemieux2024quantum} with Type I and Type III reference states discussed in App.~\ref{app:reference_states_QRS}. However, we note that the structure of this state is very similar to other states that we seek to prepare; in particular, Eq.~\eqref{eq:s_prep} is a state that looks similar to Eq.~\eqref{eq:p2_prep} for the non-local term. We reduce costs by combining these preparations; i.e. we use a single unitary state preparation routine to prepare the reference state for the non-local and local term; here, careful conditioning on the term selection register and the register containing $\kett{s}$ flags which sub-case is appropriate and prepares the correct reference state in the correct branch. See App.~\ref{sec:comp_tildeG} for a detailed construction.

\begin{table}[h!]
\begin{center}
\begin{adjustbox}{width=1.05\textwidth,center}
\begin{tabular}{|c|c|c|c|c|}
\hline
Routine & Subroutine & Toffoli gates & Ancilla qubits & Reference \\
\hline
$\PREP_2$ & $\text{PREP}_{2, \text{el}}$ & $7\lceil \log_2(\eta_{\text{val}}) \rceil + 2{\color{red}b_{\eta_{\text{el}}}} - 6$ & $\max\{{\color{red}b_Z}, \lceil \log(\eta_{\text{val}}) \rceil\}$ & App.~\ref{sec:comp_loc_prep_el} \\
& $\text{PREP}_{2, \text{ion}}$ & $6Z + \lceil \log(Z) \rceil({\color{red}b_Z}-3) + 7\lceil \log(\eta_\ion) \rceil + 2{\color{red}b_I}-6$ & ${\color{red}b_Z} + \max\{{\color{red}b_I}, \lceil \log(\eta_\ion) \rceil\}$ & App.~\ref{sec:comp_loc_prep_el} \\
$\PREP_\loc$ & $\PREP_{\loc,1}$ & $Z({2\color{red}b_s}+{\color{red} b_{\text{keep}}}+25)$ & $2{\color{red} b_{\text{keep}}} + 3$ & App.~\ref{sec:comp_loccoeff} \\
 & $\text{PREP}_{\loc, 2}$ & 0* & 0* & - \\
$\SEL_\loc$ & & $8\bar{n}$ & $\bar{n}$ & App.~\ref{sec:comp_loc_sel}\\
 $\PREP_\loc^\dagger$ & $(\PREP_{\loc, 2})^\dagger$ & 0* & 0* & - \\
 & $(\PREP_{\loc, 1})^\dagger$ & $Z({2\color{red}b_s}+{\color{red} b_{\text{keep}}}+25)$ & 0 & App.~\ref{sec:comp_loccoeff}  \\
 $\PREP_2^\dagger$ & $\text{PREP}_{2, \text{ion}}^\dagger$ & $6Z + \lceil \log(Z) \rceil({\color{red}b_Z}-3) + 7\lceil \log(\eta_\ion) \rceil + 2{\color{red}b_I}-6$ & ${\color{red}b_Z} + \max\{{\color{red}b_I}, \lceil \log(\eta_\ion) \rceil\}$ & App.~\ref{sec:comp_loc_prep_el} \\
 & $\text{PREP}_{2, \text{el}}^\dagger$ & $7\lceil \log_2(\eta_{\text{val}}) \rceil + 2{\color{red}b_{\eta_{\text{el}}}} - 6$ & $\max\{{\color{red}b_Z}, \lceil \log(\eta_{\text{val}}) \rceil\}$ & App.~\ref{sec:comp_loc_prep_el}\\
\hline
\end{tabular}
\end{adjustbox}
\caption{The resource costs for the implementation of the local interaction term. The cost labeled ``0*'' for $\PREP_{\loc, 2}$ is because the costs for this state preparation are captured in the non-local term. The parameters in red are tunable, and must be chosen to satisfy an overall error budget for the block encoding. Other parameters are defined in their linked appendices. Note $Z$ is the total number of pseudoion types.}
\label{tab:qre_loc}
\end{center}
\end{table}

\subsubsection{Electron-pseudoion interaction: the non-local term $V_{\mathrm{NL}}^{\mathrm{PI}}$}\label{subsubsec:BENonlocalPP}

We begin from Eq.~\eqref{eq:nl_legendre}. Let $X^{I,l}$ be the orthogonal matrix that diagonalizes the ($3\times3$)
real symmetric matrix $B^{I,l}$, namely $B^{I,l}=X^{I,l}D^{I,l}(X^{I,l})^{T}$
where $D^{I,l}$ is a diagonal matrix of eigenvalues $D_{\alpha}^{I,l}$. Then, we define
\begin{align}\label{eq:G_func_def}
G_{\alpha}^{I,l}(|\mathbf{k_{p}}|\bar{r}_{l}^{I}):=\sum_{b=1}^{3}[X^{I,l}]_{b\alpha}\mathrm{g}_{b}^{l}(|\mathbf{k_{p}}|\bar{r}_{l}^{I}).
\end{align}
Noting that the matrix $B^{I,l}$ (and in fact all $I$-dependent HGH parameters) actually depends on $I$ only through the ion-type $\zeta_I$,
we have that the non-local term takes a convenient form,
\small
\begin{equation}
V_{\mathrm{NL}}^{i,I}=\sum_{\substack{\mathbf{p_{1}},\mathbf{p_{2}\in G},\mathbf{P}\in \overline{G}\\
\mathbf{P}+\mathbf{p_{1}}-\mathbf{p_{2}}\in \overline{G}
}
}\sum_{\alpha=1}^{3}\sum_{l=0}^{l_{\mathrm{max}}}\frac{4\pi}{\Omega}(\bar{r}_{l}^{\zeta_I})^{3}(2l+1)D_{\alpha}^{\zeta_I,l}G_{\alpha}^{\zeta_I,l}(|\mathbf{k_{p_{2}}}|\bar{r}_{l}^{\zeta_I})G_{\alpha}^{\zeta_I,l}(|\mathbf{k_{p_{1}}}|\bar{r}_{l}^{\zeta_I})P_{l}(\mathbf{\hat{k}_{p_{1}}}\cdot\mathbf{\hat{k}_{p_{2}}})\kett{\mathbf{p_{2}},\mathbf{P}+\mathbf{p_{1}}-\mathbf{p_{2}}}\brat{\mathbf{p_{1}},\mathbf{P}}_{i,I}. \label{eq:NL_term_diagonalized}
\end{equation}
\normalsize
As mentioned in Sec.~\ref{sec:dynamic_DoF}, for practical purposes we expect $l_{\mathrm{max}} \leq 2$ and so we show how to compile the circuit for the case of $l_{\mathrm{max}} =2$, although the procedure may be easily generalized for higher $l$ if desired. The particle index register preparation routine is, \begin{align}\label{eq:NLPREP}
\PREP_{3}\kett{0}= \frac{1}{\sqrt{\eta_\mathrm{val}}} \sum^{\eta_\mathrm{val}}_{i=1} \kett{i} \otimes \frac{1}{\sqrt{\sum_{I=1}^{\eta_{\mathrm{ion}}} \lambda_{\tilde{V}^I_{\NL}}}}\sum^{\eta_\ion}_{I=1} \sqrt{ \lambda_{\tilde{V}^I_\NL}} \kett{I}
\end{align}
where
\begin{align}\label{eq:lambdatildeVINL}
\lambda_{\tilde{V}_{\mathrm{NL}}^{I}}=\sum_{l=0}^{l_{\max}}\sum_{\alpha=1}^{3}\frac{4\pi}{\Omega}(\bar{r}_{l}^{\zeta_I})^{3}(2l+1)|D_{\alpha}^{\zeta_{I},l}|\sum_{\mathbf{p_{2}}\in G}|G_{\alpha}^{\zeta_I,l}(|\v{k}_{\v{p}_2}| \bar{r}_{l}^{\zeta_I})|^2.
\end{align}

We proceed with the block-encoding of the non-local term $\mathrm{BE}_\NL$ as per the circuit structure shown in Fig.~\ref{fig:BE_NL_circuit}. We discuss the steps in detail below, noting that, as in the local term, the state preparation depends on the ion $I$ only through its type $\zeta_I$. 
\begin{figure}[h!]
\includegraphics[width=\textwidth]{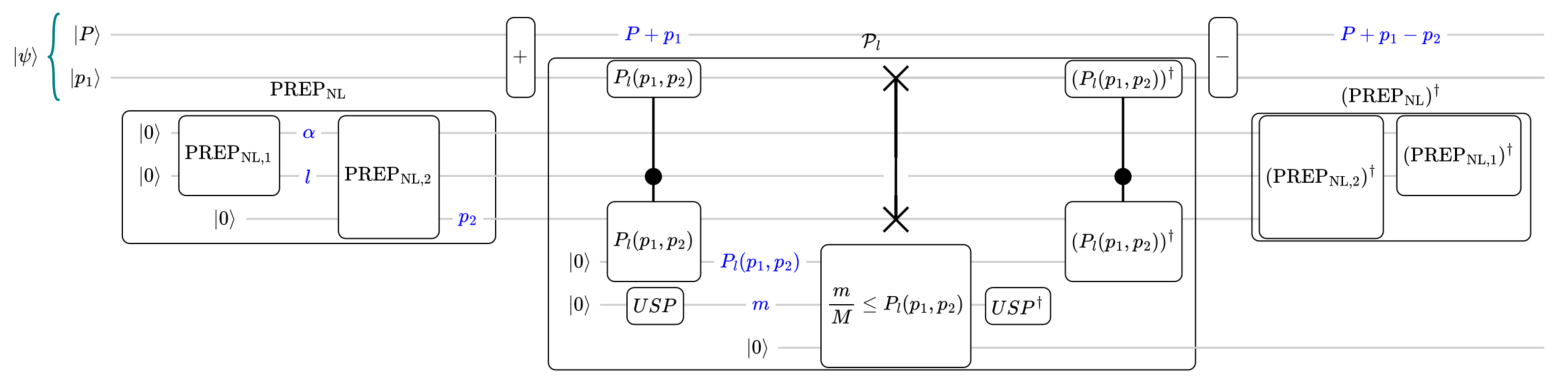}
   \caption{Circuit for block-encoding the non-local term $V^{i,I}_\NL$ in Eq.~\eqref{eq:NL_term_diagonalized}.
   The notation $P_l$ refers to the computation of the Legendre polynomial in Step 2 below. 
   Note that we have slightly modified the circuit relative to the prescription in the text so that it is manifestly self-inverse.} 
\label{fig:BE_NL_circuit}
\end{figure}

\begin{enumerate}
\item Controlled on the register encoding $\zeta_I$, we apply a unitary that prepares the following state over the angular momenta, eigenstate indices, and output indices, $l$, $\alpha$ and $\v{p}_2$ respectively:
\begin{align}
\mathrm{PREP}_{\mathrm{NL}} \kett 0 \kett{\zeta_I} =\frac{1}{\sqrt{\lambda_{\tilde{V}_{\NL}^{I}}}}\sum_{l=0}^{l_{\max}}\sum_{\alpha=1}^{3}\sqrt{\frac{4\pi}{\Omega}(\bar{r}_{l}^{\zeta_I})^{3}(2l+1)D_{\alpha}^{\zeta_{I},l}}\sum_{\mathbf{p_{2}}\in G} G_{\alpha}^{\zeta_{I},l}(|\mathbf{k}_{\mathbf{p_{2}}}|\bar{r}_l^{\zeta_I})\kett l\kett{\alpha}\kett{\mathbf{p_{2}}} \kett{\zeta_I}
\end{align}
We divide this into a sequence of two unitaries as $\mathrm{PREP}_{\mathrm{NL}}=\PREP_{\mathrm{NL},2} \cdot \PREP_{\mathrm{NL},1}$, with action
\begin{align}\label{eq:l_alpha_prep}
\PREP_{\mathrm{NL},1}\kett 0 \kett{\zeta_I} & =\frac{1}{\sqrt{\tilde{V}_{\mathrm{NL}}^{I}}}\sum_{l=0}^{l_{\max}}\sum_{\alpha=1}^{3}\sqrt{\frac{4\pi}{\Omega}(\bar{r}_{l}^{\zeta_I})^{3}(2l+1)D_{\alpha}^{\zeta_{I},l}\lambda_{G_{\alpha}^{\zeta_{I},l}}}\kett l\kett{\alpha} \kett{\zeta_I},\\ \label{eq:p2_prep}
\PREP_{\mathrm{NL},2} \kett{l}\kett{\alpha}\kett{0} \kett{\zeta_I} & =\frac{1}{\sqrt{\lambda_{G_{\alpha}^{\zeta_{I},l}}}}\sum_{\mathbf{p_{2}}\in G}G_{\alpha}^{\zeta_{I},l}(|\mathbf{k}_{\mathbf{p_{2}}}|\bar{r}_l^{\zeta_I})\kett{l}\kett{\alpha}\kett{\mathbf{p_{2}}}\kett{\zeta_I}.
\end{align}
Here $\lambda_{G_\alpha^{\zeta_I,l}} = \sum_{\v{p}_2 \in G}  |G_{\alpha}^{\zeta_I,l}(|\v{k}_{\v{p}_2}| \bar{r}_{l}^{\zeta_I})|^2$, whose value is precomputed classically, and incorporated into the quantum algorithm with appropriate data loading subroutines. 
The state in Eq.~\eqref{eq:l_alpha_prep} is prepared using coherent alias sampling. The state in Eq.~\eqref{eq:p2_prep} is prepared using rejection sampling (Ref.~\cite{lemieux2024quantum}), in combination with the equivalent reference state for the local term (see Type I reference state in App.~\ref{app:reference_states_QRS}). 
More specifically, for the non-local term we use rejection sampling to prepare a state with amplitudes proportional to $|G_\alpha^{\zeta_I,l}|$ and then we manually add signs to the parts of the domain where $G$ becomes negative (note that the functions $G$ are real-valued by definition).
The relevant piece of the quantum circuit operates in the following order:

    \begin{enumerate}
        \item Conditioned on $l, \alpha, \zeta_I$, prepare a reference state
        \begin{equation}
            \kett{\psi_{\tilde{G}}} \propto \sum_{\v{p_2} \in G} \tilde{G}_\alpha^{\zeta_I,l}(\v{k_{p_2}}\bar{r}_l^{\zeta_I})\kett{\v{p_2}},
        \end{equation}
        for some function $\tilde{G}_\alpha^{\zeta_I,l} \geq |G_\alpha^{\zeta_I,l}|$ everywhere in the domain.
        \item Conditioned on $l, \alpha, \zeta_I, \v{p_2}$, compute an approximation to the function $\kett{\bar{G}_\alpha^{\zeta_I, l}(\v{k_{p_2}}\bar{r}_l^{\zeta_I})}$, where $ \bar{G}_\alpha^{\zeta_I, l}(\v{k_{p_2}}\bar{r}_l^{\zeta_I}) = |G_\alpha^{\zeta_I,l}(|\v{k_{p_2}}|\bar{r}_l^{\zeta_I})|/\tilde{G}_\alpha^{\zeta_I,l}(\v{k_{p_2}}\bar{r}_l^{\zeta_I})$, to an ancilla register.
        \item Prepare a uniform superposition over $m=1 \ldots M$ amplitudes (we assume that $M$ is a power of two).
        \item Use an inequality test to set an ancilla to 0 when $m \leq M|G_\alpha^{\zeta_I,l}(|\v{k_{p_2}}|\bar{r}_l^{\zeta_I})|/\tilde{G}_\alpha^{\zeta_I,l}(\v{k_{p_2}}\bar{r}_l^{\zeta_I})$ (and to 1 otherwise).
        \item Use oblivious amplitude amplification to amplify the zero branch of the flag.
        \item Flip the sign of the resultant state for each $\v{k_{p_2}}$ where $G_\alpha^{\zeta_I,l}(|\v{k_{p_2}}|\bar{r}_l^{\zeta_I}) < 0$.
    \end{enumerate}

\item
Then, given that we have the registers $\v{p_1}$ (the system register) and $\v{p_2}$ (an ancilla register that is created with $\PREP$ in the first step), we apply the block-encoding of the diagonal matrix whose elements are the values of the Legendre polynomials.
This operation is controlled only on $l$. 
Namely, it is the block-encoding of
\begin{align}
\mathcal{P}_l = \sum_{\v{p}_1, \v{p}_2 \in G} P_l\left(\frac{\v{k}_{\v{p}_2}\cdot \v{k}_{\v{p}_1}}{|\v{k}_{\v{p}_1}| |\v{k}_{\v{p}_2}|}\right) \kett{\v{p}_2}\brat{\v{p}_2} \otimes \kett{\v{p}_1}\brat{\v{p}_1}.
\label{eq:legendre_diagonal_operator}
\end{align}
This contributes a rescaling factor of $1$, given that it is a diagonal operator with maximum diagonal element $1$. 
We prepare this block encoding using explicit (in the nomenclature of Ref.~\cite{lemieux2024quantum}) rejection sampling with a uniform reference state. See Fig.~\ref{fig:BE_NL_circuit}. However, we structure the arithmetic for rejection sampling in a way that we avoid having to compute a division to evaluate the argument to the Legendre polynomial. Specifically, we can prepare a uniform superposition over $m = 1 \ldots M$ basis states and then check the inequalities
\begin{equation}\label{eq:legendre_cases_main}
    \begin{cases}
        M \geq m, & l=0 \\
        (\v{k_{p_1}} \cdot \v{k_{p_{2}}}) M \geq |\v{k_{p_1}}| |\v{k_{p_2}}| m, & l=1 \\
        [3(\v{k_{p_1}} \cdot \v{k_{p_{2}}})^2 - |\v{k_{p_1}}|^2 |\v{k_{p_2}}|^2]  M \geq 2 |\v{k_{p_1}}|^2 |\v{k_{p_2}}|^2 m, & l=2.
    \end{cases}
\end{equation}
Rearranging the above inequalities recovers the correct inequality test for the Legendre polynomial. The arithmetic to prepare the input quantities to the inequality test can be broken down as follows:
    \begin{enumerate}
        \item Compute $\v{k_{p_1}} \cdot \v{k_{p_{2}}}$ and $|\v{k_{p_1}}| |\v{k_{p_2}}|$ to ancilla registers.
        \item If $l=2$, square $\v{k_{p_1}} \cdot \v{k_{p_{2}}}$ and $|\v{k_{p_1}}| |\v{k_{p_2}}|$.
        \item Prepare a uniform superposition over $m = 1 \ldots M$ basis states.
        \item Prepare the right hand side of Eq.~\eqref{eq:legendre_cases_main} by multiplying the $m$ register by $|\v{k_{p_1}}| |\v{k_{p_2}}|$ if $l=1$, and by $2|\v{k_{p_1}}|^2 |\v{k_{p_2}}|^2$ if $l=2$.
        \item Instantiate a register in computational basis state $\kett{M}$.
        \item Compute the quantity $3(\v{k_{p_1}} \cdot \v{k_{p_{2}}})^2 - |\v{k_{p_1}}|^2 |\v{k_{p_2}}|^2$.
        \item Evaluate the left hand side of Eq.~\eqref{eq:legendre_cases_main} by multiplying $M$ by $\v{k_{p_1}} \cdot \v{k_{p_{2}}}$ if $l=1$ or by $3(\v{k_{p_1}} \cdot \v{k_{p_{2}}})^2 - |\v{k_{p_1}}|^2 |\v{k_{p_2}}|^2$ if $l=2$.
        \item Carry out the inequality test.
        \item Uncompute the arithmetic in the substeps above.
    \end{enumerate}

\item We add $\v{p_1}$ to the pseudoion momentum $\v{P}$.

\item We swap the system register where $\v{p_1}$ is encoded and the ancilla register where $\v{p_2}$ is encoded.

\item We subtract $\v{p_2}$ from the pseudoion momentum $\v{P}$ and (not shown) flag an ancilla qubit with $\kett{0}$ if $\v{P}+\v{p_1}-\v{p_2} \in \overline{G}$, and with $\kett{1}$ otherwise.

\item Finally, controlled on $\zeta_I$, we apply $(\PREP_{\NL})^{\dagger}$ on the (now) ancilla register where $\v{p_1}$ is encoded.
\end{enumerate}

\noindent The costs for each of the steps and substeps above are derived in Appendix~\ref{sec:comp_nl} and collated in Table~\ref{tab:qre_nl}. The total rescaling factor is
\begin{align}\label{eq:lambdaVNL}
\lambda_{V_{\NL}}=\eta_\mathrm{val}\sum_{I=1}^{\eta_{\ion}}\lambda_{\tilde{V}_{\NL}^{I}}=\eta_\mathrm{val}\sum_{I=1}^{\eta_{\ion}}\sum_{l=0}^{l_{\max}}\sum_{\alpha=1}^{3}\frac{4\pi}{\Omega}(\bar{r}_{l}^{\zeta_I})^{3}(2l+1)|D_{\alpha}^{\zeta_{I},l}|\sum_{\mathbf{p_{2}}\in G}|G_{\alpha}^{\zeta_{I},l}(|\mathbf{k}_{\mathbf{p_{2}}}|\bar{r}_l^{\zeta_I})|^{2}
\end{align}
\begin{table}[h!]
\begin{center}
\begin{adjustbox}{width=1.05\textwidth,center}
\begin{tabular}{|c|c|c|c|c|}
\hline
Routine & Subroutine        & Toffoli gates & Ancilla qubits & Reference \\
\hline
Load $\zeta_I$ &  & $\eta_{\text{ion}}$ & $5+{\color{red}b_M}$ & App.~\ref{sec:comp_nucleardata} \\
 $\PREP_3$ & $\PREP_{3, \el}$ & $0$ & $0$ & App.~\ref{sec:comp_loc_prep_el}\\
 & $\PREP_{3, \ion}$ & $4Z + \lceil \log(Z) \rceil({\color{red}b_Z}-3) -2$ & ${\color{red}b_Z}$ & App.~\ref{sec:comp_loc_prep_el}\\
$\text{PREP}_{\NL, 1}$ & & $11Z + 3\lceil \log(9Z) \rceil + 2 {\color{red}b_{\alpha,l}} + {\color{red}b_{\text{keep}}} -8$ & $2{\color{red}b_{\text{keep}}}+\lceil \log(9Z) \rceil$ & App.~\ref{sec:comp_alprep} \\
 $\PREP_{\NL, 2}$ & $\kett{0} \rightarrow \kett{\psi_{\tilde{G}}}$ & $(1+2R)(12\tilde{n} + 74n + 4n^2 + 6n{\color{red}b_{\text{pl}}}+6n{\color{red}b_{\text{exp}}}+3{\color{red}b_{\text{rot}}}+8)$ & $2n + \max\{{\color{red} b_{\text{exp}}}, {\color{red} b_{\text{pl}}}\}$ & App.~\ref{sec:comp_tildeG} \\
 & $\kett{0} \rightarrow \kett{\bar{G}^\alpha_{l,\zeta}}$ & $(1+R)(4 \tilde{n} + 2n^2 + 7{\color{red}b}n + \frac{51}{4}{\color{red}b}^2 + 32{\color{red}b}+ 116 - 2n_{\max}(n_{\max}+{\color{red}b}))$ & $65{\color{red}b}$ & App.~\ref{sec:comp_G} \\
 & $\text{USP}_M$ & $0$ & $0$ & - \\
 & Ineq. test & $(1+R)({\color{red}b}+{\color{red}b_{\tilde{M}}})$ & ${\color{red}b}+2{\color{red}b_{\tilde{M}}}$ & App.~\ref{sec:comp_NLineq} \\
$\mathcal{P}_l$ & $\text{USP}_M$ & $0$ & $0$ & - \\
& Arithmetic & $5\tilde{n} + 5n^2 + 8{\color{red}b}n + \frac{21}{4}{\color{red}b}^2 + \frac{13}{2}{\color{red}b} - 2n_{\max}(n_{\max}+{\color{red}b}) - 6$ & & App.~\ref{sec:comp_legendre} \\
& Ineq. test & $2(\max\{{\color{red}b},\lceil \log {\color{red}M} \rceil\})^2 + \max\{{\color{red}b},\lceil \log {\color{red}M} \rceil\}$ & $2\max\{{\color{red}b},\lceil \log {\color{red}M} \rceil\}+1$ & App.~\ref{sec:comp_G} \\
& $(\text{Arithmetic})^\dagger$ & $5\tilde{n} + 5n^2 + 8{\color{red}b}n + \frac{21}{4}{\color{red}b}^2 + \frac{13}{2}{\color{red}b} - 2n_{\max}(n_{\max}+{\color{red}b}) - 6$ & 0 & App.~\ref{sec:comp_legendre} \\
& $\text{USP}^\dagger_M$ & $0$ & 0 & - \\
$\text{SWAP}_{\v{p_1}, \v{p_2}}$ & & 0 & 0 & - \\
Nucl. mom. & $\v{P} \mathrel{+}= \v{p_1} - \v{p_2}$ & $2\bar{n}$ & $\bar{n}$ & App.~\ref{sec:comp_nuclmom}\\
 & Flag if $\in G$ & 0 & 0 & App.~\ref{sec:comp_nuclmom} \\
 $\PREP_{\NL, 2}^\dagger$ & $(\text{Ineq. test})^\dagger$ & $(1+R)({\color{red}b}+{\color{red}b_{\tilde{M}}})$ & 0 & App.~\ref{sec:comp_NLineq} \\
  & $\text{USP}^\dagger_M$ & $0$ & 0 & - \\
& $\kett{\bar{G}^\alpha_{l,\zeta}} \rightarrow \kett{0}$ & $(1+R)(4 \tilde{n} + 2n^2 + 7{\color{red}b}n + \frac{51}{4}{\color{red}b}^2 + 32{\color{red}b}+ 116 - 2n_{\max}(n_{\max}+{\color{red}b}))$ & 0 & App~\ref{sec:comp_G} \\
& $\kett{\psi_{\tilde{G}}} \rightarrow \kett{0}$ & $(1+2R)(12\tilde{n} + 74n + 4n^2 + 6n{\color{red}b_{\text{pl}}}+6n{\color{red}b_{\text{exp}}}+3{\color{red}b_{\text{rot}}}+8)$ & 0 & App.~\ref{sec:comp_tildeG} \\
 $\PREP_{\NL, 1}^\dagger$ &  & $11Z + 3\lceil \log(9Z) \rceil + 2 {\color{red}b_{\alpha,l}} + {\color{red}b_{\text{keep}}} -8$ & 0 & App.~\ref{sec:comp_alprep} \\
 $\PREP^\dagger_3$ & $\PREP_{3, \ion}^\dagger$ & $4Z + \lceil \log(Z) \rceil({\color{red}b_Z}-3) -2$ & $0$ & App.~\ref{sec:comp_loc_prep_el}\\
  & $\PREP_{3, \el}^\dagger$ & $0$ & $0$ & App.~\ref{sec:comp_loc_prep_el}\\
\hline
\end{tabular}
\end{adjustbox}
\caption{The resource costs for the implementation of the non-local interaction term. The parameter $R$ quantifies the amount of repetition needed for amplitude amplification; for all elements that we explored numerically, $R \leq 3$. The parameters in red are tunable, and must be chosen to satisfy an overall error budget for the block encoding. Other parameters are defined in their linked appendices. Note $Z$ is the total number of pseudoion types.}
\label{tab:qre_nl}
\end{center}
\end{table}

\subsection{Time-evolution}\label{subsec:TimeEvolution}

We perform Hamiltonian simulation via quantum signal processing (QSP).
A summary of the method and the resource cost in terms of number of calls to the iterate is given below. 

\subsubsection{Jacobi-Angers expansion and the cost of implementing the time-evolution}\label{subsubsec:CostOfTimeEvolution}

We follow the construction in Ref.~\cite{low2019hamiltonian} and express the complex exponential with the Jacobi-Angers expansion.
For $|x|\leq 1$,
\begin{align}
e^{i x \tau} & = \cos(x \tau ) + i \sin(x \tau), \\
	\cos (\tau x) &= J_0(\tau) +2\sum_{k >0 : \; \rm{even }}^{\infty} (-1)^{k/2} J_{k}(\tau) T_{k}(x),\\
	\sin (\tau x) &= 2\sum_{k>0: \; \rm{odd}}^{\infty} (-1)^{(k-1)/2} J_{k}(\tau) T_{k}(x).
\end{align}
where $J_k(\tau)$ are the Bessel functions of the first kind and $T_k(x)=\cos(k\cos^{-1}(x))$ are the Chebyshev polynomials. For any fixed $\tau$, the above series are truncated by dropping all terms with $k \geq r$:
\begin{align}
	C_r(x) &= J_0(\tau) +2\sum_{k>0: \; \rm{even}}^{r-1} (-1)^{k/2} J_{k}(\tau) T_{k}(x),\\
	S_r(x) &= 2\sum_{k>0: \; \rm{odd}}^{r-1} (-1)^{(k-1)/2} J_{k}(\tau) T_{k}(x),
\end{align}
In Ref.~\cite{jennings2023efficient} (Lemma 6, v1) it was shown that for any $\tau \in \mathbb{R}$ we can achieve a truncation error
\begin{equation}
		 \max_{x \in [-1,1]} \left | C_r(x) - i S_r(x) - e^{-i  \tau x} \right|\leq \delta
	\end{equation}
 by setting
	\begin{equation}
		r(\tau,\delta) =\left \lceil \frac{|\tau| e}{2} +  \log \left( \frac{c}{\delta} \right) \right \rceil,
	\end{equation}
	where $c =4 ( \sqrt{2 \pi } e^{\frac{1}{13}})^{-1} \approx 1.47762$.\footnote{We note that the upper bound on the degree of the Jacobi-Angers expansion required is likely an overestimate, as an asymptotic analysis suggests that the constant prefactor of the term scaling with $\tau$ may be improved from $e/2$ to $1$~\cite{babbush2019quantum}. However, the non-asymptotic extension is nontrivial and rigorous results are not available. }
A recent result~\cite{berry2024doubling} has shown how to apply Generalized QSP~\cite{motlagh2024generalized} to construct a circuit that applies:
\begin{itemize}
 \item The controlled unitary $\ketbra{0}{0} \otimes W_{sa} + \ketbra{1}{1} \otimes W_{sa}^\dag$, $ r(\lambda t,\delta)$ many times;
\item $r(\lambda t,\delta)$ single qubit rotations, whose cost is negligible (comparatively speaking);
\item The unitary $\ketbra{0}{0} \otimes W^\dag_{sa} + \ketbra{1}{1} \otimes I$ and the unitary $\ketbra{0}{0} \otimes I + \ketbra{1}{1} \otimes W_{sa}$ once.
\end{itemize}
The resulting circuit implements a block-encoding of an operator $X_t$ satisfying
\begin{align}
    \| X_t - e^{-i H t} \| \leq \delta.
\end{align}
The circuit uses $1$ auxiliary qubit. 
Note that $W_{sa}^\dag = U_{sa} [I_s \otimes (2 \ketbra{G}{G}_a - I_a)]$, so we can control between $W_{sa}$ and $W^\dag_{sa}$ by simply controlling the reflection operator on the ancilla:
\small
\begin{align*}
\kett 0\brat 0\otimes W_{sa}+\kett 1\brat 1\otimes W_{sa}^{\dag}=(\kett 0\brat 0\otimes I_{s}\otimes R_{a}+\kett 1\brat 1\otimes I_{s}\otimes I_{a})(I\otimes U_{sa})(\kett 0\brat 0\otimes I_{s}\otimes I_{a}+\kett 1\brat 1\otimes I_{s}\otimes R_{a}),
\end{align*}
\normalsize
where $R_{a}=2\kett G\brat G_{a}-I_{a}$. Hence, the aforementioned circuit block-encoding $X_t$ requires   
\begin{align}
\label{eq:querycostHamSim}
    \left \lceil \frac{|\tau| e}{2} +  \log \left( \frac{c}{\delta} \right) \right \rceil + 2
\end{align}
applications of the unitary $U_{sa}$ (two of which are controlled on the ancilla qubit, i.e., $\kett{0}\brat{0} \otimes U_{sa}$ and $\kett{1}\brat{1} \otimes U_{sa}$, and the rest are simply $U_{sa}$), together with singly-controlled reflections on the ancilla qubits whose cost is negligible compared to the cost of $U_{sa}$.\footnote{Note also that the cost of a singly-controlled $U_{sa}$ is almost as the cost of $U_{sa}$ with no control, due to the fact that the control only needs to be implemented for the Select or the Swap parts of the block-encodings of each Hamiltonian term.}
Furthermore, there are additional gates needed for synthesizing the single-qubit rotations on the QSP's ancilla qubit, which are again negligible, although determining the angles may require a potentially challenging classical precomputation.
Hence, in our resource estimates in Sec.~\ref{sec:resource_estimates} we only account for the cost of the repeated application of $U_{sa}$.

\subsubsection{Starting from an imperfect block-encoding of $H$}
\label{subsubsec:CostOfTimeEvolutionImperfectBE}
In the discussion presented so far, we have assumed access to an exact block-encoding $U_{sa}$ of $H/\lambda$. As in Ref.~\cite{gilyen2018quantum}, let us now lift this assumption and assume that we access instead a unitary $\tilde{U}_{sa}$ which block-encodes an operator $\tilde{H}/\lambda$ with 
\begin{align}
    \| \tilde{H} - H \| \leq \delta_{\textrm{BE}}.
\end{align}
Applying the procedure discussed in the previous subsection with $U_{sa}$ replaced by $\tilde{U}_{sa}$, we block-encode an operator $\tilde{X}_t$ such that
\begin{align}
    \|\tilde{X}_t - e^{-i \tilde{H}t} \| \leq \delta'.
\end{align}
with a number of applications of $\tilde{U}_{sa}$ as in Eq.~\eqref{eq:querycostHamSim}. 
Now, using the triangle inequality, the previous inequality, and Lemma~50 in Ref.~\cite{chakraborty2018power},
\begin{align}
    \|\tilde{X}_t - e^{-i Ht} \| \leq \|\tilde{X}_t - e^{-i \tilde{H} t} \| + \|e^{-i \tilde{H} t} - e^{-i H t} \| \leq \delta' + |t| \| \tilde{H} - H \| \leq \delta' +  \delta_{\textrm{BE}} |t|.  
\end{align}
Hence, we will need $\delta' +  \delta_{\textrm{BE}} |t| = \delta$. For example, we may choose $\delta' = \delta /2$, which can be achieved with 
\begin{align}
\label{eq:querycostHamSimwithApproxBE}
    \left \lceil \frac{|\tau| e}{2} +  \log \left( 2\frac{c}{\delta} \right) \right \rceil + 2
\end{align}
applications of $\tilde{U}_{sa}$ which hence requires, 
\begin{align}
\label{eq:BEerrorrequirement}
    \delta_{\textrm{BE}} = \frac{\delta}{2|t|}.
\end{align}
This guarantees that the output block-encoding is $\delta$-close to $e^{-i H t}$, while in practice we expect much less stringent requirements will arise, due to error cancellations. 

In practice, there are multiple sources of inaccuracy in the compilation that lead to an imperfect block encoding. Indeed, some are highlighted in the tables in Section~\ref{subsec:CircuitsForBE}; all the parameters listed in red are adjustable parameters that constitute finite bit-precision for floating-point operations. In the resource estimates that follow, we try to provide an apples-to-apples comparison with~\cite{berry2023quantum} by adopting the same fixed prescription for precision of arithmetic and uniform state preparation. In principle, however, one should construct a more detailed error propagation and find the choice of parameters that minimizes the computational resources while satisfying a total error of no more than $\delta_{\BE}$. We leave this to future work.

\section{Quantum chemical identification} 
\label{sec:chemical_species_id}

Recalling from Sec.~\ref{sec:info_extraction} that the fingerprints are
described in terms of pseudoion spatial geometry, consider an expansion
of the wavefunction in the position basis,
\[
\kett{\psi}=\sum_{\mathbf{R}_{1},...,\mathbf{R}_{\eta_{\mathrm{ion}}}, \mathbf{r}}\psi(\mathbf{R}_{1},...,\mathbf{R}_{\eta_{\mathrm{ion}}}, \mathbf{r})\kett{\mathbf{R}_{1},...,\mathbf{R}_{\eta_{\mathrm{ion}}}, \mathbf{r}},
\]
with $\mathbf{r}$ collectively denoting all electron coordinates, over which we shall perform no computation.
We define a unitary $U_{X_{\alpha}}$ that performs a coherent implementation
of the fingerprint-based species counter $C_{X_{\alpha}}$
on the coordinates (see Sec.~\ref{sec:info_extraction}) and writes the result in an auxiliary register
encoding non-negative integers:
\[
U_{c}^{X_{\alpha}}\kett{\psi}\kett 0=\sum_{\mathbf{R}_{1},...,\mathbf{R}_{\eta_{\mathrm{ion}}}, \mathbf{r}}\psi(\mathbf{R}_{1},...,\mathbf{R}_{\eta_{\mathrm{ion}}}, \mathbf{r})\kett{\mathbf{R}_{1},...,\mathbf{R}_{\eta_{\mathrm{ion}}}, \mathbf{r}}\kett{C_{X_{\alpha}}(\mathbf{R}_{1},\dots,\mathbf{R}_{\eta_{\mathrm{ion}}})}.
\]
Performing the counts for all listed chemical species $U_{X}=\prod_{\alpha=1}^{M}U_{X_{\alpha}}$
provides the desired species identification.

\subsection{Validating Example}

As a concrete example, we show in this section how the quantum chemical identification (QCI) procedure is implemented for a specific subreaction that often plays a rate-limiting step for the WGS reaction -- the dissociative adsorption of $\mathrm{CO_2}$ into $\mathrm{CO}$ and atomic oxygen on a catalytic surface. We pick an Ir(100) surface for our simulation, as previous investigations based on density-functional theory (DFT) show that this surface has one of the smallest barriers for $\mathrm{CO_2}$ dissociation~\cite{liu2018theoretical}.

\subsubsection{Molecular fingerprinting}

We first identify the desired molecules -- or, more broadly, chemical species $X_\alpha$ -- that shall be identified by our fingerprinting approach, as discussed in Sec.~\ref{sec:info_extraction}. For this simple reaction, it suffices to build a fingerprinting for $\mathrm{CO_2}$ and $\mathrm{CO}$. We must first find computationally inexpensive functions which, given a set of pseudoionic positions $x=(\mathbf{R}_{1},...,\mathbf{R}_{\eta_{\mathrm{ion}}})$, output the features $F_{X_\alpha}(x)$ as defined in Eq.~\eqref{eq:feature_map}. We first estimate the potential energy $E_{X_\alpha}(x)$ of $X_\alpha$ at various ionic configurations $x$ close to the equilibrium structure. For that, we perform tight-binding calculations using the \texttt{xTB} package \cite{bannwarth2021extended} within the \texttt{GFN2-xTB} molecular parameterization~\cite{bannwarth2019gfn2}. For each molecule of interest, we sample a variety of ionic configurations $x$ by performing a molecular dynamics (MD) simulation at 5000~K employing the ASE package~\cite{larsen2017atomic} and an Andersen thermostat~\cite{andersen1980molecular} with a simulation time step of 0.1~fs. For each molecule, we construct a configurational database by extracting 5,000 structures, sampling every 10 frames from the MD simulation.

Next, we train a simple machine-learning (ML) model to predict the energy $E_{X_\alpha}(x)$. For each structure in each dataset, given by a set of ionic positions $x$ and ionic charges $z$, we compute simple feature vectors that can be easily computed by a quantum circuit.
Our approach is based on an extension of a simple molecular descriptor, the  Coulomb matrix~\cite{rupp2012fast}, defined by a matrix $M$ with entries
\begin{equation}
M_{ij}=\left\{
    \begin{matrix}
    0.5 z_i^{2.4} & \text{for } i = j \\
        \frac{z_i z_j}{|x_i - x_j|} & \text{for } i \neq j
    \end{matrix}
    \right.,
\end{equation}
where $i$ and $j$ label ions for each structure. While simple, the Coulomb matrix descriptor has one drawback for our problem: the individual entries of the matrix are not invariant as one swaps different indices associated with equivalent atomic species. Common solutions involve the usage of the eigenvectors of $M$, ordered by their respective eigenvalues, as the feature vectors, or sorting the various columns of $M$ according to their 2-norm. While it is possible to implement such approaches in a quantum circuit, they involve sorting values and incur additional overhead in circuit size and auxiliary qubits.

Here, we propose alternative Coulomb-matrix-type of descriptors that are simple to implement with quantum circuits. We first consider successively higher matrix products $M^p$, where $p \in \mathbb{Z}_{+}$ and compute the corresponding squared Frobenius norms $m_p \equiv \sum_{ij} |(M^p)_{ij}|^2$, which is invariant under atom permutation. We also compute an element-wise inverse of the Coulomb matrix $N$, $N_{ij} \equiv 1/M_{ij}$, and similarly compute the squared norms $n_p \equiv \sum_{ij} |(N^p)_{ij}|^2$. Next, we augment these scalars $m_p$ and $n_p$ with arbitrary powers $q \in \mathbb{Z}_{+}$. We define a feature vector as $u \equiv (m_1,\cdots,m_p,n_1,\cdots,n_p,\cdots,m_1^q,\cdots,m_p^q,n_1^q,\cdots,n_p^q)^T$, with dimension $2 p q$.

Next, we train a logistic regression model to classify  whether a feature vector $u$ is associated with a molecule, an unstable structure, or a different chemical compound, i.e., we build the feature indicator in Eq.~\eqref{eq:feature_indicator}. For our example, we utilize a simple criterion: we consider a structure part of a molecule if its energy per atom is within $E_\mathrm{max}=0.25$~eV of the equilibrium energy. To avoid overfitting, we add L2 regularization to our loss function and perform stratified k-fold cross-validation with $5$ splits. We find it sufficient to use $p=2$ and $q=3$ (i.e., 12 scalars to form the feature vector) to obtain an accurate prediction of various simple molecules (H$_2$O, CO, CO$_2$ and H$_2$), obtaining an area-under-the-curve (AUC) score beyond 99.9\%.

A significant advantage of this approach is that the logistic model is simple to train and tune. A logistic model fits a probability function $p(u) = [1 + \exp(w^T u + b)]^{-1}$ that the feature $u$ is associated with a desired molecule, where the training weights $w \in \mathbb{R}^{2pq}$ and the intercept $b \in \mathbb{R}$ are obtained from training the logistic regression, and where the subscript $X_\alpha$ is implied in all quantities. After we determine the weights, we seek a simpler fingerprinting function $I(u)$ that has a binary output and is easier to evaluate. We hence define $I(u) = \Theta(w^T u + b)$, where the intercept $b$ is obtained from the logistic fitting.

After the logistic fitting is performed, one may also fine-tune the decision criterion of whether a given atomic configuration is part of a molecule. Instead of changing $E_\text{max}$, generating a new dataset, and repeating the logistic regression procedure, one can simply alter the value of the intercept $b$, without changing the weights $w$, to make the fingerprinting function more or less strict. This procedure is particularly convenient because one typically needs to utilize small values of $E_\text{max}$ to ensure physically relevant fingerprinting thresholds. However, small values of $E_\text{max}$ yield a large population imbalance between configurations that are flagged as part of the molecule and those that are not. As regression techniques typically perform worse with large population imbalances, it is much more stable to utilize a larger value of $E_\text{max}$ that roughly divides the atomic configurations into equal populations that are part of and not part of a molecule during the logistic regression procedure, and simply adjust $b$ afterward to fine tune the fingerprinting function to within the desired threshold.

For more complex chemical structures, a different criterion, perhaps not based on energetics alone, can be employed to train fingerprints. Such a 
criterion depends on the desired goals of the simulation: for instance, one can identify unstable structures that are topologically similar to the reference molecule or identify a set of ionic configurations that behave, electronically and/or vibrationally, similar to the reference molecule. In other words, one can classically train fingerprints with a ground truth based on energies and/or topological similarity, depending on the simulation goals. Whatever the chosen ground truth or training strategy, the main requirement is that the results can be distilled into a relatively small number of weights to be combined in a fingerprinting function via simple coherent arithmetic.

\subsubsection{Surrogate classical molecular dynamics calculation}

To test that our fingerprinting approach can detect various molecular species throughout our quantum dynamics simulation, we perform a surrogate classical molecular dynamics for the dissociative adsorption of CO$_2$ on an Ir(100) surface. While such a surrogate classical calculation is not expected to be nearly as accurate as the quantum one we develop in this work, it is suitable to test our QCI approach. That is because at the core of the QCI is a fingerprinting that only involves simple functions of ionic coordinates. On a FTQC, our fingerprinting is applied coherently in superposition on a wavefunction, and classically we can apply fingerprinting on each element of an ensemble of classical trajectories. We simulate one such trajectory here using a simple approximation for the ionic interactions using machine-learned interatomic potentials parametrized for catalytic surfaces~\cite{chanussot2021open}.

We construct a $4\times4$ Ir(100) slab with 3 layers and initialize the atomic velocities according to a Maxwell-Boltzmann distribution at 300~K. Next, we add a CO$_2$ molecule centered around a hollow Ir site, aligning the CO$_2$ major axis parallel to the Ir(100) surface, and imprint a large population of bond-stretching vibrational modes to more quickly simulate a dissociation event. We then perform an MD simulation for $20\,000$ time steps, with a time step of $0.2$~fs. Next, we apply our QCI function on each frame of the MD trajectory, and show the resulting classification in Fig.~\ref{fig:surrogate_sim}, together with a few snapshots of the corresponding molecular configuration. Notably, the fingerprinting approach can successfully detect the breakdown of a CO$_2$ molecule into a CO molecule plus an oxygen atom. For completeness, we also include the parameter vector $w$ and the scalar $b$ obtained from our logistic regression for defining the fingerprinting function $I(u) = \Theta(w^T u + b)$ in Table~\ref{tab:featurization_params}.

\begin{figure}
\centering
\includegraphics[width=0.9\textwidth]{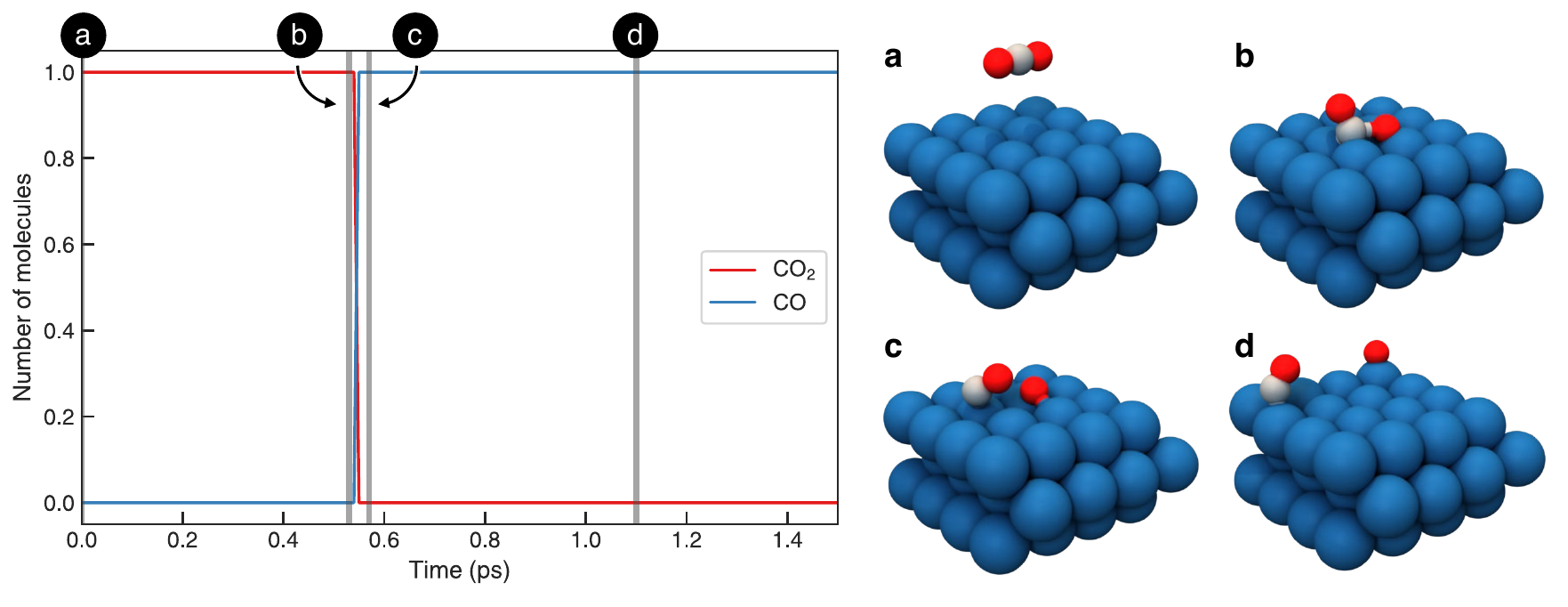}

\caption{(Left) Number of molecules per species identified as per the QCI protocol. Initially, at $t=0$, there is a single vibrationally energetic $\mathrm{CO_2}$ impinging on the Ir(100) surface and no $\mathrm{CO}$. At about $\sim0.55$~ps, the $\mathrm{CO_2}$ count vanishes and the $\mathrm{CO}$ count concurrently rises to unity, indicating the conversion of $\mathrm{CO_2}$ to $\mathrm{CO}$. (Right) Configurational snapshots at selected times (a-d) that show the conversion is indeed related to the dissociative adsorption of $\mathrm{CO_2}$ on the surface.}

\label{fig:surrogate_sim}

\end{figure}

\begin{table}
\centering
\begin{tabular}{c|c|c|c|c}
Parameter & H$_2$O & CO & CO$_2$ & H$_2$ \\
\hline
$w_{0}$ & $-2.050 \times 10^{2}$ & $-1.460 \times 10^{4}$ & $-7.193 \times 10^{4}$ & $\hphantom{-}5.254 \times 10^{-1}$ \\
$w_{1}$ & $\hphantom{-}3.993 \times 10^{1}$ & $-5.155 \times 10^{6}$ & $-3.902 \times 10^{8}$ & $-2.254 \times 10^{-2}$ \\
$w_{2}$ & $\hphantom{-}3.413 \times 10^{-1}$ & $-5.312 \times 10^{-3}$ & $-5.410 \times 10^{-1}$ & $-5.564 \times 10^{0}$ \\
$w_{3}$ & $\hphantom{-}2.696 \times 10^{-3}$ & $-2.744 \times 10^{-7}$ & $-2.894 \times 10^{-5}$ & $-7.897 \times 10^{-1}$ \\
$w_{4}$ & $-3.651 \times 10^{1}$ & $-5.503 \times 10^{6}$ & $\hphantom{-}1.195 \times 10^{8}$ & $-1.189 \times 10^{-1}$ \\
$w_{5}$ & $\hphantom{-}7.643 \times 10^{-2}$ & $-1.074 \times 10^{12}$ & $\hphantom{-}7.602 \times 10^{11}$ & $-2.355 \times 10^{-3}$ \\
$w_{6}$ & $-1.306 \times 10^{-3}$ & $-3.485 \times 10^{-7}$ & $\hphantom{-}4.229 \times 10^{-6}$ & $-4.304 \times 10^{-1}$ \\
$w_{7}$ & $\hphantom{-}8.586 \times 10^{-12}$ & $-2.210 \times 10^{-15}$ & $\hphantom{-}2.407 \times 10^{-14}$ & $\hphantom{-}3.710 \times 10^{-3}$ \\
$w_{8}$ & $-1.366 \times 10^{0}$ & $-2.387 \times 10^{9}$ & $\hphantom{-}1.099 \times 10^{10}$ & $-1.824 \times 10^{-2}$ \\
$w_{9}$ & $-1.840 \times 10^{-5}$ & $-2.350 \times 10^{17}$ & $-1.341 \times 10^{16}$ & $-3.551 \times 10^{-5}$ \\
$w_{10}$ & $-2.565 \times 10^{-7}$ & $-2.858 \times 10^{-11}$ & $\hphantom{-}3.057 \times 10^{-10}$ & $-1.053 \times 10^{-2}$ \\
$w_{11}$ & $\hphantom{-}7.613 \times 10^{-20}$ & $-2.006 \times 10^{-23}$ & $-8.579 \times 10^{-24}$ & $-7.813 \times 10^{-5}$ \\
$b$ & $-7.954 \times 10^{3}$ & $\hphantom{-}8.487 \times 10^{2}$ & $\hphantom{-}2.158 \times 10^{4}$ & $\hphantom{-}1.068 \times 10^{2}$
\end{tabular}
\caption{Parameters obtained from logistic regression for our fingerprinting function $I(u) = \Theta(w^T u + b)$ trained for a few select molecules. The input feature vectors $u$ are defined for atomic positions defined in angstroms, and do not include any scaling or centering around the mean, as is commonly done in machine-learning protocols. The values reported for the intercept $b$ for the CO and CO$_2$ molecules were increased by 500 and 2000 relative to the values directly obtained from the logistic regression to yield a sharper classification, as discussed in the text.}
\label{tab:featurization_params}
\end{table}

\subsection{Sampling and coherent sampling}
\label{sec:coherentsampling}
The initial state preparation presented earlier contains a range of adjustable parameters, including those determining the reaction kinematics (initial positions, velocities, incident angles for each reactant CoM, motifs on the catalyst surface). In many physical situations, we may be interested in knowing the reaction rate suitably averaged over chemically relevant configurations (see Sec.~\ref{sec:generalconsiderationstateprep}). We abstractly define a finite parameter space $\mathcal{S}$, which includes a subset of configuration space (e.g. initial positions/velocities) and other discrete parameters (e.g. different target motifs), over which we want to average. We also assume access to a target probability distribution $p_x$ over the parameter space $\mathcal{S}$. Let $s_x$ denote the probability that a target chemical species of interest is formed after some time $t$,\footnote{A similar argument can be made if we are specifically interested in the number of chemical species that have formed, in which case we'd have different probabilities for different product counts.} given that the system was initialized in a configuration $x\in\mathcal{S}$. The problem is to estimate 
\begin{align}
    s = \sum_{x \in \mathcal{S}} p_x s_x,
\end{align}
up to relative error $\epsilon$,  e.g., $\epsilon = 0.01$, or even $\epsilon\sim1$ if we are we are performing a cursory scan and do not care yet care about high precision.
Formally, we want an estimate $\tilde{s}$ such that 
\begin{align}
\label{eq:estimatetarget}
    | \tilde{s} - s | \leq \epsilon s.
\end{align}
We discuss two different possible solutions.
\subsubsection*{Solution 1: standard averaging}
A standard procedure is as follows:
\begin{itemize}
    \item Sample $x \in \mathcal{S}$ according to probability $p_x$.
    \item Apply the unitary $U_x$ that prepares the initial state $\kett{\psi_x(0)}$ from the all zero state.
    
    \item Run the time-evolution to obtain a state $\kett{\psi_x(t)}$. 
    Then run the chemical species identification components of the algorithm discussed in  
    Sec.~\ref{sec:chemical_species_id}, to obtain the state at time $t$,
    \begin{align}
         \sqrt{s_x} \kett{\psi^{\mathrm{identified}}_x(t)}  \kett{0} + \sqrt{1- s_x} \kett{\psi^{\perp}_x(t)} \kett{0^\perp},
    \end{align}
    where $\kett{\psi^{\textrm{identified}}_x(t)}$ is the normalized state that includes all ionic configurations in the support of $\kett{\psi_x(t)}$ identified as target species, and $\kett{\psi^{\perp}_x(t)}$ is the normalized state including all ionic configurations in the support of $\kett{\psi_x(t)}$ where the target species are identified not to exist. 
    Then, $s_x= |\braket{\psi^{\textrm{identified}}_x(t)}{\psi_x(t)}|^2$. 
    The extra register flags as $\kett{0}$ the identification of a target species and by `not zero' everything else ($\kett{0^\perp}$ is any state satisfying $\braket{0}{0^\perp}=0$). 
    Note that we absorbed phases in the definition of $\kett{0}$, $\kett{0^\perp}$.
    
    \item Measure the extra register, obtaining outcome $0$ with probability $s_x$.
    \item Repeat.
\end{itemize}
Note that we can see $s$ as the average of a Bernoulli variable that takes two values, $1$ when the target chemical species is identified at the final time (which has probability $s=\sum_x p_x s_x$) and $0$ when it is not (which has probability  $\sum_x p_x (1-s_x)$). When averaged over $N_{\textrm{sample}}$ trials, we obtain a random variable with average $s$ and variance $s(1-s)/N_{\textrm{sample}}\leq s/N_{\textrm{sample}}$. By Chebyshev's inequality, the empirical mean $\tilde{s}$ satisfies Eq.~\eqref{eq:estimatetarget} up to a constant failure probability (say, $0.1\%$) as long as
\begin{align}
    N_{\textrm{samples}} =O(s^{-1} \epsilon^{-2}). 
\end{align}
The overall cost of this procedure then scales as
\begin{align}
    O \left(\frac{C_{\textrm{init}}+C_{\textrm{algo}}}{s \epsilon^2} \right),
\end{align}
where $C_{\textrm{init}}$ is the average cost of initializing the initial state and $C_{\textrm{algo}}$ is the cost of running the rest of the algorithm. Note that the scaling with $s^{-1}$ is expected to be a limiting factor, since $s$ will be small in many instances despite our attempts to favorably bias the reaction as discussed in Sec.~\ref{sec:generalconsiderationstateprep}. This can be partially mitigated (at the cost of longer depth) by the following procedure.

\subsubsection*{Solution 2: coherent averaging by amplitude amplification}
The coherent averaging procedure is as follows:
\begin{itemize}
\item Apply a unitary on $O(\log_2 |\mathcal{S}|)$ qubit registers, preparing them in the state
\begin{align}
    \sum^{|\mathcal{S}|-1}_{x=0} \sqrt{p_x} \kett{x},
\end{align}
where $x$ is just an integer label of the configurations in $\mathcal{S}$.
    \item Apply the controlled unitary $\sum_x U_x \otimes \ketbra{x}{x}$ that prepares the initial state
    \begin{align}
       \sum_{x=0}^{|S|-1} \sqrt{p_x} \kett{\psi_x(0)} \kett{x}, 
    \end{align}
from the all zero state.
    \item Run the time-evolution and chemical species identification components of the algorithm, to obtain the state
    \begin{align}
        \sum_x \left(\sqrt{p_x s_x} \kett{\psi^{\mathrm{identified}}_x(t)} \kett{x} \kett{0} + \sqrt{p_x(1- s_x)} \kett{\psi^{\perp}_x(t)} \kett{x} \kett{0^\perp}) \right)= \sqrt{s} \kett{\phi_0} \kett{0} + \sqrt{1-s} \kett{\phi_\perp} \kett{0^\perp}, 
    \end{align}
    where $\kett{\phi_0} = \frac{1}{\sqrt{s}}  \sum_x \sqrt{p_x s_x} \kett{\psi^{\mathrm{identified}}_x(t)} \kett{x}$, $\kett{\phi_\perp} = \frac{1}{\sqrt{1-s}} \sum_x \sqrt{p_x(1- s_x)} \kett{\psi^{\perp}_x(t)} \kett{x}$, where the extra register flags as $\kett{0}$ the identification of a target species and by `not zero' as everything else. 
    \item Perform amplitude estimation to estimate the amplitude of the zero component.
\end{itemize}
Note that in the first step we need to be able to prepare a quantum state that coherently encodes the distribution $p_x$. In the worst case this has a cost $O(|\mathcal{S}|)$, but if $p_x$ has special properties (e.g., if it can be well-approximated by a much simpler reference distribution~\cite{lemieux2024quantum}), then the cost will have better scaling, up to $O(\log_2 |\mathcal{S}|)$. In the second step the initialization unitary has in the worst case a cost $O(|\mathcal{S}| C_{\textrm{init}})$, while in many cases we can expect further savings exploiting the structure of the configuration space. The third step has a cost $C_{\textrm{algo}}$, as before. 

The amplitude amplification step involves performing an algorithm in which all previous unitaries are called a number $N_{\textrm{coh}}$ of times. To bound this number we note that it suffices to take (Theorem 12, \cite{brassard2002quantum}) $\sqrt{s}/N_{\textrm{coh}} \sim \epsilon s$, which means
\begin{align}
    N_{\textrm{coh}} = \mathcal{O}(s^{-1/2} \epsilon^{-1}).
\end{align}
The overall cost of this procedure is then scaling in the worst case as
\begin{align}
    O \left(\frac{|\mathcal{S}| C_{\textrm{init}}+C_{\textrm{algo}}}{\sqrt{s} \epsilon} \right).
    \label{eq:QCI_coherent_scaling}
\end{align}
As we are expecting that $C_{\textrm{init}} \ll C_{\textrm{algo}}$, this coherent averaging procedure is expected to be much more efficient than standard in most cases, if we can afford the longer depth of the resulting circuit.

The quadratically improved scaling compared to the classical counterpart (solution 1) may still be a limiting factor in practice. If, for a reaction of interest, we have for example $s \sim 10^{-6}$, this will still give a $ 10^3 \times$ overhead to the algorithmic cost. This is why it is crucial to implement appropriate biasing techniques to avoid ballooning costs, as discussed in Sec.~\ref{sec:generalconsiderationstateprep}.

\section{Quantum Resource Estimates}\label{sec:resource_estimates}

In this section, we first present asymptotic quantum resource estimates for all parts of our algorithm in terms of the various simulation parameters.
We use $\tilde{O}$ to denote big-O notation which hides multiplicative and additive polylogarithmic factors in the basis size $|G|$ (or $|\overline{G}|$, etc.) and inverse approximation error $1/\delta$.
We then present detailed quantum resource estimates for the time-evolution part - by far the dominant cost - of our quantum algorithm for various problem instances.

\subsubsection*{Asymptotic Cost Analysis}
The cost of our initial state preparation in a worst case scenario, as analyzed in Sec.~\ref{sec:algorithm for initial state preparation}, scales as $\tilde{O}(\eta_{\mathrm{val}}|G|+\eta_{\mathrm{ion}}|\overline{G}|)$
with the two terms arising from using data lookup tables to load the classically-computed (one
to few determinants) electron state and to implement the pseudoion coordinate
transformation, respectively. In practice, however, the linear scaling in terms of $|G|$ can be improved further, especially if one can exploit structure that allows fitting the data lookup tables to more efficiently computable functions, and by employing more efficient state preparation techniques.

Time-evolution requires, as per Eq.~\eqref{eq:querycostHamSimwithApproxBE}, calling the iterate $O(|\tau| + \log(1/\delta))$
many times, where $|\tau|= \lambda |t|$ for a time $t$ (expressed as a real number in atomic units) with rescaling factor $\lambda$, and where $\delta$ is the approximation error for time-evolution.
We shall use the approximate bounds on the rescaling factors from App.~\ref{app:rescaling_bounds},
and consider a cubic lattice for ease of analysis (see relations under Eq.~\eqref{eq:ksq_formula}) such that the maximum momentum $K$ associated to basis set $G$
scales as $K\sim \left(\frac{|G|}{\Omega}\right)^{1/3}$ (and similarly for all
other sets $\overline{G},G^{0},\overline{G}_{\mathrm{trunc}}^{0}$
with maximum momenta $\bar{K},Q,Q_{\mathrm{trunc}}$ respectively). We then have that the total rescaling factor scales as,
\begin{align}
\lambda\sim\tilde{O}\p{\eta_{\mathrm{val}}\left(\frac{|G|}{\Omega}\right)^{2/3}+\eta_{\mathrm{ion}}\left(\frac{|\overline{G}|}{\Omega}\right)^{2/3}+\eta_{\mathrm{val}}^{2}\left(\frac{|G^{0}|}{\Omega}\right)^{1/3}+\eta_{\mathrm{ion}}^{2}\left(\frac{|\overline{G}_{\mathrm{trunc}}^{0}|}{\Omega}\right)^{1/3}+\eta_{\mathrm{val}}\eta_{\mathrm{ion}}}
\label{eq:rescaling_factor_scaling}
\end{align}
where the first two terms arise from $T_{\mathrm{el}},T_{\mathrm{ion}}$,
the second two terms arise from $V_{\mathrm{el}},V_{\mathrm{ion}}^{\mathrm{PI}}$,
and the last term arises from $V_{\mathrm{loc}}^\mathrm{PI},V_{\mathrm{NL}}^\mathrm{PI}$
(which have the same scaling), respectively. 
Furthermore, asymptotically, the cost of implementing the iterate is given by $\tilde{O}(\eta_{\mathrm{val}})$ and $\tilde{O}(\eta_{\mathrm{ion}})$ (e.g. due to SWUPs) with basis scaling $\sim \textrm{polylog}(|G|)$.
Hence, recalling that $\eta=\eta_{\mathrm{val}}+\eta_{\mathrm{ion}}$,
the asymptotically dominant expression for the Toffoli cost of time-evolution is given as,
\small
\begin{align}
\mathrm{Cost}(e^{-iHt}) & \sim\tilde{O}\p{(|t|+\log(1/\delta))\eta\left[\eta_{\mathrm{val}}\left(\frac{|G|}{\Omega}\right)^{2/3}+\eta_{\mathrm{ion}}(\frac{|\overline{G}|}{\Omega})^{2/3}+\eta_{\mathrm{val}}^{2}\left(\frac{|G^{0}|}{\Omega}\right)^{1/3}+\eta_{\mathrm{ion}}^{2}\left(\frac{|\overline{G}_{\mathrm{trunc}}^{0}|}{\Omega}\right)^{1/3}+\eta_{\mathrm{val}}\eta_{\mathrm{ion}}\right]} \nonumber \\
&\sim\tilde{O}\p{(|t|+\log(1/\delta))\left[\eta^{1/3}(\eta_{\mathrm{val}}|G|^{2/3}+\eta_{\mathrm{ion}}|\overline{G}|^{2/3}) + \eta^{2/3}(\eta_{\mathrm{val}}^{2}|G^{0}|^{1/3} +\eta_{\mathrm{ion}}^{2}|\overline{G}_{\mathrm{trunc}}^{0}|^{1/3})+\eta \eta_{\mathrm{val}}\eta_{\mathrm{ion}})\right]} \nonumber  \\
&\sim\tilde{O}\p{(|t|+\log(1/\delta))\left[\eta^{4/3}|G|^{2/3} + \eta^{2/3}\eta_{\mathrm{val}}^{2}|G|^{1/3}+ \eta \eta_{\mathrm{val}}\eta_{\mathrm{ion}})\right]}.
\end{align}
\normalsize
where we recall that $\delta$ is the approximation error for time-evolution. Note that in the second line we assume an extensive scaling of volume
with particle number such that $\Omega\sim\eta$ (e.g., in a slab), and in the last line we use $|G^{0}|\propto|G|$
, $|\bar{G}|\propto|G|$, $|\overline{G}_{\mathrm{trunc}}^{0}|\ll|G|,|\overline{G}|$, and $\eta_{\mathrm{ion}} \leq \eta_{\mathrm{val}}$.
Note that when the pseudoions are not dynamical quantum objects and are considered to be at fixed positions, we have $\eta_{\ion}=0$ and $\eta_{\val}= \eta$, such that one recovers the scaling found in Refs.~\cite{babbush2019sublinear, su2021fault}.

Finally, the cost of coherently averaging over initial states and performing amplitude amplification (ignoring the one-time
classical cost of developing fingerprints) for information extraction is given by Eq.~\eqref{eq:QCI_coherent_scaling}, giving the total complexity of
our algorithm,
\begin{align}
\tilde{O}\left(\frac{|\s S|\eta|G| +(\eta^{4/3}|G|^{2/3}+\eta^{2/3} \eta_{\val}^2|G|^{1/3}+\eta \eta_{\val} \eta_{\ion})(|t|+\log(1/\epsilon))}{\sqrt{s}\epsilon}\right).
\label{eq:total_complexity}
\end{align}
\noindent where we recall that $|\mathcal{S}|$ is the size of the initial parameter space (e.g. initial positions, velocities, etc.) over which we average, $s$ is the total probability that the target chemical species is formed given initial parameter space $\mathcal{S}$, $\epsilon$ is the relative error in estimating $s$, and we set $\delta\sim O(\epsilon)$ to ensure we measure the $s$ up to the desired $\epsilon$.

While it is valuable to obtain such asymptotic cost expressions, they should be used with care. Some subroutines that are omitted as asymptotically subdominant, turn out to dominate for all of the selected problem instances. In particular, while the state preparation scaling with the basis size is asymptotically dominant when generic state preparation is adopted, we will see that for the case where $|\s S| \sim 1$ and typical problem instances its cost is negligible compared to
the second term arising from time-evolution.\footnote{This is not necessarily the case if we are interested in the properties of large ensembles, where $|S|$ is large and the \emph{worst case} state preparation cost is potentially large. In that case, the use of more efficient state preparation routines becomes crucial.} Another clear illustration is seen in Fig.~\ref{fig:compilingDist}, where we observe that the block-encoding is dominated by $\PREP$ subroutines of the non-local terms rather than the $\SWUP$s, while asymptotically the cost of $\PREP$s are ignored due to polylogarithmic scaling in $|G|$.
Therefore, one requires detailed constant-factor numerical resource estimates to understand the real costs of the algorithm.

\subsubsection*{Numerical Quantum Resource Estimates for Problem Instances}

\begin{figure}[t]
    \centering
        \includegraphics[width=0.85\textwidth]{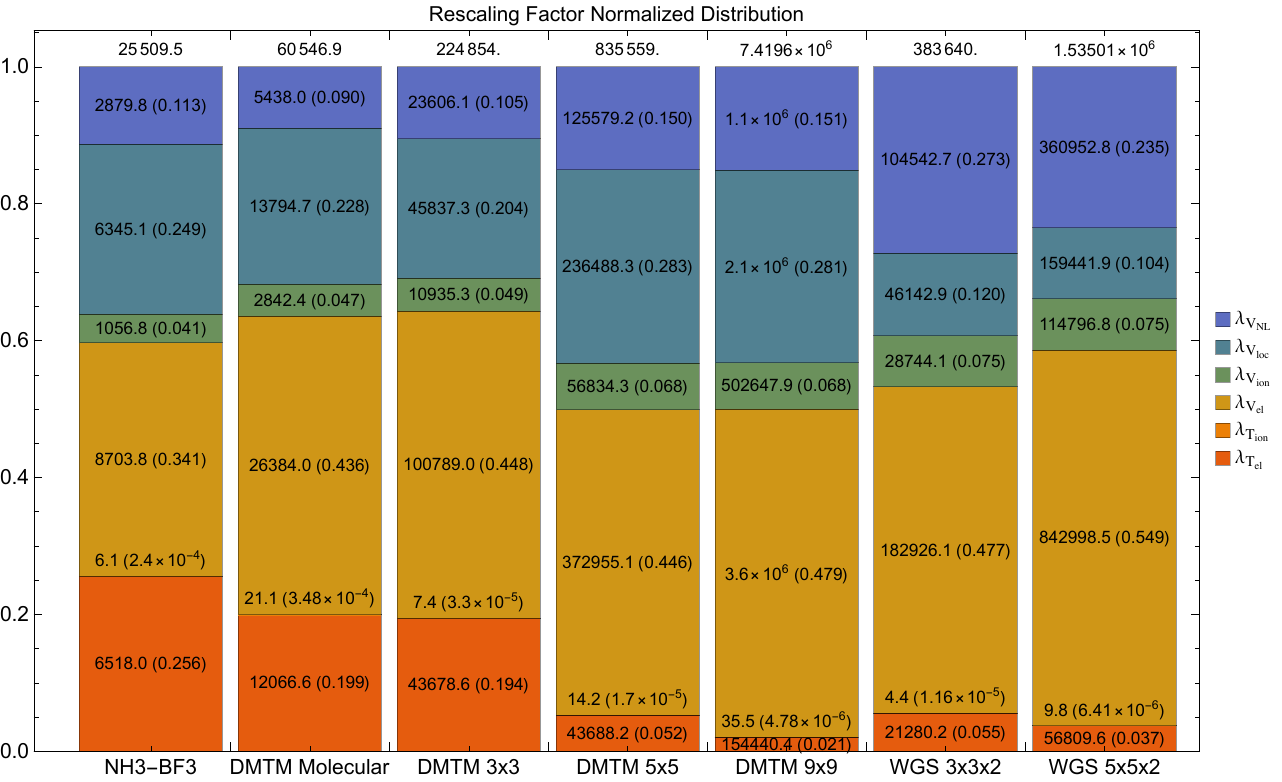}
    \caption{Normalized distribution of rescaling factors for each problem instance. Inside each color bar we show the  values of the rescaling factor for that term, and in parenthesis the relative contribution to the total rescaling (the $\lambda_{T_\mathrm{ion}}$ bar is negligible, so the corresponding number is at the bottom of the $\lambda_{V_\mathrm{el}}$ bar). The total rescaling factor is shown on top of each bar stack.}
            \label{fig:rescalingFactorDist}
\end{figure}
\begin{figure}[h!]
    \centering
    \includegraphics[width=0.85\textwidth]{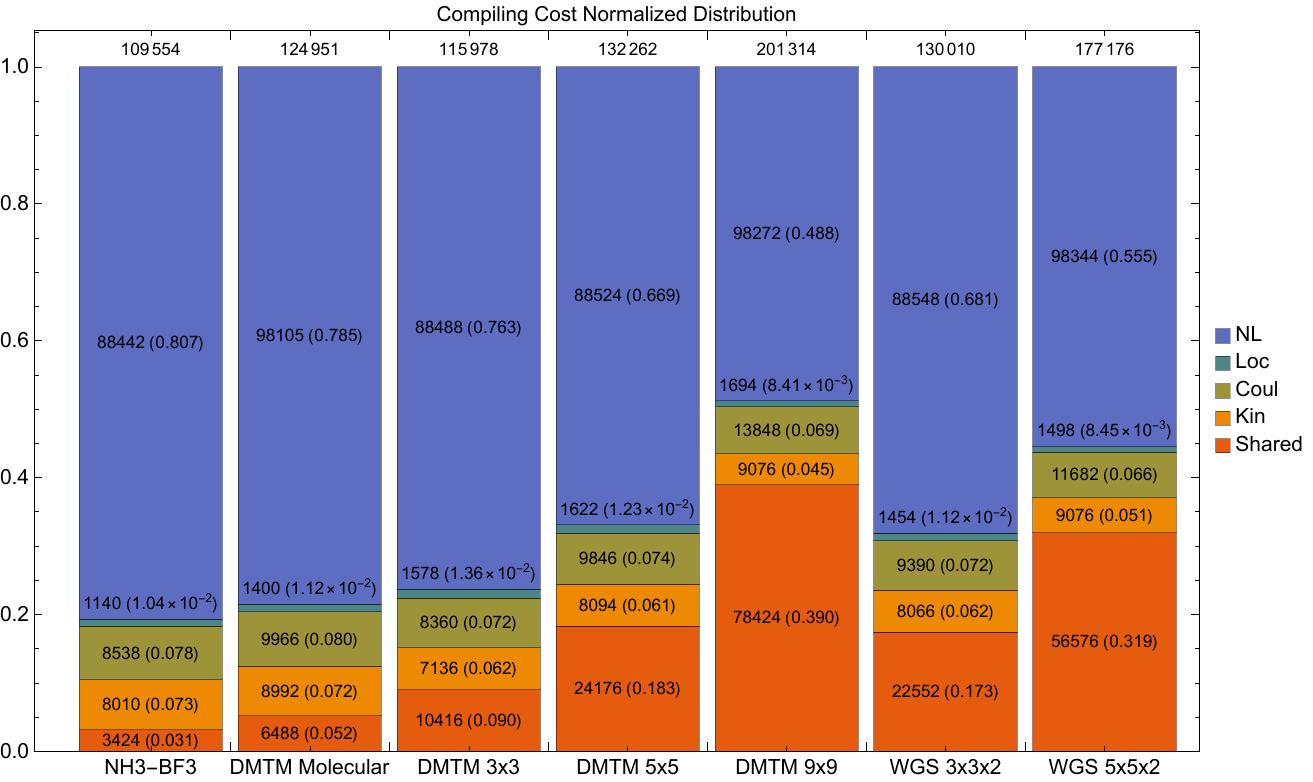}
    \caption{Normalized distribution of compiling costs (in Toffolis) for each problem instance. The numbers inside each color bar indicate the values of the cost for that term and in parenthesis, the relative fraction of the total for that term. The total compilation cost (sum of all global values per term) is shown on top of each bar stack. The numbers above the green bar are for the local term.}
            \label{fig:compilingDist}
\end{figure}
We give a detailed account of quantum resource estimates for the dominant time-evolution part of our quantum algorithm for various problem instances.
Note importantly that the resource estimates we present only refer to implementing one round of time-evolution for a given specified time $t$, and does not include the multiplicative factor $O(1/(\sqrt{s} \epsilon))$ required for amplitude amplification.
We study seven instances of the three classes of problems introduced
in Sec.~\ref{sec:sequence}, Ammonia-Boron trifluoride ($\textrm{NH}_3\textrm{BF}_3$), direct methane to methanol (DMTM) on hexagonal boron nitride with a palladium complex, and water-gas-shift (WGS) on Copper(100). 
Restating Eq.~\eqref{eq:querycostHamSimwithApproxBE} for convenience, the quantum algorithm for implementing time-evolution calls the iterate
\begin{align}
\left \lceil \frac{|\tau| e}{2} +  \log \left( \frac{2c}{\delta} \right) \right \rceil + 2
\end{align}
many times, where 
$c \approx 1.47762$ is a constant (see Sec.~\ref{subsec:TimeEvolution} for details).
Note that for a user-specified precision $\delta$, the number of calls to the iterate is dominated by the first term, which scales linearly with given time $t$, and the total rescaling factor $\lambda$. The latter, an inherent figure of merit for a block-encoding, is hence crucial in determining the overall cost and is studied in Fig.~\ref{fig:rescalingFactorDist}. It is equally as important to understand how much a single implementation of the iterate $W_{sa}$ itself costs, and so its dominant cost, given by the block-encoding $U_{sa}$, is studied in Fig.~\ref{fig:compilingDist}, with a Toffoli cost breakdown shown in detail.
As discussed in Sec.~\ref{subsec:TimeEvolution}, the total gate cost for time-evolution is the cost of the iterate times the number of calls to the iterate. Then, to good approximation, we may simply multiply the cost of $U_{sa}$ times the number of calls to the iterate. These total costs are shown in Fig.~\ref{fig:timeEvoCost}.

We first discuss our results for the rescaling factors. 
In Fig.~\ref{fig:rescalingFactorDist}, we show the distribution of the rescaling
factors, $\lambda_{V_\NL}$, $\lambda_{V_\loc}$, $\lambda_{V_{\el}}$, $\lambda_{V_{\ion}}$, $\lambda_{T_{\el}}$, and $\lambda_{T_{\ion}}$. 
The sum of these numbers determines the total rescaling factor $\lambda$.
Several observations are in order.
First, the cost
of $\lambda_{T_{\mathrm{ion}}}$ is negligible due to suppression
by the pseudoion mass which is a factor of $\sim10^{4}-10^{5}$ in the denominator
of the kinetic term. 
Second, the contributions of $\lambda_{T_{\mathrm{el}}},\lambda_{T_{\mathrm{ion}}}$ relative to
$\lambda_{V_{\mathrm{el}}},\lambda_{V_{\mathrm{ion}}},\lambda_{\mathrm{loc}}
,\lambda_{V_{\mathrm{NL}}}
$
decrease with increasing total number of particles $\eta$, since the kinetic term rescaling factors scales with a lower power of $\eta$ than the rescaling factors for the interactions (see Eq.~\eqref{eq:rescaling_factor_scaling}). 
Third, the largest contribution ($\sim40-50\%$) comes from the Coulomb interactions between valence electrons which (see App.~\ref{app:rescaling_bounds}) scales like the maximum momentum exchange $Q$
in $G^{0}$. 
In contrast, the cost from Coulomb interactions between
pseudoions is significantly smaller as it scales with the cutoff $\overline{\Lambda}_{\mathrm{trunc}}$
and, importantly, not with the maximum momentum $\overline{Q}$
in $\overline{G}^{0}$. 
Fourth, the local and non-local terms capture
a relatively modest fraction of the total rescaling, determined by the structure
of the interactions and HGH parameters, and  the scaling becomes independent
of the basis size in the limit of a large basis, due to the Gaussian decay in momentum
(see App.~\ref{app:rescaling_bounds}).
We also provide per-particle-pair
rescaling factors for the local and non-local terms in Table~\ref{tab:perPairRescalingBig}. 
While there is no steadfast rule, loosely speaking, a higher valence electron count for a pseudoion
increases its per-particle-pair rescaling factors.

In Fig.~\ref{fig:compilingDist}, we analyze the Toffoli cost of each call to the block-encoding unitary, $U_{sa}$, in terms of its constituent components corresponding to the Hamiltonian terms. Note that the quantum circuit as a whole is designed such that gates common to the block-encoding of all Hamiltonian terms are performed only once to reduce cost. These common gates are accounted for as a `shared' cost and do not necessarily perform operations relevant to any single Hamiltonian term (such as calculating the relevant matrix elements into registers), but do account for the gates which combine block-encodings of different Hamiltonian terms via LCU, and also for repeated Hamiltonian parts (e.g. SWUP operations).
Furthermore, $V_\el$ and $V_\ion^\mathrm{PI}$, and $T_{\el}$ and $T_{\ion}$ use common gates, e.g., computing $|\v{q}|^2$ and $|\v{k}|^2$, respectively, and hence the circuits block-encoding these parts are designed optimally with shared gate costs.
For the instances we studied, it is clear that the dominant cost per call is caused by the operations relevant solely to the non-local
term, attributed to arithmetic and reference state preparation (in
roughly equal proportion) for implicit quantum rejection sampling. The next leading cost contribution is from the shared circuitry as explained above, the majority
of which is attributed to SWUPs.

We note that our block-encoding, both the mathematical expressions they are based on and implementation principles considered, differ from those in the literature.
This results in substantially improved rescaling factors for the non-local term, and a mild increase in cost of block-encoding of the iterate.
Overall, in any algorithm where the iterate is called a certain number of times, such as $O(\lambda|t|)$ in a time-evolution or $O(\lambda/\epsilon)$ in energy estimation, this can lead up to an order of magnitude improvement.
We compare our rescaling factors for a selection of pseudoions in Table~\ref{tab:perPairRescalingBig}, to Ref.~\cite{berry2023quantum}'s results (see Table VII of Ref.~\cite{berry2023quantum}).
We observe a $\sim 30-180\times$ factor of improvement for the per electron-ion pair rescaling factor of the non-local terms.
While it is not straightforward to explain this improvement as a product of distinct improvements, an explanation that sheds light to it is as follows.
First, the mathematical expression we use for the operator given as in Eq.~\eqref{eq:NL_term_diagonalized} differs from the one used in Ref.~\cite{berry2023quantum} (see Eq.~(25)).
Ref.~\cite{berry2023quantum}'s block-encoding expresses the non-local term as a sum over momentum exchange in order to easily conform to the same structure as the block-encoding of the Coulomb term. Doing so then allows all of the potential terms to share the same SELECT operation. However, this comes with the cost of performing an additional LCU over momentum exchange, which does not appear in the original mathematical form.
This major difference, combined with our use of a diagonalized form of the non-local term discussed in Sec.~\ref{subsubsec:BENonlocalPP}, leads to $\sim 8-17\times$ improvement - compare the values in Table~\ref{tab:perPairRescalingBig} and Ref.~\cite{berry2023quantum}'s Table~(V) or Table~(VI).
Furthermore, we use implicit quantum rejection sampling (QRS) rather than explicit QRS in our block-encodings, and with optimal reference functions tailored to the target functions.
This can lead to an additional factor of $\sim 5-10\times$ improvement as also observed in Ref.~\cite{berry2023quantum}'s values in their Table~(VI) vs. Table~(VII).
On the other hand, our compilation per block-encoding costs approximately $\sim3-5\times$ more.
This is expected and mostly due to the fact that we perform implicit QRS in our block-encoding rather than explicit QRS as in Ref.~\cite{berry2023quantum}. 
Furthermore, our block-encoding of the local term differs from Ref.~\cite{berry2023quantum}'s only in terms of using implicit QRS rather than explicit QRS, although, in this case, our rescaling factors are similar.
These considerations eventually lead to approximately an improvement of $\sim4-6\times$ if QPE were to be implemented, for the three instances $\mathrm{Pd}_{\mathrm{CO}}3\times3,\mathrm{Pt_{CO}}2\times2,\mathrm{AlN(wurzite)}$ in Ref.~\cite{berry2023quantum}.\footnote{Note that our block-encoding includes pseudoions and electrons as compared to just electrons, which does also increase our cost. However, even with the inclusion of pseudoion degrees of freedom and their corresponding interactions with themselves and the electrons, we still achieve a significant performance gain.}

We note that it is possible that the block-encoding for the non-local terms to be further improved with different ways of partitioning the full non-local interaction or potentially improving the subroutines involved in the current block-encoding; however, we leave this for future work, especially in light
of the fact that the leading rescaling cost arises from electron-electron Coulomb
interactions. 
Finally, we reiterate that our rescaling
and overall costs are significantly lower than an all-electron block-encoding,
due to the use of pseudoions to reduce both particle number and basis
size, and by a judicious choice of cutoffs.

\begin{table}
\centering
\begin{adjustbox}{width=\textwidth,center}
\begin{tabular}{|c|c|c|c|c|c|c|c|c|c|c|c|}
\hline 
 & $\lambda_{\tilde{V}_{\mathrm{loc}}^{I}}$ & $\lambda_{\tilde{V}_{\mathrm{NL}}^{I}}$ &  & $\lambda_{\tilde{V}_{\mathrm{loc}}^{I}}$ & $\lambda_{\tilde{V}_{\mathrm{NL}}^{I}}$ &  & $\lambda_{\tilde{V}_{\mathrm{loc}}^{I}}$ & $\lambda_{\tilde{V}_{\mathrm{NL}}^{I}}$ &  & $\lambda_{\tilde{V}_{\mathrm{loc}}^{I}}$ & $\lambda_{\tilde{V}_{\mathrm{NL}}^{I}}$\tabularnewline
\hline 
\hline 
$\mathrm{H}^{1}$ & $\begin{array}{c}
8.03\\
(8.17)
\end{array}$ & $\begin{array}{c}
0\\
(0)
\end{array}$ & $\mathrm{Al}^{3}$ & $\begin{array}{c}
13.39\\
(13.81)
\end{array}$ & $\begin{array}{c}
12.74\\
(12.59)
\end{array}$ & $\mathrm{Cu}^{1}$ & $\begin{array}{c}
1.24\\
(1.38)
\end{array}$ & $\begin{array}{c}
2.93\\
(2.92)
\end{array}$ & $\mathrm{Ir}^{9}$ & $\begin{array}{c}
20.66\\
(21.92)
\end{array}$ & $\begin{array}{c}
30.61\\
(30.18)
\end{array}$\tabularnewline
\hline 
$\mathrm{B^{3}}$ & $\begin{array}{c}
10.68\\
(11.09)
\end{array}$ & $\begin{array}{c}
6.23\\
(6.23)
\end{array}$ & $\mathrm{Si}^{4}$ & $\begin{array}{c}
14.03\\
(14.59)
\end{array}$ & $\begin{array}{c}
15.39\\
(15.21)
\end{array}$ & $\mathrm{Cu}^{11}$ & $\begin{array}{c}
15.03\\
(16.56)
\end{array}$ & $\begin{array}{c}
74.18\\
(74.29)
\end{array}$ & $\mathrm{Ir}^{17}$ & $\begin{array}{c}
37.01\\
(39.38)
\end{array}$ & $\begin{array}{c}
49.88\\
(49.13)
\end{array}$\tabularnewline
\hline 
$\mathrm{C}^{4}$ & $\begin{array}{c}
17.10\\
(17.66)
\end{array}$ & $\begin{array}{c}
9.52\\
(9.52)
\end{array}$ & $\mathrm{Fe}^{8}$ & $\begin{array}{c}
9.35\\
(10.46)
\end{array}$ & $\begin{array}{c}
54.95\\
(54.87)
\end{array}$ & $\mathrm{Pd}^{10}$ & $\begin{array}{c}
17.20\\
(18.60)
\end{array}$ & $\begin{array}{c}
33.01\\
(32.85)
\end{array}$ & $\mathrm{Pt}^{10}$ & $\begin{array}{c}
22.58\\
(23.98)
\end{array}$ & $\begin{array}{c}
31.23\\
(30.73)
\end{array}$\tabularnewline
\hline 
$\mathrm{N}^{5}$ & $\begin{array}{c}
25.33\\
(26.03)
\end{array}$ & $\begin{array}{c}
13.55\\
(13.55)
\end{array}$ & $\mathrm{Fe}^{16}$ & $\begin{array}{c}
38.62\\
(40.85)
\end{array}$ & $\begin{array}{c}
89.56\\
(91.16)
\end{array}$ & $\mathrm{Pd}^{18}$ & $\begin{array}{c}
49.08\\
(51.59)
\end{array}$ & $\begin{array}{c}
34.16\\
(33.88)
\end{array}$ & $\mathrm{Pt}^{18}$ & $\begin{array}{c}
38.60\\
(41.11)
\end{array}$ & $\begin{array}{c}
46.72\\
(45.91)
\end{array}$\tabularnewline
\hline 
$\mathrm{O^{6}}$ & $\begin{array}{c}
35.08\\
(35.91)
\end{array}$ & $\begin{array}{c}
18.23\\
(18.27)
\end{array}$ & $\mathrm{Ni^{10}}$ & $\begin{array}{c}
12.85\\
(14.25)
\end{array}$ & $\begin{array}{c}
68.71\\
(68.71)
\end{array}$ & $\mathrm{W}^{6}$ & $\begin{array}{c}
9.88\\
(10.72)
\end{array}$ & $\begin{array}{c}
19.63\\
(19.29)
\end{array}$ &  &  & \tabularnewline
\hline 
$\mathrm{F}^{7}$ & $\begin{array}{c}
45.89\\
(46.87)
\end{array}$ & $\begin{array}{c}
23.41\\
(23.58)
\end{array}$ & $\mathrm{Ni^{18}}$ & $\begin{array}{c}
44.83\\
(47.34)
\end{array}$ & $\begin{array}{c}
76.04\\
(76.61)
\end{array}$ & $\mathrm{W}^{14}$ & $\begin{array}{c}
28.94\\
(30.90)
\end{array}$ & $\begin{array}{c}
45.44\\
(44.83)
\end{array}$ &  &  & \tabularnewline
\hline 
\end{tabular}
\end{adjustbox}
\caption{Rescaling factors per pair of electron-pseudoion with the first (second) column for the local (non-local) terms. In each cell, the top number is the exact value and the bottom number in parenthesis is an approximate bound computed as per the discussion in App.~\ref{app:rescaling_bounds}. Note that the numbers technically vary for a given problem instance but the differences in practice are often negligible (a few percent or less).}
\label{tab:perPairRescalingBig}
\end{table}

\section{Conclusion} \label{sec:conclusion}
In this work, we developed a general and practical end-to-end framework to simulate the real-time quantum dynamics of chemical systems, including fully-interacting (beyond Born-Oppenheimer approximation) electronic and nuclear degrees of freedom, in first quantization with a plane wave basis. To achieve this, we first introduced the notion of a pseudoion, which combines chemically inactive (core) electrons and the nucleus into a single point-like ionic object with an effective charge, lifting the well-known $1-$body pseudopotentials from quantum chemistry literature to $2$-body interaction terms.
Second, we provide an initial state preparation protocol, taking into account physical accuracy and algorithmic costs, in order to efficiently and flexibly initialize the quantum state across all motional degrees of freedom - molecular translations, rotations, vibrations, and electronics. Third, to evolve the initial state in time, we construct an efficient block-encoding of a Hamiltonian comprising interacting pseudoions and valence (chemically active) electrons, heavily utilizing quantum rejection sampling methods~\cite{lemieux2024quantum}. 
Finally, we develop a new paradigm for information extraction via Quantum Chemical Identification (QCI), wherein we classically develop and validate chemical fingerprints for molecular identification based on a combination of physical intuition and machine-learning techniques, and show how to implement the fingerprints coherently on the quantum computer utilizing amplitude estimation for efficient readout. 
We stress that we make no physical approximations in our time-evolution besides the use of a finite plane-wave basis and the general usage of pseudoions.\footnote{Other spatially-localized or hybrid bases are certainly directions to consider in the future, although this might require careful consideration as to how localized bases will evolve in time during the evolution.}

The substantial methodological and technical effort to develop this general framework via a detailed analysis of chemical physics, as well as construction and optimization of the quantum circuits involved, is illustrated through
quantum resource estimates for 3 classes (a total of 7 specific problem instances) of problems. These involve both molecular/cluster systems and extended systems discussed in Sec.~\ref{sec:sequence}, with a detailed cost analysis presented in Sec.~\ref{sec:resource_estimates}.
We find costs in the range $10^{11}-10^{14}$ Toffolis per femtosecond of evolution, depending on the use-case  (recall that the timescale for bond reconfiguration in chemistry can range from femtoseconds to picoseconds). 
To these, one needs to add a multiplicative overhead of approximately the inverse square-root of the reaction yield (probability of desired chemical products) - hence we stress the importance of porting classical biasing techniques to the quantum algorithm. These resource estimates are encouraging, since they are not entirely different from those expected from energy estimation in first quantization~\cite{berry2023quantum}, but tackle real-time quantum evolution, a problem for which a more limited suite of classical methods is available and the threshold of classical simulability occurs at smaller problem sizes.\footnote{Specifying a direct classical comparison is difficult since this type of exact dynamics simulation is seldom, if ever, considered in conventional quantum chemistry. However, as discussed at the end of Sec.~\ref{subsec:QCforChemicalDynamics}, we note that full configuration-interaction real-time simulations with additional constraints such as the BO approximation as in ~\cite{lars-hendrik2019ultrafast} are still limited to just a few atoms and for a few femtoseconds. Furthermore, Ref.~\cite{lars-hendrik2019ultrafast} remarks that allowing joint electron and nuclear motion would allow the study of charge transfer and chemical reactions, as we consider in this work.} Nonetheless, we expect that further algorithmic, compiling, and physics optimizations will be needed to determine scientifically interesting and cost-efficient dynamical problems for early-generation FTQC.
For example, one may apply architectural optimizations based on active volume~\cite{litinski2022active} in order to reduce compilation costs that are dominated by arithmetic, as was recently shown in the context of energy estimation~\cite{caesura2025faster}.

The present work is the most in-depth study of the potential of quantum computers to probe entirely new physical regimes via fully quantum mechanical calculations of a molecular system dynamics, with the aim 
to understand (and eventually better control) catalytic chemical processes at a fundamental level.
We consider this as a milestone in a broader research effort,  opening up a range of new venues that we broadly categorize into three themes:  simulating other elementary processes, handling the computational issue of rare events, and the incorporation of additional physics. 

Regarding other elementary processes, in this manuscript, we have focused on studying bond reconfiguration/reaction mechanisms. However, the same concepts may be applied towards studying adsorption and desorption processes. Note that a number of techniques and heuristics coming from less accurate theories may be needed to accelerate the process so that the likelihood of an event is not too low, while retaining the ability to extrapolate the true rates and/or infer binding energies.  A combination of adsorption, bond reconfiguration, and desorption forms the basic set of elementary processes that produces surface chemistry. Our exact quantum dynamical simulations on short timescales might aid in discovering the types of structures that naturally occur due to the fully-coupled motion of electrons and pseudoions, although significant work must be done on how to best synergize our methods with existing classical methods.

Concerning rare events, we indicated that if we simply initialize states according to thermodynamic considerations our quantum algorithm may display low reaction rates, i.e. small wavefunction amplitudes of states corresponding to non-trivial bond reconfigurations, since reaction events rely on fairly unlikely thermodynamic fluctuations. Luckily, the challenge of having to sample from rare events in not an issue specific to quantum algorithms, so one can take inspiration from classical techniques to mitigate this problem. We have already taken steps to avoid inactive configurations in Sec. \ref{sec:generalconsiderationstateprep}, but more can be done to lift classical techniques into the quantum algorithm.

Finally, in terms of additional physics, in this work we have only considered isolated quantum systems, with an `environment' provided only by a limited number catalyst atoms  mimicking the catalyst surface. Real chemical reactions, however, critically involve thermalization and require notions of temperature and pressure, thus necessitating further discussion of how to handle open system effects for realistic modeling. Furthermore, there is increasing scientific interest in photocatalysis and plasmonic catalysis (and also electrochemistry), which additionally require the inclusion of electromagnetic fields. Many reactions also involve radicals and open-shell species that play a crucial role in determining reaction rates. Some of these settings can be tackled within the framework described in this work or simple extensions thereof, but others (e.g. inclusion of open system dynamics) require considerable further modeling and algorithmic development that we leave to future work.

\section*{Author contributions and acknowledgments}
Author and contribution lists are alphabetical.
ML, B\c{S}, KS developed the initial state preparation, block-encoding algorithm, and quantum chemical identification (QCI) protocol.
FJ developed the initial state preparation, provided guidance with chemical physics, developed the QCI protocol, and performed the classical simulations to validate QCI.
SP performed the compilation of the algorithm.
B\c{S}, KS performed the numerical resource estimation.
KS envisioned and managed the project.
All authors contributed to discussing, reading, and writing the manuscript.
We particularly acknowledge helpful discussions with Arvin Kakekhani and Sohrab Ismail-Beigi, and are thankful for collaboration with our colleagues at PsiQuantum on related and other topics.

\bibliographystyle{unsrt}
\bibliography{Bibliography}

\newpage

\appendix 
\part*{Appendices}
\addcontentsline{toc}{part}{III - Appendices}

\section{Basis change from real space to plane waves}
\label{app:basis_change}
We discuss the keys steps required in changing basis from real space
in Eq.~\eqref{eq:H_PI}  to plane waves in Eqs.~\eqref{eq:Tel_plane-wave-elements-valence}, \eqref{eq:Tion_plane-wave-elements-repeated}, \eqref{eq:Vel_plane-wave-elements-valence}, \eqref{eq:VPI_ion_plane_waves}, \eqref{eq:VPI_el-ion_plane_waves_total}, \eqref{eq:local_term_plane_waves}, \eqref{eq:nl_legendre}.
The kinetic terms $T_{\mathrm{el}},T_{\mathrm{ion}}$ are straightforwardly
diagonal in plane waves. The Coulomb terms $V_{\mathrm{el}},V_{\mathrm{ion}}^\mathrm{PI}$
utilize the textbook result $\int_{\mathbf{r}}e^{-i\mathbf{k}\cdot\mathbf{r}}\frac{1}{|\mathbf{r}|}=\frac{4\pi}{|\mathbf{k}|^{2}}$
to change basis. The local and non-local
terms are more involved and, for completeness, we walk through the
calculations carefully.

\subsection*{Local term}

Using the matrix elements given in Eqs.~\eqref{eq:localPP_term_position} and~\eqref{eq:local_term_position},
\small
\[
\mel{\mathbf{r_{2}},\mathbf{R_{2}}}{V_{\mathrm{loc}}^{i,I}}{\mathbf{r_{1}},\mathbf{R_{1}}}=\p{\frac{-Z_{I}}{|\mathbf{r_{1}}-\mathbf{R_{1}}|}\mathrm{erf}(\bar{\lambda}_{\mathrm{loc}}^{I}|\mathbf{r_{1}}-\mathbf{R_{1}}|)+e^{-(\bar{\lambda}_{\mathrm{loc}}|\mathbf{r_{1}}-\mathbf{R_{1}}|)^{2}}\sum_{c=1}^{4}C_{c}^{I}(\sqrt{2}\bar{\lambda}_{\mathrm{loc}}^{I}|\mathbf{r_{1}}-\mathbf{R_{1}}|)^{2(c-1)}}\delta_{\mathbf{r_{1}}\mathbf{r_{2}}}\delta_{\mathbf{R_{1}}\mathbf{R_{2}}},
\]
\normalsize
we move to the plane wave basis in electrons,

\begin{align}
V_{\mathrm{loc}}^{i,I} & =\int_{\mathbf{r_{1}},\mathbf{r_{2}}}\int_{\mathbf{R_{1}},\mathbf{R_{2}}}\mel{\mathbf{r_{2}},\mathbf{R_{2}}}{V_{\mathrm{loc}}^{i,I}}{\mathbf{r_{1}},\mathbf{R_{1}}}\kett{\mathbf{r_{2}},\mathbf{R_{2}}}\brat{\mathbf{r_{1}},\mathbf{R_{1}}}_{i,I}\nonumber \\
 & =\int_{\mathbf{R}}\sum_{\substack{\mathbf{p}\in G,\mathbf{q}\in G^{0}\\
\mathbf{p}-\mathbf{q}\in G
}
}(V_{\mathrm{loc},1}+V_{\mathrm{loc},2})e^{-i\mathbf{k_{q}}\cdot\mathbf{R}}\kett{\mathbf{p}-\mathbf{q},\mathbf{R}}\brat{\mathbf{p},\mathbf{R}}_{i,I},\label{eq:Local_term_mixed-basis_kR-level}
\end{align}
where the two terms are (recalling that $\bar{\lambda}_{\mathrm{loc}}^{I}:=\frac{1}{\sqrt{2}\bar{r}_{\mathrm{loc}}^{I}}$),
\small
\begin{align*}
V_{\mathrm{loc},1} & =\frac{1}{\Omega}\int_{\mathbf{r}}e^{-i\mathbf{k_{q}}\cdot\mathbf{r}}(\frac{-Z_{I}^{\mathrm{PI}}}{|\mathbf{r}|}\mathrm{erf}(\bar{\lambda}_{\mathrm{loc}}^{I}|\mathbf{r}|))=\frac{4\pi}{\Omega}(-\frac{Z_{I}^{\mathrm{PI}}}{|\mathbf{k_{q}}|^{2}}e^{-(|\mathbf{k_{q}}|\bar{r}_{\mathrm{loc}}^{I})^{2}/2}),\\
V_{\mathrm{loc},2} & =\frac{1}{\Omega}\sum_{c=1}^{4}C_{c}^{I}2^{c-1}\int_{\mathbf{r}}e^{-i\mathbf{k_{q}}\cdot\mathbf{r}}e^{-(\bar{\lambda}_{\mathrm{loc}}^{I}|\mathbf{r}|)^{2}}(\bar{\lambda}_{\mathrm{loc}}^{I}|\mathbf{r}|)^{2(c-1)}=\frac{4\pi}{\Omega}\sum_{c=1}^{4}C_{c}^{I}2^{c-1}\frac{1}{|\mathbf{k_{q}}|}\int_{0}^{\infty}drr\sin(|\mathbf{k_{q}}|r)e^{-(\bar{\lambda}_{\mathrm{loc}}^{I}r)^{2}}(\bar{\lambda}_{\mathrm{loc}}^{I}r)^{2(c-1)}.
\end{align*}
\normalsize
For each summand $c=1,2,3,4$ separately,
\small
\begin{align*}
\frac{C_{1}^{I}}{|\mathbf{k_{q}}|}\int_{0}^{\infty}drr\sin(|\mathbf{k_{q}}|r)e^{-(\bar{\lambda}_{\mathrm{loc}}^{I}r)^{2}} & =C_{1}^{I}\sqrt{\frac{\pi}{2}}\bar{r}_{\mathrm{loc}}^{3}e^{-(|\mathbf{k_{q}}|\bar{r}_{\mathrm{loc}})^{2}/2},\\
\frac{2C_{2}^{I}}{|\mathbf{k_{q}}|}\int_{0}^{\infty}drr\sin(|\mathbf{k_{q}}|r)e^{-(\bar{\lambda}_{\mathrm{loc}}^{I}r)^{2}}(\bar{\lambda}_{\mathrm{loc}}^{I}r)^{2} & =C_{2}^{I}\sqrt{\frac{\pi}{2}}e^{-(|\mathbf{k_{q}}|\bar{r}_{\mathrm{loc}})^{2}/2}(3\bar{r}_{\mathrm{loc}}^{3}-\bar{r}_{\mathrm{loc}}^{5}|\mathbf{k_{q}}|^{2}),\\
\frac{4C_{3}^{I}}{|\mathbf{k_{q}}|}\int_{0}^{\infty}drr\sin(|\mathbf{k_{q}}|r)e^{-(\bar{\lambda}_{\mathrm{loc}}^{I}r)^{2}}(\bar{\lambda}_{\mathrm{loc}}^{I}r)^{4} & =C_{3}^{I}\sqrt{\frac{\pi}{2}}e^{-(|\mathbf{k_{q}}|\bar{r}_{\mathrm{loc}})^{2}/2}(15\bar{r}_{\mathrm{loc}}^{3}-10\bar{r}_{\mathrm{loc}}^{5}|\mathbf{k_{q}}|^{2}+\bar{r}_{\mathrm{loc}}^{7}|\mathbf{k_{q}}|^{4}),\\
\frac{8C_{4}^{I}}{|\mathbf{k_{q}}|}\int_{0}^{\infty}drr\sin(|\mathbf{k_{q}}|r)e^{-(\bar{\lambda}_{\mathrm{loc}}^{I}r)^{2}}(\bar{\lambda}_{\mathrm{loc}}^{I}r)^{6} & =C_{4}^{I}\sqrt{\frac{\pi}{2}}e^{-(|\mathbf{k_{q}}|\bar{r}_{\mathrm{loc}})^{2}/2}(105\bar{r}_{\mathrm{loc}}^{3}-105\bar{r}_{\mathrm{loc}}^{5}|\mathbf{k_{q}}|^{2}+21\bar{r}_{\mathrm{loc}}^{7}|\mathbf{k_{q}}|^{4}-\bar{r}_{\mathrm{loc}}^{9}|\mathbf{k_{q}}|^{6}),
\end{align*}
\normalsize
where above we have suppressed the superscript $I$ on $\bar{r}_{\mathrm{loc}}^{I}$
on the RHS for brevity. Consolidating both terms in the overall expression
and noting that $\int_{\mathbf{R}}e^{-i\mathbf{k_{q}}\cdot\mathbf{R}}\kett{\mathbf{R}}\brat{\mathbf{R}}=\sum_{\mathbf{P}}\kett{\mathbf{P}+\mathbf{q}}\brat{\mathbf{P}}$
as a result of $\frac{1}{\Omega}\int_{\mathbf{r}}e^{-i(\mathbf{k}-\mathbf{k}')\cdot\mathbf{r}}=\delta_{\mathbf{k}\mathbf{k}'}$,
we obtain,

\begin{align}
V_{\mathrm{loc}}^{i,I} & =\sum_{\substack{\mathbf{p},\mathbf{P}\in G,\mathbf{q}\in G^{0}\\
\mathbf{p-q},\mathbf{P}+\mathbf{q}\in G
}
}h_{I}^{\mathrm{loc}}(|\mathbf{k_{q}}|\bar{r}_{\mathrm{loc}}^{I})\kett{\mathbf{p-q},\mathbf{P}+\mathbf{q}}\brat{\mathbf{p},\mathbf{P}}_{i,I},\nonumber \\
h_{I}^{\mathrm{loc}}(|\mathbf{k_{q}}|\bar{r}_{\mathrm{loc}}^{I}) & =e^{-(|\mathbf{k_{q}}|\bar{r}_{\mathrm{loc}}^{I})^{2}/2}(-\frac{4\pi(\bar{r}_{\mathrm{loc}}^{_{I}})^{2}Z_{I}^{\mathrm{PI}}}{\Omega(|\mathbf{k_{q}}|\bar{r}_{\mathrm{loc}}^{_{I}})^{2}}\nonumber \\
 & +\frac{4\pi(\bar{r}_{\mathrm{loc}}^{I})^{3}}{\Omega}\sqrt{\frac{\pi}{2}}(C_{1}^{I}+C_{2}^{I}(3-(\bar{r}_{\mathrm{loc}}^{I}|\mathbf{k_{q}}|)^{2})+C_{3}^{I}(15-10(\bar{r}_{\mathrm{loc}}^{I}|\mathbf{k_{q}}|)^{2}+(\bar{r}_{\mathrm{loc}}^{I}|\mathbf{k_{q}}|)^{4})\nonumber \\
 & +C_{4}^{I}(105-105(\bar{r}_{\mathrm{loc}}^{I}|\mathbf{k_{q}}|)^{2}+21(\bar{r}_{\mathrm{loc}}^{I}|\mathbf{k_{q}}|)^{4}-(\bar{r}_{\mathrm{loc}}^{I}|\mathbf{k_{q}}|)^{6}))).\label{eq:Local_term_C_coeffs}
\end{align}
Reorganizing the expression by polynomial degree,

\begin{equation}
h_{I}^{\mathrm{loc}}(|\mathbf{k_{q}}|\bar{r}_{\mathrm{loc}}^{I})=\frac{4\pi(\bar{r}_{\mathrm{loc}}^{I})^{3}}{\Omega}\sqrt{\frac{\pi}{2}}e^{-(|\mathbf{k_{q}}|\bar{r}_{\mathrm{loc}}^{I})^{2}/2}\sum_{s=-1}^{3}c_{s}^{I}(|\mathbf{k_{q}}|\bar{r}_{\mathrm{loc}}^{I})^{2s},\label{eq:Local_term_c_coeffs}
\end{equation}
we have the coefficients,

\begin{align*}
c_{-1}^{I} & =-\sqrt{\frac{2}{\pi}}\frac{Z_{I}^{\mathrm{PI}}}{\bar{r}_{\mathrm{loc}}^{_{I}}},\;c_{0}^{I}=C_{1}^{I}+3C_{2}^{I}+15C_{3}^{I}+105C_{4}^{I},\;c_{1}^{I}=-C_{2}^{I}-10C_{3}^{I}-105C_{4}^{I},\;c_{2}^{I}=C_{3}^{I}+21C_{4}^{I},\;c_{3}^{I}=-C_{4}^{I},
\end{align*}
which is the same as Eq.~\eqref{eq:local_term_plane_waves} with coefficients in Eq.~\eqref{eq:c_coeffs}.
The local term has the simple physical intuition of momentum exchange
between the electron and the pseudoion with elements that dependent
only on the magnitude of the electron momentum of the state (similar
to the Coulomb interaction), with both a Coulomb-like term ($s=-1$)
dominating at $|\mathbf{k_{q}}|<1$ and polynomial terms ($s\ge0$)
dominating at $|\mathbf{k_{q}}|>1$, all under a Gaussian envelope.
Note that for simulation purposes on a finite plane wave basis set,
the pseudoion momentum after exchange is required to be present in
the basis.

\subsection*{Non-local term }\label{subsec:Non-local-term_plane-waves_app}

Using the matrix elements given in Eqs.~\eqref{eq:NLPP_term_position} and \eqref{eq:NL_term_position},

\[
\mel{\mathbf{r_{2}},\mathbf{R_{2}}}{V_{\mathrm{NL}}^{i,I}}{\mathbf{r_{1}},\mathbf{R_{1}}}=\sum_{l=0}^{l_{\mathrm{max}}}\sum_{m=-l}^{l}\sum_{a,b=1}^{3}\qb{\mathbf{r_{2}},\mathbf{R_{2}}}{\zeta_{a}^{I,l,m}}B_{a,b}^{I,l}\qb{\zeta_{b}^{I,l,m}}{\mathbf{r_{1}},\mathbf{R_{1}}}\delta_{\mathbf{R_{1}},\mathbf{R_{2}}},
\]
we move to the plane wave basis in electrons,

\begin{align}
V_{\mathrm{NL}}^{i,I} & =\int_{\mathbf{r_{1}},\mathbf{r_{2}}}\int_{\mathbf{R_{1}},\mathbf{R_{2}}}\kett{\mathbf{r_{2}},\mathbf{R_{2}}}\mel{\mathbf{r_{2}},\mathbf{R_{2}}}{V_{\mathrm{NL}}^{i,I}}{\mathbf{r_{1}},\mathbf{R_{1}}}\brat{\mathbf{r_{1}},\mathbf{R_{1}}}_{i,I}\nonumber \\
 & =\sum_{\mathbf{p_{1}},\mathbf{p_{2}}\in G}\int_{\mathbf{R}}\sum_{l=0}^{l_{\mathrm{max}}}\sum_{m=-l}^{l}\sum_{a,b=1}^{3}B_{a,b}^{I,l}\qb{\mathbf{p_{2}},\mathbf{R}}{\zeta_{a}^{I,l,m}}\qb{\zeta_{b}^{I,l,m}}{\mathbf{p_{1}},\mathbf{R}}\kett{\mathbf{p_{2}},\mathbf{R}}\brat{\mathbf{p_{1}},\mathbf{R}}_{i,I}. \label{eq:NL_term_mixed-basis_zetakR-level}
\end{align}
Using the plane wave expansion $e^{i\mathbf{k_{p}}\cdot\mathbf{r}}=4\pi\sum_{l=0}^{\infty}\sum_{m=-l}^{l}i^{l}j_{l}(|\mathbf{k_{p}}|r)Y_{l}^{m}(\hat{\mathbf{k_{p}}})Y_{l}^{m}(\hat{\mathbf{r}})^{*}$
with $j_l$ denoting the spherical Bessel functions, and the orthonormality of spherical harmonics $\int_{\theta=0}^{\pi}\int_{\phi=0}^{2\pi}\sin\theta d\theta d\phi Y_{l}^{m}(\theta,\phi)Y_{l'}^{m'}(\theta,\phi)^{*}=\delta_{ll'}\delta_{mm'}$,\footnote{We use the convention for spherical harmonics as,

\begin{align*}
Y_{l}^{m}(\theta,\phi) & =(-1)^{m}\sqrt{\frac{(2l+1)}{4\pi}\frac{(l-m)!}{(l+m)!}}P_{l}^{m}(\cos\theta)e^{im\phi}\\
P_{l}^{m}(x) & =2^{l}(1-x^{2})^{m/2}\sum_{k=m}^{l}\frac{k!}{(k-m)!}x^{k-m}\left(\begin{array}{c}
l\\
k
\end{array}\right)\left(\begin{array}{c}
\frac{l+k-1}{2}\\
l
\end{array}\right)
\end{align*}
where the second line defines the associated Legendre polynomials
without the Condon-Shortley phase to avoid double-counting the phase
($m=0$ yields the standard Legendre polynomials).} 
\small
\begin{align}
\qb{\mathbf{p},\mathbf{R}}{\zeta_{a}^{lm}} & :=\int_{\mathbf{r}}\qb{\mathbf{p}}{\mathbf{r}}\qb{\mathbf{r},\mathbf{R}}{\zeta_{a}^{lm}}=\int_{\mathbf{r}}\frac{1}{\sqrt{\Omega}}e^{i\mathbf{k_{p}}\cdot\mathbf{r}}\zeta_{a}^{l}(|\mathbf{r}-\mathbf{R}|)Y_{l}^{m}(\widehat{\mathbf{r}-\mathbf{R}})=\frac{4\pi}{\sqrt{\Omega}}(\bar{r}_{l}^{I})^{\frac{3}{2}}i^{l}e^{i\mathbf{k_{p}}\cdot\mathbf{R}}Y_{l}^{m}(\mathbf{\hat{k}_{p}})\mathrm{g}_{a}^{l}(|\mathbf{k_{p}}|\bar{r}_{l}^{I})\label{eq:zeta_kR_elements}
\end{align}
\normalsize
where we define a real ``radial'' function $\mathrm{g}_{a}^{l}(|\mathbf{k_{p}}|\bar{r}_{l}^{I})$
and explicitly evaluate the integral,

\begin{align}
\mathrm{g}_{a}^{l}(|\mathbf{k_{p}}|\bar{r}_{l}^{I}) & :=\frac{1}{(\bar{r}_{l}^{I})^{\frac{3}{2}}}A_{a,I}^{l}\int_{0}^{\infty}drr^{2}r^{l+2(a-1)}e^{-\frac{1}{2}(\frac{r}{\bar{r}_{l}^{I}})^{2}}j_{l}(|\mathbf{k_{p}}|r)\nonumber \\
 & =\frac{1}{(\bar{r}_{l}^{I})^{\frac{3}{2}}}A_{a,I}^{l}\p{\sqrt{\pi}2^{a-\frac{3}{2}}|\mathbf{k_{p}}|^{l}(\bar{r}_{l}^{I})^{2(a+l+\frac{1}{2})}\frac{\Gamma[a+l+\frac{1}{2}]}{\Gamma[l+\frac{3}{2}]}\;_{1}F_{1}(a+l+\frac{1}{2};l+\frac{3}{2};-\frac{1}{2}(|\mathbf{k_{p}}|\bar{r}_{l}^{I})^{2})}\nonumber \\
 & =e^{-\frac{1}{2}(|\mathbf{k_{p}}|\bar{r}_{l}^{I})^{2}}(|\mathbf{k_{p}}|\bar{r}_{l}^{I})^{l}\p{\frac{\sqrt{\pi}2^{a-1}(a-1)!}{\sqrt{\Gamma(l+2a-\frac{1}{2})}}\mathrm{L}_{a-1}^{l+\frac{1}{2}}(\frac{1}{2}(|\mathbf{k_{p}}|\bar{r}_{l}^{I})^{2})},\label{eq:g_function_2}
\end{align}
where $\Gamma(z)$ is the gamma function and $\;_{1}F_{1}(i;j;z)$
is Kummer's confluent hypergeometric function that can be expressed
in terms of generalized Laguerre polynomials as

\[
\frac{\Gamma(1-i)\Gamma(j)}{\Gamma(j-i)}\mathrm{L}_{-i}^{j-1}(z)=\;_{1}F_{1}(i;j;z).
\]
Employing Kummer's transformation $\;_{1}F_{1}(i;j;z)=e^{z}\;_{1}F_{1}(j-i;j;-z)$
with $i=(l+\frac{1}{2})+a$, $j=(l+\frac{1}{2})+1$, $z=-\frac{1}{2}(k\bar{r}_{l}^{I})^{2}$
yields the final result where the term in parenthesis is a degree
$2(a-1)$ polynomial of $|\mathbf{k_{p}}|\bar{r}_{l}^{I}$. The normalization
of the $\mathrm{g}$-function is $\int_{0}^{\infty}dkk^{2}\mathrm{g}_{a}^{l}(k)^{2}=\frac{\pi}{2}$.
Putting everything together and moving to plane waves in pseudoions
yields,\footnote{The original Ref.~\cite{hartwigsen1998relativistic} (and subsequent work in Ref.~\cite{berry2023quantum}) shows a factor $(-1)^{l}$. We believe this is
erroneous since the factor $i^{l}$ in Eq.~\eqref{eq:zeta_kR_elements}
cancels in Eq.~\eqref{eq:NL_term_mixed-basis_zetakR-level} due to conjugation
since both bra and ket present.}
\small
\[
V_{\mathrm{NL}}^{i,I}=\sum_{\substack{\mathbf{p_{1}},\mathbf{p_{2}},\mathbf{P}\in G\\
\mathbf{P}+\mathbf{p_{1}}-\mathbf{p_{2}}\in G
}
}\sum_{l=0}^{l_{\mathrm{max}}}\sum_{m=-l}^{l}\sum_{a,b=1}^{3}\frac{(4\pi)^{2}(\bar{r}_{l}^{I})^{3}}{\Omega}Y_{l}^{m}(\mathbf{\hat{k}_{p_{2}}})\mathrm{g}_{a}^{l}(|\mathbf{k_{p_{2}}}|\bar{r}_{l}^{I})B_{a,b}^{I,l}\mathrm{g}_{b}^{l}(|\mathbf{k_{p_{1}}}|\bar{r}_{l}^{I})Y_{l}^{m}(\mathbf{\hat{k}_{p_{1}}})^{*}\kett{\mathbf{p_{2}},\mathbf{P}+\mathbf{p_{1}}-\mathbf{p_{2}}}\brat{\mathbf{p_{1}},\mathbf{P}}_{i,I}.
\]
\normalsize
Using the addition theorem $\frac{2l+1}{4\pi}P_{l}(\hat{\mathbf{x}}\cdot\hat{\mathbf{y}})=\sum_{m=-l}^{l}Y_{l}^{m}(\hat{\mathbf{y}})Y_{l}^{m}(\hat{\mathbf{x}})^{*}$,
we obtain Eq.~\eqref{eq:nl_legendre}
(reproduced below),
\small
\begin{align}
V_{\mathrm{NL}}^{i,I} & =\sum_{\substack{\mathbf{p_{1}},\mathbf{p_{2}},\mathbf{P}\in G\\
\mathbf{P}+\mathbf{p_{1}}-\mathbf{p_{2}}\in G
}
}\sum_{a,b=1}^{3}\sum_{l=0}^{l_{\mathrm{max}}}\frac{4\pi}{\Omega}(\bar{r}_{l}^{I})^{3}(2l+1)\mathrm{g}_{a}^{l}(|\mathbf{k_{p_{2}}}|\bar{r}_{l}^{I})B_{a,b}^{I,l}\mathrm{g}_{b}^{l}(|\mathbf{k_{p_{1}}}|\bar{r}_{l}^{I})P_{l}(\mathbf{\hat{k}_{p_{1}}}\cdot\mathbf{\hat{k}_{p_{2}}})\kett{\mathbf{p_{2}},\mathbf{P}+\mathbf{p_{1}}-\mathbf{p_{2}}}\brat{\mathbf{p_{1}},\mathbf{P}}_{i,I}.
\end{align}
\normalsize
The intuition is that, under the constraint of total momentum conservation
between pseudoion and electron, the matrix element connecting electron states
$\mathbf{p_{1}}\rightarrow\mathbf{p_{2}}$ has ``radial'' dependence
on momenta through the overlap of their respective $\mathrm{g}$-functions
($G$-functions in the diagonalized version in Eq.~\eqref{eq:NL_term_diagonalized})
and ``angular'' dependence through the Legendre polynomials $P_{l}$,
i.e. $\mathbf{\hat{k}_{p_{1}}}\cdot\mathbf{\hat{k}_{p_{2}}}$ is the
cosine of the angle between the momenta of the two states. Our block-encoding procedure in Sec.~\ref{subsubsec:BENonlocalPP} exploits this structure.

\section{Detailed physical justification for initial states} \label{app:detailed_physical_justification}

We utilize well-known results in physics to further justify our choice of initial state.

\subsubsection*{Solids}
Many metals are qualitatively described by Fermi liquid theory (Ch. 15 of \cite{girvin2019modern}), which, in short, states that electrons in solids can be treated as effective non-interacting fermionic quasiparticles. For metals, typical Fermi temperatures $T_\mathrm{F}$ are
between $10^{4}-10^{5}K$ (Table 2.1 of \cite{ashcroft1976solid}). With common operating temperatures of many
chemical reactions on the order of a few hundred $K$, we are often in
the regime $T\ll T_\mathrm{F}$. Employing the Sommerfeld expansion (Ch. 7 of \cite{kardar2007statistical}) for a metal with $N$ ions, the electronic energy and heat capacity at constant volume becomes,
\begin{align}
E_{\mathrm{el}}(T\ll T_\mathrm{F}) & \approx\frac{3}{5}Nk_{B}T_\mathrm{F}\p{1+\frac{5}{12}\pi^{2}(\frac{T}{T_\mathrm{F}})^{2}+...},\qquad C_{V,\mathrm{el}}(T\ll T_\mathrm{F})\approx\frac{\pi^{2}}{2}Nk_{B}(\frac{T}{T_\mathrm{F}}).
\end{align}
Only electron quasiparticles with $k_BT$ around $\sim k_BT_\mathrm{F}$
are thermally excited and this represents only a small fraction $\sim\frac{T}{T_\mathrm{F}}\ll1$
of the electron quasiparticles. Hence, to very good approximation,
we can assume the metal is in the ground state of non-interacting
electron quasiparticles, i.e. in an anti-symmetrized product state
of the $N$ lowest quasiparticle modes of the metal. For semiconductors and insulators, the typical electronic band gaps are $E_{\mathrm{gap}}\gtrsim0.5-10\mathrm{eV}$ \cite{strehlow1973compilation} and so the thermal population in the conduction band above the ground state is exponentially small and therefore carries a small fraction of the energy in the system.\footnote{One might comment that several useful properties of semiconductors
come from thermally excited electronic carriers since the bandgaps
are smaller than insulators and molecules. However, the thermal energy
of these carriers compared to the ground state energy is still negligible
at typical reaction temperatures and so our
approximation should hold well for the initial state.} Hence, again to good approximation, the electrons are in the ground state.

The ionic vibrations, aka phonons, for most solids are well-described by the Debye model (Ch. 6 of \cite{girvin2019modern}), which, in short, states that phonons can be treated as non-interacting bosonic quasiparticles with a linear dispersion up to a cutoff known as the Debye energy $\hbar\omega_{D}$. Typical Debye temperatures $T_{D}=\frac{\hbar\omega_{D}}{k_{B}}$ are between $10^{2}-10^{3}K$ (Ch. 23 and front cover of \cite{ashcroft1976solid}).\footnote{For example,
bulk copper has $T_{D}=315K$ and $T_{F}=81600K$.} Hence, for the ionic motion, we have that $T\sim T_{D}$ and therefore a
significant thermal population of (non-interacting) phonons exists.
Therefore, we create a truncated thermal state, i.e. a thermal
state with suitable occupation cutoffs, over phonon modes which have energies in the range of $1-70\mathrm{meV}$ for typical solids \cite{ioffeSemi,NIMS2023phonon}.
A practical requirement is that we do \emph{not} have $T\gg T_{D}$,
since this would imply a very high truncation order and/or signal
the breakdown of the applicability of the Debye model itself, e.g.
due to significant anharmonic phonon interactions.
Exact integral expressions for the energy and specific heat at constant
volume for the Debye model are common knowledge (Ch. 6 in \cite{kardar2007statistical}), but we show the results for $T>T_{D}$ (omitting the zero point energy) which is often the
typical situation,
\begin{align}
E_{\mathrm{ph}}(T>T_{D}) & \approx3Nk_{B}T,\qquad C_{V,\mathrm{ph}}(T>T_{D})\approx3Nk_{B}.
\end{align}
For metals, we see that the ratio of electronic and phononic thermal energies (ignoring energy of the electronic ground state) and specific heats scales as $\sim T/T_\mathrm{F}$. Hence, the electrons contain negligible thermal energy compared to the phonons and any amount of additional thermal energy will almost entirely appear in the phonons. For semiconductors and insulators, due to the sizable band gap, the electronic thermal energy is even more strongly negligible relative to the phonon thermal energy. Implicit in treating electrons and phonons via a pair of non-interacting
quasiparticle models is the underlying assumption that electrons and
phonons are, to good approximation, uncorrelated.
This is consistent with the BO approximation since the bare electron
masses are replaced by electron quasiparticle effective masses which
still continue to be much smaller than the masses of the pseudoions,\footnote{Here, effective mass refers to either the interaction-renormalized mass in Fermi liquid theory or the band effective mass for semiconductors/insulators. Typical effective masses for these quasiparticles, which informs their motional properties,
are between $m^{*}\sim0.01-10$ (in units of the electron mass) (Table 2.3 in \cite{ashcroft1976solid} for metals and Appendix F in \cite{sze2007physics} for semiconductors).} while
pseudoion masses are still $10^{3}-10^{5}$ larger. These facts reemphasize that taking a product state of electron quasiparticles
in the ground state and a truncated thermal state of phonons is an
energetically reasonable approximation to the exact thermal state
of the solid at $t=0$, before the reaction takes place.  Our initial state construction for vibrations and electrons can be
unjustified when the model of electron quasiparticles
breaks down and/or strong electron/phonon correlations arise; these
usually occur in exotic condensed matter phases that are not relevant
to the present scope, or at time $t>0$ when the reaction occurs, which is consistent with our choice of evolving the initial state under the full (beyond-BO) interacting Hamiltonian.

\subsubsection*{Molecules}

Many small to medium-sized molecules have HOMO-LUMO gaps of $E_{\mathrm{gap}}\gtrsim5\mathrm{eV}$.
For example, electronic structure calculations yield gaps of $\mathrm{CO_{2}}:\sim16-20\mathrm{eV}$
and $\mathrm{H_{2}O}:14-18\mathrm{eV}$. Similarly, many small molecules
have a small discrete set vibrational modes of $E_{\mathrm{vib}}\gtrsim50\mathrm{meV}$.
For example, we have the vibrational modes $\mathrm{CO_{2}}:83,165,291\mathrm{meV}$
and $\mathrm{H_{2}O}:198,453,466\mathrm{meV}$ \cite{NISTCCCDB2022}.
We have again that for typical operating temperatures $E_{\mathrm{vib}}\sim k_{B}T\ll E_{\mathrm{gap}}$, and so we keep the same product state form as before.

\section{Reference states for quantum rejection sampling}
\label{app:reference_states_QRS}

We provide the 3 classes of reference states used in quantum rejection
sampling throughout the algorithm. To compute success probabilities
of the target states below (excluding the 1d quantum harmonic oscillator
case), we use the $\mathrm{NH_3BF_3}$ instance, although the results are fairly insensitive to the exact problem instance.

\subsection{Type I (Constant with exponential tail)}

This reference state is used for the state preparation routine $\mathrm{PREP}_\mathrm{loc,2}$
(Eq.~\eqref{eq:s_prep}) for $s\geq0$ in the local term block-encoding,
for the state preparation routine $\PREP_\mathrm{NL,2}$
(Eq.~\eqref{eq:p2_prep}) in the non-local term block-encoding,
and a 1-dimensional version is used for preparation of the quantum
harmonic oscillator eigenstates (Eq.~\eqref{eq:QHO_discretized_eigenstate}). We
construct a piecewise function with two regions, wherein the function
is uniform in the inner region and an exponential tail in the outer
region.

\subsubsection*{Non-local and local terms}

In the case of the non-local term (Eq.~\eqref{eq:reference_func} reproduced here),

\begin{align}
\tilde{G}_{\alpha}^{\zeta,l}(\mathbf{k_{p}}\bar{r}_{l}^{\zeta_{I}})=\begin{cases}
\max_{\mathbf{p}}|G_{\alpha}^{\zeta,l}(|\mathbf{k_{p}}|\bar{r}_{l}^{\zeta})| & \mathrm{for}\;\mathbf{p}\in\lozenge:=\{\mathbf{p}:|k_{\mathbf{p}}^{(j)}\bar{r}_{l}^{\zeta}|\leq(k^{*})_{\alpha}^{\zeta,l}\forall j\}\\
d_{\alpha}^{\zeta,l}e^{-\gamma_{\alpha}^{\zeta,l}||\mathbf{k_{p}}\bar{r}_{l}^{\zeta}||_{1}} & \mathrm{for}\;\mathbf{p}\in G\backslash\lozenge
\end{cases}
\end{align}
where $||\mathbf{k_{p}}||_{1}=\sum_{j}|k_{\mathbf{p}}^{(j)}|$ is the
1-norm of the vector $\mathbf{k_{p}}$ in Cartesian components, i.e.
$k_{\mathbf{p}}^{(j)}=\sum_{\alpha=1}^{3}p_{\alpha}b_{\alpha}^{(j)}$
where $b_{\alpha}^{(j)}$ is the $j$-th Cartesian component of reciprocal
vector $\mathbf{b}_{\alpha}$. Note that this form is indeed a product over
each Cartesian direction making it to easy prepare. For the $s\geq0$
terms, instead of parameters $l,\alpha$, we have parameter $s$ such
that $\tilde{G}_{\alpha}^{\zeta,l}(\mathbf{k_{p}}\bar{r}_{l}^{\zeta})\rightarrow\tilde{G}_{s}^{\zeta}(\mathbf{k_{p}}\bar{r}_{\mathrm{loc}}^{\zeta})$
as seen in Eq.~\eqref{eq:reference_func_loc}, and we have the simple
formula that $\max_{\mathbf{p}}|\tilde{G}_{s}^{\zeta_{}}(\mathbf{k_{p}}\bar{r}_{\mathrm{loc}}^{\zeta})|=(\frac{2s}{e})^{s/2}$
for $s=1,2,3$ and $1$ for $s=0$. For the non-local (local) term,
we choose parameters $k^{*},d$ that have explicit dependence solely
on $l$ ($s$), while $\gamma$ is chosen to be constant. The parameters $k^{*},\gamma$
have implicit dependence on $\zeta$ through $\bar{r}_{l}^{\zeta}$
($\bar{r}_{\mathrm{loc}}^{\zeta}$) for the non-local (local) term.
The parameters are summarized in Table \ref{tab:qrs_reference_params},
where we drop all the parameter labels since they are clear from
context.
\begin{table}
\centering

\begin{tabular}{|r@{\extracolsep{0pt}.}l|r@{\extracolsep{0pt}.}l|r@{\extracolsep{0pt}.}l|r@{\extracolsep{0pt}.}l|r@{\extracolsep{0pt}.}l|r@{\extracolsep{0pt}.}l|r@{\extracolsep{0pt}.}l|r@{\extracolsep{0pt}.}l|r@{\extracolsep{0pt}.}l|}
\cline{3-18}
\multicolumn{2}{c|}{} & \multicolumn{6}{c|}{$l$} & \multicolumn{10}{c|}{$s$}\tabularnewline
\cline{3-18}
\multicolumn{2}{c|}{} & \multicolumn{2}{c|}{0} & \multicolumn{2}{c|}{1} & \multicolumn{2}{c|}{2} & \multicolumn{2}{c|}{-1} & \multicolumn{2}{c|}{0} & \multicolumn{2}{c|}{1} & \multicolumn{2}{c|}{2} & \multicolumn{2}{c|}{3}\tabularnewline
\hline 
\multicolumn{2}{|c|}{$\frac{k^{*}}{(\frac{\pi}{6})^{1/3}\bar{r}}$} & 1&4 & 1&8 & 2&2 & \multicolumn{2}{c|}{1} & \multicolumn{2}{c|}{1} & 2&1 & 2&7 & 3&1\tabularnewline
\hline 
\multicolumn{2}{|c|}{$\sqrt{3}\bar{r}\gamma$} & 1&5 & 1&5 & 1&5 & 1&69 & \multicolumn{2}{c|}{1} & \multicolumn{2}{c|}{1} & \multicolumn{2}{c|}{1} & \multicolumn{2}{c|}{1}\tabularnewline
\hline 
\multicolumn{2}{|c|}{$d$} & \multicolumn{2}{c|}{7} & \multicolumn{2}{c|}{10} & \multicolumn{2}{c|}{12} & \multicolumn{2}{c|}{$e^{1.6}$} & \multicolumn{2}{c|}{$e^{1.02}$} & \multicolumn{2}{c|}{$e^{1.9}$} & \multicolumn{2}{c|}{$e^{3.0}$} & \multicolumn{2}{c|}{$e^{4.25}$}\tabularnewline
\hline 
\end{tabular}

\caption{Parameters for the Type I and Type III reference states used for rejection
sampling inside the block-encodings of the non-local and local terms
Note that $\bar{r}$ indicates $\bar{r}_{l}^{\zeta_{I}},\bar{r}_{\mathrm{loc}}^{\zeta_{I}}$
for the non-local and local terms respectively.}
\label{tab:qrs_reference_params}
\end{table}
\begin{figure}
    \centering
    \begin{subfigure}{0.55\textwidth} 
        \centering
        \includegraphics[width=\textwidth]{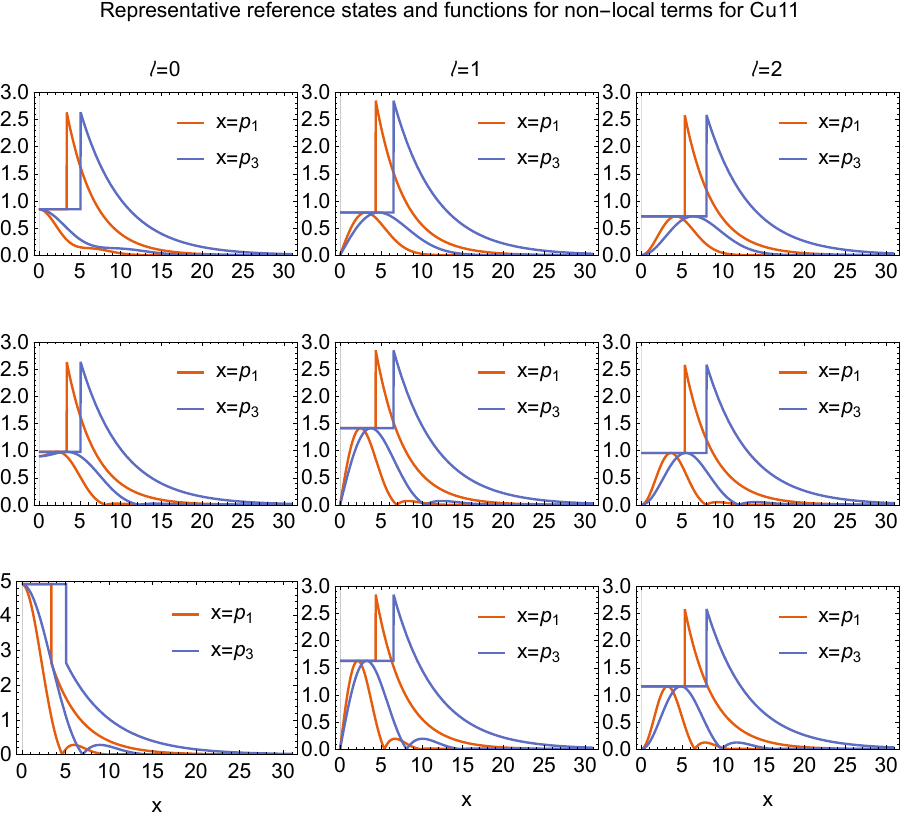}
        \caption{Non-local terms}
        \label{fig:exampleRefStatesBENonlocal}
    \end{subfigure}
    \hfill
    \begin{subfigure}{0.4\textwidth} 
        \centering
        \includegraphics[width=\textwidth]{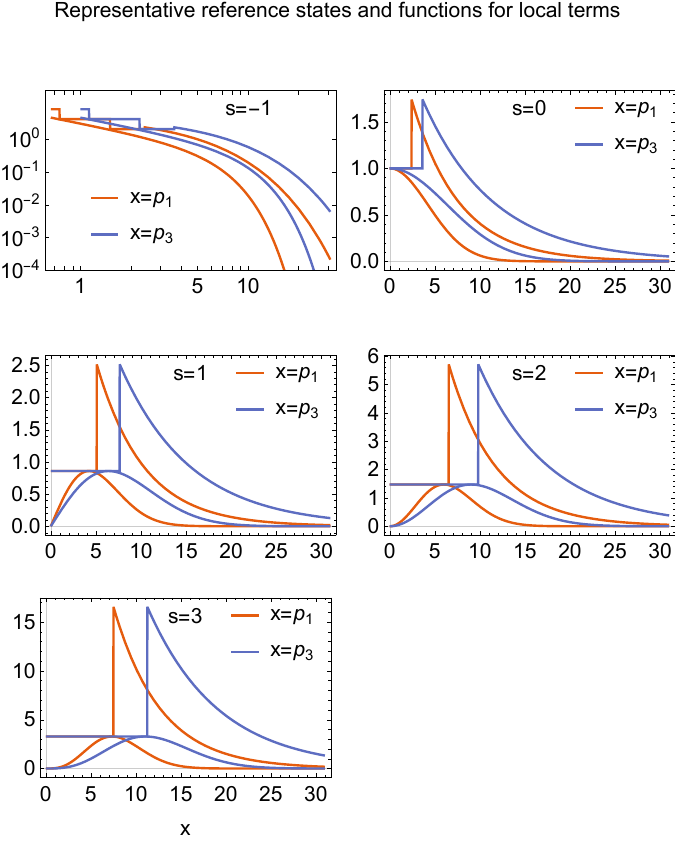}
        \caption{Local terms}
        \label{fig:exampleRefStatesBELocal}
    \end{subfigure}

    \caption{Examples of the unnormalized Type I (and Type III for $s=-1$) reference states and modulus of the target functions. The non-local terms (a) and local terms (b) are shown for the $\mathrm{NH_3BF_3}$ simulation instance. Specifically, we show the reference states along $(x,0,0)\bar{r}$ and $(0,0,x)\bar{r}$, i.e., along the $\bar{r}p_1$ (blue) and $\bar{r}p_3$ (red) axes (for ease, we set $\bar{r}=1$), respectively, for $l=0,1,2$ for the non-local terms for $\mathrm{Cu^{11}}$ and for the local terms $s=-1,0,1,2,3$. For each $l$, the columns are the 3 eigenstates labeled by $\alpha=1,2,3$.}
    \label{fig:exampleRefStatesBE}
\end{figure}
In Fig.~\ref{fig:exampleRefStatesBE}, we show the (unnormalized) Type I
reference states (for non-local and $s\geq0$) bounding the target
functions along the $p_{1}$ and $p_{3}$ axes for the pseudoion $\zeta=\mathrm{Cu^{11}}$.
On the left hand side, columns represent $l$-values and rows represent
the eigenstate values $\alpha=1,2,3$. On the right hand side, the
same cuts for $s\geq0$ are shown. While the reference function
may not always appear to tightly bind the target function along any
specific cut for a given set of parameters, the overall probability
of success, approximately given by the ratio of the volume integrals
of the squared functions,
\begin{align}
p_{\mathrm{succ}} & =\frac{\sum_{\mathbf{p}\in G}G_{\alpha}^{\zeta,l}(|\mathbf{k_{p}}|\bar{r}_{l}^{\zeta_{}})^{2}}{\sum_{\mathbf{p}\in G}\tilde{G}_{\alpha}^{\zeta,l}(\mathbf{k_{p}}\bar{r}_{l}^{\zeta_{}})^{2}}\approx\frac{\int_{\mathbf{p}\in G}G_{\alpha}^{\zeta,l}(|\mathbf{k_{p}}|\bar{r}_{l}^{\zeta})^{2}}{\int_{\mathbf{p}\in G}\tilde{G}_{\alpha}^{\zeta,l}(\mathbf{k_{p}}\bar{r}_{l}^{\zeta})^{2}}
\end{align}
\begin{figure}
\centering
\includegraphics[width=0.8\textwidth]{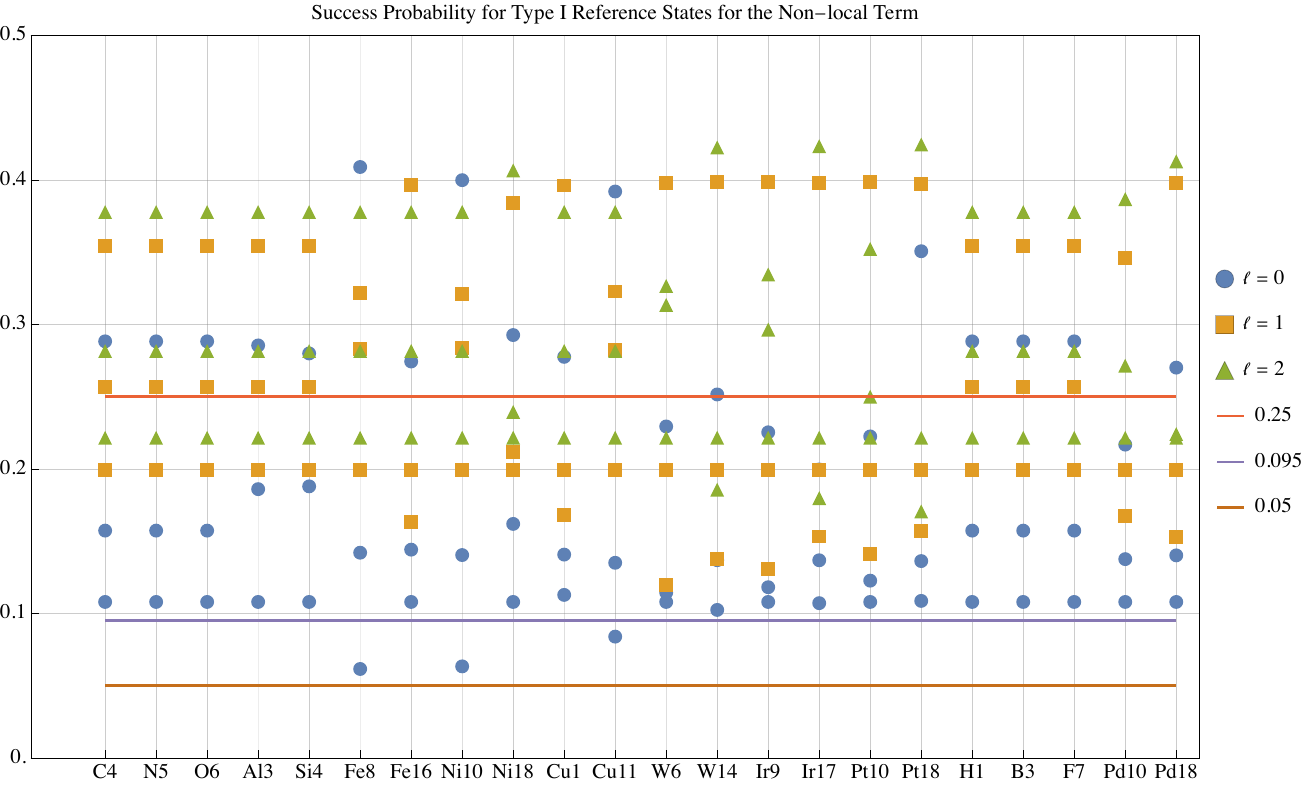}
\caption{Success probabilities for the Type I reference state in the non-local term $\tilde{G}^{\zeta,l}_\alpha$ for a selection of HGH pseudoions (for the $\mathrm{NH_3BF_3}$ instance, although results are insensitive to exact problem instance). Colors indicate the parameter $l$ and the 3 markers per color indicate the 3 eigenstates denoted by parameter $\alpha$. The successively lower horizontal lines indicate the threshold probabilities for $1,2,3$ rounds of amplification. Note that very few cases fall below the 2 round threshold.}
\label{fig:succProbNL}
\end{figure}
are sufficiently large to ensure that for most pseudoions $\zeta$
and cases $l,\alpha$, only 1-2 rounds of amplification are required
to restore the full norm of the target function. This is apparent
in Fig.~\ref{fig:succProbNL}, where we show the success probabilities
of preparing the target function for a substantial selection of pseudoions
(though not an exhaustive list of all HGH-based pseudoions), where the $0.25,0.095,0.05$ lines denote
the threshold probabilities for $1,2,3$ rounds of amplifications
respectively, and where the blue, orange, green markers denote the
$l=0,1,2$ cases respectively, with the 3 points per color denoting
the 3 eigenstates. Only in a few cases does the success probability
fall below the 2 round threshold. The results suggest that with a
little fine-tuning of the reference state parameters (by incorporating
some $\alpha$-dependence) to address low probability cases, one can
always construct a Type I reference state above the 2 round threshold
for all HGH pseudoions. This is what we assume in resource estimation. For the $s\geq0$ local terms, the
success probabilities are $0.45,0.41,0.36,0.31$ for $s=0,1,2,3$
respectively, and so only require 1 round of amplification.

\subsubsection*{Quantum harmonic oscillator}

In non-dimensionalized shape coordinates $\bar{q}^{s,k}=\sqrt{\omega_{s,k}}q^{s,k}$,
the single-mode eigenstates of a quantum harmonic oscillator in Eq.~\eqref{eq:QHO_discretized_eigenstate} has the target function,
\begin{equation}
\phi_{l}^{s,k}(\bar{q}^{s,k})=e^{-\frac{1}{2}(\bar{q}^{s,k})^{2}}\mathrm{H}_{l}(\bar{q}^{s,k})
\end{equation}
where $\mathrm{H}_{l}(x)$ is the $l$-th Hermite polynomial. We provide
a 1-dimensional Type I reference function $\tilde{\phi}_{l}^{s,k}(\bar{q}^{s,k})$
to explicitly construct the unitary $U_{l}^{s,k}$ below Eq.~\eqref{eq:QHO_discretized_eigenstate} using rejection sampling,

\begin{equation}
\tilde{\phi}_{l}^{s,k}(\bar{q}^{s,k})=\begin{cases}
\max_{\bar{q}_{k}}|\phi_{l}^{s,k}(\bar{q}^{s,k})| & \mathrm{for}\;|\bar{q}^{s,k}|\leq\bar{q}_{l}^{*}\\
d_{l}e^{-\gamma_{l}|\bar{q}^{s,k}|} & \mathrm{for}\;|\bar{q}^{s,k}|>\bar{q}_{l}^{*}
\end{cases}
\end{equation}
where, for $l\geq1$, the parameters $\gamma_{l}=-\frac{1}{\phi_{l}^{s,k}(\bar{q}_{l}^{*})}\frac{\d\phi_{l}^{s,k}}{\d\bar{q}^{s,k}}|_{\bar{q}^{s,k}=\bar{q}_{l}^{*}}$
and $d_{l}=e^{\alpha_{l}\bar{q}_{l}^{*}}\phi_{l}^{s,k}(\bar{q}_{l}^{*})$
are given, respectively, by matching derivatives and values of the
exponential tail $d_{l}e^{-\gamma_{l}\bar{q}^{s,k}}$ and the target
function $\phi_{l}^{s,k}(\bar{q}^{s,k})$ at the classical turning
point $\bar{q}_{l}^{*}=\sqrt{2l+1}$. For $l=0$, we use the same
reference state with $\gamma_{0}=-\frac{1}{2},d_{0}=\pi^{-1/4},\bar{q}_{0}^{*}=1$.

\begin{figure}
    \centering
    \begin{subfigure}{0.58\textwidth} 
        \centering
        \includegraphics[width=\textwidth]{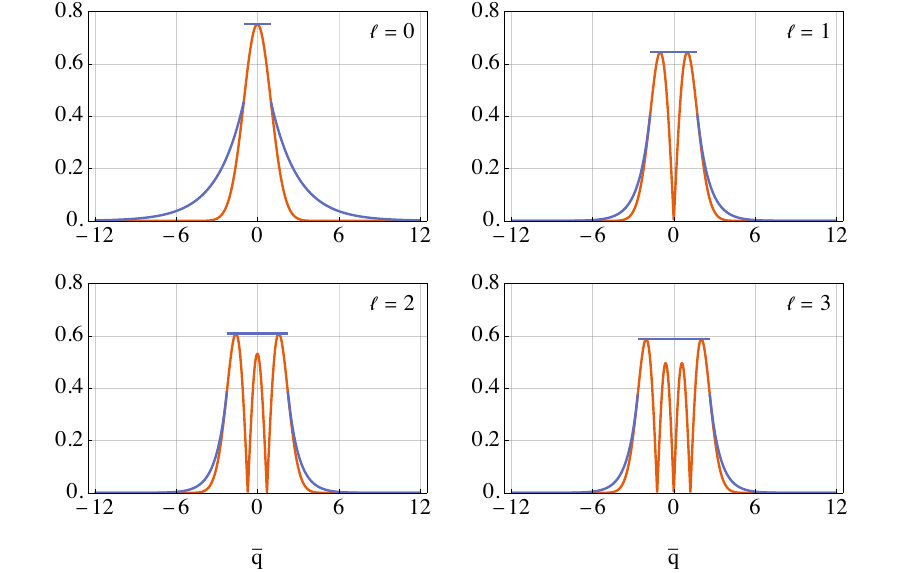}
        \caption{Reference states $\tilde{\phi}_{l}^{s,k}(\bar{q}^{s,k})$ (blue) and functions $|\phi_{l}^{s,k}(\bar{q}^{s,k})|$ (red) for the first four harmonic oscillator harmonics $l=0,1,2,3$.}
        \label{fig:QHO_harmonic_refs}
    \end{subfigure}
    \hfill
    \begin{subfigure}{0.38\textwidth} 
        \centering
        \includegraphics[width=\textwidth]{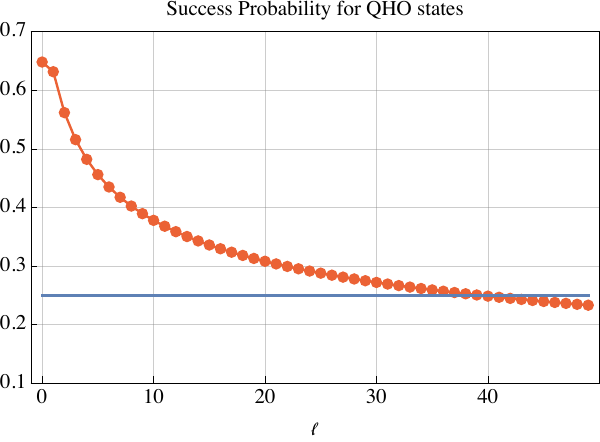}
        \caption{Success probability using references $\tilde{\phi}_{l}^{s,k}(\bar{q}^{s,k})$ for functions $|\phi_{l}^{s,k}(\bar{q}^{s,k})|$ as a function of the harmonic index $l$. The blue line is at $0.25$, indicating the threshold for one round of amplification.}
        \label{fig:QHO_succ_prob}
    \end{subfigure}

    \caption{Comparison of reference states and success probabilities for quantum harmonic oscillators.}
    \label{fig:QHO_harmonics}
\end{figure}

The success probability is shown in Fig.~\ref{fig:QHO_harmonics} where
we see that the first $40$ harmonics $l=0,...,39$ all require $1$
round of amplitude amplification, which suffices for practical purposes
relevant to this work. Should higher harmonics be desired, either
one may use the same reference state with more rounds of amplification
or alternatively construct a more refined reference, e.g. with more
piecewise constant levels below the max value which works well for
the internal region inside the classical turning points, to bring
the success probability back above the $1$ round threshold. As with
all initial state preparation subroutines, this preparation only occurs
once to initialize the time-evolution for a given problem instance,
and hence the resource cost is negligible compared to the time-evolution
and we omit precise compilation and numerical resource estimation
here.

\subsection{Type II (Power-law ladder)}

This reference state is used for the state preparation routine $\PREP_\mathrm{coul,el},\PREP_\mathrm{coul,ion}$
(Eq.~\eqref{eq:PREP_V}) in the Coulomb term block-encoding.
Defining $R_{\mathbf{p}}:=\max_{j}|k_{\mathbf{p}}^{(j)}|$,

\[
\tilde{G}(\mathbf{k_{p}})=\begin{cases}
0 & \mathrm{for}\;R_{\mathbf{p}}\in[0,\Lambda_{\mathrm{IR}})\\
2^{1-\mu(\mathbf{k_{p}})} & \mathrm{for}\;R_{\mathbf{p}}\geq\Lambda_{\mathrm{IR}}
\end{cases}
\]
where $\mu(\mathbf{k_{p}})=1+\left\lfloor \log_{2}(R_{\mathbf{p}})\right\rfloor $
is an index label for all piecewise constant levels that form a ``ladder''
bounding the power law amplitudes of the target state $\frac{1}{|\mathbf{k_{p}}|}$,
and $\Lambda_{\mathrm{IR}}=\min_{\mathbf{p},\mathbf{p}\ne0}|\mathbf{k_{p}}|$
is a cutoff to avoid the singularity at $\mathbf{p}=0$. The success
probability is approximately $0.31$ for the $\mathrm{NH_3BF_3}$ simulation instance,
indicating only 1 round of amplification.\footnote{In the limit of large basis sizes, footnote 2 in Ref.~\cite{lemieux2024quantum}
with $d=3$ and $x=3$ gives an analytical integral estimate using
box integrals for the success probability as $0.274$, close to the
numerical value found. We note that this
analytical value further differs from the $0.239$ value computed
in the earlier Ref.~\cite{babbush2019sublinear}  which is lower by factor
of $\frac{1}{(1-\frac{1}{2^{3}})}$ due to an imperfect (but efficient)
preparation of the reference state in their circuit construction.}

\subsection{Type III (Power-law ladder with exponential tail)}

This reference state is used for the state preparation routine $\mathrm{PREP}_\mathrm{loc,2}$
(Eq.~\eqref{eq:s_prep}) for $s=-1$ in the local term block-encoding.
Defining $R_{\mathbf{p}}^{\zeta}:=\max_{j}|k_{\mathbf{p}}^{(j)}\bar{r}_{\mathrm{loc}}^{\zeta}|$
(similar to the Type II state),

\[
\tilde{G}_{s=-1}^{\zeta}(\mathbf{k_{p}}\bar{r}_{\mathrm{loc}}^{\zeta_{}})=\begin{cases}
0 & R_{\mathbf{p}}^{\zeta_{}}\in[0,\Lambda_{\mathrm{IR}})\\
2^{1-\mu(\mathbf{k_{p}}\bar{r}_{\mathrm{loc}}^{\zeta})} & \mathrm{for}\;R_{\mathbf{p}}^{\zeta}\in[\Lambda_{\mathrm{IR}},k_{s=-1}^{*})\\
d_{s=-1}e^{-\gamma_{\alpha}^{\zeta}||\mathbf{k_{p}}\bar{r}_{\mathrm{loc}}^{\zeta}||_{1}} & \mathrm{for}\;R_{\mathbf{p}}^{\zeta}\geq k_{s=-1}^{*}
\end{cases}
\]
where as before, $\mu(\mathbf{k_{p}}\bar{r}_{\mathrm{loc}}^{\zeta_{}})=1+\left\lfloor \log_{2}(R_{\mathbf{p}}^{\zeta_{}})\right\rfloor $
is an index label for all piecewise constant levels that form the
``ladder,'' and $\Lambda_{\mathrm{IR}}=\min_{\mathbf{p},\mathbf{p}\ne0}|\mathbf{k_{p}}|$
is the lower cutoff. The same cuts as before are shown on the right
hand side of Fig.~\ref{fig:exampleRefStatesBE}. The success probability
for the $\mathrm{NH_3BF_3}$ instance discussed earlier is approximately
$0.29$, also indicating only 1 round of amplification (and that the
power law in the interior region dominates the success probability
relative to the tail, hence giving a similar number to the Type II
case).

\section{Approximate rescaling factor bounds}
\label{app:rescaling_bounds}
We find easily-computable approximate upper bounds on the rescaling factors for each
of the pairwise interaction block-encoding terms to obtain a rough estimate of query complexity in the time-evolution.
Recall that $\mathbf{k}=\s B\mathbf{p}$ where $\s B$ is an affine
transformation built from the reciprocal lattice vectors and that
$\Omega\det\s B=(2\pi)^{3}$ where $\Omega$ is the real-space simulation
cell volume. In general, as we will see below, the
rescaling factor bounds on the local and non-local terms are independent
of the simulation cell and basis size while the rescaling factor bounds
on the Coulomb term (and exact rescaling factor for the kinetic term)
only depend on the maximum norm of the momentum exchange (momentum
for the kinetic term). This makes the bounds easy to compute as rough
estimates for the total rescaling factor without having to generate
a large basis and numerically sum many terms as is required in the
exact formulas.

\subsubsection*{Coulomb Term}

We approximately bound the below sum with an integral, and further
circumscribe the integration region $\diamond^{0}:=\{\mathbf{k}:\mathbf{k}\in\s BG^{0}\}$
with a sphere of radius $Q=\max_{\mathbf{q}\in\diamond^{0}}|\mathbf{q}|$.

\begin{equation}
\frac{1}{\Omega}\sum_{\mathbf{q}\in G^{0}}\frac{1}{|\mathbf{k_{q}}|^{2}}\lesssim\frac{1}{\Omega\det\s B}\int_{\mathbf{q}\in\diamond^{0}}d^{3}q\frac{1}{|\mathbf{q}|^{2}}\leq\frac{4\pi}{\Omega\det\s B}\int_{0}^{Q}dqq^{2}\frac{1}{q^{2}}=\frac{4\pi}{\Omega\det\s B}Q=\frac{1}{2\pi^{2}}Q.
\end{equation}
The same computation applies for the momentum exchange in $\overline{G}^0_\mathrm{trunc}$ with bounding sphere radius $Q_\mathrm{trunc}$. Hence, we have,

\begin{equation}
\lambda_{V_{\mathrm{el}}+V_{\mathrm{ion}}}\leq\eta_{\mathrm{val}}(\eta_{\mathrm{val}}-1)\frac{Q}{\pi}+\sum_{I\ne J=1}^{\eta_{\mathrm{ion}}}Z_{I}^{\mathrm{PI}}Z_{J}^{\mathrm{PI}}\frac{Q_\mathrm{trunc}}{\pi},\label{eq:rescaling_coulomb_bound}
\end{equation}
where the two terms are the bounds for $\lambda_{V_\mathrm{el}},\lambda_{V_\mathrm{ion}}$, respectively. This bound is easy to compute with only knowledge of the maximum momentum exchanges.

\subsubsection*{Local term}

We perform the same bounding procedure as the Coulomb term on the
below sum,
\small
\begin{align}
\frac{(\bar{r}_{\mathrm{loc}}^{\zeta_I})^{3}}{\Omega}\sum_{\mathbf{q}\in G^{0}}e^{-(|\mathbf{k_{q}}|\bar{r}_{\mathrm{loc}}^{\zeta_I})^{2}/2}(|\mathbf{k_{q}}|\bar{r}_{\mathrm{loc}}^{\zeta_I})^{2s} & \lesssim\frac{(\bar{r}_{\mathrm{loc}}^{\zeta_I})^{3}}{\Omega\det\s B}\int_{\mathbf{q}\in\diamond^{0}}d^{3}qe^{-(|\mathbf{q}|\bar{r}_{\mathrm{loc}}^{\zeta_I})^{2}/2}(|\mathbf{q}|\bar{r}_{\mathrm{loc}}^{\zeta_I})^{2s}\nonumber \\
 & \leq\frac{4\pi(\bar{r}_{\mathrm{loc}}^{\zeta_I})^{3}}{\Omega\det\s B}\int_{0}^{Q}dqq^{2}e^{-(|\mathbf{q}|\bar{r}_{\mathrm{loc}}^{\zeta_I})^{2}/2}(|\mathbf{q}|\bar{r}_{\mathrm{loc}}^{\zeta_I})^{2s}\nonumber \\
 & =\frac{4\pi(\bar{r}_{\mathrm{loc}}^{\zeta_I})^{3}}{\Omega\det\s B}\p{\frac{2^{s}\sqrt{2}}{(\bar{r}_{\mathrm{loc}}^{\zeta_I})^{3}}(\Gamma(s+\frac{3}{2})-\Gamma[s+\frac{3}{2},\frac{(Q\bar{r}_{\mathrm{loc}}^{\zeta_I})^{2}}{2}])}\nonumber \\
 & \leq\frac{4\pi(\bar{r}_{\mathrm{loc}}^{\zeta_I})^{3}}{\Omega\det\s B}\p{\frac{2^{s}\sqrt{2}}{(\bar{r}_{\mathrm{loc}}^{\zeta_I})^{3}}\Gamma(s+\frac{3}{2})}\nonumber \\
 & =\frac{\sqrt{2}}{2\pi^{2}}2^{s}\Gamma(s+\frac{3}{2}),
\end{align}
where the second gamma function rapidly decays as a function of $Q$,
i.e. for $Q>5$, this term is almost vanishing which is true for all
physical situations under consideration, and hence the second-to-last line is almost an equality. This leads to per-particle-pair
rescaling factor bound,

\begin{align}
\lambda_{\tilde{V}_{\mathrm{loc}}^{I}} & \leq\sum_{s=-1}^{3}\sum_{c\in\{0,1\}}2\pi\sqrt{\frac{\pi}{2}}|c_{s}^{\zeta_I}|\p{\frac{\sqrt{2}}{2\pi^{2}}2^{s}\Gamma(s+\frac{3}{2})}=\frac{2}{\sqrt{\pi}}\sum_{s=-1}^{3}|c_{s}^{\zeta_I}|2^{s}\Gamma(s+\frac{3}{2})=|c_{-1}^{\zeta_I}|+|c_{0}^{\zeta_I}|+3|c_{1}^{\zeta_I}|+15|c_{2}^{\zeta_I}|+105|c_{3}^{\zeta_I}|,
\end{align}
\normalsize
and therefore we have,

\begin{equation}
\lambda_{V_\mathrm{loc}}\leq\eta_{\mathrm{val}}\sum_{I=1}^{\eta_{\mathrm{ion}}}\p{|c_{-1}^{\zeta_I}|+|c_{0}^{\zeta_I}|+3|c_{1}^{\zeta_I}|+15|c_{2}^{\zeta_I}|+105|c_{3}^{\zeta_I}|}\label{eq:rescaling_loc_bound}
\end{equation}
which is trivial to estimate directly from the HGH parameters.

\subsubsection*{Non-local term}

We again perform the same bounding procedure (but with $\diamond:=\{\mathbf{k}:\mathbf{k}\in\s BG\}$)
until we reach a radial integral with bound $K=\max_{\mathbf{k}\in\diamond}|\mathbf{k}|$,

\begin{align*}
\frac{(\bar{r}_{l}^{\zeta_I})^{3}}{\Omega}\sum_{\mathbf{\mathbf{p}}\in G}G_{\alpha}^{\zeta_I,l}(|\mathbf{k_{p}}|\bar{r}_{l}^{\zeta_I})^{2}\lesssim\frac{(\bar{r}_{l}^{\zeta_I})^{3}}{(\bar{r}_{l}^{\zeta_I})^{3}\Omega\det\s B}\int_{\mathbf{k}\in\diamond}d^{3}kG_{\alpha}^{\zeta_I,l}(|\mathbf{k_{p}}|)^{2}\leq\frac{4\pi}{\Omega\det\s B}\int_{0}^{K}dkk^{2}G_{\alpha}^{\zeta_I,l}(k)^{2}\leq\frac{1}{4\pi}\tilde{C}_{\alpha}^{\zeta_I,l},
\end{align*}
where in the last inequality, we can extend $K\rightarrow\infty$ with almost equality since the integrand is exponentially decaying, and define the positive constants $\tilde{C}_{\alpha}^{\zeta_I,l}:=\frac{2}{\pi}\int_{0}^{\infty}dkk^{2}G_{\alpha}^{\zeta_I,l}(k)^{2}$.
This leads to per-particle-pair rescaling factor bound,

\begin{equation}
\lambda_{\tilde{V}_{\mathrm{NL}}^{I}}\leq\sum_{l=0}^{l_{\max}}\sum_{\alpha=1}^{3}(2l+1)|D_{\alpha}^{\zeta_I,l}|\tilde{C}_{\alpha}^{\zeta_I,l},
\end{equation}
and therefore we have,
\begin{equation}
\lambda_{V_\mathrm{NL}}\leq\eta_{\mathrm{val}}\sum_{I=1}^{\eta_{\mathrm{ion}}}\sum_{l=0}^{l_{\max}}\sum_{\alpha=1}^{3}(2l+1)|D_{\alpha}^{\zeta_I,l}|\tilde{C}_{\alpha}^{\zeta_I,l}.\label{eq:rescaling_NL_bound}
\end{equation}
The key here is that $\tilde{C}_{\alpha}^{\zeta_I,l}$
are near unity constants as we see in Fig.~\ref{fig:NLboundscoeffs}
for a wide selection of pseudoions and are easy to compute directly
from the HGH parameters, i.e. diagonalize the matrix $B^{\zeta_I,l}$ to
obtain eigenvalues $D_{\alpha}^{\zeta_I,l}$ and eigenvector matrix $X^{\zeta_I,l}$,
construct $G_{\alpha}^{\zeta_I,l}$ from $\mathrm{g}_{a}^{l}$, and
then numerically integrate to find $\tilde{C}_{\alpha}^{\zeta_I,l}$.

\bigskip

Table~\ref{tab:rescalingRatiosBounds} shows the ratio of the approximate rescaling factor bounds to their corresponding exact values for all of the problem instances considered. Note that the approximate bounds are close to the exact values within a few percent for the local and non-local terms, but overestimate the Coulomb interactions by $\sim40\%-90\%$. In net, the total rescaling factor is within $\sim15\%-50\%$ which is overall quite accurate for a quick estimate.

\begin{figure}
\centering
\includegraphics[width=0.8\textwidth]{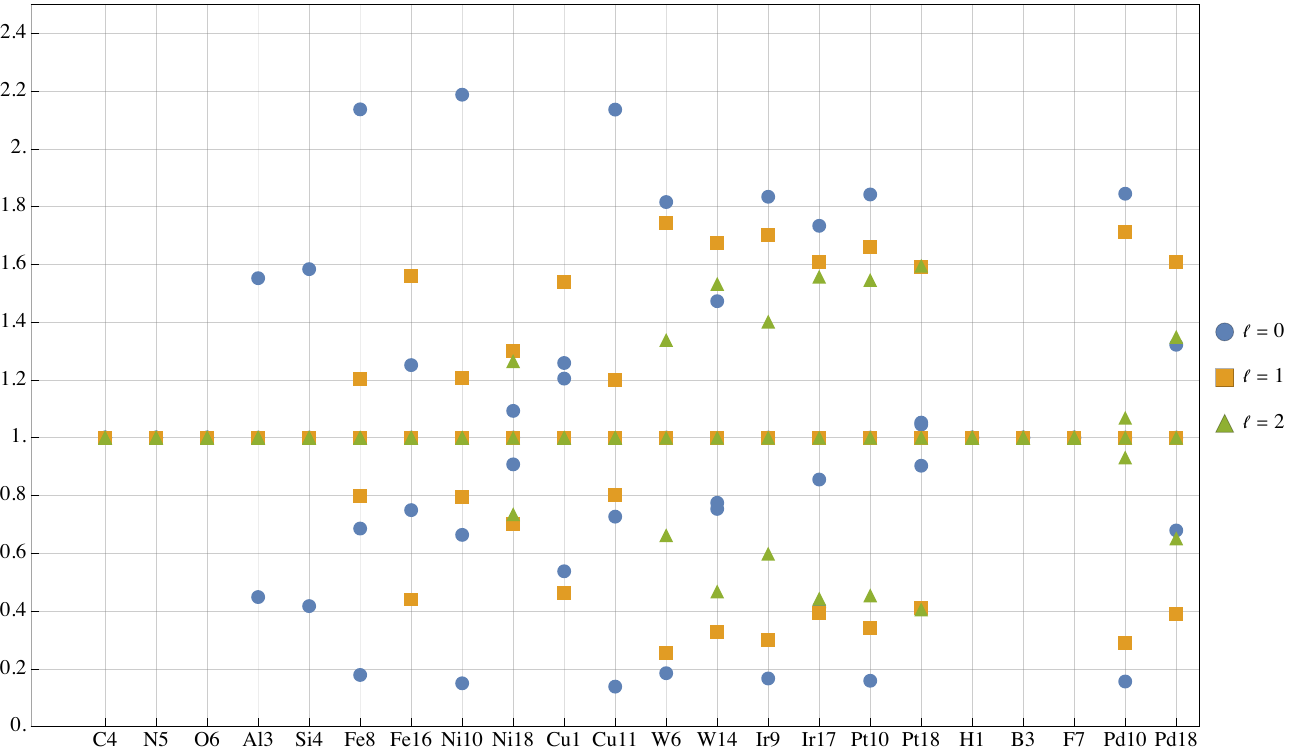}
\caption{The coefficients $\tilde{C}^{\zeta_I,l}_\alpha$ in the rescaling factor bound for the non-local term. Note that most of the coefficients are near unity for the selection of HGH pseudoions shown.}
\label{fig:NLboundscoeffs}
\end{figure}

\begin{table}
\centering
\begin{adjustbox}{width=\textwidth,center}
\begin{tabular}{|c|c|c|c|c|c|c|c|}
\hline 
 & NH3BF3 & DMTM Molecular & DMTM $3 \times 3$ & DMTM $5 \times 5$ & DMTM $9 \times 9$ & WGS $3 \times 3 \times 2$ & WGS $5 \times 5 \times 2$\tabularnewline
\hline 
\hline 
$\lambda_{T_{\mathrm{el}}}$ & 1.00 & 1.00 & 1.00 & 1.00 & 1.00 & 1.00 & 1.00\tabularnewline
\hline 
$\lambda_{T_{\mathrm{ion}}}$ & 1.00 & 1.00 & 1.00 & 1.00 & 1.00 & 1.00 & 1.00\tabularnewline
\hline 
$\lambda_{V_{\mathrm{el}}}$ & 1.46 & 1.42 & 1.88 & 1.74 & 1.75 & 1.42 & 1.44\tabularnewline
\hline 
$\lambda_{V_{\mathrm{ion}}}$ & 1.43 & 1.43 & 1.69 & 1.75 & 1.76 & 1.41 & 1.42\tabularnewline
\hline 
$\lambda_{V_\mathrm{loc}}$ & 1.02 & 1.02 & 1.02 & 1.02 & 1.02 & 1.05 & 1.05\tabularnewline
\hline 
$\lambda_{V_\mathrm{NL}}$ & 1.01 & 1.00 & 1.00 & 1.00 & 1.00 & 1.00 & 1.00\tabularnewline
\hline 
$\lambda$ & 1.18 & 1.21 & 1.43 & 1.39 & 1.42 & 1.24 & 1.28\tabularnewline
\hline 
\end{tabular}
\end{adjustbox}
\caption{Ratios of the approximate rescaling factor bounds to their corresponding exact values for each of the Hamiltonian terms. Note that the kinetic terms shown for completeness are all exactly unity since no bounding procedure is needed and the exact formulas in Eq.~\eqref{eq:lambdaT} are easy to use.}
\label{tab:rescalingRatiosBounds}
\end{table}

\section{Compilation of the block encoding}\label{sec:comp}

There are two routines that we will use significantly throughout the block encoding: (i) preparation of a uniform superposition over an arbitrary number of amplitudes (``USP''), and (ii) quantum rejection sampling (``QRS''). 

For USP, we reuse the construction from Ref.~\cite{babbush2018encoding} for preparing a uniform superposition over $d$ amplitudes. In the worst case, $d$ has no factors that are a power of two; in this case, the subroutine consists of two inequality tests over $\lceil \log(d) \rceil$ qubits and two rotations with a $d$-dependent angle. If these rotations are carried out using phase gradient state addition (with a $b$-bit phase gradient state), the total cost is

\begin{equation}\label{eq:comp_usp}
    2(\lceil \log(d) \rceil - 1) + 2 (b - 3) = 2\lceil \log(d) \rceil + 2 b - 8.
\end{equation}
We assume that the cost of preparing the phase gradient state is amortized across the entire algorithm and therefore has negligible cost per evocation. The USP subroutine uses $\lceil \log(d) \rceil + 1$ ancillae for the arithmetic for the inequality test and $b$ ancillae for the rotation, but these ancillae are not instantiated concurrently.
USP has tunable parameters that dictate the quality of the state that is prepared. These parameters will be left as free parameters in the resource estimation sections below, and we defer discussion of fixing them when we consider specific problem instances.

For QRS, we use the framework from~\cite{lemieux2024quantum}. In particular, QRS proceeds as a coherent analogue of rejection sampling; in order to prepare a state with amplitudes proportional to some target function $f(x)$, we instead prepare a state with amplitudes corresponding to some bounding function $g(x)$, chosen such that it is easier to prepare than $f(x)$ itself. We can then use coherent arithmetic to ``flag'' the portion of the state over $f(x)$, versus that over the remainder; amplitude amplification then amplifies the part corresponding to $f(x)$ and dampens the rest. Judicious choice of $g(x)$ often leads to low resource counts in practice. In extending this preparation technique to block encodings, we also inherit the distinction between \emph{explicit} and \emph{implicit} rejection sampling; in explicit rejection sampling, the QRS occurs inside $\PREP$, in implicit rejection sampling, the QRS occurs inside $\SEL$. See~\cite{lemieux2024quantum}, Sec. 5.3 for detail.

\subsection{Subroutines shared across the block encoding}\label{sec:comp_shared}

\subsubsection{Selecting between terms}\label{sec:comp_tu_rotn}

The selection between terms is achieved by preparing the state

\begin{align}
\PREP_{\textrm{terms}} \kett{0}= \frac{1}{\sqrt{\lambda}} \left(\sqrt{\lambda_{T_\el+T_\ion}} \kett{0} + \sqrt{\lambda_{V_\el + V_\ion}} \kett{1} + \sqrt{\lambda_{V_\loc}} \kett{2} + \sqrt{\lambda_{V_{\NL}}} \kett{3}\right),
\end{align}
encoded in unary. We prepare this state by first preparing the encoding in binary, using at most three rotations, then converting from binary to unary using three CSWAPS. If the rotations are performed over $b_P$ bits, the total cost of this preparation is $3b_P + 3$. We assume the same cost for the uncomputation.

\subsubsection{Swapping particle registers}\label{sec:comp_swaps}

For every term in the block encoding, SELECT requires carrying out some arithmetic on particle registers, conditioned on the index of the particle.
As noted in~\cite{su2021fault} (Sec.II B), it is typically cheaper to ``swap up'' the particle to an ``active'' register conditioned on the index register once, and then to flag which term in the block encoding to carry out based on the state of flag qubits, rather than copying particles up and down for each term of the block encoding. 
In particular, we must swap up four registers: an electron indexed by $i$, an electron indexed by $j$, an ion indexed by $I$ and an ion indexed by $J$. 
Each electron register contains $n$ qubits, each ion register contains $\bar{n}$ qubits, and must be indexed over $\eta_{\text{val}}$ and $\eta_{\text{ion}}$ indices for electron and ion terms, respectively. 
Given two indices each for electrons and ions ($i$ and $j$ for electrons and $I$ and $J$ for ions), the cost for swapping up $\eta$ particles where each particle encodes its momentum in $n$ qubits controlled on their index, is
\begin{equation}
    n\eta + \eta - 1,
\end{equation}
where the first term comes from the SWAPs and the rest from the indexing. Given that we must swap up pairs of electrons and ions, the total cost is
\begin{equation}
    2n\eta_{\text{val}}+2\bar{n}\eta_{\text{ion}}+2(\eta_{\text{val}}-1) + 2(\eta_{\text{ion}}-1),
\end{equation}
We consider an uncompute with the same cost. The additional cost of controlling SELECTs to indicate which term in the block encoding to apply is included in the costs in the sections below.

\subsection[Kinetic term]{The kinetic term $T = T_\el + T_\ion$}\label{sec:comp_T}

\subsubsection{$\PREP_{0}$ (Kinetic)}\label{sec:comp_T_prep_eta}

The initial preparation is of the state

\begin{align}
\PREP_0\kett 0=\sqrt{\frac{\lambda_{T_{\mathrm{el}}}}{\lambda_{T_{\mathrm{el}}+T_{\mathrm{ion}}}}}\p{ \sum_{i=1}^{\eta_{\mathrm{val}}}\kett i+\sqrt{\frac{\lambda_{\tilde{T}_{\mathrm{ion}}}}{\lambda_{T_{\mathrm{el}}}}} \sum_{I=1}^{\eta_{\mathrm{ion}}}\sqrt{\frac{1}{M_{I}}}\kett{I}}.
\end{align}

Let the state $\kett{\phi^T_\eta} \propto \sum_{\bar{I}=1}^\eta \sqrt{\frac{1}{\bar{M}_{\bar{I}}}} \kett{\bar{I}}$, where $\bar{I}$ is a composite index that runs over electrons and ions $\bar{I} = 1, \ldots \eta_\val, \eta_\val + 1, \ldots \eta_\val + \eta_\ion$ and where $\bar{M}_{\bar{I}}$ is the normalized mass of particle $\bar{I}$ (defined so that $\bar{M}_{\bar{I}}=1$ for indices corresponding to electrons, and that $\bar{M}_{\bar{I}} = \frac{\lambda_{\tilde{T}_{\mathrm{el}}}}{\lambda_{T_{\mathrm{ion}}}} M_I$ for ions). 
We assume that $b_M$-bit approximants to the quantities $\frac{1}{\sqrt{\bar{M}_{\bar{I}}}}$ have been loaded, indexed on $\bar{I}$, using the QROM in App.~\ref{sec:comp_alprep}. 
The preparation of $\kett{\phi^T_\eta}$ is then carried out by rejection sampling against a uniform reference state $\frac{1}{\sqrt{\eta}}\sum_{\bar{I}=1}^\eta \kett{\bar{I}}$. 
The cost of preparing the reference state from Eq.~\eqref{eq:comp_usp} is $2\lceil \log(\eta) \rceil + 2 b_T - 8$, for some user-specified precision $b_T$. 
The cost of the inequality test is dictated by the precision with which one requires to prepare the target state. It is shown in~\cite{lemieux2024quantum}, Lemma 2.11 that preparing a sampling state that is uniform with $M_T \geq \frac{2}{\epsilon_T}$ amplitudes suffices to prepare a target to within precision $\epsilon_T$ (we also assume that $M_T$ is chosen to be a power of two to simplify preparation costs). The inequality test then requires $\lceil \log \frac{2}{\epsilon_T} \rceil$ Toffolis and an equal number of ancillae. The output of the inequality test must be amplified; the success probability is

\begin{equation}
    P_{\text{succ}} = \sqrt{\frac{\sum_{\bar{I}=1}^\eta \frac{1}{M_{\bar{I}}}}{\eta}} = \sqrt{\frac{\eta_\val + \sum_{I=1}^{\eta_{\text{ion}}} \frac{1}{M_{I}}}{\eta}} \geq \sqrt{\frac{\eta_\val}{\eta}},
\end{equation}
and so we can guarantee one round of amplification provided at least half the particles are electrons. This requires three calls to both the inequality test and the USP. We must also add a single control for term selection. This can be achieved by (i) controlling the inequality test (doubling its cost); and (ii) adding a controlled USP to uncompute the reference state (with cost $7\lceil \log(\eta) \rceil + 2b_T - 6$). The total cost is therefore $6\lceil \log \frac{2}{\epsilon_T} \rceil + 13\lceil \log(\eta) \rceil + 8 b_T - 30$. 

\subsubsection{Preparing the reference state for $|\mathbf{k_p}|^2$}\label{sec:comp_T_usp}

We consider the same approach given as in Ref.~\cite{berry2023quantum}; i.e. to prepare a uniform reference state (discussed in Sec. VII C therein). 
The cost of USP is given in Eq.~\eqref{eq:comp_usp}, with $d=\max_\mathbf{p} |\mathbf{k_p}|^2$ for electrons (or equivalently $d=\max_\mathbf{P} |\mathbf{k_P}|^2$ for ions). The reference state for ions should be encoded over a larger number of amplitudes than for electrons. To achieve this, we check whether $\bar{I} \leq \eta_\val$ by applying an inequality test to the particle index register (the output of this inequality test flags whether we are encoding an electron or an ion). The inequality test has cost $\lceil \log(\eta) \rceil$, and can be uncomputed using only Cliffords. We then apply two different USPs controlled on this flag. Assuming that the register encoding $|\mathbf{k_p}|^2$ is of size $b$, and that of $|\mathbf{k_P}|^2$ is of size $\bar{b}$, the cost for each controlled USP is $7b + 2b_k - 6$ and $7\bar{b} + 2b_k - 6$, respectively. The total cost is then $\lceil \log(\eta) \rceil + 7b + 7\bar{b} + 4b_k - 12$, for some precision $b_k$. 

\subsubsection{$\text{SEL}_T$}\label{sec:comp_T_sel}

We compute the quantity $|\mathbf{k_p}|^2$ (and likewise $|\mathbf{k_P}|^2$) once to an ancilla register, to be shared across both the kinetic and interaction terms (for the latter, see App.~\ref{sec:comp_G}). The cost of this preparation is included in the interaction term. Given the preparation of a uniform reference state in App.~\ref{sec:comp_T_usp}, $\text{SEL}_T$ then reduces to a single inequality test as discussed in Sec.~\ref{subsubsec:BEKinetic}. The register encoding $|\mathbf{k_P}|^2$ (and $|\mathbf{k_p}|^2$) is of size $\bar{b}$ and $b$, respectively, hence the cost is $b+\bar{b}$ Toffolis.

\subsection[Coulomb terms]{The Coulomb terms $V_{\text{el}}$, $V_{\text{ion}}^\mathrm{PI}$}\label{sec:comp_Coul}

\subsubsection{$\PREP_1$ (Coulomb)}\label{sec:comp_Coul_V}

We aim to prepare a state $\kett{\phi_V}$ (where we drop the superscript $\mathrm{PI}$ for brevity on the ion charges $Z_I$), defined as

\begin{align}
\kett{\phi_{V}} \propto \left( \sqrt{\lambda_{\tilde{V}_\el}} \sum^{\eta_\val}_{i\neq j=1} \kett{i,j} + \sqrt{\lambda_{\tilde{V}_\ion}} \sum^{\eta_\val + \eta_\ion}_{I\neq J= \eta_\val +1} \sqrt{Z_I Z_J}\kett{I,J} \right).
\end{align}

\noindent To prepare $\kett{\phi_{V}}$ (up to entanglement with junk), we first prepare a state

\begin{equation}
    \kett{\tilde{\phi}_V} \propto \left(  \sum_{i=1}^{\eta_\val} \kett{i} + \frac{\lambda^{1/4}_{V_\ion}}{\lambda^{1/4}_{V_\el}} \sum_{I=\eta_\val+1}^{\eta_\val + \eta_{\text{ion}}} \sqrt{Z_I} \kett{I}\right) \otimes \left(\sum_{j=1}^{\eta_\val} \kett{j} + \frac{\lambda^{1/4}_{V_\ion}}{\lambda^{1/4}_{V_\el}} \sum_{J=\eta_\val+1}^{\eta_\val + \eta_{\text{ion}}} \sqrt{Z_J} \kett{J}\right).
\end{equation} 
For readability in the proceeding compilation, we absorb the prefactor $\frac{\lambda^{1/2}_{V_\ion}}{\lambda^{1/2}_{V_\el}}$ into the coefficients $Z_I$ and $Z_J$. For atomic numbers $Z_I$ let $\kappa = \eta_\val + \sum_{I=1}^{\eta_{\text{ion}}} Z_I = \sum_{\bar{I}=1}^{\eta} Z_{\bar{I}}$, where

\begin{align}
Z_{\bar{I}}=\begin{cases}
1 & \bar{I}=1,..,\eta_\val, \\
Z_{\bar{I}-\eta_\val} & \bar{I}=\eta_\val+1,...,\eta_\val+\eta_{\mathrm{ion}}.
\end{cases}
\end{align}

Note that $\kappa = 2\eta_\val$ for charge-neutral systems. Then we (i) first prepare a uniform superposition over values $z = 1 \ldots \kappa$ (with cost $2\lceil \log\kappa \rceil + 2b_\kappa - 8$); (ii) using a QROM, load $\bar{I}$ for all $Z_{\bar{I}-1} < z \leq Z_{\bar{I}}$. 
This generates the state we want, up to entanglement with junk. 
To see this note that the state after the USP is transformed to

\begin{equation}
    \frac{1}{\sqrt{\kappa}} \sum_{z=1}^\kappa \kett{z} \mapsto \frac{1}{\sqrt{\kappa}} \sum_{z=1}^\kappa \kett{z} \kett{\bar{I}(z)} = \sum_{\bar{I}=1}^{\eta}\sum_{z={Z_{\bar{I}-1}}}^{Z_{\bar{I}}} \kett{z} \kett{\bar{I}(z)}.
\end{equation}
\noindent But $\bar{I}(z)$ is a constant for all $Z_{\bar{I}-1} < z \leq Z_{\bar{I}}$, and so rearranging and dropping the $z$ argument gives
\begin{equation}
    \sum_{\bar{I}=1}^{\eta}\sum_{z=Z_{\bar{I}-1}}^{Z_{\bar{I}}} \kett{z} \kett{\bar{I}(z)} = \sum_{\bar{I}=1}^{\eta} \sqrt{Z_{\bar{I}}} \kett{\bar{I}} \left(\frac{1}{\sqrt{Z_{\bar{I}}}}\sum_{z=Z_{\bar{I}-1}}^{Z_{\bar{I}}} \kett{z} \right) = \sum_{\bar{I}=1}^{\eta} \sqrt{Z_{\bar{I}}} \kett{\bar{I}} \kett{\text{junk}_{\bar{I}}}.
\end{equation}

Of course, it would be unnecessarily costly to use a QROM that indexes over all values $z=1 \ldots \kappa$ with cost $\kappa$ given that the majority of the loaded values are just equal to the index itself (because $Z_{\bar{I}}=1$ provided that $\bar{I} \leq \eta_\val$). 
Instead, we (i) copy the index value to the output register using CNOTs (with zero cost); (ii) subtract $\eta_\val$ from the index register (with cost $\lceil \log\eta \rceil - 2$); (iii) use a QROM over $\kappa-\eta_\val$ indices to fix up the incorrect values by loading $z \oplus \bar{I}(z)$ (with cost $\kappa - \eta_\val$); (iv) add $\eta_\val$ back into the index register (with cost $\lceil \log\eta \rceil - 2$). The total cost of this part is then $\kappa - \eta_\val + 2\lceil \log \eta \rceil - 4$. Note that this must be repeated twice: once for $\bar{I}$ and once for $\bar{J}$. 

The simplest way to add a control to the preparation of $\kett{\phi_V}$ for the term selection in Eq.~\eqref{eq:CombiningPREP} is to control the QROM here. In the case where the control is off and the QROM is not applied, the $\kett{\bar{I}}$ register stays in the state $\kett{0}$ and the rest of the subroutine only prepares junk in ancilla registers. The added cost of controlling the QROM is a small constant number of Toffolis, which we omit from the cost.

Once $\kett{\tilde{\phi}_V}$ is prepared (up to junk), we must flag the part of the state that we wish to amplify in order to yield $\kett{\phi_V}$. First, we flag whether $\bar{I}=\bar{J}$ with cost $\lceil \log\eta \rceil$; we then check whether $\bar{I}$ or $\bar{J}$ is greater than $\eta_{\text{el}}$, with cost $2 \lceil \log\eta \rceil$. The subspace that we wish to suppress is the one in which:

\begin{equation}
    (\bar{I} \neq \bar{J}) \vee (\bar{I} > \eta_\val \wedge \bar{J} \leq \eta_\val) \vee (\bar{I} \leq \eta_\val \wedge \bar{J} > \eta_\val).
\end{equation}

This logic can be executed with at most four Toffolis and four ancillae (and the cost of uncomputation is free). This subroutine costs $2(\kappa-\eta_\val +5 \lceil \log \eta \rceil+2\lceil \log\kappa \rceil + 2b_\kappa - 8)$, counting costs from the initial USP using Eq.~\eqref{eq:comp_usp}, from the QROM (assuming no free clean ancillae) and the flagging logic. 
The factor of two outside the expression accounts for the fact that the subroutine must be repeated twice, once for $\bar{I}$ and once for $\bar{J}$. The success probability is

\begin{equation}
    P_{\text{succ}} = \frac{\eta_\val(\eta_\val-1)+\sum_{I \neq J} Z_I Z_J}{(\eta_\val +\sum_I Z_I)(\eta_\val+\sum_J Z_J)}.
\end{equation}

\noindent For charge-neutral systems, $\sum_I Z_I = \eta_\val$ and so

\begin{equation}
    P_{\text{succ}} = \frac{\eta_\val(\eta_\val-1)+\sum_{I \neq J} Z_I Z_J}{4\eta^2_\val} = \frac{1}{4} + \frac{\left(\sum_{I \neq J} Z_I Z_J\right)-\eta_\val}{4\eta^2_\val} \geq \frac{1}{4},
\end{equation}

\noindent where in the last inequality we have used that $\sum_{I \neq J} Z_I Z_J \geq \sum_I Z_I = \eta_\val$. One round of amplification therefore suffices. The cost is then multiplied by three (two computations, plus one uncomputation) yielding a total cost

\begin{equation}
    6(\kappa-\eta_\val +5 \lceil \log \eta \rceil + 2\lceil \log\kappa \rceil + 2b_\kappa - 8).
\end{equation}
Note that we must also perform a single rotation to amplify appropriately, but we assume that the cost is negligible in comparison to the costs above. Assuming that the system is charge-neutral, the total cost is 

\begin{equation}
    6(\eta_\val +5 \lceil \log \eta \rceil + 2\lceil \log 2\eta_\val \rceil + 2b_\kappa - 8).
\end{equation}
In most cases, the most significant contribution to the ancilla count is the $\lceil \log(\zeta) \rceil = \lceil \log(2\eta_\val) \rceil$ qubits necessary for the QROM.

\subsubsection{$\PREP_{\text{coul, \el}}$, $\PREP_{\text{coul, \ion}}$}\label{sec:comp_Coul_q}

The subroutine $\PREP_{\text{coul}, \el}$ prepares the state $\kett{\phi_{q}^{V}} \propto\sum_{\mathbf{q}\in G^{0}}\sqrt{\frac{\pi}{\Omega|\mathbf{k_{q}}|^2}}\kett{\mathbf{q}}$. The compilation for this state proceeds identically to the construction in~\cite{berry2023quantum} (Sec. VII A and App. B) and discussed in the context of quantum rejection sampling in~\cite{lemieux2024quantum}. 
In particular, we must (i) compute $|\mathbf{k_q}|^2$ to a register; (ii) prepare an appropriate reference state (provided in App.~\ref{app:reference_states_QRS}); (iii) multiply the reference by a uniform state; (iv) carry out an inequality test; (v) reflect and uncompute steps (i)-(iii). 
The procedure must also be invoked in reverse for the uncomputation of $\kett{\phi_{q}^{V}}$. The breakdown in~\cite{berry2023quantum} considers that $|\mathbf{k_q}|^2$ is computed to a register three times; twice in the preparation (in order to amplify once), and once in the unpreparation (because the final invocation can be uncomputed using measurement-based uncomputation). 
Additionally, the register containing $|\mathbf{k_q}|^2$ must be multiplied by a uniform superposition register in order to carry out rejection sampling. However, we assume that $M$ is chosen to be a power of two and the multiplication has no cost (it consists only of padding the register containing $|\mathbf{k_q}|^2$ by $\log M$ bits).
We will keep the computation and uncomputation costs separate in the tabulation below.

As for the cost of computing $|\mathbf{k_q}|^2$, the arithmetic is given in~\cite{berry2023quantum}, App. C.
While the arithmetic is simple, the details of the cost depend somewhat on the structure of the Gramian derived from the Bravais vectors. 
We will assume the worst-case cost for a number of plane waves in each dimension $(n_1, n_2, n_3)$ which is given by
\begin{equation}
    \frac{5}{2}(n_1^2 + n_2^2 + n_3^2) + 2(n_1 + n_2 + n_3)^2 + 4b_g(n_1+n_2+n_3),
\end{equation}
where $b_g$ is a finite bit-precision for the Gramian elements, but note that the cost can be reduced given information about the lattice geometry. For the sake of readability, let $n_1^2 + n_2^2 + n_3^2 = \tilde{n}$ and $n_1 + n_2 + n_3=n$. Then the leading order cost for the computation of $\kett{\phi_{q}^{V}}$ is

\begin{equation}
    5\tilde{n} + 4n^2 + 8b_g n,
\end{equation}

\noindent and for the uncomputation

\begin{equation}
    \frac{5}{2}\tilde{n} + 2n^2 + 4b_g n.
\end{equation}

Assuming that the number of bits to encode momenta is large relative to the bit precision with which we carry out arithmetic, the overwhelming cost to the ancilla count are the $\tilde{n}$ qubits necessary to encode $|\mathbf{k_q}|^2$. Likewise for the pseudoion term, the subroutine $\PREP_{\text{coul}, \ion}$ prepares the state $\kett{\phi_Q^V} \propto \sum_{\v{Q} \in \bar{G}^0_{\text{trunc}}} \sqrt{\frac{\pi}{\Omega|\mathbf{k_{Q}}|^2}}\kett{\mathbf{Q}}$. This preparation proceeds identically as above, but now we span over $\bar{G}^0_{\text{trunc}}$ momenta. In general, this quantity will be small given the ion-ion cutoff. However, we make the conservative assumption that there is no truncation and we must prepare a state over all momenta in $\bar{G}^0$. In this case, we have the same cost as the electronic case but with $n$ replaced with $\bar{n}$ and $\tilde{n}$ replaced with $\tilde{\bar{n}}$ to account for the increased number of plane waves spanning the ion momentum register.

\subsubsection{$\SEL_{\text{coul}, \el}$, $\SEL_{\text{coul}, \ion}$}\label{sec:comp_Coul_sel}

The SELECT for the electronic Coulomb term is given by

\begin{align}
 \SEL_{\text{coul}, \el} &= \sum_{\mathbf{q}\in G^{0}}\sum_{c\in\{0,1\}} \kett{\mathbf{q},c}\brat{\mathbf{q},c}\otimes U_{(\mathbf{q},c)}^{\text{coul, \el}} \\
U_{(\mathbf{q},c)}^{\text{coul}, \el} & =\sum_{\substack{\mathbf{p},\mathbf{p'}\in G}
}(-1)^{c([\mathbf{p}-\mathbf{q}\notin G]\vee[\mathbf{p'}+\mathbf{q}\notin G])}\kett{\mathbf{p-q}}\brat{\mathbf{p}}\otimes\kett{{\bf p'+q}}\brat{{\bf p'}}.
\end{align}

The construction is identical to that in Ref.~\cite{su2021fault}, Sec II D (except we iterate over the composite index $\bar{I}$), and so we describe it only briefly. 
The addition of the transferred momentum $q$ is complicated by the fact that the momenta are not encoded in two's complement, but in a signed representation. 
Conversion, addition and back-conversion costs $8n$ Toffolis (Ref.~\cite{su2021fault}, Eq. (94)).
Checking whether the added/subtracted momenta $\mathbf{p}-\mathbf{q}$ and $\mathbf{p}'+\mathbf{q}$ are in $G$ is not strictly necessary, given that this branch flips high bits in these registers to 1 if outside the prerequisite range, and we can select on $\kett{0}$ for these bits, implying that that branch lies outside the block encoding. 
However, if the checking is explicitly included then it adds an additional four Toffolis (one each to check $\mathbf{p}-\mathbf{q}$ and $\mathbf{p}'+\mathbf{q}$, one to OR them, and one to AND them with the bit $c$). 
We do not include the additional Toffolis in the resource count in the main body. 
As for ancilla qubits, $\max_i n_i$ qubits are needed for the arithmetic and four are needed for the checking.

The SELECT for the ion-ion term is almost identical:

\begin{align}
 \SEL_{\text{coul}, \ion} &= \sum_{\mathbf{Q}\in \bar{G}_{\text{trunc}}^{0}}\sum_{c\in\{0,1\}} \kett{\mathbf{Q},c}\brat{\mathbf{Q},c}\otimes U_{(\mathbf{Q},c)}^{\text{coul, \ion}} \\
U_{(\mathbf{Q},c)}^{\text{coul}, \ion} & =\sum_{\substack{\mathbf{P},\mathbf{P'}\in \bar{G}_{\text{trunc}}}
}(-1)^{c([\mathbf{P}-\mathbf{Q}\notin \bar{G}_{\text{trunc}}]\vee[\mathbf{P'}+\mathbf{Q}\notin \bar{G}_{\text{trunc}}])}\kett{\mathbf{P-Q}}\brat{\mathbf{P}}\otimes\kett{{\bf P'+Q}}\brat{{\bf P'}}.
\end{align}
It therefore incurs an equivalent cost, but with $n$ replaced by $\bar{n}$ to account for the larger registers needed to store the pseudoion momenta.

\subsection[Local interaction term]{The local interaction term, $V^{\text{PI}}_{\text{loc}}$}\label{sec:comp_loc}

\subsubsection{$\PREP_2$ (interaction term)}\label{sec:comp_loc_prep_el}

The subroutine $\PREP_2$ prepares a state

\begin{equation}
    \PREP_2\kett{0} = \frac{1}{\sqrt{\eta_\val}} \sum_{i=1}^{\eta_\val} \kett{i} \otimes \frac{1}{\sqrt{\lambda_{V_\loc}/\eta_\val}} \sum_{I=1}^{\eta_\ion} \sqrt{\lambda_{{\tilde{V}}^I_\loc}} \kett{I} = \kett{\Phi_\el} \otimes \kett{\Phi_\ion}.
\end{equation}

The state $\kett{\Phi_{\mathrm{el}}}$ is a uniform state prepared with USP. However, we must include the cost of the control on the term selection register. The cost of a controlled USP is $7\lceil \log(\eta_\val) \rceil + 2b_{\eta_\val} - 6$ and uses $\lceil \log(\eta_\val) \rceil$ additional ancillae. For the non-local term, we must also prepare $\kett{\Phi_{\mathrm{el}}}$ controlled on the term selection register. However, this cost can be shared; we encode the local and non-local terms in the term selection register with the bases $\kett{0100}$ and $\kett{1000}$, respectively, and so if we wish to prepare $\kett{\Phi_{\mathrm{el}}}$ conditioned on either of these terms, we only need to CNOT the high bit conditioned on the second-highest bit to flip to $\kett{1100}$ and $\kett{1000}$, and then prepare $\kett{\Phi_{\mathrm{el}}}$ once conditioned on the high bit (and then undo the CNOT).

We have the state
\begin{equation}
   \kett{\Phi_{\mathrm{ion}}} = \sum^{\eta_\ion}_{I=1} \sqrt{ \frac{\lambda_{\tilde{V}^I_\loc}}{{\lambda_{V_{\loc}}/{\eta_\val}}}} \kett{I}.
\end{equation}
However, the coefficient $\lambda_{\tilde{V}^I_\loc}$ only depends on ion $I$ insofar as it depends on the ionic species $\zeta_I$. Instead of preparing $\kett{\Phi_{\mathrm{ion}}}$ directly, we first prepare a state

\begin{equation}
    \kett{\Phi_\zeta} = \frac{1}{\sqrt{\sum_{\zeta=1}^Z \lambda_{\tilde{V}^\zeta_\loc}m_\zeta}} \sum_{\zeta=1}^Z \sqrt{\lambda_{\tilde{V}^\zeta_\loc}m_\zeta} \kett{\zeta},
\end{equation}
where $m_\zeta$ is the multiplicity of the ion species $\zeta$. We then use a QROM over $\zeta$ to load $m_\zeta$, use USP to prepare a uniform superposition over $m_\zeta$ amplitudes, and then uncompute the QROM:

\begin{align}
    \kett{\Phi_\zeta} &\mapsto \frac{1}{\sqrt{\sum_{\zeta=1}^Z \lambda_{\tilde{V}^\zeta_\loc}m_\zeta}} \sum_{\zeta=1}^Z \sqrt{\lambda_{\tilde{V}^\zeta_\loc}m_\zeta} \kett{\zeta}\kett{m_\zeta} \\
    &\mapsto \frac{1}{\sqrt{\sum_{\zeta=1}^Z \lambda_{\tilde{V}^\zeta_\loc}m_\zeta}} \sum_{\zeta=1}^Z \sqrt{\lambda_{\tilde{V}^\zeta_\loc}m_\zeta} \kett{\zeta}\kett{m_\zeta} \left(\frac{1}{\sqrt{m_\zeta}}\sum_{I=1}^{m_\zeta} \kett{I}\right) \\
    &\mapsto \frac{1}{\sqrt{\sum_{\zeta=1}^Z \lambda_{\tilde{V}^\zeta_\loc}m_\zeta}} \sum_{\zeta=1}^Z \sqrt{\lambda_{\tilde{V}^\zeta_\loc}m_\zeta} \kett{\zeta} \left(\frac{1}{\sqrt{m_\zeta}}\sum_{I=1}^{m_\zeta} \kett{I}\right).
\end{align}
Rearranging yields
\begin{align}
    \frac{1}{\sqrt{\sum_{\zeta=1}^Z \lambda_{\tilde{V}^\zeta_\loc}m_\zeta}} \sum_{\zeta=1}^Z \sqrt{\lambda_{\tilde{V}^\zeta_\loc}m_\zeta} \kett{\zeta} \left(\frac{1}{\sqrt{m_\zeta}}\sum_{I=1}^{m_\zeta} \kett{I}\right) &= \frac{1}{\sqrt{\sum_{\zeta=1}^Z \lambda_{\tilde{V}^\zeta_\loc}m_\zeta}} \sum_{\zeta=1}^Z \sum_{I=1}^{m_\zeta} \sqrt{\lambda_{\tilde{V}^\zeta_\loc}} \kett{\zeta}\kett{I} \\
    &= \frac{1}{\sqrt{\sum_{I=1}^{\eta_\ion} \lambda_{\tilde{V}^I_\loc}}} \sum^{\eta_\ion}_{I=1} \sqrt{\lambda_{\tilde{V}^\zeta_\loc}} \kett{\zeta}\kett{I},
\end{align}
which is $\kett{\Phi_{\mathrm{ion}}}$ up to entanglement with junk in the $\kett{\zeta}$ register (that we can ignore given that this is a state preparation routine inside a block encoding).

The cost of preparing a state $\kett{\Phi_\zeta}$ just using arbitrary state preparation to $b_Z$ bits of precision using routines like~\cite{low2018trading} is $4Z+\lceil \log Z \rceil (b_Z - 3) - 2$~\cite{zini2023quantum}. The QROM to load $m_\zeta$ has cost $Z$ (and we assume the same cost for its uncomputation). Adding a control (for the term selection register) to each of these subroutines only requires a small constant number of Toffolis, which we omit. The controlled USP has cost at most $7\lceil \log (\eta_\ion)\rceil + 2b_I - 4$ (i.e. in the case where all the ions are the same species). The total cost, therefore, is $6Z+\lceil \log Z \rceil (b_Z - 3) + 7\lceil \log (\eta_\ion)\rceil + 2b_I - 6 $. We require $b_Z$ ancillas for the arbitrary state prep and $\max\{b_I, \lceil \log (\eta_\ion)\rceil\}$ for the USP.

As for the non-local term, the QROM to load $m_\zeta$ and the USP is shared with the local term, and we can therefore compute those steps once (using the trick above). The only retained cost is that for the preparation of $\kett{\Phi_\zeta}$ (where now we are loading analogous coefficients that are constructed from $V_\NL^\mathrm{PI}$ rather than $V_{\loc}^\mathrm{PI}$).

\subsubsection{$\PREP_{\loc, 1}$}\label{sec:comp_loccoeff}

\begin{equation}
    \PREP_{\loc, 1}\kett 0 \kett{\zeta_I} =\frac{1}{\sqrt{\lambda_{\tilde{V}_{\mathrm{loc}}^{I}}}}\sum_{s=-1}^{3}\sum_{c\in\{0,1\}}\sqrt{\frac{2\pi(\bar{r}_{\mathrm{loc}}^{\zeta_{I}})^{3}}{\Omega}\sqrt{\frac{\pi}{2}}|c_{s}^{\zeta_{I}}|\lambda^{\zeta_{I},s}_\loc}\kett{s,\mathrm{sgn}(c_{s}^{\zeta_{I}}),c}\kett{\zeta_I}.
\end{equation}
Given that the index $s$ is only over five values, we compute this state using a routine for preparing states with arbitrary amplitudes. For arbitrary state preparation, we use coherent alias sampling. Note, however, that the preparation is multiplexed over the ion index $I$ (or more specifically, the ion species index $\zeta_I$). Given that $\zeta_I = 1 \ldots Z$ is loaded in unary using the QROM in App.~\ref{sec:comp_nucleardata}, the total cost for this subroutine is the cost of a singly-controlled alias sampler, multiplied by the total number of nuclear species $Z$. The steps for the alias sampling are as follows:

\begin{enumerate}
    \item Controlled on the $\zeta_I$ bit, prepare a uniform superposition over 5 amplitudes with cost $15+2b_{s}$ for some precision $b_{s}$.
    \item Controlled on the $\zeta_I$ bit, use a QROM to load alt values and keep probabilities with a cost of 7 Toffolis. We also use this QROM to load $b$-bit approximants to the precomputed coefficients $\theta_\alpha^{\zeta,l}$ as needed in Sec.~\ref{sec:comp_tildeG}, and the coefficients $[Y^\alpha_{l,\zeta}]_a$, $\frac{1}{b^{\zeta,l}_\alpha}$, $\gamma^{\zeta,l}_\alpha$ and $\frac{1}{\max_{\mathbf{p}}|G_{\alpha}^{\zeta,l}(|\mathbf{k_p}|\bar{r}^\zeta_l)|}$ as needed in Sec.~\ref{sec:comp_G}.
    \item For keep probabilities with $b_{\text{keep}}$ bits, apply an inequality test with cost $b_{\text{keep}}$.
    \item Conditioned on the flag from the inequality test, swap the alt values using 3 Toffolis.
\end{enumerate}
The total cost for the alias sampler is $Z(2b_s + b_{\text{keep}} + 25)$. The subroutine is cheap enough that we just assume doubling the cost to include uncomputation. As for qubits, $3+b_{\text{keep}}$ are necessary for the QROM and $b_{\text{keep}}$ for the inequality test.

\subsubsection{$\SEL_\loc$}\label{sec:comp_loc_sel}

SELECT for the local interaction term is given by

\begin{align*}
\mathrm{SEL}_\loc & = \sum_{s=-1}^{3}\sum_{\mathbf{q}\in G^{0}}\sum_{c\in\{0,1\}}\kett{s,\mathrm{sgn}(c_{s}^{\zeta_I}),\mathbf{q},c}\brat{s,\mathrm{sgn}(c_{s}^{\zeta_I}),\mathbf{q},c}\otimes U_{(s,\mathrm{sgn}(c_{s}^{\zeta_I}),\mathbf{q},c)}^{\loc}\\
U_{(s,\mathrm{sgn}(c_{s}^{\zeta_I}),\mathbf{q},c)}^{\loc} & =\sum_{\mathbf{p},\mathbf{P}\in G}(-1)^{c(\mathbf{p}-\mathbf{q}\notin G\lor\mathbf{P}+\mathbf{q}\notin G)+\mathrm{sgn}(c_{s}^{\zeta_I})}\kett{\mathbf{p}-\mathbf{q},\mathbf{P}+\mathbf{q}}\brat{\mathbf{p},\mathbf{P}}.
\end{align*}
This has an identical structure to the Coulomb term in App.~\ref{sec:comp_Coul_sel}, but where the momentum transfer is between an electron-ion pair. 
It therefore requires a slightly larger cost of $8\bar{n}$ Toffolis (because the register encoding $\mathbf{P}$ is larger than that of $\mathbf{p}$). The cost for checking overflow of the output momenta is free, as noted above and in~\cite{su2021fault} Sec. II D (as the carry-out bits for the arithmetic already flag overflow).

\subsection[Non-local interaction term]{The non-local interaction term, $V^{\text{PI}}_{\text{NL}}$}\label{sec:comp_nl}

\subsubsection{Nuclear data loading}\label{sec:comp_nucleardata}

We use a single QROM over the pseudoion index $I$ to (i) load the corresponding nuclear species $\zeta_I$ for the interaction term; and (ii) to load a function of the pseudoion masses $M_I^{-\frac{1}{2}}$ for state preparation in the kinetic term. In the following section, we will often drop the subscript $I$ in $\zeta_I$ for the sake of readability. The Toffoli cost of this operation is just $\eta_{\text{ion}}$. Given how few species we consider, we will load $\zeta$ in unary. The total ancilla cost is then $5+b_M$.

\subsubsection{$\text{PREP}_{\NL, 1}$}\label{sec:comp_alprep}

The state $\PREP_{\NL, 1} \kett{0}$ is given by

\begin{equation}
    \mathrm{PREP}_{\NL, 1}\kett 0 \kett{\zeta} =\frac{1}{\sqrt{\tilde{V}_{\mathrm{NL}}^{I}}}\sum_{l=0}^{l_{\max}}\sum_{\alpha=1}^{3}\sqrt{\frac{4\pi}{\Omega}(\bar{r}_{l}^{\zeta})^{3}(2l+1)D_{\alpha}^{\zeta,l}\lambda_{G_{\alpha}^{\zeta,l}}}\kett l\kett{\alpha} \kett{\zeta}.
\end{equation}
We note that the coefficients $D_{\alpha}^{{\zeta}, l}$ and $\lambda_{G_{\alpha}^{\zeta,l}}$ only depend on $I$ insofar as they depend on the nuclear species $\zeta$. Note in principle that  $D_{\alpha}^{{\zeta}, l}$ can have negative values and therefore the amplitudes of this state can be complex. However, we make the simplifying assumption that the entries are positive; negative entries can be remedied by including phase gates at the end of the preparation with negligible cost. Given this assumption, we prepare this state using coherent alias sampling. Given that $\zeta$, $l$ and $\alpha$ are in unary, we first have to convert to binary and contiguize into a single index. For a system with $Z$ ion species, the contiguized index runs from 1 to $9Z$ in the worst case. The conversion can be done with just Cliffords, and the contiguization requires no more than $2Z$ Toffolis (for two $Z$-bit adders). The steps for the alias sampling then proceed as follows:

\begin{enumerate}
    \item Prepare a uniform superposition over $9Z$ amplitudes with cost $2\lceil \log(9Z) \rceil + 2b_{\alpha,l}-8$ for some precision $b_{\alpha,l}$.
    \item Use a QROM to load alt values and keep probabilities with a cost of $9Z$ Toffolis. We also use this QROM to load $b$-bit approximants to the precomputed coefficients $\theta_\alpha^{\zeta,l}$ as needed in Sec.~\ref{sec:comp_tildeG}, and the coefficients $[Y^\alpha_{l,\zeta}]_a$, $\frac{1}{b^{\zeta,l}_\alpha}$, $\gamma^{\zeta,l}_\alpha$ and $\frac{1}{\max_{\mathbf{p}}|G_{\alpha}^{\zeta,l}(|\mathbf{k_p}|\bar{r}^\zeta_l)|}$ as needed in Sec.~\ref{sec:comp_G}.
    \item For keep probabilities with $b_{\text{keep}}$ bits, apply an inequality test with cost $b_{\text{keep}}$.
    \item Conditioned on the flag from the inequality test, swap the alt values using $\lceil \log(9Z) \rceil$ Toffolis.
\end{enumerate}
The total cost for the alias sampler is $11Z + 3\lceil \log(9Z) \rceil + 2 b_{\alpha,l} + b_{\text{keep}} -8$. The subroutine is cheap enough that we just assume doubling the cost to include uncomputation. As for qubits, $\lceil \log(9Z) \rceil+b_{\text{keep}}$ are necessary for the QROM and $b_{\text{keep}}$ for the inequality test.

\subsubsection{Preparation of the reference state $\kett{\psi_{\tilde{G}}}$}\label{sec:comp_tildeG}

We aim to use rejection sampling to prepare the spherically-symmetric 3D state $\kett{\psi_{G}}=\sum_{\mathbf{p}}G_{\alpha}^{\zeta,l}(|\mathbf{k_{p}}|\bar{r}_{l}^{\zeta})\kett{\mathbf{p}}$
with amplitudes $G_{\alpha}^{\zeta,l}(|\mathbf{k_{p}}|\bar{r}_{l}^{\zeta})=\sum_{a=1}^{3}X_{\alpha a}^{\zeta,l} \mathrm{g}_{a}^{l}(|\mathbf{k_{p}}|\bar{r}_{l}^{\zeta})$.
Fixing the signs $\mathrm{sgn}(G_{\alpha}^{I,l}(|\mathbf{k_{p}}|\bar{r}_{l}^{\zeta}))$ is done at a later stage and discussed in Sec.~\ref{sec:comp_G}. We therefore define a reference state 
$\kett{\psi_{\tilde{G}}}=\sum_{\mathbf{p}}\tilde{G}_{\alpha}^{\zeta,l}(\mathbf{k_{p}})\kett{\mathbf{p}}$ such that, for all $\mathbf{p}$, $\tilde{G}_{\alpha}^{\zeta,l} \geq |G_{\alpha}^{\zeta,l}(|\mathbf{k_{p}}|\bar{r}_{l}^{\zeta})|$ (but chosen such that the state with amplitudes $\tilde{G}_{\alpha}^{\zeta,l}$ is easier to prepare than $G_{\alpha}^{\zeta,l}$ itself). Specifically, for the non-local interaction term we choose $\tilde{G}_{\alpha}^{\zeta,l}$ as

\begin{align}\label{eq:reference_func}
\tilde{G}_{\alpha}^{\zeta,l}(\mathbf{k_{p}})=\begin{cases}
\max_{\mathbf{p}}|G_{\alpha}^{\zeta,l}(|\mathbf{k_{p}}|\bar{r}_{l}^{\zeta})| & \mathrm{for}\;\mathbf{p}\in\lozenge:=\{\mathbf{p}:|k_{\mathbf{p}}^{(j)}\bar{r}_{l}^{\zeta}|\leq(k^{*})_{\alpha}^{\zeta,l}\forall j\}\\
d_{\alpha}^{\zeta,l}e^{-\gamma_{\alpha}^{\zeta,l}||\mathbf{k_{p}}\bar{r}_{l}^{\zeta}||_{1}} & \mathrm{for}\;\mathbf{p}\in G\backslash\lozenge
\end{cases}
\end{align}
where the parameters $\max_{\mathbf{p}}|G_{\alpha}^{\zeta,l}(|\mathbf{k_{p}}|\bar{r}_{l}^{\zeta}))|,(k^{*})^{\zeta,l}_\alpha$, $d^{\zeta,l}_\alpha$ and $\gamma^{\zeta,l}_\alpha$ are classically precomputed (see Table~\ref{tab:qrs_reference_params}). We have used the notation $\v{k}_{\v{p}}^{(j)} = \sum_{a=1}^3 p_a b_a^{(j)}$, where $b_a^{(j)}$ is the $j^{th}$ component of the $a^{th}$ reciprocal lattice vector, and the index-free symbol $G$ to denote the space of all valid momenta. Note also that, while the target function is a function of the 2-norm of $\v{k_p}$ (and throughout we have used the shorthand $|\v{k_p}| = \Vert \v{k_p} \Vert_2$), the reference state's dependence comes in the exponent and rather depends on $\Vert \v{k_p} \Vert_1 = \sum_j |k_{\mathbf{p}}^{(j)}|$. Example reference functions are shown in Fig.~\ref{fig:exampleRefStatesBE}. We combine the reference state preparation here with the local interaction term. For the local term with $s \geq 0$, we choose

\begin{equation}\label{eq:reference_func_loc}
\tilde{G}^\zeta_{s \geq 0}(\mathbf{k_{p}})=
\begin{cases}
c_s^\zeta, & \textrm{for} \; \v{p} \in \lozenge := \{\v{p}: |k_{\mathbf{p}}^{(j)}\bar{r}_{l}^{\zeta}| \leq (k^*)^\zeta_s \; \forall j\}\\
d_s^\zeta e^{-\gamma^\zeta_s \Vert\mathbf{k_{p}}\bar{r}_{l}^{\zeta}\Vert_1}, & \textrm{for}\; \v{p} \in G\setminus\lozenge,
\end{cases}
\end{equation}
and for the local term with $s=-1$ we choose
\begin{equation}\label{eq:reference_func_loc_s-1}
\tilde{G}^\zeta_{s=-1}(\mathbf{k_{p}})=
\begin{cases}
2^{1-\mu}, & \textrm{for} \; \v{p} \in \lozenge := \{\v{p}: |k_{\mathbf{p}}^{(j)}\bar{r}_{l}^{\zeta}| \leq (k^*)_{-1} \; \forall j\}\\
d_{-1}^\zeta e^{-\gamma^\zeta_{-1} \Vert\mathbf{k_{p}}\bar{r}_{l}^{\zeta}\Vert_1}, & \textrm{for}\; \v{p} \in G\setminus\lozenge,
\end{cases}
\end{equation}
where $\mu = 1 + \lfloor \log(\max_j\{|\v{k_p}^{(j)}|\}) \rfloor$. Specifically, for the non-local term conditioned on $l$, $\zeta$ and $\alpha$ we will prepare a state:

\begin{equation}
    \kett{\psi_{\tilde{G}}} = \frac{1}{\mathcal{N}_\text{NL}} \left[d^{\zeta,l}_\alpha \left(\sum_{\v p \in G \setminus \lozenge} e^{-\gamma^{\zeta,l}_\alpha \Vert \mathbf{k_{p}}\bar{r}_{l}^{\zeta} \Vert_1}\kett{\v p}\right) + \max_{\v p} |G_\alpha^{l,\zeta}| \left(\sum_{\v p \in \lozenge} \kett{\v p}\right)\right].
\end{equation}
Likewise for the local term with $s \geq 0$:
\begin{equation}
    \kett{\psi_{\tilde{G}}} = \frac{1}{\mathcal{N}_{\text{loc}, s \geq 0}} \left[d_s^\zeta \left(\sum_{\v p \in G \setminus \lozenge} e^{-\gamma_s^\zeta \Vert \mathbf{k_{p}}\bar{r}_{l}^{\zeta} \Vert_1}\kett{\v p}\right) + c_s^\zeta \left(\sum_{\v p \in \lozenge} \kett{\v p}\right)\right].
\end{equation}
and the local term with $s=-1$:
\begin{equation}
    \kett{\psi_{\tilde{G}}} = \frac{1}{\mathcal{N}_{\text{loc}, s=-1}} \left[d_{-1}^\zeta \left(\sum_{\v p \in G \setminus \lozenge} e^{-\gamma_s^\zeta \Vert \mathbf{k_{p}}\bar{r}_{l}^{\zeta} \Vert_1}\kett{\v p}\right) + \left(\sum_{\v p \in \lozenge} 2^{1-\mu} \kett{\v p}\right)\right]
\end{equation}

For readability, we drop the $\zeta$, $l$ and $\alpha$ dependence (and the $s$ dependence for the local term) in the compilation below and reintroduce them when needed for resource estimates. We will also absorb the parameter $\bar{r}_l^\zeta$ in the 1-norm into the parameter $\gamma_\alpha^{\zeta, l}$. Preparing superpositions over the points in $\lozenge$ is nontrivial, and instead we opt to prepare superpositions over simpler domains and then use amplitude amplification to remove parts that we don't want. In particular, let $\square_\text{ins} = \{\v p : |p_a| \leq p_{\text{ins}}, \forall a\}$ and $\square_\text{circ} = \{\v p : |p_a| \leq p_{\text{circ}}, \forall a\}$ be the squares that inscribe and circumscribe $\lozenge$, respectively. Also, let $\square_{\text{out}} = \{\v p : |p_a| \leq p_{\text{max}} - p_{\text{ins}}, \forall a\}$. As written, $\Vert \mathbf{k_{p}} \Vert_1 = \sum_j |\v{k_p}^{(j)}|$ is not immediately separable into its $p_a$ components and so is not simply preparable as a product state over $a$ (which would complicate the compilation). However, define the indicator

\begin{equation}
    I_j = \begin{cases}
        0 & \textrm{if} \; \sum_a p_a b_a^{(j)} \geq 0 \\
        1 & \textrm{if} \; \sum_a p_a b_a^{(j)} < 0.
    \end{cases}
\end{equation}
Then $\Vert \v{k_p} \Vert_1 = \sum_j \left| \sum_a p_a b_a^{(j)} \right| = \sum_j (-1)^{I_j} \sum_a p_a b_a^{(j)} = \sum_a p_a \left(\sum_j (-1)^{I_j}b_a^{(j)}\right)$, which is separable. For each component $p_a$, we will load rotation angles that, if implemented appropriately, approximate a rotation by $\gamma \sum_j (-1)^{I_j}b_a^{(j)}$.

To prepare the state, we first initialize and rotate an ancilla to match the relative weight of the two branches of the reference, $\kett{\psi_0} = \cos\theta^{\zeta,l}_\alpha\kett{0} + \sin\theta^{\zeta,l}_\alpha\kett{1}$ for some appropriately chosen $\theta^{\zeta,l}_\alpha$ (and likewise for some appropriately chosen $\theta_s$ for the local term). This has cost $b_{\text{rot}}$ if implemented up to $b_{\text{rot}}$ bits. Let $f(\mu) = 2^{1-\mu}$ for the $s=-1$ local term and $f(\mu)=1$ otherwise. The next steps are elucidated below:

\begin{align}
    \kett{\psi_0} \xrightarrow{(i)} \frac{1}{\mathcal{N}_1} &\left[ \cos\theta \left(\sum_{\v p \in \square_{\text{out}}} \kett{\v p}\right)\kett{0} + \sin\theta\left(\sum_{\v p \in \square_{\text{circ}}} \kett{\v p}\right)\kett{1} \right] \\
    \xrightarrow{(ii)} \frac{1}{\mathcal{N}_1} &\left[ \cos\theta \left(\sum_{\v p \in \square_{\text{out}}} \kett{\v p}\kett{\v{k_p}}\right)\kett{0} + \sin\theta\left(\sum_{\v p \in  \square_{\text{circ}} } \kett{\v p}\kett{\v{k_p}}\right)\kett{1} \right] \\
    \xrightarrow{(iii)} \frac{1}{\mathcal{N}_1} &\left[ \cos\theta \left(\sum_{\v p \in \square_{\text{out}}} \kett{\v p}\kett{\v{k_p}}\kett{\mu(\v{k_p})}\right)\kett{0} + \sin\theta\left(\sum_{\v p \in  \square_{\text{circ}} } \kett{\v p}\kett{\v{k_p}}\kett{\mu(\v{k_p})}\right)\kett{1} \right] \\
    \xrightarrow{(iv)} \frac{1}{\mathcal{N}_1} &\left[ \cos\theta \left(\sum_{\v p \in \square_{\text{out}}} \kett{\v p}\kett{\v{k_p}}\kett{\mu}\right)\kett{0} + \sin\theta\left(\sum_{\v p \in  \square_{\text{circ}} } f(\mu) \kett{\v p}\kett{\v{k_p}}\kett{\mu}\right)\kett{1} \right] \\
        \end{align}
    \begin{align}
    \xrightarrow{(v)} \frac{1}{\mathcal{N}_2} &\left[ \cos\theta \left(\sum_{\v p \in \square_{\text{out}}} e^{-\gamma \sum_a p_a \left( \sum_j (-1)^{I_j} b_a^{(j)} \right)}\kett{\v p}\kett{\v{k_p}}\kett{\mu}\right)\kett{0} + \sin\theta\left(\sum_{\v p \in \square_{\text{circ}}} f(\mu)\kett{\v p}\kett{\v{k_p}}\kett{\mu}\right)\kett{1} \right] \\
    = \frac{1}{\mathcal{N}_2} &\left[ \cos\theta \left(\sum_{\v p \in \square_{\text{out}}} e^{-\gamma \Vert \v{k_p} \Vert_1}\kett{\v p}\kett{\v{k_p}}\kett{\mu}\right)\kett{0} + \sin\theta\left(\sum_{\v p \in \square_{\text{circ}}} f(\mu)\kett{\v p}\kett{\v{k_p}}\kett{\mu}\right)\kett{1} \right] \\
    \xrightarrow{(vi)} \frac{1}{\mathcal{N}_2} &\left[ \cos\theta \left(\sum_{\v p \in G \setminus \square_{\text{ins}}} e^{-\gamma \Vert \v{k_p} \Vert_1}\kett{\v p}\kett{\v{k_p}}\kett{\mu}\right)\kett{0} + \sin\theta\left(\sum_{\v p \in \square_{\text{circ}}} f(\mu)\kett{\v p}\kett{\v{k_p}}\kett{\mu}\right)\kett{1} \right] \\
    \xrightarrow{(vii)} \frac{1}{\mathcal{N}_2} &\Bigg[ \cos\theta \left(\sum_{\v p \in \lozenge \setminus \square_{\text{ins}}} e^{-\gamma \Vert \v{k_p} \Vert_1}\kett{\v p}\kett{\v{k_p}}\kett{\mu}\kett{0}+\sum_{\v p \in G \setminus \lozenge} e^{-\gamma \Vert \v{k_p} \Vert_1}\kett{\v p}\kett{\v{k_p}}\kett{\mu}\kett{1}\right)\kett{0} \\
    &+ \sin\theta\left(\sum_{\v p \in \lozenge} f(\mu)\kett{\v p}\kett{\v{k_p}}\kett{\mu}\kett{0} +\sum_{\v p \in \square_{\text{circ}} \setminus \lozenge} f(\mu)\kett{\v p}\kett{\v{k_p}}\kett{\mu}\kett{1}\right)\kett{1} \Bigg] \\
    \xrightarrow{(viii)} \frac{1}{\mathcal{N}_2} &\Bigg[ \cos\theta \left(\sum_{\v p \in \lozenge \setminus \square_{\text{ins}}} e^{-\gamma \Vert \v{k_p} \Vert_1}\kett{\v p}\kett{\v{k_p}}\kett{\mu}\kett{1}+\sum_{\v p \in G \setminus \lozenge} e^{-\gamma \Vert \v{k_p} \Vert_1}\kett{\v p}\kett{\v{k_p}}\kett{\mu}\kett{0}\right)\kett{0} \\
    &+ \sin\theta\left(\sum_{\v p \in \lozenge} f(\mu)\kett{\v p}\kett{\v{k_p}}\kett{\mu}\kett{0} +\sum_{\v p \in \square_{\text{circ}} \setminus \lozenge} f(\mu)\kett{\v p}\kett{\v{k_p}}\kett{\mu}\kett{1}\right)\kett{1} \Bigg] \\
    \xrightarrow{(ix)} \frac{1}{\mathcal{N}_3} &\Bigg[ \cos\theta \left(\sum_{\v p \in G \setminus \lozenge} e^{-\gamma \Vert \v{k_p} \Vert_1}\kett{\v p}\kett{\v{k_p}}\kett{\mu}\kett{0}\right)\kett{0} + \sin\theta\left(\sum_{\v p \in \lozenge} f(\mu)\kett{\v p}\kett{\v{k_p}}\kett{\mu}\kett{0}\right)\kett{1} \Bigg] \\
    \xrightarrow{(x)} \frac{1}{\mathcal{N}_3} &\Bigg[ \cos\theta \left(\sum_{\v p \in G \setminus \lozenge} e^{-\gamma \Vert \v{k_p} \Vert_1}\kett{\v p}\kett{\v{k_p}}\kett{0}\right)\kett{0} + \sin\theta\left(\sum_{\v p \in \lozenge} f(\mu)\kett{\v p}\kett{\v{k_p}}\kett{0}\right)\kett{1} \Bigg] \\
    \xrightarrow{(xi)} \frac{1}{\mathcal{N}_3} &\Bigg[ \cos\theta \left(\sum_{\v p \in G \setminus \lozenge} e^{-\gamma \Vert \v{k_p} \Vert_1}\kett{\v p}\kett{\v{k_p}}\right) + \sin\theta\left(\sum_{\v p \in \lozenge} f(\mu)\kett{\v p}\kett{\v{k_p}}\right) \Bigg] \\
    \xrightarrow{(xii)} \frac{1}{\mathcal{N}_3} &\Bigg[ \cos\theta \left(\sum_{\v p \in G \setminus \lozenge} e^{-\gamma \Vert \v{k_p} \Vert_1}\kett{\v p}\right) + \sin\theta\left(\sum_{\v p \in \lozenge} f(\mu)\kett{\v p}\right) \Bigg].
\end{align}

\begin{enumerate}[label=(\roman*)]
    \item The uniform state preparation on both branches can be carried out by controlled USP, but given that we will flag out unwanted subspaces using amplitude amplification, we ``overprepare'' a uniform superposition over $2^{\lceil \log(p_\text{circ}) \rceil}$ amplitudes rather than $p_{\text{circ}}$ using controlled Hadamards and we combine the flagging of the superfluous amplitudes with the flagging for $\lozenge$. In the worst case, this step requires 2n controlled Hadamards, each costing a single Toffoli. 
    \item Evaluating $\sum_a p_a b_a^{(j)}$ to an ancilla register requires three multiplications and two additions, with cost $\tilde{n} + 2n$. This must be evaluated for each value of $j$, so the total cost is $3\tilde{n} + 6n$.
    \item We compute $\mu(\v{k_p}) = 1 + \lfloor \log(\max_j |\v{k_p}^{(j)}|) \rfloor$ in unary. Computing $\max_j |\v{k_p}^{(j)}|$ requires three inequality tests to find the maximum component, each with cost $n$, and then a ladder of $3n$ Toffolis to copy the state of the maximum component to an ancilla register. The function $1+\lfloor \log(x) \rfloor$ corresponds to finding the most-significant non-zero digit of $x$. If $x$ has $n$ bits, then $1+\lfloor \log(x) \rfloor$ is also encoded in $n$ bits in unary. If we label the bits of $x$ from least-significant to most-significant, this function can be achieved by initializing an output register in the state $\kett{0 \ldots 01}$ and then flipping the $j^{th}$ zero conditioned on whether $x \geq 2^{j}$. Naively checking each inequality via inequality test has cost $n$, and there are $n-1$ inequalities to check. Flipping bits requires CNOTs only. The total cost of this step is therefore $n^2 + 5n$.
    \item We wish to effect a transformation on the branch of the wavefunction inside $\square_{\text{circ}}$, and for the $s=-1$ part of the local pseudopotential only. Let $\square_{\text{circ}} = \lozenge \cup (\square_{\text{circ}} \setminus \lozenge)$, and decompose the region $\lozenge$ as a set of annuli indexed by $\mu$: $\lozenge = \bigcup_{\mu=1}^{\mu_{\max}} \lozenge_\mu$. Then the branch of the wavefunction we are interested in is in the (un-normalized) state $\sum_{\v p \in  \square_{\text{circ}} } \kett{\v p}\kett{\v{k_p}}\kett{\mu} = \sum_\mu \sum_{\v p \in \lozenge_\mu} \kett{\v p}\kett{\v{k_p}}\kett{\mu} + \sum_{\v p \in (\square{\text{circ}} \setminus \lozenge)} \kett{\v p}\kett{\v{k_p}}\kett{\mu}$. The target state given $f(\mu)$ and conditioned on the flag for the $s=-1$ part of the local pseudopotential is  $\sum_\mu \sum_{\v p \in \lozenge_\mu} 2^{1-\mu} \kett{\v p}\kett{\v{k_p}}\kett{\mu} + \sum_{\v p \in (\square{\text{circ}} \setminus \lozenge)} 2^{-\mu_{\max}} \kett{\v p}\kett{\v{k_p}}\kett{\mu} = \sum_\mu 2^{1-\mu} |\lozenge_{\mu}| \sum_{\v p \in \lozenge_\mu} \kett{\v p}\kett{\v{k_p}}\kett{\mu} + \sum_{\v p \in (\square{\text{circ}} \setminus \lozenge)} 2^{-\mu_{\max}} \kett{\v p}\kett{\v{k_p}}\kett{\mu}$. Given that $\mu$ is encoded in unary, this transformation can be implemented using a ladder of controlled rotations like in the power law reference state preparation in~\cite{lemieux2024quantum}, Sec. 3.3 (and also in \cite{babbush2019quantum}). However, in these examples the multiplicities of the points in each pair of adjacent annuli are a constant fraction and so the preparation reduces to a ladder of singly-controlled Hadamards (i.e. with target on qubit $j$ and control on qubit $j-1$ in the register encoding $\mu$). In this instance, the multiplicities of points in $\lozenge_\mu$ and $\lozenge_{\mu+1}$ are not so obviously related; and so instead we assume arbitrary angle rotations (to a finite bit precision $b_{\text{pl}}$). For each rotation, we must add an additional two controls: one on the branch ancilla, and one on the flag qubit indicating the $s=-1$ local term. Each rotation can be implemented with two Toffolis and two uncontrolled rotations, with cost $2b_{\text{pl}} + 2$. There are $\mu_{\max} = n$ rotations, so the cost of this step is $2b_{\text{pl}} n + 2n$.
    \item To load the rotation angles for the tail, we must first compute $I_j$; however, this is free because it just corresponds to the carry bit of the register containing $\v{k}_{\v{p}}^{(j)}$. There are naively eight different choices for the summation $\sum_j (-1)^{I_j} b_a^{(j)}$ given the possible evaluations of $(-1)^{I_j}$. However, we can pull out a ``global'' sign difference and cut down the loading by half (e.g. we load the case where $I_j=1, \forall j$, but the case where $I_j = -1, \forall j$ is generated from the first case by a CNOT on the sign bit conditioned on, say, the value of $I(1)$). For the four remaining cases, we load all possible rotation angles and then swap up the correct ones based on the evaluation of $I_j$. If the rotation angles are loaded to $b_{\text{exp}}$ bits of precision, then swapping up the correct angles for the $a^{th}$ component requires $3b_{\text{exp}}$ CSWAPS (or an equal number of Toffoli gates), and so the cost is $9b_{\text{exp}}$ once we have summed over $a$. We carry out a $b_{\text{exp}}$-bit rotation on every qubit in the $\kett{\v p}$ register, conditioned on the ancilla being in the state $\kett{1}$. This has total cost $2b_{\text{exp}}n$.
    \item To switch from a domain $\square_{\text{out}}$ to $G \setminus \square_{\text{ins}}$ for the branch on which we prepared the exponential state, we shift each component $p_a$ by $p_{ins}$. This requires three controlled additions, with total cost $2n$.  
    \item We next compute whether $\v p \in \lozenge$ to an additional ancilla qubit (call this qubit the ``$\lozenge$ ancilla''). Given that we have already computed the components of $\v{k_p}$, this step requires only a inequality test for each of the three components plus two Toffolis to AND the results together. This step therefore has cost $3n+2$.
    \item Given that we want to keep the case when $\v p \in \lozenge$ for the uniform part and $\v p \in G \setminus \lozenge$ for the exponential part, we must flip the output of the $\lozenge$ ancilla controlled on the branch ancilla. This requires only a single CNOT, with no non-Clifford cost. 
    \item The subspace that we wish to amplify is now encoded in the $\kett{0}$ subspace of the $\lozenge$ ancilla. Naively, the amount of amplification depends on the input parameters to the problem. However, we note that both the subspace and its amplitude are known in advance, and so we can incorporate the same ``partial reflection'' trick as in USP to amplify this branch to unit relative amplitude with a single round. This requires three calls to the preparation above, plus some small overhead to carry out the reflections. We omit this overhead in the resource estimate.
    \item We assume that the cost to uncompute $\mu$ is the same as the cost of computation; i.e. $n^2 + 5n$.
    \item The $\lozenge$ ancilla is already in the $\kett{0}$ state and can be discarded. Uncomputation of the branch ancilla can be carried out by rechecking whether $\v p \in \lozenge$. If we have retained the outputs of the inequality tests in step (v), this requires only two Toffolis. The inequality tests themselves are uncomputed with cost $3n+2$.
    \item Finally, the calculation of $\v k_p$ is uncomputed with cost $3\tilde{n}+6n$.
\end{enumerate}

The total cost is therefore $12\tilde{n} + 74n + 4n^2 + 6 b_{\text{pl}} n + 6b_{\text{exp}}n + 3b_{\text{rot}} + 8$. For $R$ rounds of amplification, this must be repeated $1+R$ times (and the dagger $R$ times). We assume that both have the same cost, and therefore that the complexity above must be multiplied by $1+2R$. As for ancillas, the rotations require $b_{\text{exp}}$ qubits and the $\v{k_p}$ and $\mu$ registers are over $n$ qubits, and the intermediate arithmetic requires $n$ qubits. Other constant-factor costs (such as additional ancillae for amplitude amplification and checking inequality tests) are omitted.

\subsubsection{Computing $G_\alpha^{\zeta,l}(|\v{k_{p}}|\bar{r}_{l}^{\zeta})/\tilde{G}_\alpha^{\zeta, l}(\v{k_{p}})$ and $G_s^{\zeta}(|\v{k_{p}}|\bar{r}_{\loc}^{\zeta})/\tilde{G}_s^{\zeta}(\v{k_{p}})$ to an ancilla register}\label{sec:comp_G}

In order to carry out rejection sampling, for the non-local term we must coherently evaluate the inequality $G_\alpha^{\zeta,l} M \geq \tilde{G}_\alpha^{\zeta,l} m$ for a uniformly-prepared set of amplitudes $m$. In practice, we rearrange this inequality to check $\frac{G_\alpha^{\zeta,l}}{\tilde{G}_\alpha^{\zeta, l}} M \geq m$. 

Let $\bar{G}_\alpha^{\zeta, l}:= \frac{G_\alpha^{\zeta, l}}{\tilde{G}_\alpha^{\zeta, l}}$.
The preparation of the state $\kett{\bar{G}_\alpha^{\zeta, l}(\v{k})}$ largely proceeds analogously to the construction in~\cite{berry2023quantum}, Sec VI D, for $\kett{F_a^{\zeta, l}(\v{k})}$ (given that they are both a polynomial multiplied by a Gaussian) but with a few minor modifications; in particular that we are concerned with a decomposition into radial functions $\mathrm{g}_a^l(|\v{k}| r^\zeta_l)$ as in Eqs.~\eqref{eq:G_func_def} and~\eqref{eq:g_function_2}. 
We assume that the registers are encoded as signed integers. To first do the conversion from $g$ to $G$, we write

\begin{align}
\label{eq:gtoG}
G_{\alpha}^{\zeta,l}(|\mathbf{k_{p}}|\bar{r}_{l}^{\zeta}) &= \sum_{a=1}^{3}[X^{\zeta,l}]_{a\alpha}\mathrm{g}_{a}^{l}(|\mathbf{k_{p}}|\bar{r}_{l}^{\zeta}) \\
&= \sum_{a=1}^{3}[X^{\zeta,l}]_{a\alpha} e^{-(|\v{k}|\bar{r}_{l}^{\zeta})^2/2}(|\v{k}|\bar{r}_{l}^{\zeta})^{l}\frac{\sqrt{\pi}2^{a-1}(a-1)!}{\sqrt{\Gamma(l+2a-\frac{1}{2})}}\mathrm{L}_{a-1}^{l+\frac{1}{2}}((|\v{k}|\bar{r}_{l}^{\zeta})^2/2)\\
&= e^{-(|\v{k}|\bar{r}_{l}^{\zeta})^2/2}(|\v{k}|\bar{r}_{l}^{\zeta})^{l} \sum_{a=1}^{3}[X^{\zeta,l}]_{a\alpha} \frac{\sqrt{\pi}2^{a-1}(a-1)!}{\sqrt{\Gamma(l+2a-\frac{1}{2})}}\mathrm{L}_{a-1}^{l+\frac{1}{2}}((|\v{k}|\bar{r}_{l}^{\zeta})^2/2)\\
&= e^{-(|\v{k}|\bar{r}_{l}^{\zeta})^2/2}(|\v{k}|\bar{r}_{l}^{\zeta})^{l} \sum_{a=1}^{3} [Y^{\alpha}_{l,\zeta}]_a \mathrm{L}_{a-1}^{l+\frac{1}{2}}((|\v{k}|\bar{r}_{l}^{\zeta})^2/2) \\
&= e^{-(|\v{k}|\bar{r}_{l}^{\zeta})^2/2}(|\v{k}|\bar{r}_{l}^{\zeta})^{l} \sum_{a=1}^{3} [Y^{\alpha}_{l,\zeta}]_a \sum_{x=0}^{a-1} c_{x,la} (|\v{k}| \bar{r}_{l}^{\zeta})^{2x},
\end{align}
where we have defined $[Y^{\alpha}_{l,\zeta}]_a = [X^{\zeta,l}]_{a\alpha} \frac{\sqrt{\pi}2^{a-1}(a-1)!}{\sqrt{\Gamma(l+2a-\frac{1}{2})}}$, and implicitly defined $c_{x, la}$ through the polynomial expansion of the Laguerre polynomial $L$. To convert from $G$ to $\bar{G}$, we use the definition of $\tilde{G}$ in Eq.~\eqref{eq:reference_func}. Explicitly,

\begin{equation}\label{eq:barG}
    \bar{G}_\alpha^{\zeta, l}(\v{k}) = \begin{cases}
\frac{1}{\max_{\mathbf{p}}|G_{\alpha}^{\zeta,l}(\mathbf{k})|}|\v{k}|^l e^{-(\bar{r}_{l}^{\zeta} |\v{k}|)^2/2} \sum_{a=1}^3 [Y^{\alpha}_{l,\zeta}]_a \sum^{a-1}_{x=0} c_{x,la} (\bar{r}_{l}^{\zeta}|\v{k}|)^{2x} & |\mathbf{k}^{(j)}| \leq k_{l}^{*} \; \forall j\\
\frac{1}{d_{l}}|\v{k}|^l e^{\gamma\Vert\mathbf{k}\Vert_1-(\bar{r}_{l}^{\zeta} |\v{k}|)^2/2} \sum_{a=1}^3 [Y^{\alpha}_{l,\zeta}]_a \sum^{a-1}_{x=0} c_{x,la} (\bar{r}_{l}^{\zeta}|\v{k}|)^{2x} & \text{otherwise.}
\end{cases}
\end{equation}
Likewise, the equivalent definition for the local term for $s \geq 0$ is given by Eq.~\eqref{eq:reference_func_loc}:
\begin{equation}\label{eq:barGloc}
    \bar{G}_s^{\zeta}(\v{k}) = \begin{cases}
\frac{1}{c_s} e^{-(\bar{r}^\zeta_{\loc} |\v{k}|)^2/4} (\bar{r}^{\zeta}_\loc|\v{k}|)^{s} & |\mathbf{k}^{(j)}| \leq k_{s}^{*} \; \forall j\\
\frac{1}{d_{s}} e^{\gamma\Vert\mathbf{k}\Vert_1-(\bar{r}^\zeta_\loc |\v{k}|)^2/4} (\bar{r}^{\zeta}_\loc|\v{k}|)^{s} & \text{otherwise,}
\end{cases}
\end{equation}
and the definition for the local term for $s=-1$ is given by Eq.~\eqref{eq:reference_func_loc_s-1}:
\begin{equation}\label{eq:barGloc_s-1}
    \bar{G}_{-1}^{\zeta}(\v{k}) = \begin{cases}
2^{1-\mu} e^{-(\bar{r}^\zeta_{\loc} |\v{k}|)^2/4} (\bar{r}^{\zeta}_\loc|\v{k}|)^{s} & |\mathbf{k}^{(j)}| \leq k_{s}^{*} \; \forall j\\
\frac{1}{d_{s}} e^{\gamma\Vert\mathbf{k}\Vert_1-(\bar{r}^\zeta_\loc |\v{k}|)^2/4} (\bar{r}^{\zeta}_\loc|\v{k}|)^{s} & \text{otherwise.}
\end{cases}
\end{equation}

We seek to carry out these state preparations jointly; i.e. to prepare a state proportional to $\kett{\varsigma=0, \varphi = 0}\kett{\bar{G}_\alpha^{\zeta, l}(\v{k_{p_2}})} + \kett{\varsigma=1, \varphi = 0}\kett{\bar{G}_s^{\zeta}(\v{k_{p_2}})} + \kett{\varsigma=1, \varphi = 1}\kett{\bar{G}_{-1}^{\zeta}(\v{k_{p_2}})}$, conditioned on registers containing $l$, $\alpha$, $\zeta$ and $s$ (and where the flag $\varsigma$ indicates whether we are implementing the nonlocal term or the local term, and the flag $\varphi$ indicates whether we are implementing the $s=-1$ or $s \geq 0$ piece of the local term). We outline the steps and corresponding costs to construct this state here:

\begin{enumerate}
    \item From $\v{p_2}$, compute $|\v{k_{p_2}}|^2$ to an ancilla register. If we take the worst-case from~\cite{berry2023quantum}, Appendix C then the cost is $\frac{5}{2}\tilde{n} + 2n^2 + 4bn$ for a $b$-bit approximation. However, we can pull out one of the multiplications (it would be most prudent to pick the biggest multiplicand) by a real number from this calculation and absorb it into the coefficient preparation of $\bar{r}^l_\zeta$ (or $\bar{r}^\zeta_\loc$). The cost for this step is then estimated to be $\frac{5}{2}\tilde{n} + 2n^2 + 4bn - 2n_{\max}(n_{\max}+b)$, where $n_{\max}=\max(n_1, n_2, n_3)$. This is also the cost that we assume for the preparation of $|\v{k_P}|^2$ for the kinetic term (but with $n$ replaced with $\bar{n}$).
    \item Using the coefficients loaded in Sec.~\ref{sec:comp_alprep}, compute $(\bar{r}^l_\zeta)^2|\v{k_{p_2}}|^2$ to an ancilla register if $\varsigma=0$ and $(\bar{r}_\loc^\zeta)^2|\v{k_{p_2}}|^2$ if $\varsigma=1$ and $s \neq 0$ with leading-order cost $2b^2$.
    \item If $\varsigma=0$, square the previous step out-of-place to evaluate $(\bar{r}^l_\zeta)^4|\v{k_{p_2}}|^4$, with cost $\frac{b^2}{2}$. The addition of a control introduces a cost linear in $b$, which we omit.
    \item To evaluate $\sum_{x=0}^{a-1} c_{x,la}(\bar{r}^l_\zeta \v{k_{p_2}})^{2x}$, we make the following remarks (in analogy to~\cite{berry2023quantum}, Sec VI D):
        \begin{enumerate}
            \item For $a=1$, we only require the constant $c_{0, l1}$ which can be copied down with zero cost.
            \item For $a=2$, we require $c_{0, l2} + c_{1, l2}(\bar{r}^l_\zeta \v{k_{p_2}})^2 = c_{0, l2} - (\bar{r}^l_\zeta \v{k_{p_2}})^2$. This is a single controlled addition with cost $2b$.
            \item For $a=3$, we require $c_{0, l3} + c_{1, l3}(\bar{r}^l_\zeta \v{k_{p_2}})^2 + c_{2, l3}(\bar{r}^l_\zeta \v{k_{p_2}})^4 = c_{0, l3} + c_{1, l3}(\bar{r}^l_\zeta \v{k_{p_2}})^2 + (\bar{r}^l_\zeta \v{k_{p_2}})^4$. For the systems we consider, we only have $\alpha=3$ when $l=0$. In this case, we must evaluate $c_{0, 03} + c_{1, 03}(\bar{r}^l_\zeta \v{k_{p_2}})^2 + (\bar{r}^l_\zeta \v{k_{p_2}})^4 = 15 - 10(r^l_\zeta \v{k_{p_2}})^2 + (\bar{r}^l_\zeta \v{k_{p_2}})^4$. The cost is that of two controlled bit-shifts (to give $2(\bar{r}^l_\zeta \v{k_{p_2}})^2$ and $8(\bar{r}^l_\zeta \v{k_{p_2}})^2$, respectively) and three controlled $b$-bit additions, with total cost $8b$.
        \end{enumerate}
    The total cost of this step is therefore $10b$. If $\varsigma=1$, the coefficients $c_{x,la}=0$, and this summation evaluates to zero.
    \item The constants $[Y^{\alpha}_{l,\zeta}]_a$ were loaded using the QROM in Sec.~\ref{sec:comp_alprep}, multiplexed over $\zeta$, $l$ and $\alpha$. Multiplying by this constant to produce $[Y^{\alpha}_{l,\zeta}]_a \sum^{a-1}_{x=0} c_{x,la} (\bar{r}^{\zeta}_l|\v{k}|)^{2x}$ has cost $b^2$. Naively, we would need to evaluate this multiplication three times (once for each $a$). However, we can absorb a common factor into the preparation of the constants outside the summation in the expression for $\bar{G}$, and we therefore only have to evaluate two multiplications with total cost $2b^2$. If $\varsigma=1$, the multiplicand generated in the previous step is zero and therefore the output is zero.
    \item Two additions are needed to complete the sum over $a$ and to evaluate $\sum_{a=1}^3 [Y^{\alpha}_{l,\zeta}]_a \sum^{a-1}_{x=0} c_{x,la} (\bar{r}^{\zeta}_l|\v{k}|)^{2x}$, with cost $2b$.
    \item Unlike~\cite{berry2023quantum}, we seek to prepare a state directly approximating the quantity $G_\alpha^{\zeta, l}$, rather than an upper bound to it. Unfortunately, this requires multiplying by $|\v{k_{p_2}}|^l$ for the nonlocal term. We have already prepared this for $l=2$ in Step 1, but we have not prepared $|\v{k_{p_2}}|$ in the case where $l=1$. We compute this from the register containing $|\v{k_{p_2}}|^2$ using a controlled square-root circuit from~\cite{munoz2018tcount}. However, we make two adjustments: we substitute more efficient (i.e. Gidney~\cite{gidney2018halving}) base adders, and we count resources in Toffolis rather than Ts. Specifically for $b$-bit input, the algorithm in~\cite{munoz2018tcount} has three parts: (i) an ``initial subtraction'' consisting of one adder with cost $b$; a ``conditional addition'' that consists of additions of $2(j+1)$-bit addends for $j=2 \ldots \frac{b}{2}-1$; a ``remainder restoration'' that uses a single controlled addition with cost $2b$. The total cost is therefore $3b + \sum_{j=2}^{\frac{b}{2}-1} 2(j+1) = \frac{b^2}{4}+\frac{7b}{2}-6$. 
    \item Multiplication by the prefactor $|\v{k_{p_2}}|^l$ requires a single controlled multiplication. The cost of the control is linear in $b$, which we omit. The primary contribution has cost $b^2$. This completes the state preparation for the polynomial part of $\bar{G}_\alpha^{\zeta, l}$ in Eq.~\eqref{eq:barG}.
    \item For the local term, we must construct $(\bar{r}_\loc^\zeta)^s|\v{k_{p_2}}|^s$ from $(\bar{r}_\loc^\zeta)^2|\v{k_{p_2}}|^2$. Controlled on $\varsigma=1$ and $s \neq 2$, we square-root this register out-of-place with cost given in Step 7. We then multiply $(\bar{r}_\loc^\zeta)|\v{k_{p_2}}|$ with $(\bar{r}_\loc^\zeta)^2|\v{k_{p_2}}|^2$ in the case where $s=3$, using a single multiplication with leading-order cost $b^2$. For the $s=-1$ part, rather than compute the reciprocal we instead multiply the alternate side of the inequality test for rejection sampling by $(\bar{r}_\loc^\zeta)|\v{k_{p_2}}|$ (i.e. we carry out a single multiplication conditioned on whether both $\varphi = 1$ and $s=1$). The total cost for this step is therefore $\frac{5b^2}{4}+\frac{7b}{2}-6$. This completes the state preparation for the polynomial part of $\bar{G}_s^\zeta$ in Eqs.~\eqref{eq:barGloc} and~\eqref{eq:barGloc_s-1}.
    \item For the non-local term, we must check whether $|\v{k}_{\v{p_2}}^{(j)}| = |\sum_\alpha p_\alpha b_\alpha^{(j)}| \leq k^*_l$, for all $j$ (where here, $b_\alpha$ is the $\alpha^{th}$ reciprocal lattice vector). Likewise for the local term, we must check the same condition against the precomputed coefficient $k^*_s$. The coefficients $k^*_l$ and $k^*_s$ are precomputed and loaded using the QROM in Sec.~\ref{sec:comp_alprep}. Evaluating $|\v{k}_{\v{p_2}}^{(j)}|$ requires three multiplications of an $n_j$-bit component with a $b$-bit classical multiplicand (plus two additions). The cost for each multiplication is $\frac{n_j^2}{2}+bn_j$. The total cost of evaluating $|\v{k}_{\v{p_2}}^{(j)}|$ is therefore $\frac{1}{2}\tilde{n} + bn + 2b$. This must be carried out three times (one for each $j$), plus the cost of three inequality tests. The total cost is $\frac{3}{2}\tilde{n} + 3bn + 9b$.
    \item Dependent on both $\varsigma$ and the outcomes of the inequality tests in the previous step, we must multiply by an $l$- or $s$-dependent constant. Both of these families of coefficients are assumed to be precomputed classically and loaded using the QROM in Sec.~\ref{sec:comp_alprep}. In the worst case this step requires two controlled multiplications. Including the controls introduces a complexity linear in $b$, which we omit. The cost for this step is then $2 b^2$.
    \item We must evaluate $\gamma\Vert \v{k_{p_2}} \Vert_1 = \sum_j |\v{k}_{\v{p_2}}^{(j)}|$. Each of these components have already been computed; the cost of this step is therefore the cost of two additions: $2b$.
    \item If the flag in Step 9 indicates that the criterion is not satisfied $\forall j$, we must modify the argument to the exponential such that we evaluate $\gamma\Vert\v{k_{p_2}}\Vert_1-(\bar{r}^\zeta_l |\v{k_{p_2}}|)^2/2$ rather than $-(\bar{r}^\zeta_l |\v{k_{p_2}}|)^2/2$ for the nonlocal term (and likewise $\gamma\Vert\v{k_{p_2}}\Vert_1-(\bar{r}^\zeta_\loc |\v{k_{p_2}}|)^2/4$ rather than $-(\bar{r}^\zeta_\loc |\v{k_{p_2}}|)^2/4$ for the local term). This requires a single controlled subtraction with cost $2b$.
    \item The preparation of the exponential part in all cases is achieved with QROM interpolation as described in~\cite{berry2023quantum}, Sec. VI C. We assume the pessimistic cost derived there of $\frac{11}{4}b^2 + 128$.
    \item Multiplication of the exponential and polynomial parts to yield $\bar{G}_\alpha^{\zeta, l}$ and $\bar{G}_s^{\zeta}$ has cost $b^2$.

\end{enumerate}

The reference function $\tilde{G}$ is guaranteed to upper bound $|G|$, not just $G$, and the rejection sampling (in both success probability and circuit compilation) assumes that one is comparing $\tilde{G}$ with $|G|$ (and that the appropriate phases are added after the rejection sampling procedure). Given that $\kett{G_\alpha^{\zeta, l}}$ is stored as a signed integer, $\kett{|G_\alpha^{\zeta, l}|}$ is extracted just by ignoring the sign bit. The inequality test then proceeds on $\kett{|G_\alpha^{\zeta, l}|}$ and $\kett{\tilde{G}_\alpha^{\zeta, l}}$, producing a state with amplitudes proportional to $|G_\alpha^{\zeta, l}|$. To modify the resultant state so that it encodes amplitudes proportional to $G_\alpha^{\zeta, l}$ itself, we only need to apply a single Pauli Z to the sign bit.

The total Toffoli cost of Steps 1-15 is then $4\tilde{n} + 2n^2 + 7bn + \frac{51}{4} b^2 + 32 b -2n_{\max}(n_{\max} + b) + 116$. Note that, in the rejection sampling procedure outlined in~\cite{lemieux2024quantum}, in order to amplify we must call this subroutine (including its dagger) multiple times. In Table~\ref{tab:qre_nl}, we denote the number of amplification rounds by $R$. In~\cite{lemieux2024quantum}, we require $1+R$ queries to the routine and $1+R$ to its dagger (see Table 1 therein). However, given that the steps above only consist of coherent arithmetic we can retain the intermediate ancillas and apply the dagger at no Toffoli cost. As shown in Fig.~\ref{fig:succProbNL}, for all elements that we have tested, we can safely take $R \leq 2$ with small tweaks to the reference function.

We also require ancillas to store the intermediate arithmetic quantities. Specifically, we require ancillas to store $|\v{k_{p_2}}|^2$, $(\bar{r}^\zeta_l|\v{k_{p_2}}|)^2$, $(\bar{r}^\zeta_l|\v{k_{p_2}}|)^4$, $\sum_x c_{x, la}(\bar{r}^\zeta_l|\v{k_{p_2}}|)^{2x}$, $[Y^{\alpha}_{l,\zeta}]_a \sum^{a-1}_{x=0} c_{x,la} (\bar{r}^{\zeta}_l|\v{k}|)^{2x}$ (for each $a$), $\sum_{a=1}^3 [Y^{\alpha}_{l,\zeta}]_a \sum^{a-1}_{x=0} c_{x,la} (\bar{r}^{\zeta}_l|\v{k}|)^{2x}$, $\gamma \Vert \v{k_{p_2}} \Vert_1$, the exponential term, and $\bar{G}_\alpha^{\zeta, l}$ itself. If we assume that $|\v{k_{p_2}}|^2$ is computed to $b$ bits, we require roughly $2b$, $4b$, $8b$, $8b$, $24b$, $8b$, $2b$, $b$ and $8b$ qubits for each register, respectively (for a total of $65b$ qubits). While we expect that we can truncate without significantly affecting the quality of the output state, we do not carry out that analysis in this work.

\subsubsection{Inequality tests}\label{sec:comp_NLineq}

Inequality tests are necessary for rejection sampling for in the nonlocal term. For the steps in Sec.~\ref{subsubsec:BENonlocalPP}, this requires checking whether $\frac{|G|}{\tilde{G}} M \geq m$, for some target state $|G|$, reference state $\tilde{G}$, number of samples $M$ and a uniform superposition over $m=1, \ldots, M$. 
Given that the quantity $\frac{|G|}{\tilde{G}}$ has already been computed in previous steps, we only need to cost the two multiplications to prepare each side of the inequality test and the test itself. However, we make the simplifying assumption that $M$ is a power of two (or more precisely, for any target $M$ as specified by the error budget, we prepare over $\tilde{M} = 2^{\lceil \log M \rceil}$ amplitudes). This can only increase the fidelity with which the target state has prepared, and makes the preparation for the inequality test trivial. In particular, both preparing the uniform state over $m$ and multiplying by $M$ has no non-Clifford cost. The only cost, therefore, is for the inequality test itself, which has cost $b+ b_{\tilde{M}}$ (where $b_{\tilde{M}} = \log{\tilde{M}}$). For $R$ rounds of amplification, this inequality test must be repeated $1+R$ times (and its dagger $R$ times). However, the dagger can be implemented with no Toffoli cost if ancillas are retained. The total cost is therefore $(1+R)(b+ b_{\tilde{M}})$.

We require an extra $b_{\tilde{M}}$ qubits to pad for the multiplication and $b+ b_{\tilde{M}}$ for the inequality test.

\subsubsection{$\mathcal{P}_l$: block encoding the Legendre polynomial}\label{sec:comp_legendre}

We aim to prepare the block encoding

\begin{align}
\sum_{\v{p_1}, \v{p_2} \in G} P_l\left(\frac{\v{k_{p_2}}\cdot \v{k_{p_1}}}{|\v{k_{p_1}}| |\v{k_{p_2}}|}\right) \kett{\v{{p_2}}}\brat{\v{{p_2}}} \otimes \kett{\v{{p_1}}}\brat{\v{{p_1}}}.
\end{align}

\noindent The steps to prepare this block encoding are given in Sec.~\ref{subsubsec:BENonlocalPP}, Step 3, but we collate the costs here. 
As in Sec.~\ref{sec:comp_G}, intermediate quantities are computed in signed integer representation. 
As noted in the main body, we can prepare a uniform superposition over $m = 1, \ldots, M$ basis states and then check the inequalities
\begin{equation}\label{eq:legendre_cases}
    \begin{cases}
        M \geq m, & l=0, \\
        (\v{k_{p_1}} \cdot \v{k_{p_{2}}}) M \geq |\v{k_{p_1}}| |\v{k_{p_2}}| m, & l=1, \\
        [3(\v{k_{p_1}} \cdot \v{k_{p_{2}}})^2 - |\v{k_{p_1}}|^2 |\v{k_{p_2}}|^2]  M \geq 2 |\v{k_{p_1}}|^2 |\v{k_{p_2}}|^2 m, & l=2.
    \end{cases}
\end{equation}
Rearranging the above inequalities recovers the correct inequality test for the Legendre polynomial. We already have $|\v{k_{p_2}}|$ stored as an intermediate quantity from the previous step, which we can reuse. The substeps are as follows:
\begin{enumerate}
    \item Compute $|\v{k_{p_1}}|$ equivalently to the steps for $\v{k_{p_2}}$. This has cost $\frac{5}{2}\tilde{n}+2n^2 + 4bn-2n_{\max}(n_{\max} + b) + \frac{b^2}{4} + \frac{7}{2}b - 6$ (this is the combined cost to evaluate the norm and then take the square root). Note that we can apply the square root out-of-place and then reuse the calculation of $|\v{k_{p_1}}|^2$ for the kinetic term.
    \item Compute $\v{k_{p_1}} \cdot \v{k_{p_{2}}}$ and $|\v{k_{p_1}}| |\v{k_{p_2}}|$ to ancilla registers. 
    For the former, we inherit the procedure and costs from Ref.~\cite{berry2023quantum}, App. C. 
    Specifically, the cost of computing the dot product is $\frac{5}{2}\tilde{n} + 3n^2 + 4bn$. The cost of calculating $|\v{k_{p_1}}| |\v{k_{p_2}}|$ is just the cost of a single multiplication, $b^2$, as both multiplicands have already been computed previously.
    \item If $l=2$, square $\v{k_{p_1}} \cdot \v{k_{p_{2}}}$ and $|\v{k_{p_1}}| |\v{k_{p_2}}|$. 
    The cost of two squarings is $b^2$, using Eq. D37 from Ref.~\cite{sanders2020compilation}. 
    The inclusion of a single control introduces a complexity linear in $b$, which we omit.
    \item Prepare a uniform superposition over $m = 1, \ldots, M$ basis states. 
    We assume that $M$ is a power of two, and therefore this step has no cost.
    \item Prepare the righthand side of Eq.~\eqref{eq:legendre_cases} by multiplying the $m$ register by $|\v{k_{p_1}}| |\v{k_{p_2}}|$ if $l=1$ and by $|\v{k_{p_1}}|^2 |\v{k_{p_2}}|^2$ if $l=2$. The correct power is computed for $l=1$ and $l=2$ given step 3. We therefore need only a single multiplication, controlled on $l \neq 0$. The multiplication has cost $b^2$, and the inclusion of a single control introduces a complexity linear in $b$, which we omit.
    \item Compute the quantity $3(\v{k_{p_1}} \cdot \v{k_{p_{2}}})^2 - |\v{k_{p_1}}|^2 |\v{k_{p_2}}|^2$ to an ancilla register, when $l=2$. Note that we don't care what is computed to this ancilla register for other values of $l$, so we can carry out an uncontrolled evaluation. The subtraction has cost $b$. Multiplication by the constant 3 has cost $2b$.
    \item Evaluate the lefthand side of Eq.~\eqref{eq:legendre_cases} by multiplying $\v{k_{p_1}} \cdot \v{k_{p_{2}}}$ by $M$ if $l=1$ or $3(\v{k_{p_1}} \cdot \v{k_{p_{2}}})^2 - |\v{k_{p_1}}|^2 |\v{k_{p_2}}|^2$ by $M$ if $l=2$. Two multiplications has cost $2b^2$. The inclusion of a single control introduces a complexity linear in $b$, which we omit.
        \item Carry out the inequality test. This has cost as given in App.~\ref{sec:comp_NLineq}.
        \item Uncompute the arithmetic in the substeps above.
\end{enumerate}

The total cost of the arithmetic (substeps 1, 2, 3, 5, 6, 7) is 
\begin{equation}
    5\tilde{n}+ 5n^2 + 8bn - 2n_{\max}(n_{\max} + b) + \frac{21}{4}b^2 + \frac{13}{2}b - 6.
\end{equation}
While we could maintain intermediate ancilla registers and uncompute with reduced cost, we make the pessimistic assumption that the uncomputation is as expensive as the computation. Note that, as in Sec.~\ref{sec:comp_G}, the rejection sampling is strictly with respect to $|P_l|$ rather than $P_l$. However, we compute $P_l$, carry out the inequality test without the sign bit, and then fix up the phase by applying a Pauli Z to the sign bit after the rejection sampling. We require ancilla registers for $(\v{k_{p_1}} \cdot \v{k_{p_{2}}})^2$, $|\v{k_{p_1}}| |\v{k_{p_2}}|$, $|\v{k_{p_1}}|^2 |\v{k_{p_2}}|^2$, $m|\v{k_{p_1}}|^2 |\v{k_{p_2}}|^2$, $3(\v{k_{p_1}} \cdot \v{k_{p_{2}}})^2 - |\v{k_{p_1}}|^2 |\v{k_{p_2}}|^2$, and $[3(\v{k_{p_1}} \cdot \v{k_{p_{2}}})^2 - |\v{k_{p_1}}|^2 |\v{k_{p_2}}|^2]M$. These have cost $2b$, $2b$, $3b$, $2b+2$ and $3b+2$, respectively. The total ancilla count for the arithmetic is therefore $11b+4$.

\subsubsection{Nuclear momentum}\label{sec:comp_nuclmom}

Sec.~\ref{subsubsec:BENonlocalPP} requires that we add to the nuclear momenta $\v P \rightarrow \v P + \v{p_1} - \v{p_2}$. We carry this out with one addition and one subtraction (which costs the same as an addition). Register $\kett{\v P}$,is the largest, with $\bar{n}$ qubits, so we incur a total cost of $2\bar{n}$. Each addition requires $\bar{n}$ ancillas. We also explicitly flag if  $\v P + \v{p_1} - \v{p_2} \in G$. As noted in~\cite{su2021fault}, Sec. II D, the carry-out bits from the arithmetic above already encode this overflow (and so no additional checking is required).

\end{document}